\documentstyle[preprint,aps,eqsecnum,epsfig]{revtex}
\tightenlines
\begin{document}
\draft

\newcommand{\beq}{\begin{equation}}
\newcommand{\eeq}{\end{equation}}
\newcommand{\bea}{\begin{eqnarray}}
\newcommand{\eea}{\end{eqnarray}}
\newcommand{\cir}{{\buildrel \circ \over =}}

\title{Hamiltonian Linearization of the Rest-Frame Instant Form of
Tetrad Gravity in a Completely Fixed 3-Orthogonal Gauge:  a
Radiation Gauge for Background-Independent Gravitational Waves in
a Post-Minkowskian Einstein Spacetime.}

\author{Juri Agresti}

\address{Dipartimento di Fisica\\
Universita' di Firenze\\Polo Scientifico, via Sansone 1\\
 50019 Sesto Fiorentino, Italy\\
 E-mail agresti@fi.infn.it}

\author{and}

\author{Roberto De Pietri}

\address{Dipartimento di Fisica\\ Campus Universitario,
Universita' di Parma\\ Viale delle Scienze, 43100 Parma, Italy\\
depietri@pr.infn.it}

\author{and}

\author{Luca Lusanna}

\address{
Sezione INFN di Firenze\\Polo Scientifico, via Sansone 1\\
 50019 Sesto Fiorentino, Italy\\
E-mail lusanna@fi.infn.it}

\author{and}

\author{Luca Martucci}

\address{Dipartimento di Fisica and INFN Sezione di Milano\\
Universita' di Milano I\\
 via G.Celoria  16\\
 20133 Milano\\
 E-mail luca.martucci@mi.infn.it}

\maketitle
\begin{abstract}

In the framework of the rest-frame instant form of tetrad gravity,
where the Hamiltonian is the weak ADM energy ${\hat E}_{ADM}$, we
define a special completely fixed 3-orthogonal Hamiltonian gauge,
corresponding to a choice of {\it non-harmonic} 4-coordinates, in
which the independent degrees of freedom of the gravitational
field are described by two pairs of canonically conjugate Dirac
observables (DO) $r_{\bar a}(\tau ,\vec \sigma )$, $\pi_{\bar
a}(\tau ,\vec \sigma )$, $\bar a = 1,2$. We define a Hamiltonian
linearization of the theory, i.e. gravitational waves, {\it
without introducing any background 4-metric}, by retaining only
the linear terms in the DO's in the super-hamiltonian constraint
(the Lichnerowicz equation for the conformal factor of the
3-metric) and the quadratic terms in the DO's in ${\hat E}_{ADM}$.
{\it We solve all the constraints} of the linearized theory: this
amounts to work in a well defined post-Minkowskian
Christodoulou-Klainermann space-time. The Hamilton equations imply
the wave equation for the DO's $r_{\bar a}(\tau ,\vec \sigma )$,
which replace the two polarizations of the TT harmonic gauge, and
that {\it linearized Einstein's equations are satisfied}    .
Finally we study the geodesic equation, both for time-like and
null geodesics, and the geodesic deviation equation.

\vskip 1truecm
\noindent \today
\vskip 1truecm

\end{abstract}
\pacs{}

\newpage

\vfill\eject

\section
{Introduction}

In a series of papers \cite{1,2,3} the rest-frame instant form of
the Hamiltonian ADM formulation of both metric and tetrad gravity
was given. The aim of these papers was to arrive at a formulation
of general relativity with matter such that the switching off of
the Newton constant $G$ would produce the description of the same
matter in the rest-frame instant form of parametrized Minkowski
theories \cite{4,5,1}, with the general relativistic covariance
deparametrizing to the special relativistic one of these theories,
and to study its Hamiltonian formulation consistently with
Dirac-Bergmann theory of constraints.
\bigskip

To implement this program we must select the following family of
non-compact space-times:

i) {\it globally hyperbolic}, so that the ADM Hamiltonian
formulation \cite{6} is well defined if we start from the ADM
action;

ii) {\it topologically trivial}, so that they can be foliated with
space-like hyper-surfaces $\Sigma_{\tau}$ diffeomorphic to $R^3$
(3+1 splitting of space-time with $\tau$, the scalar parameter
labeling the leaves, as a {\it mathematical time});

iii) {\it asymptotically flat at spatial infinity} and with
boundary conditions at spatial infinity independent from the
direction, so that the {\it spi group} of asymptotic symmetries is
reduced to the Poincare' group with the ADM Poincare' charges as
generators \footnote{When we switch off the Newton constant $G$,
the ADM Poincare' charges, expressed in $\Sigma_{\tau}$-adapted
4-coordinates, become the generators of the {\it internal
Poincare' group} of parametrized Minkowski theories \cite{1}.}. In
this way we can eliminate the {\it super-translations}, namely the
obstruction to define angular momentum in general relativity, and
we have the same type of boundary conditions which are needed to
get well defined non-Abelian charges in Yang-Mills theory, opening
the possibility of a unified description of the four interactions
with all the fields belonging to same function space \cite{5}. All
these requirements imply that the {\it allowed foliations} of
space-time must have the space-like hyper-planes tending in a
direction-independent way to Minkowski space-like hyper-planes at
spatial infinity, which moreover must be orthogonal there to the
ADM 4-momentum. But these are the conditions satisfied by the
singularity-free Christodoulou-Klainermann space-times \cite{7},
in which, in presence of matter, the allowed hyper-surfaces define
the {\it rest frame of the universe} and naturally become the
Wigner hyper-planes of the rest-frame instant form of the
parametrized Minkowski theories describing the same matter when $G
\mapsto 0$. Therefore there are {\it preferred asymptotic inertial
observers}, which can be identified with the {\it fixed stars}.
These allowed hyper-surfaces are called {\it Wigner-Sen-Witten
(WSW) hyper-surfaces}, because it can be shown that the
Frauendiener re-formulation \cite{8} of Sen-Witten equations
\cite{9} for triads allows (after the restriction to the solutions
of Einstein's equations) to transport the asymptotic tetrads of
the asymptotic inertial observers in each point of the
hyper-surface, generating a {\it local compass of inertia} to be
used to define {\it rotations with respect to the fixed stars}
\footnote{Instead the standard Fermi-Walker transport of the
tetrads of a time-like observer is a standard of non-rotation with
respect to a local observer in free fall.}. Besides the existence
of a realization of the Poincare' group, the only other needed
property is that the admissible WSW hyper-surfaces must admit an
involution \cite{10} allowing the definition of a generalized
Fourier transform with its associated concepts of positive and
negative energy, so to avoid the claimed impossibility to define
particles in curved space-times \cite{11}.

iv) All the fields have to belong to suitable {\it weighted
Sobolev spaces} so that the allowed space-like hyper-surfaces are
Riemannian 3-manifolds without Killing vectors: in this way we
avoid the analogue of the Gribov ambiguity in general relativity
and we can get a unification of the function spaces of gravity and
particle physics.\bigskip

After all these preliminaries it is possible to study the
Hamiltonian formulation of both ADM metric \cite{1} and tetrad
\cite{2,3} gravity \footnote{More natural for the coupling to the
fermions of the standard model of particles. Moreover tetrad
gravity is naturally a theory of time-like accelerated observers,
generalizing the ones of parametrized Minkowski theories. } with
their (8 and 14 respectively) first class constraints as
generators of the Hamiltonian gauge transformations. Then it is
possible to look, at least at a heuristic level, for
Shanmugadhasan canonical transformations \cite{12,4,5}
implementing  {\it the separation between the gauge variables and
the Dirac observables} (DO) for the gravitational field
\footnote{See Refs.\cite{13,14,15} for the interpretation of the
gauge variables as {\it generalized inertial effects} and of the
DO's (the non-local deterministically predictable physical degrees
of freedom of the gravitational field) as {\it generalized tidal
effects}. The non-locality of DO's (all 3-space is needed for
their determination) may be interpreted as a form of Mach
principle, even if our boundary conditions are not compatible with
the standard interpretation of such principle given by Einstein
and Wheeler, who choose spatially compact space-times without
boundary \cite{16}. }. A complete exposition of these topics is in
Refs.\cite{1,2,3}, where it is shown that it is possible to define
a {\it rest frame instant form of gravity} in which the effective
Hamiltonian for the evolution is the {\it weak ADM energy}
$E_{ADM}$ \footnote{Therefore the formulations with a frozen
reduced phase space are avoided. The super-hamiltonian constraint
generates {\it normal} deformations of the space-like
hyper-surfaces, which are {\it not} interpreted as a time
evolution (like in the Wheeler-DeWitt approach) but as the
Hamiltonian gauge transformations ensuring that the description of
gravity is independent from the 3+1 splittings of space-time like
it happens in parametrized Minkowski theories.} \cite{17}.

\medskip

Let us consider tetrad gravity. In Refs. \cite{2,3} there is a new
parametrization of tetrad gravity, still utilizing the ADM action,
which allows to solve 13 of the 14 first class constraints. After
an allowed 3+1 splitting of space-time with space-like
hyper-surfaces $\Sigma_{\tau}$, we introduce adapted coordinates
\footnote{Instead of local coordinates $x^{\mu}$ for $M^4$, we use
local coordinates $\sigma^A$ on $R\times \Sigma \approx M^4$ with
$\Sigma \approx R^3\,\,\,$ [$x^{\mu}=z^{\mu}(\sigma )$ with
inverse $\sigma^A= \sigma^A(x)$], i.e. a {\it
$\Sigma_{\tau}$-adapted holonomic coordinate basis} for vector
fields $\partial_A={{\partial}\over {\partial \sigma^A}}\in
T(R\times \Sigma ) \mapsto b^{\mu}_A(\sigma )
\partial_{\mu} ={{\partial z ^{\mu}(\sigma )}\over {\partial
\sigma^A}} \partial_{\mu} \in TM^4$, and for differential
one-forms $dx^{\mu}\in T^{*}M^4 \mapsto d\sigma^A=b^A
_{\mu}(\sigma )dx^{\mu}={{\partial \sigma^A(z)}\over {\partial
z^{\mu}}} dx ^{\mu} \in T^{*}(R\times \Sigma )$. The 4-metric has
Lorentz signature $\epsilon\, (+,-,-,-)$ with $\epsilon = \pm 1$
according to particle physics and general relativity conventions
respectively. The induced 4-metric and inverse 4-metric become in
the new basis

\begin{eqnarray*}
 {}^4g_{AB} &=& \lbrace {}^4g_{\tau\tau}= \epsilon
(N^2-{}^3g_{rs}N^rN^s); {}^4g_{\tau r}=- \epsilon \,
{}^3g_{rs}N^s; {}^4g_{rs}=-\epsilon \, {}^3g_{rs}\rbrace =
\nonumber \\
 &=& \epsilon\, [l_Al_B - {}^3g_{rs}\, (\delta^r_A + N^r\,
\delta^{\tau}_A)(\delta^s_B+ N^s\delta^{\tau}_B)],\nonumber \\
 {}^4g^{AB} &=& \lbrace {}^4g^{\tau\tau}= {{\epsilon}\over {N^2}};
{}^4g^{\tau r}=-{{\epsilon \, N^r} \over {N^2}};
{}^4g^{rs}=-\epsilon ({}^3g^{rs} - {{N^rN^s}\over {N^2}}) \rbrace
=\epsilon [l^Al^B -{}^3g^{rs}\delta^A_r\delta^B_s].
 \end{eqnarray*}

\noindent For the unit normals to $\Sigma_{\tau}$ we have
$l^A={{\epsilon}\over N} (1; -N^r)$ and $l_A = N \partial_A \tau
=N \delta^{\tau}_A = (N; \vec 0)$. We introduce the 3-metric  of
$\Sigma_{\tau}$: $\, {}^3g_{rs}=-\epsilon \, {}^4g_{rs}$ with
signature (+++). If ${}^4\gamma^{rs}$ is the inverse of the
spatial part of the 4-metric (${}^4\gamma^{ru}\,
{}^4g_{us}=\delta^r_s$), the inverse of the 3-metric is
${}^3g^{rs}=-\epsilon \, {}^4\gamma^{rs}$ (${}^3g^{ru}\,
{}^3g_{us}=\delta^r_s$). We have the following form for the line
element in $M^4$:

\begin{eqnarray*}
 ds^2 &=& \epsilon (N^2-{}^3g_{rs}N^rN^s) (d\tau )^2-2\epsilon \,
{}^3g_{rs}N^s d\tau d\sigma^r -\epsilon \, {}^3g_{rs}
d\sigma^rd\sigma^s=\nonumber \\
 &=& \epsilon \Big[ N^2(d\tau )^2 -{}^3g_{rs}(d\sigma^r+N^rd\tau )(d\sigma^s+
N^sd\tau )\Big].
 \end{eqnarray*} }.
The arbitrary cotetrads ${}^4E^{(\alpha )}_{\mu}$, appearing in
the 4-metric of the ADM action principle, are rewritten in adapted
coordinates, ${}^4E^{(\alpha )}_A$, and replaced at the
Hamiltonian level by the lapse $N$ and shift $N_{(a)} =
{}^3e_{(a)r}\, N^r$ functions,  cotriads ${}^3e_{(a)r}$ and boost
parameters $\varphi^{(a)}$

\bea
 \left( \begin{array}{l} {}^4E^{(o)}_A \\ {}^4E^{(a)}_A \end{array} \right) &=&
\left( \begin{array}{cc}  \sqrt{1+\sum_{(c)} \varphi^{(c) 2}}
&-\epsilon \varphi_{(b)}\\  \varphi^{(a)} &
\delta^{(a)}_{(b)}-\epsilon {{\varphi^{(a)}\varphi_{(b)} }\over
{1+ \sqrt{1+\sum_{(c)}\varphi^{(c) 2}} }} \end{array} \right)
\times\nonumber \\ &&\left( \begin{array}{l} {}^4_{(\Sigma
)}{\check {\tilde E}}^{(o)}_A=(N;\vec 0)\\
 {}^4_{(\Sigma )}{\check {\tilde
E}}^{(b)}_A=(N^{(b)}={}^3e^{(b)}_rN^r; {}^3e^{(b)}_r) \end{array}
\right) ,\nonumber \\
 &&{}\nonumber \\
 &&{}\nonumber \\
  \Rightarrow {}^4g_{AB}&=&{}^4E^{(\alpha )}_A\,
{}^4\eta_{(\alpha )(\beta )}\, {}^4E^{(\beta )}_B= {}^4_{(\Sigma
)}{\check E}^{(\alpha )}_A\, {}^4\eta_{(\alpha )(\beta )}\,
{}^4_{(\Sigma )}{\check E}^{(\beta )}_B=\nonumber \\
 &=&\epsilon \left( \begin{array}{cc} (N^2- {}^3g_{rs}N^rN^s) &
-{}^3g_{st}N^t\\ -{}^3g_{rt}N^t & -{}^3g _{rs} \end{array} \right)
,\nonumber \\
 &&{}\nonumber \\
  &&{}\nonumber \\
\left( \begin{array}{l} {}^4E^{\mu}_{(o)}\\
{}^4E^{\mu}_{(a)}\end{array} \right) &=&\left( \begin{array}{cc}
\sqrt{1+\sum_{(c)} \varphi^{(c) 2}} &- \varphi^{(b)}\\ \epsilon
\varphi_{(a)} & \delta_{(a)}^{(b)}-\epsilon
{{\varphi_{(a)}\varphi^{(b)} }\over {1+
\sqrt{1+\sum_{(c)}\varphi^{(c) 2}} }} \end{array} \right) \,
\left( \begin{array}{l} l^{\mu} \\ b^{\mu}_s\, {}^3e^s_{(b)}
\end{array} \right) , \nonumber \\
 &&{}\nonumber \\
 &&{}\nonumber \\
  \left( \begin{array}{l} {}^4E^A_{(o)} \\ {}^4E^A_{(a)} \end{array} \right)
&=&\left( \begin{array}{cc}  \sqrt{1+\sum_{(c)} \varphi^{(c) 2}}
&- \varphi^{(b)}\\ \epsilon \varphi^{(a)} &
\delta_{(a)}^{(b)}-\epsilon {{\varphi_{(a)}\varphi^{(b)} }\over
{1+ \sqrt{1+\sum_{(c)}\varphi^{(c) 2}} }} \end{array} \right) \,
\left( \begin{array}{l} {}^4_{(\Sigma )}{\check {\tilde
E}}^A_{(o)}=(1/N; -N^r/N) \\ {}^4_{(\Sigma )}{\check {\tilde
E}}^A_{(b)}=(0; {}^3e^r_{(b)})
\end{array} \right) ,\nonumber \\
 &&{}\nonumber \\
 &&{}\nonumber \\
  \Rightarrow {}^4g^{AB}&=&{}^4E^A_{(\alpha )}\, {}^4\eta^{(\alpha
)(\beta )}\, {}^4E^B_{(\beta )}= {}^4_{(\Sigma )}{\check
E}_{(\alpha )}^A\, {}^4\eta_{(\alpha )(\beta )}\, {}^4_{(\Sigma
)}{\check E}_{(\beta )}^B=\nonumber \\
 &=&\epsilon \left( \begin{array}{cc} {1\over {N^2}} & -
{{N^s}\over {N^2}} \\ - {{N^r}\over {N^2}} & -
({}^3g^{rs}-{{N^rN^s}\over {N^2}})\end{array} \right).
 \label{I1}
 \eea

\bigskip

The conjugate momenta are ${\tilde \pi}^N(\tau ,\vec \sigma )$,
${\tilde \pi}^{\vec N}_{(a)}(\tau ,\vec \sigma )$, ${}^3{\tilde
\pi}^r_{(a)}(\tau ,\vec \sigma )$, ${\tilde \pi}_{(a)}^{\vec
\varphi}(\tau ,\vec \sigma )$, respectively. There are ten primary
constraints ${\tilde \pi}^N(\tau ,\vec \sigma ) \approx 0$,
${\tilde \pi}^{\vec N}_{(a)}(\tau ,\vec \sigma ) \approx 0$,
${\tilde \pi}^{\vec \varphi}_{(a)}(\tau ,\vec \sigma ) \approx 0$,
${}^3{\tilde M}_{(a)}(\tau ,\vec \sigma ) = \epsilon_{(a)(b)(c)}\,
{}^3e_{(b)r}(\tau ,\vec \sigma )\, {}^3{\tilde \pi}^r_{(c)}(\tau
,\vec \sigma ) \approx 0$ and four secondary ones: the
super-hamiltonian constraint ${\cal H}(\tau ,\vec \sigma ) \approx
0$ and the super-momentum constraints ${\cal H}^r(\tau ,\vec
\sigma ) \approx 0$. Therefore there are {\it 14 arbitrary gauge
variables}, four of which are the lapse and shift functions. All
the constraints are first class and the Dirac Hamiltonian depends
on 10 Dirac multipliers.
\bigskip

As shown in Ref.\cite{1}, a consistent treatment of the boundary
conditions at spatial infinity requires the explicit separation of
the {\it asymptotic} part of the lapse and shift functions from
their {\it bulk} part: $N(\tau ,\vec \sigma ) = N_{(as)}(\tau
,\vec \sigma ) + n(\tau , \vec \sigma )$, $N_r(\tau ,\vec \sigma )
= N_{(as)r}(\tau ,\vec \sigma ) + n_r(\tau , \vec \sigma )$, with
$n(\tau ,\vec \sigma )$ and $n_r(\tau ,\vec \sigma )$ tending to
zero at spatial infinity in a direction-independent way. On the
contrary, $N_{(as)}(\tau ,\vec \sigma ) = - \lambda_{\tau}(\tau )
- {1\over 2}\, \lambda_{\tau u}(\tau )\, \sigma^u$ and
$N_{(as)r}(\tau ,\vec \sigma ) = - \lambda_{r}(\tau ) - {1\over
2}\, \lambda_{r u}(\tau )\, \sigma^u$. The {\it
Christodoulou-Klainermann space-times} \cite{7}, with their {\it
rest-frame} condition of zero ADM 3-momentum and absence of
super-translations, are singled out by these considerations. As
already said the allowed foliations of these space-times tend
asymptotically to Minkowski hyper-planes in a
direction-independent way and are asymptotically orthogonal to the
ADM four-momentum. The leaves $\Sigma_{\tau}$ are the  WSW
hyper-surfaces. They have $N_{(as)}(\tau ,\vec \sigma ) = -
\epsilon$, $N_{(as) r}(\tau ,\vec \sigma ) = 0$. As in
Refs.\cite{1,2,3}, from now on the lapse and shift functions $N$
and $N_r$ will be replaced by $-\epsilon + n$ and $n_r$.

\medskip

It is pointed out in the papers \cite{1,3} that in order to have
well defined asymptotic {\it weak and strong ADM Poincare'
charges} (generators of the asymptotic Poincare' group) all fields
must have a suitable direction-independent limit at spatial
infinity. Recall that the {\it strong} ADM energy is the flux
through the surface at spatial infinity of a function of the
3-metric only, and it is weakly equal to the {\it weak} ADM energy
(volume form) which contains all the dependence on the ADM
momenta. This implies \cite{1} that the super-hamiltonian
constraint must be interpreted as the equation ({\it Lichnerowicz
equation}) that uniquely determines the {\it conformal factor}
$\phi (\tau ,\vec \sigma )= (det\, {}^3g(\tau ,\vec \sigma
))^{1/12}$ of the 3-metric as a functional of the other variables.
This means that the associated gauge variable is the {\it
canonical momentum $\pi_{\phi}(\tau ,\vec \sigma )$ conjugate to
the conformal factor}: it carries information about the extrinsic
curvature of $\Sigma_{\tau}$. It is just this variable, and {\it
not} York's time, which parametrizes the {\it normal} deformation
of the embeddable space-like hyper-surfaces $\Sigma_{\tau}$. Since
different forms of $\Sigma_{\tau}$ correspond to different 3+1
splittings of $M^4$, the gauge transformations generated by the
super-hamiltonian constraint correspond to the transition from an
allowed 3+1 splitting to another one: this is the gauge orbit in
the phase space over super-space. The theory is therefore
independent of the choice of the 3+1 splitting like in
parametrized Minkowski theories. As a matter of fact, a gauge
fixing for the super-hamiltonian constraint is a choice of a
particular 3+1 splitting and this is done by fixing the momentum
$\pi_{\phi}(\tau ,\vec \sigma )$ conjugate to the conformal
factor. This shows the dominant role of the conformal 3-geometries
in the determination of the physical degrees of freedom, just as
in the Lichnerowicz-York conformal approach \cite{18,19,16}.
\bigskip

As explained in Refs.\cite{1,3}, following a suggestion of Dirac
\cite{20}, we restrict our space-times to {\it admit asymptotic
Minkowski Cartesian coordinates}, namely the admissible
4-coordinate systems $x^{\mu} = z^{\mu} (\sigma^A)$  have the
property $x^{\mu} \rightarrow \delta^{\mu}_{(\mu )}\, z^{(\mu
)}_{(\infty )}(\tau ,\vec \sigma )$ at spatial infinity with
$z^{(\mu )}_{(\infty )}(\tau ,\vec \sigma ) = x^{(\mu )}_{(\infty
)}(\tau ) + b^{(\mu )}_{(\infty )r}(\tau )\, \sigma^r$ [$(\mu )$
are flat indices]. Therefore the leaves of the admissible 3+1
splittings of our space-times tend to Minkowski space-like
hyper-planes at spatial infinity in a direction-independent way.
While $x^{(\mu )}_{(\infty )}(\tau )$ denotes the world-line of a
point (centroid) arbitrarily chosen as origin of the 3-coordinate
systems on each $\Sigma_{\tau}$, the $b^{(\mu )}_{(\infty )A}(\tau
)$'s are flat asymptotic tetrads with $b^{(\mu )}_{(\infty )\tau}$
equal to the normal $l^{(\mu )}_{(\infty )}$ to the asymptotic
Minkowski hyper-planes. To force the existence of these asymptotic
coordinates, we must add to tetrad gravity  ten extra
configuration degrees of freedom, namely the centroid $x^{(\mu
)}_{(\infty )}(\tau )$ and the six independent components of the
tetrads $b^{(\mu )}_{(\infty ) A}(\tau )$, and of the conjugate
ten momenta, $p^{(\mu )}_{(\infty )}$ and a spin tensor
$S_{(\infty )}^{(\mu )(\nu )}$. As shown in Refs.\cite{1,3}, this
increase of variables is balanced by adding ten extra first class
constraints determining the ten extra momenta: $\chi^A =
p^A_{(\infty )} - {\hat P}^A_{ADM} \approx 0$, $\chi^{AB} =
J^{AB}_{(\infty )} - {\hat J}^{AB}_{ADM} \approx 0$, where the
weak (volume form) ADM Poincare' charges ${\hat P}^A_{ADM}$,
${\hat J}^{AB}_{ADM}$ and the quantities $p^A_{(\infty )} =
b^A_{(\infty )(\mu )}\, p^{(\mu )}_{(\infty )}$, $J^{AB}_{(\infty
)} = b^A_{(\infty )(\mu )}\, b^B_{(\infty )(\nu )}\, S^{(\mu )(\nu
)}_{(\infty )}$ are expressed in $\Sigma_{\tau}$-adapted
4-coordinates.

\medskip

In this way the ten extra configuration degrees of freedom become
arbitrary {\it gauge variables}. After the splitting of the lapse
and shift functions in the asymptotic and bulk parts, this
approach \cite{1,3} implies the replacement of the canonical
Hamiltonian $\int d^3\sigma\, \Big[ n\, {\cal H} + n_r\, {\cal
H}^r\Big] (\tau ,\vec \sigma ) + \lambda_A(\tau )\, {\hat
P}^A_{ADM} + {1\over 2}\, \lambda_{AB}(\tau )\, {\hat
J}^{AB}_{ADM}$ with the Hamiltonian $\int d^3\sigma\, \Big[ n\,
{\cal H} + n_r\, {\cal H}^r + \lambda_n\, \pi_n + \lambda_{\vec
n\, r}\, \pi^r_{\vec n} + \mu_{(a)}\, M_{(a)} + \rho_{(a)}\,
\pi^{\vec \varphi}_{(a)}\Big] (\tau ,\vec \sigma ) +
\lambda_A(\tau )\, [p^A_{(\infty )} -  {\hat P}^A_{ADM}] + {1\over
2}\, \lambda_{AB}(\tau )\, [J^{AB}_{(\infty )} - {\hat
J}^{AB}_{ADM}]$.

\medskip

The following boundary conditions (compatible with
Christodoulou-Klainermann space-times) ensure that the Hamiltonian
gauge transformations preserve these asymptotic properties
defining  our class of space-times

\bea
 {}^3e_{(a)r}(\tau ,\vec \sigma )\,&& \rightarrow_{r\,
\rightarrow \infty }\, (1+{M\over {2r}}) \delta_{(a)r}+
O(r^{-3/2}),\nonumber \\
 &&{}\nonumber \\
 \Rightarrow && {}^3g_{rs}(\tau ,\vec \sigma
)=[{}^3e_{(a)r}\, {}^3e_{(a)s}] (\tau ,\vec \sigma )\,
\rightarrow_{r\, \rightarrow \infty}\, (1+{M\over r}) \delta_{rs}+
O(r^{-3/2}),\nonumber \\
 &&{}\nonumber \\
  {}^3{\tilde \pi}^{r}_{(a)}(\tau ,\vec \sigma)\, && \rightarrow_{r\,
  \rightarrow \infty}\, O(r^{-5/2}),\nonumber \\
 n(\tau ,\vec \sigma )\,&& \rightarrow_{r\, \rightarrow \infty}\,
O(r^{-(2+\epsilon )}),\qquad \epsilon > 0,\nonumber \\
  n_{r}(\tau ,\vec \sigma )\,&& \rightarrow_{r\,
\rightarrow \infty}\, O(r^{-\epsilon}),\nonumber \\
 \alpha_{(a)}(\tau ,\vec \sigma )\,&& \rightarrow_{r\, \rightarrow \infty}\,
O(r^{-(1+\epsilon )}),\nonumber \\
 \varphi_{(a)}(\tau ,\vec \sigma )\,&& \rightarrow_{r\, \rightarrow \infty}\,
O(r^{-(1+\epsilon )}).
 \label{I2}
 \eea

\bigskip

As shown in Refs.\cite{1,3}, to get the rest-frame instant form of
tetrad gravity, with its WSW hyper-surfaces as leaves of the
admissible 3+1 splittings (when $p^{(\mu )}_{(\infty )}$ is
time-like) the gauge fixings must be added with the following
procedure :\bigskip

i) Add three gauge fixings on the boost parameters, namely
$\varphi^{(a)}(\tau ,\vec \sigma ) \approx 0$: in this way we
choose  cotetrads adapted to $\Sigma_{\tau}$. The time constancy
of this gauge fixing determines the 3 Dirac multipliers
$\rho_{(a)}(\tau ,\vec \sigma )$.

ii) Add three gauge fixings, $\alpha_{(a)}(\tau ,\vec \sigma )
\approx 0$, to the rotation constraints ${}^3{\tilde M}_{(a)}(\tau
,\vec \sigma ) \approx 0$: in this way we fix a reference
orientation of the cotriads. The time constancy of these gauge
fixings determine the 3 Dirac multipliers $\mu_{(a)}(\tau ,\vec
\sigma )$.

iii) Add three gauge fixings $\chi_r(\tau ,\vec \sigma ) \approx
0$ to the secondary super-momentum constraints: this amounts to a
choice of 3-coordinates on $\Sigma_{\tau}$. The requirement of
time constancy of the constraints $\chi_r(\tau ,\vec \sigma )
\approx 0$ will generate three gauge fixings $\varphi_r(\tau ,\vec
\sigma ) \approx 0$ for the primary constraints $\pi^r_n(\tau
,\vec \sigma ) \approx 0$, which determine the shift functions
$n_r(\tau ,\vec \sigma )$ (and therefore the coordinate-dependent
gravito-magnetic aspects and the eventual anisotropy of light
propagation). The time constancy of the $\varphi_r(\tau ,\vec
\sigma )$'s will determine the 3 Dirac multipliers $\lambda_{\vec
n r}(\tau ,\vec \sigma )$.

iv) Add a gauge fixing $\chi (\tau ,\vec \sigma ) \approx 0$ to
the secondary super-hamiltonian constraint, which determines the
form of the space-like hyper-surface $\Sigma_{\tau}$. Its time
constancy produces the gauge fixing $\varphi (\tau ,\vec \sigma )
\approx 0$ for the primary constraint $\pi_n(\tau ,\vec \sigma )
\approx 0$, which determines the lapse function $n(\tau ,\vec
\sigma )$, i.e. how the surfaces $\Sigma_{\tau}$ are packed in the
foliation. Now the 3+1 splitting of space-time is completely
determined and the time constancy of $\varphi (\tau ,\vec \sigma )
\approx 0$ determines the last Dirac multiplier $\lambda_n(\tau
,\vec \sigma )$. A posteriori after having solved the Hamilton
equations one could find the embedding $z^{\mu}(\tau ,\vec \sigma
)$ of the WSW hyper-surfaces into the space-time.

v) Add 6  suitable gauge fixings on the gauge variables $b^{(\mu
)}_{(\infty )A}(\tau )$, so that, after having gone to Dirac
brackets, we get $J^{AB}_{(\infty )} \equiv {\hat J}^{AB}_{ADM}$
and $\lambda_{AB}(\tau ) = 0$. As a consequence of this gauge
fixing, the gauge variables $x^{(\mu )}_{(\infty )}(\tau )$ must
be replaced by the {\it canonical non-covariant 4-center of mass}
${\tilde x}^{(\mu )}_{(\infty )}(\tau )$ of the universe. All this
implies that the asymptotic Minkowski space-like hyper-planes
become orthogonal to the weak ADM 4-momentum ${\hat P}^A_{ADM}$
and that the canonical Hamiltonian becomes $\int d^3\sigma\, \Big[
n\, {\cal H} + n_r\, {\cal H}^r + \lambda_n\, \pi_n +
\lambda_{\vec n\, r}\, \pi^r_{\vec n} + \mu_{(a)}\, M_{(a)} +
\rho_{(a)}\, \pi^{\vec \varphi}_{(a)}\Big] (\tau ,\vec \sigma ) -
\lambda_{\tau}(\tau )\, [\epsilon_{(\infty )} - {\hat
P}^{\tau}_{ADM}] + \lambda_r(\tau )\, {\hat P}^r_{ADM}$ with
$\epsilon_{(\infty )} = -\epsilon\, \sqrt{\epsilon\, p^2_{(\infty
)}}$. Namely only the four first class constraints
$\epsilon_{(\infty )} - {\hat P}^{\tau}_{ADM} \approx 0$ and
${\hat P}^r_{ADM} \approx 0$ are left, With the constraints ${\hat
P}^r_{ADM} \approx 0$ being {\it the rest frame condition for the
universe}.

vi) Add the gauge fixing $\tau - T_{(\infty )} \approx 0$,
implying $\lambda_{\tau}(\tau ) = 0$ and identifying the
mathematical time $\tau$ with the rest frame time $T_{(\infty )} =
p_{(\infty )(\mu )}\, {\tilde x}^{(\mu )}_{(\infty )} /
\epsilon_{(\infty )} = p_{(\infty )(\mu )}\, x^{(\mu )}_{(\infty
)} / \epsilon_{(\infty )}$ of the universe. The canonical 4-center
of mass ${\tilde x}^{(\mu )}_{(\infty )}$ becomes a {\it decoupled
point particle clock}.

vii) Either add the natural gauge fixings ${\hat J}^{\tau r}_{ADM}
\approx 0$ to the constraints ${\hat P}^r_{ADM} \approx 0$ or work
by choosing $\lambda_r(\tau ) = 0$ as a pre-gauge-fixing (see the
discussion at the beginning of Section V) .

\medskip

At this stage the canonical reduction is completed by going to
Dirac brackets and it can be shown  \cite{1,3} that the Dirac
Hamiltonian for the rest-frame instant form of tetrad gravity is
the {\it weak} (volume form) ADM energy: $H_D = {\hat E}_{ADM} = -
\epsilon\, {\hat P}^{\tau}_{ADM}$. It becomes the effective
Hamiltonian for the gauge invariant observables parametrizing the
reduced phase space in the rest-frame instant form.\bigskip

To find a canonical basis of Dirac observables for the
gravitational field in absence of known solutions of the
super-hamiltonian constraint, we can perform a
quasi-Shanmugadhasan canonical transformation adapted to only 13
of the constraints and utilize the information  that this
constraint has to be interpreted as the {\it Lichnerowicz
equation} for the conformal factor $\phi (\tau ,\vec \sigma ) =
\gamma^{1/12}(\tau ,\vec \sigma ) = (det\, {}^3g(\tau ,\vec \sigma
) )^{1/ 12} = e^{q(\tau ,\vec \sigma )/2}$ of the 3-metric. The
result of this {\it point} canonical transformation is ($\bar a =
1,2$)

\begin{equation}
\begin{minipage}[t]{3cm}
\begin{tabular}{|l|l|l|l|} \hline
$\varphi^{(a)}$ & $n$ & $n_r$ & ${}^3e_{(a)r}$ \\ \hline $\approx
0$ & $\approx 0$ & $  \approx 0 $ & ${}^3{\tilde \pi}^r_{(a)}$
\\ \hline
\end{tabular}
\end{minipage} \hspace{1cm} {\longrightarrow \hspace{.2cm}} \
\begin{minipage}[t]{4 cm}
\begin{tabular}{|lllll|l|l|} \hline
$\varphi^{(a)}$ & $n$ & $n_r$ & $\alpha_{(a)}$ & $\xi^{r}$ &
$\phi$ & $r_{\bar a}$\\ \hline $\approx0$ &
 $\approx 0$ & $\approx 0$ & $\approx 0$
& $\approx 0$ &
 $\pi_{\phi}$ & $\pi_{\bar a}$ \\ \hline
\end{tabular}
\end{minipage}.
 \label{I3}
 \end{equation}

\noindent where $\alpha_{(a)}(\tau ,\vec \sigma )$ are angles and
$\xi^r(\tau ,\vec \sigma )$ are a parametrization of the group
manifold of the passive 3-diffeomorphisms of $\Sigma_{\tau}$,
describing its changes of 3-coordinates \footnote{Since there is
no canonical choice of an origin in the 3-diffeomorphism group
manifold, the choice of a 3-coordinate system is done in two
steps: i) by adding the gauge fixings $\xi^r(\tau ,\vec \sigma ) -
\sigma^r \approx 0$, declaring that this 3-coordinate system $\{
\sigma^r \}$ on $\Sigma_{\tau}$ is conventionally chosen as origin
of the 3-coordinate systems; ii) by choosing a parametrization of
the reduced cotriads, and therefore of the 3-metric associated
with the chosen 3-coordinate system, only in terms of the 3
functions $\phi (\tau ,\vec \sigma )$, $r_{\bar a}(\tau ,\vec
\sigma )$.}. The entries $\approx 0$ are the new momenta
corresponding to 13 Abelianized first class constraints: besides
the lapse, shift and boost momenta, there are three Abelianized
rotation constraints ${\tilde \pi}^{\vec \alpha}_{(a)}(\tau ,\vec
\sigma ) \approx 0$ and three Abelianized super-momentum
constraints ${\tilde \pi}^{\vec \xi}_r(\tau ,\vec \sigma ) \approx
0$.

\medskip

The Hamiltonian {\it gauge variables} are the 13 configuration
variables $\varphi^{(a)}(\tau ,\vec \sigma )$, $n(\tau ,\vec
\sigma )$, $n_r(\tau ,\vec \sigma )$, $\alpha_{(a)}(\tau ,\vec
\sigma )$, $\xi^r(\tau ,\vec \sigma )$ (they depend on the
cotetrads and its space gradients) and the momentum
$\pi_{\phi}(\tau ,\vec \sigma )$ conjugate to the conformal factor
(it depends also on the time derivative of the cotetrads). The
variables $\xi^r(\tau ,\vec \sigma )$ and $\pi_{\phi}(\tau ,\vec
\sigma )$ can be thought as a possible 4-coordinate system with
the Lorentz signature given by the pattern {\it 3 configuration +
1 momentum} variables.\bigskip

Even if we do not know the expression of the final variables
$\alpha_{(a)}$, $\xi^r$, $\pi_{\phi}$, $r_{\bar a}$, $\pi_{\bar
a}$ in terms of the original variables, the point nature of the
canonical transformation allows to write the following inverse
relations (the form of the cotriad was determined by solving the
multi-temporal equations for the gauge transformations generated
by the first class constraints \cite{3})

\bea
 {}^3e_{(a)r}(\tau ,\vec \sigma ) &=&
 {}^3R_{(a)(b)}(\alpha_{(e)}(\tau ,\vec \sigma ))\, {{\partial
 \xi^s(\tau ,\vec \sigma )}\over {\partial \sigma^r}}\, {}^3{\hat
 e}_{(b)s}(\phi(\tau , \vec \xi (\tau ,\vec \sigma )), r_{\bar a}
(\tau , \vec \xi (\tau ,\vec \sigma ))\, ),\nonumber \\
 &&{}\nonumber \\
 {}^3{\tilde \pi}^r_{(a)}(\tau ,\vec \sigma ) &=& \sum_s\, \int
 d^3\sigma_1\, {\cal K}^r_{(a)s}(\vec \sigma ,{\vec \sigma}_1,
 \tau | \alpha_{(e)}, \xi^u, \phi ,r_{\bar a}]\, \Big( \phi^{-2}\,
 e^{-{1\over {\sqrt{3}}}\, \sum_{\bar a}\, \gamma_{\bar as}\,
 r_{\bar a}}\Big)(\tau ,{\vec \sigma}_1)\nonumber \\
 &&\Big[ {{\pi_{\phi}}\over {6 \phi}} + \sqrt{3}\, \sum_{\bar b}\,
 {\gamma}_{\bar bs}\, \pi_{\bar b}\Big](\tau ,{\vec \sigma}_1)
 +\nonumber \\
 &+& \int d^3\sigma_1\, F^r_{(a)(b)}(\vec \sigma ,{\vec \sigma}_1,
 \tau | \alpha_{(e)}, \xi^u, \phi ,r_{\bar a}]\, {\tilde
 \pi}^{\vec \alpha}_{(b)}(\tau ,{\vec \sigma}_1) +\nonumber \\
 &+& \sum_u\, \int d^3\sigma_1\, G^{ru}_{(a)}(\vec \sigma ,{\vec \sigma}_1,
 \tau | \alpha_{(e)}, \xi^u, \phi ,r_{\bar a}]\, {\tilde
 \pi}^{\vec \xi}_u(\tau ,{\vec \sigma}_1),\nonumber \\
 &&{}\nonumber \\
 &&\Downarrow \nonumber \\
 &&{}\nonumber \\
 {}^3g_{rs}(\tau ,\vec \sigma ) &=& \sum_a\, {}^3e_{(a)r}(\tau
 ,\vec \sigma )\, {}^3e_{(a)s}(\tau ,\vec \sigma ) =\nonumber \\
 &=& {{\partial
 \xi^u(\tau ,\vec \sigma )}\over {\partial \sigma^r}}\,
 {{\partial \xi^v(\tau ,\vec \sigma )}\over {\partial \sigma^s}}\,
 \sum_a\, [{}^3e_{(a)u}\, {}^3e_{(a)v}](\phi(\tau , \vec \xi (\tau
 ,\vec \sigma )), r_{\bar a}(\tau , \vec \xi (\tau ,\vec \sigma ))\, ).
 \label{I4}
 \eea

\noindent Here ${}^3R$ are arbitrary rotation matrices and
${}^3{\hat e}_{(a)r}$ are reduced cotriads depending only on the
three functions $\phi$ and $r_{\bar a}$. The kernels ${\cal
K}^r_{(a)u}$, $F^r_{(a)(b)}$ and $G^{ru}_{(a)}$ are not known
explicitly: they are the solution of  quasi-linear partial
differential equations determined by the canonicity of the point
transformation and by the fact that the rotation constraints do
not have vanishing Poisson brackets with the super-momentum
constraints \cite{3}. Once we have found the general solution of
the quasi-linear equations for the kernels and we have done a
definite choice of gauge, the six rotation and super-momentum
constraints give further restrictions on the kernels, which amount
to restrict the general solution to a particular one.

\medskip

The family of {\it 3-orthogonal gauges} (all having a diagonal
3-metric, ${}^3g_{rs} = {}^3e_{(a)r}\, {}^3e_{(a)s} = 0\quad
for\,\,\, r \not= s$), which is  parametrized by the last gauge
variable $\pi_{\phi}(\tau ,\vec \sigma )$, is determined by the
gauge fixings $\xi^r(\tau ,\vec \sigma ) - \sigma^r \approx 0$ and
by the following parametrization of the reduced cotriads

\bea
 {}^3{\hat e}_{(a)r} &=& \delta_{(a)r}\, \phi^2\, e^{{1\over
{\sqrt{3}}}\, \sum_{\bar a}\, \gamma_{\bar ar}\, r_{\bar
a}},\nonumber \\
 &&{}\nonumber \\
 &&\Downarrow \nonumber \\
 &&{}\nonumber \\
 {}^3{\hat g}_{rs} &=& \sum_a\, {}^3{\hat e}_{(a)r}\, {}^3{\hat
 e}_{(a)s} = \delta_{rs}\, \phi^4\,  e^{{2\over
{\sqrt{3}}}\, \sum_{\bar a}\, \gamma_{\bar ar}\, r_{\bar a}}.
 \label{I5}
 \eea

The numerical constants $\gamma_{\bar ar}$, satisfying $\sum_{\bar
a}\, \gamma_{\bar ar} = 0$, $\sum_u\, \gamma_{\bar au}\,
\gamma_{\bar bu} = \delta_{\bar a\bar b}$, $\sum_{\bar a}\,
\gamma_{\bar u}\, \gamma_{\bar av} = \delta_{uv} - {1\over 3}$,
define a one-parameter family of quasi-Shanmugadhasan canonical
bases (\ref{I3}). In the gauge $\alpha_{(a)}(\tau ,\vec \sigma )
\approx 0$, $\xi^r(\tau ,\vec \sigma ) - \sigma^r \approx 0$, we
have ${}^3e_{(a)r}(\tau ,\vec \sigma ) \approx {}^3{\hat
e}_{(a)r}(\tau ,\vec \sigma )$.

\medskip

The physical deterministically predictable degrees of freedom of
the gravitational field are the non-local  DO's (their expression
in terms of the original variables is not known) $r_{\bar a}(\tau
,\vec \sigma )$, $\pi_{\bar a}(\tau ,\vec \sigma )$, $\bar a=1,2$.
In general they are not Bergmann observables, i.e. coordinate
independent quantities, being non-tensorial and
coordinate-dependent.

\medskip

Even if we do not know the solution $\phi = \phi [\xi^r,
\pi_{\phi}, r_{\bar a}, \pi_{\bar a}]$ of the Lichnerowicz
equation, {\it the class of Hamiltonian gauges defined by the
gauge fixing $\chi (\tau ,\vec \sigma ) = \pi_{\phi}(\tau ,\vec
\sigma ) \approx 0$ has the special property that the DO's
$r_{\bar a}(\tau ,\vec \sigma )$, $\pi_{\bar a}(\tau ,\vec \sigma
)$ remain canonical also at the level of Dirac brackets}, so that
these gauges can be named {\it radiation gauges}. When in a
radiation gauge there is no other residual gauge freedom, we can
express every tensor over the space-time only in terms of the two
pairs of canonically conjugate DO's.

\medskip

This allows {\it for the first time} to arrive at a completely
fixed Hamiltonian gauge of the 3-orthogonal type \footnote{Namely
with ${}^3g_{rs}(\tau ,\vec \sigma )$ diagonal and with
${}^3g_{rr}(\tau ,\vec \sigma ) = f_r(r_{\bar a}(\tau ,\vec \sigma
))$ after the solution of the Lichnerowicz equation. The
3-orthogonal class of gauges seems to be the nearest one to the
physical laboratories on the Earth. Let us remember that the
standards of length and time are {\it coordinate units} and not
Bergmann observables \cite{21}. }, which when restricted to the
solutions of Einstein's equations (i.e. {\it on-shell}) is
equivalent to a well defined choice of 4-coordinates for
space-time.
\bigskip

It is evident that the Hamiltonian gauge variables of canonical
gravity carry an information about observers in space-time, so
that they are not inessential variables like in electromagnetism
and Yang-Mills theory but take into account the fact that in
general relativity global inertial reference frames do not exist
\footnote{The equivalence principle only allows the existence of
local inertial frames along time-like geodesics describing the
world-line of a scalar test particle in free fall. }. The
separation between gauge variables and DO's is an extra piece of
(non-local) information \cite{13,14,15}, which has to be added to
the equivalence principle, asserting the local impossibility to
distinguish gravitational from inertial effects, to visualize
which of the local forces acting on test matter are {\it
generalized inertial (or fictitious) forces depending on the
Hamiltonian gauge variables} and which are {\it genuine
gravitational forces depending on the DO's, which are absent in
Newtonian gravity} \footnote{When dynamical matter will be
introduced, this Hamiltonian procedure will lead to distinguish
among action-at-a-distance, gravitational and inertial effects.}.
Both types of forces have a {\it different appearance in different
4-coordinate systems}.

\medskip

In every 4-coordinate system (on-shell completely fixed
Hamiltonian gauge)\bigskip

i) the {\it genuine tidal gravitational forces} appearing in the
geodesic deviation equation will be well defined gauge-dependent
functionals only of the DO's associated to that gauge, so that
DO's can be considered {\it generalized non-local tidal degrees of
freedom} ;

ii) the geodesics will have a different geometrical form which
again is functionally dependent on the DO's in that gauge;

iii) the description of the relative 3-acceleration of a free
particle in free fall given in the local rest frame of an observer
will generated various terms identifiable with the general
relativistic extension of the non-relativistic inertial
accelerations and again these terms will depend on both the DO's
and the Hamiltonian gauge variables \footnote{Note that the
coordinate-dependent definition of gravito-magnetism as the
effects induced by ${}^4g_{\tau r}$ is a pure inertial effect,
because it is determined by the shift gauge variables.}.\bigskip

Therefore the Hamiltonian gauge variables, which change value from
a gauge to another one, {\it describe the change in the
appearance} of both the physical and apparent gravitational forces
going (on-shell) from a coordinate system to another one.
\bigskip

In this paper we shall start from tetrad gravity in the preferred
completely fixed 3-orthogonal gauge $\pi_{\phi}(\tau ,\vec \sigma
) \approx 0$ \footnote{The other gauge fixings are
$\varphi_{(a)}(\tau ,\vec \sigma ) \approx 0$, $\alpha_{(a)}(\tau
,\vec \sigma ) \approx 0$, $\xi^r(\tau ,\vec \sigma ) - \sigma^r
\approx 0$ and Eqs.(\ref{I5}). } and we shall define a
background-independent Hamiltonian linearization of the theory
after this gauge fixation, which, on-shell, corresponds to a
definite choice of a 4-coordinate system.\bigskip

The {\it standard linearization} \cite{22,23,24} of Einstein's
equations selects those space-times which admit nearly Lorentzian
4-coordinate systems and whose 4-metric is well approximated by
the {\it splitting} ${}^4g_{\mu\nu} = {}^4\eta_{\mu\nu} +
{}^4h_{\mu\nu}$, $|{}^4h_{\mu\nu}| << 1$ ({\it weak field}
approximation) with ${}^4\eta_{\mu\nu}$ the {\it background}
Minkowski 4-metric in Cartesian coordinates. Then it is assumed
that $\partial_{\alpha}\, {}^4h_{\mu\nu} ={1\over L}
O({}^4h_{\mu\nu})$, where for the length $L$ one can take the {\it
reduced wavelength $\lambda /2\pi$ of the resulting gravitational
waves} \footnote{With the flat Minkowski background, the
background Riemann tensor vanishes and the background radius of
curvature ${\cal R}$ is infinite (${\cal R}^{-2}$ is of the order
of the Riemann tensor). If ${\cal A}$ is the amplitude of the
gravitational wave, the {\it weak field} approximation is valid if
${\cal A} << 1$. In the more general {\it short-wave}
approximation \cite{22} the background is a vacuum Einstein
space-time with 4-metric ${}^4g^{(B)}_{\mu\nu}$ and typical
background radius of curvature ${\cal R}$. The splitting
${}^4g_{\mu\nu} = {}^4g^{(B)}_{\mu\nu} + {}^4h_{\mu\nu}$ is done
in {\it steady} 4-coordinates, where, if ${\cal A}$ is the
amplitude of the perturbation, one has: 1) ${}^4h_{\mu\nu} =
O({\cal A})$; 2) $\partial_{\alpha}\, {}^4g^{(B)}_{\mu\nu} =
O({\cal R}^{-1})$; 3) $\partial_{\alpha}\, {}^4h_{\mu\nu} =
O(({{\lambda}\over {2\pi}})^{-1})$. The short-wave approximation
is valid if ${\cal A} << 1$ and ${{\lambda}\over {2\pi}} << {\cal
R}$. }. Namely, one {\it assumes} the existence of solutions of
Einstein's equations admitting nearly Cartesian 4-coordinates and
{\it split them in a background}  and {\it a perturbation}, with a
residual gauge freedom in the choice of 4-coordinates. Then, after
the restriction to the family of {\it harmonic gauges}, one {\it
replaces the residual gauge freedom on the harmonic 4-coordinates
with the gauge freedom of a manifestly covariant spin-2 theory
over Minkowski space-time with Cartesian coordinates}.
Gravitational waves are usually analyzed in the special {\it TT
harmonic gauge}, a special case of the Lorentz gauges of the
spin-2 theory. Therefore, there is a {\it discontinuity in the
conceptual interpretation} and one gets that a special
relativistic theory with its associated absolute space-time
chrono-geometric structure and theory of measurement replaces
Einstein space-time with its problematic concerning the
identification of point-events (general covariance and the Hole
Argument) and with a theory of measurement in curved space-times
still to be developed (we have only an axiomatic theory employing
{\it test} non-dynamical objects) \cite{13,14,15}.
\bigskip

Instead, by working in the preferred completely fixed Hamiltonian
gauge with well determined 3-coordinates on the WSW hyper-surfaces
and well determined lapse and shift functions (so that the
reconstruction of Einstein space-time and of its 4-coordinates can
be done by using the embedding of WSW hyper-surfaces), we have the
possibility of making the linearization of the theory on the WSW
hyper-surfaces by approximating the Lichnerowicz equation and the
Hamiltonian, namely the weak ADM energy, with their linear and
quadratic parts in the DO's respectively, {\it without never
introducing a background}. In this way we may evaluate the
linearized conformal factor of the 3-metric and the linearized
lapse and shift functions and to obtain a linearized Einstein
space-time with 3-orthogonal coordinates on the WSW hyper-surfaces
(with a linearized embedding into the space-time), namely a {\it
post-Minkowskian Einstein space-time} linearization of a
Christodoulou-Klainermann space-time. It turns out that this gauge
is {\it not} a harmonic gauge (and, as a consequence, not a TT
harmonic gauge), but rather it is a {\it radiation gauge} for the
gravitational field without any residual gauge freedom. The DO's
$r_{\bar a}(\tau ,\vec \sigma )$ replace the two polarizations
$e_{+}$ and $e_{\times}$ of the TT harmonic gauge. Moreover,
besides the absence of background, there is also {\it no
post-Newtonian expansion}: our approximate solution of Einstein
equations in the chosen 4-coordinate system describes a linearized
Einstein manifold with Lorentz signature near to the Minkowski
space-time. It can be called a {\it post-Minkowskian space-time}
with a linearized gravitational field dynamically modifying
special relativity.
\bigskip

In the special 3-orthogonal gauge with $\pi_{\phi}(\tau,\vec
\sigma )=0$, the canonical variables for the gravitational field
are $r_{\bar a}(\tau ,\vec \sigma )$, $\pi_{\bar a}(\tau ,\vec
\sigma )$, and $\phi (\tau ,\vec \sigma )= e^{q(\tau ,\vec \sigma
)/2} = \phi [r_{\bar a}(\tau ,\vec \sigma ),\pi_{\bar a}(\tau
,\vec \sigma )]$ is the conformal factor solution of the
Lichnerowicz equation in this gauge. Since the linearization
consists in taking only the terms quadratic in the DO's inside the
weak ADM energy and only linear terms in the Lichnerowicz
equation, we will do the following assumptions: \hfill\break

A) We assume $| r_{\bar a}(\tau ,\vec \sigma ) | << 1$ on each WSW
hyper-surface and $|\partial_u r_{\bar a}(\tau ,\vec \sigma )|
\sim {1\over L} O(r_{\bar a})$, $|\partial_u\partial_v r_{\bar
a}(\tau ,\vec \sigma )| \sim {1\over {L^2}} O(r_{\bar a})$, where
$L$ is a {\it big enough characteristic length interpretable as
the reduced wavelength $\lambda / 2\pi$ of the resulting
gravitational waves}. Since the conjugate momenta $\pi_{\bar
a}(\tau ,\vec \sigma )$ have the dimensions of ${{action}\over
{L^3}}$, i.e. of ${k\over L}$ with $k={{c^3}\over {16\pi G}}$, we
assume $|\pi_{\bar a}(\tau ,\vec \sigma )| \sim {k\over L}
O(r_{\bar a})$, $|\partial_u \pi_{\bar a}(\tau ,\vec \sigma )|
\sim {k\over {L^2}} O(r_{\bar a})$, $|\partial_u\partial_v
\pi_{\bar a}(\tau ,\vec \sigma )| \sim {k\over {L^3}} O(r_{\bar
a})$. Therefore, $r_{\bar a}(\tau ,\vec \sigma )$ and $\pi_{\bar
a}(\tau ,\vec \sigma )$ are {\it slowly varying over the length L}
(for $r_{\bar a}, \pi_{\bar a}\, \rightarrow \, 0$ we get the void
space-times of Ref.\cite{3}). From Eq.(\ref{d7}) we get that the
Riemann tensor of our space-time is of order ${1\over {L\, k}}\,
O(\pi_{\bar a}) = {1\over {L^2}}\, O(r_{\bar a}) \approx {\cal
R}^{-2}$, where ${\cal R}$ is the mean radius of curvature.
Therefore we get that the {\it requirements of the weak field
approximation are satisfied}: i) ${\cal A} = O(r_{\bar a})$, if
${\cal A}$ is the amplitude of the gravitational wave; ii)
${L\over {{\cal R}}} = O(r_{\bar a})$, namely ${{\lambda}\over
{2\pi}} << {\cal R}$.

\bigskip

B) We also assume $q(\tau ,\vec \sigma ) \sim O(r_{\bar a})$,
$\partial_u q(\tau ,\vec \sigma ) \sim {1\over L} O(r_{\bar a})$,
$\partial_u\partial_vq(\tau ,\vec \sigma ) \sim {1\over {L^2}}
O(r_{\bar a})$, so that we get $\phi (\tau ,\vec \sigma ) =
e^{q(\tau ,\vec \sigma )/2} \approx 1+ {1\over 2} q(\tau ,\vec
\sigma ) +O(r^2_{\bar a})$ for the conformal factor. The
Lichnerowicz equation becomes a linear partial differential
equation for $q(\tau ,\vec \sigma )$. The linearization is done by
systematically discarding terms of order $O(r^2_{\bar a})$ in the
Lichnerowicz equation and $O(r^3_{\bar a})$ in the weak ADM
energy.\bigskip

With these positions we can, for the first time, {\it solve all
the constraints} of tetrad gravity, Lichnerowicz equation
included, and find explicitly the kernels in Eq.(\ref{I4}). We can
check that our gauge is not a harmonic gauge. Notwithstanding this
fact, the Hamilton equations imply the {\it wave equation} for the
DO's $r_{\bar a}(\tau ,\vec \sigma )$. We can check explicitly
that the linearized Einstein equations in this 4-coordinate system
are satisfied. {\it Therefore, for the first time we get a
definition of gravitational wave in a linearized post-Minkoskian
Einstein space-time without introducing any background} and
independently from the post-Newtonian approximation.

\medskip

After a comparison with other approaches, in which it is
emphasized the coordinate-dependent nature of effects like
gravito-magnetism, we make a study of the time-like geodesics and
of the geodesic deviation equation. The study of null geodesics
allows to show the post-Minkowskian modification of the light-cone
in each point of space-time. Then we analyze the embedding of the
WSW hyper-surfaces of our gauge in the post-Minkowskian space-time
and the associated congruence of time-like observers. Moreover,
anticipating the introduction of matter to be done in a future
paper, the comparison of our gauge with the post-Newtonian
approach shows that it is the equation determining the bulk part
$n(\tau ,\vec \sigma )$ of the lapse function which tends to the
Poisson equation for the Newton potential for $c \rightarrow
\infty$.

\bigskip

In Section II we solve the linearized Lichnerowicz equation and
the rotation and super-momentum constraints in our gauge. In
Section III we determine the quadratic part of the weak ADM energy
in terms of the DO's and we evaluate the lapse and shift functions
of our gauge. Then we find the linearized 4-metric and the tetrads
of our gauge and we study the Landau-Lifschiz energy-momentum
pseudo-tensor in our 4-coordinates. Section IV contains a
comparison of our results with the Lichnerowicz-York conformal
approach, with the standard harmonic gauges (and their associated
gravito-electric-magnetic analogy) and with the post-Newtonian
approximation. The Hamilton equations for the DO's and their
solution, included the verification that Einstein's equations are
satisfied, are presented in Section V. In Section VI we study the
time-like geodesics and the geodesic deviation equation of our
space-time. Section VII contains the determination of the
embedding of the WSW hyper-surfaces of our gauge into the
space-time and the study of the associated congruence of time-like
observers and of the null geodesics. Final remarks and the future
perspectives for the introduction of matter are drawn in the
Conclusions.

In Appendix A we reproduce the results about tetrad gravity
\cite{3} needed for this paper. In Appendix B we give the
determination of some kernels connected with the Shanmugadhasan
transformation. Appendix C contains the material on the Fourier
transform on the WSW hyper-surfaces of our gauge needed for the
study of the Hamilton equations. Finally in Appendix D there is
the linearized form of the relevant 3-tensors on the WSW
hyper-surfaces of our gauge and of the relevant 4-tensors over our
space-time. In this Appendix there is the explicit verification
that Einstein's equations are satisfied.

\vfill\eject

\section{Linearization in the 3-Orthogonal Gauge with
$\pi_{\phi}(\tau ,\vec \sigma ) =0$ and solution of the
constraints.}

In this Section we study the linearization of tetrad gravity and
the solution of its constraints in the completely fixed
3-orthogonal gauge with $\pi_{\phi} (\tau ,\vec \sigma ) =0$
defined in the Introduction.

\subsection{The Super-Hamiltonian Constraint.}

Let us first consider  the linearization of the {\it Lichnerowicz
equation}, see Eq.(\ref{a1}), which is the super-hamiltonian
constraint of Eq.(\ref{a2}), interpreted as an equation for the
conformal factor of the 3-metric.

Since the super-hamiltonian constraint of Eq.(\ref{a2})becomes
${\tilde {\cal H}}_R(\tau ,\vec \sigma )={{\epsilon c^3}\over
{16\pi G}} \Big( -8\triangle \phi (\tau,\vec \sigma )+{2\over
{\sqrt{3}}} \sum_{\bar au} \gamma_{\bar au} \partial^2_u r_{\bar
a}(\tau ,\vec \sigma )\Big) +{1\over {L^2}}O(r^2_{\bar a}) \approx
0$, where $\triangle = {\vec \partial}^2$ is the flat Laplacian
[$\triangle {1\over {4\pi |\vec \sigma |}} =- \delta^3(\vec \sigma
)$], only the term $(-\tilde \triangle +{1\over 8}\, {}^3\tilde
R)\phi$ in Eq.(\ref{a1})  gives a contribution of order $O(r_{\bar
a})$. As a consequence the linearized Lichnerowicz equation for
$\phi = e^{q/2} = 1+{1\over 2}q+O(r^2_{\bar a})$ becomes

\beq
 \triangle q(\tau ,\vec \sigma ) = {1\over {2\sqrt{3}}}
\sum_{u\bar a} \gamma_{\bar au} \partial^2_ur_{\bar a}(\tau ,\vec
\sigma ) +{1\over {L^2}} O(r_{\bar a}^2),
 \label{II1}
  \eeq

\noindent whose solution vanishing at spatial infinity is

\beq
 q(\tau ,\vec \sigma ) = -{1\over {2\sqrt{3}}} \sum_{u\bar a}
\gamma_{\bar au} \int d^3\sigma_1 {{\partial^2_{1u}r_{\bar a}(\tau
,{\vec \sigma}_1)} \over {4\pi |\vec \sigma -{\vec \sigma}_1|}}
+O(r^2_{\bar a}).
 \label{II2}
 \eeq

\subsection{The Rotation and Super-Momentum Constraints.}

After having solved the super-hamiltonian constraint we have to
solve the six rotation and super-momentum constraints. As said in
the Introduction, this is equivalent \cite{3} to find the kernels
appearing in the linearization of Eqs.(\ref{I4}). After putting
equal to zero the Abelianized rotation and super-momentum
constraints in Eqs.(\ref{I4}), the old momenta ${}^3{\check
{\tilde \pi}}^r_{(a)}(\tau ,\vec \sigma )\, {\buildrel {def} \over
=}\, {}^3{\tilde \pi}^r_{(a)}(\tau ,\vec \sigma
){|}_{3-O,\pi_{\phi} =0}$ \footnote{${|}_{3-0}$ means in the
family of 3-orthogonal gauges.} are given by  Eqs.(\ref{a4}),
whose linearization is

\bea
 {}^3{\check {\tilde \pi}}^r_{(a)}(\tau ,\vec \sigma ) &=&
\sqrt{3} \sum_{s\bar a} \gamma_{\bar as} \int d^3\sigma_1 {\tilde
{\cal K}}^r_{(a)s}(\vec \sigma ,{\vec \sigma}_1,\tau | \phi ,
r_{\bar a}] \Big( \phi^{-2}e^{-{1\over {\sqrt{3}}}\sum_{\bar
b}\gamma_{\bar bs}r_{\bar b}}\, \pi_{\bar a}\Big) (\tau ,{\vec
\sigma}_1) =\nonumber \\
 &=&\sqrt{3} \sum_{s\bar a} \gamma_{\bar as} \int d^3\sigma_1
 {\cal K}^{(o) r}_{(a)s}(\vec \sigma ,{\vec \sigma}_1)
 \pi_{\bar a}(\tau ,{\vec \sigma}_1)+{k\over {L^2}}O(r^2_{\bar a})=\nonumber \\
 &=&\sqrt{3} \sum_{s\bar a} \gamma_{\bar as} \int d^3\sigma_1 \Big[
 \delta^r_s\delta_{(a)s} \delta^3(\vec \sigma ,{\vec \sigma}_1)
 +{\cal T}^{(o) r}_{(a)s}(\vec \sigma ,{\vec \sigma}_1)\Big]
 \pi_{\bar a}(\tau ,{\vec \sigma}_1)+\nonumber \\
  &+&{k\over {L^2}}O(r^2_{\bar a}),\nonumber \\
  &&{}\nonumber \\
  &&with\nonumber \\
 &&{}\nonumber \\
 {\cal K}^{(o) r}_{(a)s}(\vec \sigma ,{\vec \sigma}_1) &=&
 {\tilde {\cal K}}^r_{(a)s}(\vec \sigma ,{\vec \sigma}_1,\tau |1,0]=
 \delta^r_s\delta_{(a)s}\delta^3(\vec \sigma ,{\vec \sigma}_1)+
 {\cal T}^{(o)r}_{(a)s}(\vec \sigma ,{\vec \sigma}_1)=\nonumber \\
 &=& \delta^r_s\delta_{(a)s}\delta^3(\vec \sigma ,{\vec
 \sigma}_1)-{{\partial G^{(o)rs}_{(a)}(\vec \sigma ,{\vec
 \sigma}_1)}\over {\partial \sigma_1^s}}.
 \label{II3}
 \eea

\noindent Therefore, the linearized kernels ${\cal
K}^{(o)r}_{(a)s}$'s are determined by the linearized kernels
$G^{(o)ru}_{(a)}$'s.\bigskip

As a consequence, the linearization implies that the partial
differential equations (\ref{a5}) for the kernels $G^{ru}_{(a)}$'s
have to be restricted to $r_{\bar a}(\tau ,\vec \sigma )=
\pi_{\bar a}(\tau ,\vec \sigma )=0$ (zero curvature limit), namely
the linearized kernels $G^{(o)ru}_{(a)}$'s are the same as for
void space-times \cite{3}. Their general solutions $G^{(o)
ru}_{(a)}(\vec \sigma ,{\vec \sigma}_1)$'s will determine the
${\cal K}^{(o) r}_{(a)u}$'s [Eq.(\ref{II3})] and the $F^{(o)
r}_{(a)(b)}$'s [see Appendix B]. Moreover, to satisfy
simultaneously the rotation and super-momentum constraints , the
${\cal K}^{(o) r}_{(a)u}$'s must satisfy the linearized version of
Eqs.(\ref{a6}). These linearized equations, which restrict the
general solution, will be given in Eqs.(\ref{II8}) and are
equivalent to the statement that the old momenta ${}^3{\tilde
\pi}^r_{(a)}$ satisfy the linearized form of the six rotation and
super-momentum constraints if we have

 \bea
 &&{}^3{\check {\tilde \pi}}^b_{(a)}(\tau ,\vec \sigma ) =
 {}^3{\check {\tilde \pi}}^a_{(b)}(\tau ,\vec \sigma ), \qquad a \not= b,\nonumber \\
  &&{}\nonumber \\
  &&\partial_r\, {}^3{\check {\tilde \pi}}^r_{(a)}(\tau ,\vec \sigma ) =0.
  \label{II4}
  \eea

\noindent Eqs.(\ref{II4}) also correspond to the linearization of
the three Einstein equations associated with the super-momentum
constraints of ADM metric gravity.
\bigskip

The zero curvature limit of Eqs.(\ref{a5}) implies that the
$G^{(o)ru}_{(a)}$'s are determined by the following linear partial
differential equations \footnote{$r_1, r_2 \not= b$, $r_1 \not=
r_2$;
$\epsilon_{(a)(r_1)(b)}=-\delta_{(a)r_2}\epsilon_{(r_1)(r_2)(b)}$,
$\epsilon_{(a)(r_2)(b)}=\delta_{(a)r_1}\epsilon_{(r_1)(r_2)(b)}$,
$\epsilon_{(u)(r_1)(b)}=-\delta_{(u)r_2}\epsilon_{(r_1)(r_2)(b)}$,
$\epsilon_{(u)(r_2)(b)}=\delta_{(u)r_1}\epsilon_{(r_1)(r_2)(b)}$.}

\bea
 1)&&\,\, s=a\quad homogeneous\, equations:\nonumber \\
 &&{}\nonumber \\
 && {{\partial G^{(o)a2}_{(a)}
 ({\vec \sigma} ,{\vec \sigma}_1)}\over {\partial \sigma_1^1}} +
 {{\partial G^{(o)a1}_{(a)}
 ({\vec \sigma},{\vec \sigma}_1)}\over {\partial \sigma_1^2}} =\nonumber \\
 &&= {{\partial G^{(o)a3}_{(a)}
 ({\vec \sigma},{\vec \sigma}_1)}\over {\partial \sigma_1^2}} +
 {{\partial G^{(o)a2}_{(a)}
 ({\vec \sigma},{\vec \sigma}_1)}\over {\partial \sigma_1^3}} =\nonumber \\
 &&= {{\partial G^{(o)a1}_{(a)}
 ({\vec \sigma},{\vec \sigma}_1)}\over {\partial \sigma_1^3}} +
 {{\partial G^{(o)a3}_{(a)}
 ({\vec \sigma},{\vec \sigma}_1)}\over {\partial \sigma_1^1}} = 0,
 \qquad a=1,2,3;\nonumber \\
 &&{}\nonumber \\
 &&\Downarrow \nonumber \\
 &&{}\nonumber \\
 &&{{\partial^2 G^{(o)a1}_{(a)}(\vec \sigma ,{\vec
 \sigma}_1)}\over {\partial \sigma_1^2 \partial \sigma_1^3}}=
{{\partial^2 G^{(o)a2}_{(a)}(\vec \sigma ,{\vec
 \sigma}_1)}\over {\partial \sigma_1^3 \partial \sigma_1^1}}=
 {{\partial^2 G^{(o)a3}_{(a)}(\vec \sigma ,{\vec
 \sigma}_1)}\over {\partial \sigma_1^1 \partial \sigma_1^3}}=0,\nonumber \\
 &&{}\nonumber \\
 2)&&\,\, s\not= a\,\, s\not= r,\, r\not= a\quad inhomogeneous\, equations:\nonumber \\
 &&{}\nonumber \\
 &&{{\partial G^{(o)sr}_{(a)}
 ({\vec \sigma},{\vec \sigma}_1)}\over {\partial \sigma_1^s}} +
 {{\partial G^{(o)ss}_{(a)}
 ({\vec \sigma},{\vec \sigma}_1)}\over {\partial \sigma_1^r}} =\nonumber \\
  &&={{\partial G^{(o)sa}_{(a)}
 ({\vec \sigma},{\vec \sigma}_1)}\over {\partial \sigma_1^r}} +
 {{\partial G^{(o) sr}_{(a)}
 ({\vec \sigma},{\vec \sigma}_1)}\over {\partial \sigma_1^a}} =0,\nonumber \\
 &&{{\partial G^{(o)ss}_{(a)}
 ({\vec \sigma},{\vec \sigma}_1)}\over {\partial \sigma_1^a}} +
 {{\partial G^{(o)sa}_{(a)}
 ({\vec \sigma},{\vec \sigma}_1)}\over {\partial \sigma_1^s}}
 = \delta^3({\vec \sigma};{\vec \sigma}_1).
 \label{II5}
 \eea

Each set of homogeneous equations, considered as equations for
functions of $\vec \sigma$, is of the form $\partial_2\,
u_{(1)}(\vec \sigma )+\partial_1\, u_{(2)}(\vec \sigma ) =
\partial_3\, u_{(2)}(\vec \sigma )+\partial_2\, u_{(3)}(\vec \sigma ) =
\partial_1\, u_{(3)}(\vec \sigma ) +
\partial_3\, u_{(1)}(\vec \sigma ) =0$. This is a
system of three linear partial differential equations for the
three unknown functions $u_{(i)}(\vec \sigma )$ of {\it elliptic}
type, since the determinant of its characteristic matrix \cite{25}
is $2\xi_1\xi_2\xi_3 \not= 0$. Moreover it is integrable, since
$u_{(r)}(\vec \sigma ) = f_{(r)}(\sigma^r)$ with arbitrary
$f_{(r)}$ are solutions of the system. We do not know whether they
exhaust all the possible solutions. Therefore,
$G^{(o)ar}_{(a)}({\vec \sigma},{\vec
\sigma}_1)=h^{(o)ar}_{(a)}({\vec \sigma},{\sigma}_1^r)$, with
$h^{(o)ar}_{(a)}$ arbitrary functions, are solutions of the
homogeneous equations.

As a consequence, if ${\bar G}^{(o)ru}_{(a)}({\vec \sigma},{\vec
\sigma}_1)$, $r\not= a$, is a particular solution of each set of
inhomogeneous equations (\ref{II5}), then the general solution is
$G^{(o)ru}_{(a)}({\vec \sigma},{\vec \sigma}_1)={\bar
G}^{(o)ru}_{(a)}({\vec \sigma},{\vec \sigma}_1
)+g^{(o)ru}_{(a)}({\vec \sigma},{\vec \sigma}_1)$, $r\not= a$,
with the $g^{(o)ru}_{(a)}$'s arbitrary homogeneous solutions
(again with $g^{(o)ru}_{(a)}({\vec \sigma},\sigma_1^r)$, if this
is the most general solution of the associated homogeneous
equations).

\bigskip

Therefore the general solution of Eqs.(\ref{II5}) for the kernels
$G^{(o)ru}_{(a)}({\vec \sigma},{\vec \sigma}_1)$  can be written
in the following form

\beq
 G^{(o)ru}_{(a)}({\vec \sigma},{\vec \sigma}_1) =
\delta^r_{(a)} h^{(o)au}_{(a)}({\vec \sigma},{\vec \sigma}_1)+
(1-\delta^r_{(a)})[{\bar G}^{(o)ru}_{(a)}({\vec \sigma},{\vec
\sigma}_1)+g^{(o)ru}_{(a)}({\vec \sigma},{\vec \sigma}_1)],
 \label{II6}
 \eeq

\noindent with arbitrary $h^{(o)au}_{(a)}$'s and
$g^{(o)ru}_{(a)}$'s. Then, Eq(\ref{II3}) gives the following
expression for the kernels ${\cal K}^{(o)r}_{(a)u}$'s

\bea
 &&{\cal K}^{(o)r}_{(a)u}({\vec \sigma},{\vec \sigma}_1)=\delta^r_{(a)}
 \delta_{(a)u} \delta^3({\vec \sigma},{\vec \sigma}_1)-\nonumber \\
 &-& \Big( \delta^r_{(a)} {{\partial
 h^{(o)au}_{(a)}({\vec \sigma},{\vec \sigma}_1)}\over {\partial
 \sigma^u_1}} +(1-\delta^r_{(a)})[ {{\partial
 {\bar G}^{(o)ru}_{(a)}({\vec \sigma},{\vec \sigma}_1)}\over {\partial
 \sigma^u_1}}+{{\partial
 g^{(o)ru}_{(a)}({\vec \sigma},{\vec \sigma}_1)}\over {\partial
 \sigma^u_1}}]\Big) .
 \label{II7}
 \eea

\medskip

The solutions of Eqs.(\ref{II5}) for the $G^{(o)ru}_{(a)}$'s are
restricted by the requirement that the ${\cal K}^{(o)r}_{(a)u}$'s
of Eqs.(\ref{II7}) satisfy the zero curvature limit of Eqs.
(\ref{a6}), which in the 3-orthogonal gauges become \footnote{At
zeroth order ($\phi (\tau ,\vec \sigma ) = 1$)  in the
3-orthogonal gauges  the spin connection vanishes, ${}^3{\hat
\omega}_{r(a)}(\tau ,\vec \sigma ) =0 + O(r_{\bar a})$, see
Eq.(\ref{a7}).}

\bea
 \sum_r \Big[ \delta_{(a)r} {\cal K}^{(o)r}_{(b)u} -
\delta_{(b)r} {\cal K}^{(o)r}_{(a)u} \Big] (\vec \sigma ,{\vec
\sigma}_1)
 &=& \Big[ {{\partial G^{(o)au}_{(b)}
(\vec \sigma ,{\vec \sigma}_1;\tau )}\over {\partial
\sigma^u_1}}-{{\partial G^{(o)bu}_{(a)} (\vec \sigma ,{\vec
\sigma}_1;\tau )}\over {\partial \sigma^u_1}} \Big] = 0,\qquad
a\not= b,\nonumber \\
 &&{}\nonumber \\
 \partial_r\,  {\cal K}^{(o)r}_{(a)u}
 (\vec \sigma ,{\vec \sigma}_1) &=& \partial_r\,
\Big[ \delta_{(a)}^r\delta_{(a)u} \delta^3(\vec \sigma ,{\vec
\sigma}_1) -  {{\partial G^{(o)ru}_{(a)}(\vec \sigma ,{\vec
\sigma}_1;\tau )} \over {\partial \sigma^u_1}}\Big]
 = 0.
 \label{II8}
 \eea

The first set of Eqs.(\ref{II8}) becomes the following set of
three linear partial differential equations to get the
$g^{(o)ru}_{(a)}$'s of Eq.(\ref{II6}) with $a \not= b$ in terms of
the ${\bar G}^{(o)ru}_{(a)}$'s

\bea
 {{\partial f^u_{ab}({\vec \sigma},{\vec \sigma}_1)}\over
 {\partial \sigma^u}} &=& {{\partial }\over {\partial
 \sigma^u_1}}\, \Big( g^{(o)au}_{(b)}({\vec \sigma},{\vec \sigma}_1) -
 g^{(o)bu}_{(a)}({\vec \sigma},{\vec \sigma}_1)\Big) =\nonumber \\
 &=&-\Big(  {{\partial
 {\bar G}^{(o)au}_{(b)}({\vec \sigma},{\vec \sigma}_1)}\over {\partial
 \sigma_1^u}} - {{\partial
 {\bar G}^{(o)bu}_{(a)}({\vec \sigma},{\vec \sigma}_1)}\over {\partial
 \sigma^u_1}} \Big)\,
 {\buildrel {def} \over =}\, m^u_{ab}({\vec \sigma}_1,\vec \sigma
 ).
 \label{II9}
 \eea

 \noindent For each pair $a\not= b$, this is a system of three
 elliptic linear partial differential equations for the
 $f^u_{ab}$'s. Each choice of the $g^{(o)au}_{(b)}$'s, $a\not= b$,
 which gives a solution of this system, implies that the
 associated kernels ${\cal K}^{(o)r}_{(a)u}$'s satisfy the rotation
 constraints.\bigskip

 Having found a solution for the $g^{(o)au}_{(b)}$'s, $a\not= b$, the
 second set of Eqs.(\ref{II8}) becomes the following set of
 equations for the $h^{(o)au}_{(a)}$'s of Eq.(\ref{II6}) in terms of the ${\bar
 G}^{(o)ru}_{(a)}$'s and $g^{(o)ru}_{(a)}$'s

 \bea
  &&\sum_{r}\, \partial_r \delta^r_{(a)}\Big[
  {{\partial h^{(o)au}_{(a)}({\vec \sigma},{\vec \sigma}_1
  )}\over {\partial \sigma_1^u}}\Big] = \sum_{r}  \partial_r\,
   \Big[ \delta^r_{(a)}\delta_{(a)u} \delta^3({\vec
  \sigma},{\vec \sigma}_1)-\nonumber \\
  &&-(1-\delta^r_{(a)}) [ {{\partial
  {\bar G}^{(o)ru}_{(a)}({\vec \sigma},{\vec \sigma}_1
  )}\over {\partial \sigma_1^u}}+ {{\partial g^{(o)ru}_{(a)}({\vec \sigma},{\vec
  \sigma}_1)}\over {\partial \sigma_1^u}}]\Big)\Big] .
  \label{II10}
  \eea

By using the linearization of the Green function of the covariant
divergence given in Eqs.(\ref{a8}) and (\ref{a9}), we get
($f^{su}_{(T)}$ are solutions of the homogeneous equation)

  \bea
  &&{{\partial h^{(o)su}_{(s)}({\vec \sigma},{\vec \sigma}_1
  )}\over {\partial \sigma_1^u}} =  f^{su}_{(T)}({\vec \sigma},{\vec \sigma}_1)-\nonumber \\
  &&-\int d^3\sigma_2 \sum_{(a)}\delta_{(s)(a)} c^s({\vec \sigma},{\vec
  \sigma}_2)\sum_{r} {{\partial}\over {\partial \sigma^r_2}}
   \Big[ \delta^r_{(a)}\delta_{(a)u}
   \delta^3({\vec \sigma}_2,{\vec \sigma}_1)-\nonumber \\
  &&-(1-\delta^r_{(a)}) [ {{\partial
  {\bar G}^{(o)ru}_{(a)}({\vec \sigma}_2,{\vec \sigma}_1
  )}\over {\partial \sigma^u_1}}+ {{\partial g^{(o)ru}_{(a)}({\vec \sigma}_2,{\vec
  \sigma}_1)}\over {\partial \sigma_1^u}}]\Big] .
  \label{II11}
  \eea

\noindent Again this is a system of elliptic linear partial
differential equations for the $h^{(o)au}_{(a)}$'s with fixed $a$.
\bigskip

After having found the solutions of Eqs. (\ref{II5}), (\ref{II9})
and (\ref{II11}) for the kernels $G^{(o)ru}_{(a)}$'s, ${\cal
K}^{(o)r}_{(a)u}$'s and $F^{(o)r}_{(a)(b)}$'s, every remaining
arbitrariness will be fixed by the boundary conditions at spatial
infinity for the momenta ${}^3{\tilde \pi}^r_{(a)}(\tau ,\vec
\sigma )$,  given in Eqs.(\ref{I2}). The final solutions are
equivalent not only to the solution of the rotation and
diffeomorphisms constraints, but also to their Abelianization in
the 3-orthogonal gauges with $\alpha_{(a)}(\tau ,\vec \sigma )=0$.

\subsection{A Solution of the Rotation and Super-Momentum Constraints.}

We have found the following particular solution $G^{(o)ru}_{(a)} =
{\bar G}^{(o)ru}_{(a)}$ of Eqs.(\ref{II5}) with $g^{(o)au}_{(b)} =
h^{(o)au}_{(a)}=0$ and vanishing for $|\vec \sigma | \rightarrow
\infty$ (we use $\delta(\sigma^r,\sigma_1^r\rightarrow -\infty )
=0$ for finite $\sigma^r$),

\bea
G^{(o) ar}_{(a)}({\vec \sigma} ,{\vec \sigma}_1) &=& 0,\qquad
a=1,2,3,\nonumber \\
 &&{}\nonumber \\
 G^{(o) 21}_{(1)}({\vec \sigma} ,{\vec \sigma}_1)&=&
 G^{(o) 11}_{(2)}({\vec \sigma} ,{\vec \sigma}_1)
  =G^{(o) 33}_{(2)}({\vec \sigma} ,{\vec \sigma}_1)=
 G^{(o) 23}_{(3)}({\vec \sigma} ,{\vec \sigma}_1)=\nonumber \\
 &=&{1\over 2} \int_{-\infty}^{\sigma_1^2} dw_1^2\, \delta^3(\vec \sigma
 ,\sigma_1^1w_1^2\sigma_1^3)=
  {1\over 2}\, \delta(\sigma^1,\sigma^1_1)\, \theta
  (\sigma_1^2,\sigma^2)\,
 \delta (\sigma^3,\sigma^3_1),\nonumber \\
 &&{}\nonumber \\
 G^{(o) 22}_{(1)}({\vec \sigma} ,{\vec \sigma}_1)&=&
 G^{(o) 12}_{(2)}({\vec \sigma} ,{\vec \sigma}_1)
  =G^{(o) 33}_{(1)}({\vec \sigma} ,{\vec \sigma}_1)=
 G^{(o) 13}_{(3)}({\vec \sigma} ,{\vec \sigma}_1)= \nonumber \\
 &=&{1\over 2} \int_{-\infty}^{\sigma_1^1} dw_1^1\, \delta^3(\vec \sigma
 ,w^1_1\sigma^2_1\sigma^3_1)=
 {1\over 2}\, \theta (\sigma_1^1,\sigma^1)\, \delta
 (\sigma^2,\sigma^2_1)\,
 \delta (\sigma^3,\sigma^3_1),\nonumber \\
 &&{}\nonumber \\
 G^{(o) 31}_{(1)}({\vec \sigma} ,{\vec \sigma}_1)&=&
 G^{(o) 11}_{(3)}({\vec \sigma} ,{\vec \sigma}_1)
 =G^{(o) 32}_{(2)}({\vec \sigma} ,{\vec \sigma}_1)=
 G^{(o) 22}_{(3)}({\vec \sigma} ,{\vec \sigma}_1)=\nonumber \\
 &=&{1\over 2} \int_{-\infty}^{\sigma^3_1} dw_1^3\, \delta^3(\vec \sigma
 ,\sigma_1^1\sigma_1^2w^3_1)=
 {1\over 2}\, \delta (\sigma^1,\sigma^1_1)\, \delta
 (\sigma^2,\sigma^2_1)\,
 \theta (\sigma^3_1,\sigma^3),\nonumber \\
 &&{}\nonumber \\
 G^{(o) 23}_{(1)}({\vec \sigma} ,{\vec \sigma}_1)&=&
 G^{(o) 13}_{(2)}({\vec \sigma} ,{\vec \sigma}_1)=
 -{1\over 2}{{\partial}\over {\partial \sigma_1^3}} \int_{-\infty}^{\sigma_1^1} dw_1^1
 \int_{-\infty}^{\sigma_1^2} dw_1^2\,   \delta^3(\vec \sigma ,w_1^1w_1^2\sigma^3_1)=
 \nonumber \\
 &=& - {1\over 2}\, \theta (\sigma^1_1,\sigma^1)\, \theta
 (\sigma^2_1,\sigma^2)\,
 {{\partial \delta (\sigma^3,\sigma^3_1)}\over {\partial \sigma^3_1}},\nonumber \\
 &&{}\nonumber \\
 G^{(o) 32}_{(1)}({\vec \sigma} ,{\vec \sigma}_1)&=&
 G^{(o) 12}_{(3)}({\vec \sigma} ,{\vec \sigma}_1)=
 - {1\over 2}\, {{\partial}\over {\partial \sigma_1^2}} \int_{-\infty}^{\sigma_1^1} dw_1^1
 \int_{-\infty}^{\sigma_1^3} dw_1^3\, \delta^3(\vec \sigma ,w_1^1\sigma_1^2w^3_1)=
 \nonumber \\
 &=& - {1\over 2}\, \theta (\sigma^1_1,\sigma^1)\, {{\partial \delta (\sigma^2,\sigma^2_1)}\over
 {\partial \sigma^2_1}}\, \theta (\sigma^3_1,\sigma^3),\nonumber \\
 &&{}\nonumber \\
 G^{(o) 31}_{(2)}({\vec \sigma} ,{\vec \sigma}_1)&=&
 G^{(o) 21}_{(3)}({\vec \sigma} ,{\vec \sigma}_1)=
 - {1\over 2}\, {{\partial}\over {\partial \sigma_1^1}} \int_{-\infty}^{\sigma_1^2} dw_1^2
 \int_{-\infty}^{\sigma_1^3} dw_1^3\, \delta^3(\vec \sigma ,\sigma_1^1w^2_1w^3_1)=
 \nonumber \\
 &=& - {1\over 2}\, {{\partial \delta (\sigma^1,\sigma^1_1)}\over {\partial \sigma_1^1}}
 \, \theta (\sigma^2_1,\sigma^2)\, \theta (\sigma^3_1,\sigma^3),\nonumber \\
 &&{}\nonumber \\
 &&or\nonumber \\
 &&{}\nonumber \\
 G^{(o) ru}_{(a)}({\vec \sigma} ,{\vec \sigma}_1) &=& - {1\over 2}\,
 [1-\delta^r_{(a)}]\,
  [1-2(\delta_{ru}+\delta_{au})]\nonumber \\
 &&{{\partial}\over {\partial \sigma_1^u}} \int_{-\infty}^{\sigma_1^r} dw_1^r
 \int_{-\infty}^{\sigma_1^a} dw^a_1\, \delta^3(\vec \sigma ,w_1^rw_1^a\sigma_1^{k\not= (r,a)})=
 \nonumber \\
 &=& - {1\over 2}\, [1-\delta^r_{(a)}]\,
  [1-2(\delta_{ru}+\delta_{au})]\nonumber \\
 && {{\partial}\over {\partial \sigma_1^u}}\, \Big[ \theta
 (\sigma_1^r,\sigma^r)\,
 \theta(\sigma_1^a,\sigma^a)\, \delta(\sigma^{k\not= (r,a)},\sigma^{k\not= (r,a)}_1)\Big].
\label{II12}
 \eea

\noindent where $\theta (x)$ is the step function [${d\over
{dx}}\theta (x)=\delta (x)$]. This implies the following
expression for the kernels ${\cal K}^{(o)r}_{(a)u}$'s

\bea {\cal K}^{(o) r}_{(a)u}({\vec \sigma} ,{\vec \sigma}_1) &=&
\delta^r_u\delta_{(a)u} \delta^3(\vec \sigma ,{\vec \sigma}_1) +
{\cal T}^{(o) r}_{(a)u}({\vec \sigma} ,{\vec \sigma}_1) =
\delta_{(a)}^r\delta_{(a)u} \delta^3(\vec \sigma ,{\vec \sigma}_1)
- {{\partial G^{(o) ru}_{(a)}({\vec \sigma},{\vec \sigma}_1)}
 \over {\partial \sigma^u_1}},\nonumber \\
 &&{}\nonumber \\
 &&with\nonumber \\
 &&{}\nonumber \\
 {\cal T}^{(o) a}_{(a)u}({\vec \sigma} ,{\vec \sigma}_1) &=& 0, \qquad a=1,2,3,\nonumber \\
 &&{}\nonumber \\
 {\cal T}^{(o) 2}_{(1)1}({\vec \sigma} ,{\vec \sigma}_1)&=&
 {\cal T}^{(o) 1}_{(2)1}({\vec \sigma} ,{\vec \sigma}_1)= - {1\over
 2}\, {{\partial}\over {\partial \sigma_1^1}}\, \int_{-\infty}^{\sigma_1^2} dw_1^2
 \, \delta^3(\vec \sigma ,\sigma^1_1w^2_1\sigma^3_1)=\nonumber \\
 &=& - {1\over 2}\, {{\partial \delta (\sigma^1,\sigma^1_1)}\over {\partial \sigma_1^1}}
 \, \theta (\sigma_1^2,\sigma^2)\, \delta (\sigma^3,\sigma^3_1)=\nonumber \\
 &=& - {1\over 2}\, {{\partial^2}\over {(\partial
 \sigma_1^1)^2}}\,
 \int_{-\infty}^{\sigma_1^1} dw_1^1 \int_{-\infty}^{\sigma_1^2} dw_1^2
 \, \delta^3(\vec \sigma ,w_1^1w_1^2\sigma^3_1) ,\nonumber \\
 &&{}\nonumber \\
 {\cal T}^{(o) 2}_{(1)2}({\vec \sigma} ,{\vec \sigma}_1)&=&
 {\cal T}^{(o) 1}_{(2)2}({\vec \sigma} ,{\vec \sigma}_1)= - {1\over
 2}\, {{\partial}\over {\partial \sigma_1^2}}\, \int_{-\infty}^{\sigma_1^1} dw_1^1
 \, \delta^3(\vec \sigma ,w_1^1\sigma^2_1\sigma^3_1)=\nonumber \\
 &=& - {1\over 2}\, \theta (\sigma_1^1,\sigma^1)\, {{\partial \delta (\sigma^2,\sigma^2_1)}\over
 {\partial \sigma^2_1}}\, \delta (\sigma^3,\sigma^3_1)=\nonumber \\
 &=& - {1\over 2}\, {{\partial^2}\over {(\partial
 \sigma_1^2)^2}}\,
 \int_{-\infty}^{\sigma_1^1} dw_1^1 \int_{-\infty}^{\sigma_1^2} dw_1^2
 \, \delta^3(\vec \sigma ,w_1^1w_1^2\sigma^3_1),\nonumber \\
 &&{}\nonumber \\
 {\cal T}^{(o) 2}_{(1)3}({\vec \sigma} ,{\vec \sigma}_1)&=&
 {\cal T}^{(o) 1}_{(2)3}({\vec \sigma} ,{\vec \sigma}_1)= {1\over 2}
\, {{\partial^2}\over {(\partial \sigma_1^3)^2}}\,
\int_{-\infty}^{\sigma_1^1} dw_1^1
 \int_{-\infty}^{\sigma_1^2} dw_1^2\,
 \delta^3(w_1^1w_1^2\sigma^3_1)=\nonumber \\
 &=& {1\over 2}\, \theta (\sigma_1^1,\sigma^1)\, \theta
 (\sigma_1^2,\sigma^2)\,
 {{\partial^2\delta (\sigma^3,\sigma^3_1)}\over {(\partial \sigma_1^3)^2}},\nonumber \\
 &&{}\nonumber \\
 {\cal T}^{(o) 3}_{(1)1}({\vec \sigma} ,{\vec \sigma}_1)&=&
 {\cal T}^{(o) 1}_{(3)1}({\vec \sigma} ,{\vec \sigma}_1)= - {1\over 2}
 \, {{\partial}\over {\partial \sigma^1_1}}\, \int_{-\infty}^{\sigma_1^3} dw_1^3
 \, \delta^3(\vec \sigma ,\sigma_1^1\sigma_1^2w^3_1)=\nonumber \\
 &=& - {1\over 2}\, {{\partial \delta (\sigma^1,\sigma^1_1)}\over {\partial \sigma_1^1}}
 \, \delta (\sigma^2,\sigma^2_1)\, \theta (\sigma^3_1,\sigma^3)=\nonumber \\
 &=& - {1\over 2}\, {{\partial^2}\over {(\partial \sigma_1^1)^2}}
 \, \int_{-\infty}^{\sigma_1^1} dw_1^1 \int_{-\infty}^{\sigma_1^3} dw_1^3
 \, \delta^3(\vec \sigma ,w_1^1\sigma_1^2w^3_1),\nonumber \\
 &&{}\nonumber \\
 {\cal T}^{(o) 3}_{(1)2}({\vec \sigma} ,{\vec \sigma}_1)&=&
 {\cal T}^{(o) 1}_{(3)2}({\vec \sigma} ,{\vec \sigma}_1)= {1\over 2}
 \, {{\partial^2}\over {(\partial \sigma_1^2)^2}}\, \int_{-\infty}^{\sigma_1^1} dw_1^1
 \int_{-\infty}^{\sigma_1^3} dw_1^3
 \, \delta^3(w_1^1\sigma_1^2w^3_1)=\nonumber \\
 &=& {1\over 2}\, \theta (\sigma_1^1,\sigma^1)\, {{\partial^2\delta (\sigma^2,\sigma^2_1)}\over
 {(\partial \sigma_1^2)^2}}\, \theta (\sigma_1^3,\sigma^3),\nonumber \\
 &&{}\nonumber \\
 {\cal T}^{(o) 3}_{(1)3}({\vec \sigma} ,{\vec \sigma}_1)&=&
 {\cal T}^{(o) 1}_{(3)3}({\vec \sigma} ,{\vec \sigma}_1)= - {1\over 2}
 \, {{\partial}\over {\partial \sigma^3_1}}\, \int_{-\infty}^{\sigma_1^1} dw_1^1
 \, \delta^3(\vec \sigma ,w_1^1\sigma_1^2\sigma^3_1)=\nonumber \\
 &=& - {1\over 2}\, \theta (\sigma_1^1,\sigma^1)\, \delta (\sigma^2,\sigma^2_1)
 \, {{\partial \delta (\sigma^3,\sigma^3_1)}\over {\partial \sigma_1^3}}=\nonumber \\
 &=& - {1\over 2}\, {{\partial^2}\over {(\partial \sigma_1^3)^2}}
 \, \int_{-\infty}^{\sigma^1_1} dw_1^1 \int_{-\infty}^{\sigma_1^3} dw_1^3
 \, \delta^3(\vec \sigma ,w_1^1\sigma_1^2w^3_1),\nonumber \\
 &&{}\nonumber \\
 {\cal T}^{(o) 2}_{(3)1}({\vec \sigma} ,{\vec \sigma}_1)&=&
 {\cal T}^{(o) 3}_{(2)1}({\vec \sigma} ,{\vec \sigma}_1)= {1\over 2}
 \, {{\partial^2}\over {(\partial \sigma_1^1)^2}} \int_{-\infty}^{\sigma_1^2} dw_1^2
 \int_{-\infty}^{\sigma_1^3} dw_1^3\,
 \delta^3(\vec \sigma ,\sigma_1^1w_1^2w^3_1)=\nonumber \\
 &=& {1\over 2}\, {{\partial^2\delta (\sigma^1,\sigma^1_1)}\over {(\partial \sigma_1^1)^2}}
 \, \theta (\sigma_1^2,\sigma^2)\, \theta (\sigma_1^3,\sigma^3),\nonumber \\
 &&{}\nonumber \\
 {\cal T}^{(o) 2}_{(3)2}({\vec \sigma} ,{\vec \sigma}_1)&=&
 {\cal T}^{(o) 3}_{(2)2}({\vec \sigma} ,{\vec \sigma}_1)= - {1\over 2}
 \, {{\partial}\over {\partial \sigma_1^2}}\, \int_{-\infty}^{\sigma_1^3} dw_1^3
 \, \delta^3(\vec \sigma ,\sigma_1^1\sigma_1^2w^3_1)=\nonumber \\
 &=& - {1\over 2}\, \delta (\sigma^1,\sigma^1_1)\, {{\partial \delta (\sigma^2,\sigma^2_1)}\over
 {\partial \sigma_1^2}}\, \theta (\sigma_1^3,\sigma^3)=\nonumber \\
 &=& - {1\over 2}\, {{\partial^2}\over {(\partial \sigma_1^2)^2}}
 \, \int_{-\infty}^{\sigma_1^2} dw_1^2 \int_{-\infty}^{\sigma_1^3} dw_1^3
 \, \delta^3(\vec \sigma ,\sigma_1^1w_1^2w^3_1),\nonumber \\
 &&{}\nonumber \\
 {\cal T}^{(o) 2}_{(3)3}({\vec \sigma} ,{\vec \sigma}_1)&=&
 {\cal T}^{(o) 3}_{(2)3}({\vec \sigma} ,{\vec \sigma}_1)= - {1\over 2}
 \, {{\partial}\over {\partial \sigma_1^3}} \int_{-\infty}^{\sigma_1^2} dw_1^2
 \, \delta^3(\vec \sigma ,\sigma_1^1w_1^2\sigma^3_1)=\nonumber \\
 &=& - {1\over 2}\, \delta (\sigma^1,\sigma^1_1)\, \theta (\sigma_1^2,\sigma^2)
 \, {{\partial \delta (\sigma^3,\sigma^3_1)}\over {\partial \sigma_1^3}}=\nonumber \\
 &=& - {1\over 2}\, {{\partial^2}\over {(\partial \sigma_1^3)^2}}
 \, \int_{-\infty}^{\sigma_1^2} dw_1^2 \int_{-\infty}^{\sigma_1^3} dw_1^2
 \, \delta^3(\vec \sigma ,\sigma_1^1w_1^2w^3_1),\nonumber \\
  &&{}\nonumber \\
  &&or\nonumber \\
  &&{}\nonumber \\
{\cal T}^{(o) r}_{(a)u}({\vec \sigma} ,{\vec \sigma}_1) &=&
{1\over 2}\, [1-\delta^r_{(a)}]\,
[1-2(\delta_{ru}+\delta_{au})]\nonumber \\
 &&{{\partial^2}\over {(\partial \sigma_1^u)^2}}\,  \int_{-\infty}^{\sigma_1^r} dw_1^r
 \int_{-\infty}^{\sigma_1^a} dw_1^a\, \delta^3(\vec \sigma ,w_1^rw_1^a\sigma_1^{k\not= (r,a)})=
 \nonumber \\
 &=& {1\over 2}\, [1-\delta^r_{(a)}]\, [1-2(\delta_{ru}+\delta_{au})]\nonumber \\
 &&{{\partial^2}\over {(\partial \sigma_1^u)^2}}\, \Big[ \theta (\sigma_1^r,\sigma^r)
 \, \theta (\sigma_1^a,\sigma^a)\, \delta(\sigma^{k\not= (r,a)},\sigma^{k\not= (r,a)}_1)\Big].
 \label{II13}
 \eea

The kernels $F^{(o)r}_{(a)(b)}$'s are given in Appendix B.

\subsection{The Old Cotriad Momenta from the Solution of the Constraints.}

Eqs.(\ref{II3})  imply that the cotriad momenta, solution of both
the linearized rotation and super-momentum constraints, have the
following expression in terms of the DO momenta $\pi_{\bar a}(\tau
,\vec \sigma )$

\bea
 {}^3{\check {\tilde \pi}}^1_{(1)}(\tau ,\vec \sigma ) &=&
\sqrt{3} \sum_{\bar a}\gamma_{\bar a1} \pi_{\bar a}(\tau ,\vec
\sigma ),\nonumber \\ {}^3{\check {\tilde \pi}}^2_{(1)}(\tau ,\vec
\sigma )&=&-{{\sqrt{3}}\over 2} \sum_{\bar a}\Big[ \gamma_{\bar
a1}\int d^3\sigma_1 {{\partial \delta (\sigma^1,\sigma_1^1)}\over
{\partial \sigma^1_1}} \theta (\sigma_1^2,\sigma^2)\delta
 (\sigma^3,\sigma_1^3)\pi_{\bar a}(\tau ,{\vec \sigma}_1)+\nonumber \\
 &+&\gamma_{\bar a2}\int d^3\sigma_1 \theta (\sigma^1_1,\sigma^1)
 {{\partial \delta (\sigma^2,\sigma^2_1)}\over {\partial
 \sigma_1^2}} \delta (\sigma^3,\sigma^3_1) \pi_{\bar a}(\tau
 ,{\vec \sigma}_1)-\nonumber \\
 &-&\gamma_{\bar a3} \int d^3\sigma_1 \theta (\sigma_1^1,\sigma^1)
 \theta (\sigma_1^2,\sigma^2) {{\partial^2 \delta
 (\sigma^3,\sigma^3_1)}\over {(\partial \sigma_1^3)^2}}\pi_{\bar
 a}(\tau ,{\vec \sigma}_1)\Big] =\nonumber \\
&=& {{\sqrt{3}}\over 2} \sum_{\bar a} \Big[ \gamma_{\bar a1}
\int^{\infty}_{\sigma^2} d\sigma_1^2 {{\partial \pi_{\bar a}(\tau,
 \sigma^1\sigma_1^2\sigma^3)}\over {\partial \sigma^1}} +\nonumber \\
 &+&\gamma_{\bar a2}  \int^{\infty}_{\sigma^1} d\sigma_1^1 {{\partial \pi_{\bar a}(\tau
 ,\sigma_1^1\sigma^2\sigma^3)}\over {\partial \sigma^2}} +\nonumber \\
 &-& \gamma_{\bar a3}  \int^{\infty}_{\sigma^1} d\sigma_1^1
 \int^{\infty}_{\sigma^2} d\sigma_1^2 {{\partial^2\pi_{\bar a}(\tau
 ,\sigma_1^1\sigma_1^2\sigma^3)}\over {(\partial \sigma^3)^2}}\Big],\nonumber \\
 {}^3{\check {\tilde \pi}}^3_{(1)}(\tau ,\vec \sigma )&=& -{{\sqrt{3}}\over 2}
 \sum_{\bar a} \Big[ \gamma_{\bar a1} \int d^3\sigma_1 {{\partial
 \delta (\sigma^1,\sigma_1^1)}\over {\partial \sigma_1^1}} \delta
 (\sigma^2,\sigma_1^2) \theta (\sigma_1^3,\sigma^3) \pi_{\bar
 a}(\tau ,{\vec \sigma}_1)-\nonumber \\
 &-&\gamma_{\bar a2} \int d^3\sigma_1 \theta (\sigma_1^1,\sigma^1)
 \delta (\sigma^2,\sigma_1^2) {{\partial \delta
 (\sigma^3,\sigma_1^3)}\over {\partial \sigma_1^3}} \pi_{\bar
 a}(\tau ,{\vec \sigma}_1)\Big] =\nonumber \\
 &=& -{{\sqrt{3}}\over 2}
 \sum_{\bar a} \Big[ \gamma_{\bar a1}
 \int^{\infty}_{\sigma^3} d\sigma_1^3 {{\partial \pi_{\bar a}(\tau ,\sigma^1\sigma^2\sigma_1^3)}
 \over {\partial \sigma^1}} +\nonumber \\
 &+&\gamma_{\bar a2}
 \int^{\infty}_{\sigma^1} d\sigma_1^1 \int^{\infty}_{\sigma^3} d\sigma_1^3
 {{\partial^2\pi_{\bar a}(\tau ,\sigma_1^1\sigma^2\sigma_1^3)}\over {(\partial
 \sigma^2)^2}} +\nonumber \\
 &+&\gamma_{\bar a3}  \int^{\infty}_{\sigma^1} d\sigma_1^1
 {{\partial \pi_{\bar a}(\tau ,\sigma_1^1\sigma^2\sigma^3)}\over {\partial \sigma^3}}
  \Big] ,\nonumber \\
 &&{}\nonumber \\
 {}^3{\check {\tilde \pi}}^1_{(2)}(\tau ,\vec \sigma ) &=&
 {}^3{\check {\tilde \pi}}^2_{(1)}(\tau ,\vec \sigma ),\nonumber \\
 {}^3{\check {\tilde \pi}}^2_{(2)}(\tau ,\vec \sigma ) &=& \sqrt{3}
 \sum_{\bar a} \gamma_{\bar a2} \pi_{\bar a}(\tau ,\vec \sigma ),\nonumber \\
 {}^3{\check {\tilde \pi}}^3_{(2)}(\tau ,\vec \sigma ) &=& -{{\sqrt{3}}\over 2}
 \sum_{\bar a} \Big[ \gamma_{\bar a1}\int d^3\sigma_1 {{\partial^2
 \delta (\sigma^1,\sigma_1^1)}\over {(\partial \sigma_1^1)^2}}
 \theta (\sigma_1^2,\sigma^2) \theta (\sigma_1^3,\sigma^3)
 \pi_{\bar a}(\tau ,{\vec \sigma}_1)+\nonumber \\
 &+&\gamma_{\bar a2} \int d^3\sigma_1 \delta (\sigma^1,\sigma_1^1)
 {{\partial \delta (\sigma^2,\sigma_1^2)}\over {\partial
 \sigma_1^2}} \theta (\sigma_1^3,\sigma^3) \pi_{\bar a}(\tau
 ,{\vec \sigma}_1)+\nonumber \\
 &+&\gamma_{\bar a3} \int d^3\sigma_1 \theta (\sigma_1^1,\sigma^1)
 \delta (\sigma^2,\sigma_1^2) {{\partial \delta
 (\sigma^3,\sigma_1^3)}\over {\partial \sigma_1^3}} \pi_{\bar
 a}(\tau ,{\vec \sigma}_1)\Big] =\nonumber \\
 &=& {{\sqrt{3}}\over 2}
 \sum_{\bar a} \Big[ \gamma_{\bar a1}
 \int^{\infty}_{\sigma^2} d\sigma_1^2 \int^{\infty}_{\sigma^3} d\sigma_1^3
 {{\partial^2\pi_{\bar a}(\tau ,\sigma^1\sigma_1^2\sigma_1^3)}\over {(\partial \sigma^1)^2}}
  +\nonumber \\
 &+& \gamma_{\bar a2}  \int^{\infty}_{\sigma^3} d\sigma_1^3
 {{\partial\pi_{\bar a}(\tau ,\sigma^1\sigma^2\sigma_1^3)}\over {\partial \sigma^2}}+\nonumber \\
 &+& \gamma_{\bar a3}
 \int^{\infty}_{\sigma^2} d\sigma_1^2 {{\partial \pi_{\bar a}(\tau
 ,\sigma^1\sigma_1^2\sigma^3)}\over {\partial \sigma^3}}\Big],\nonumber \\
 {}^3{\check {\tilde \pi}}^1_{(3)}(\tau ,\vec \sigma ) &=&
 {}^3{\check {\tilde \pi}}^3_{(1)}(\tau ,\vec \sigma ),\nonumber \\
 {}^3{\check {\tilde \pi}}^2_{(3)}(\tau ,\vec \sigma ) &=&
 {}^3{\check {\tilde \pi}}^3_{(2)}(\tau ,\vec \sigma ),\nonumber \\
 {}^3{\check {\tilde \pi}}^3_{(3)}(\tau ,\vec \sigma ) &=& \sqrt{3}
 \sum_{\bar a} \gamma_{\bar a3} \pi_{\bar a}(\tau ,\vec \sigma ),\nonumber \\
 &&{}\nonumber \\
 &&or\nonumber \\
 &&{}\nonumber \\
 {}^3{\check {\tilde \pi}}^r_{(a)}(\tau ,\vec \sigma ) &=&\sqrt{3}
 \sum_{\bar a}\gamma_{\bar ar} \delta_{(a)}^r \pi_{\bar a}(\tau
 ,\vec \sigma )+
  {{\sqrt{3}}\over 2}[1-\delta^r_{(a)}] \sum_{\bar au} \gamma_{\bar au}
  [1-2(\delta_{ru}+\delta_{au})]\nonumber \\
  && {{\partial^2}\over {(\partial \sigma^u)^2}}
  \int^{\infty}_{\sigma^r} d\sigma_1^r
  \int^{\infty}_{\sigma^a} d\sigma^a_1\, \pi_{\bar a}(\tau ,\sigma_1^r
  \sigma_1^a\sigma^{k\not= r,a}).
\label{II14}
 \eea

Clearly Eqs.(\ref{II4}) are satisfied.\bigskip

The solution (\ref{II12}) for the $G^{(o)ru}_{(a)}$'s is such that
the momenta ${}^3{\check {\tilde \pi}}^r_{(a)}(\tau ,\vec \sigma
)$ of Eq.(\ref{II14}) tend to zero for $|\vec \sigma | \rightarrow
\infty$, as required by Eqs.(\ref{I2}), {\it if the momenta
$\pi_{\bar a}(\tau ,\vec \sigma )$ satisfy the restrictions}

\beq
  \int_{-\infty}^{+\infty}
d\sigma^r\, \pi_{\bar a}(\tau ,\vec \sigma ) = 0.
 \label{II15}
 \eeq

\noindent For instance these restrictions are satisfied if
$\pi_{\bar a}(\tau ,\vec \sigma )={{\partial^3{\tilde \pi}_{\bar
a}(\tau ,\vec \sigma )}\over {\partial \sigma^1\partial
\sigma^2\partial \sigma^3}}$ with ${{\partial^2{\tilde \pi}_{\bar
a}(\tau ,\vec \sigma )}\over {\partial \sigma^i\partial
\sigma^j}}\, {\rightarrow}_{\sigma^k\rightarrow \infty}\, 0$
[$i,j,k$ cyclic] in a direction-independent way.
\bigskip

We have not succeeded in finding a solution without these
restrictions. Eqs.(\ref{II15}) can be thought of as 6 additional
constraints defined on 2-dimensional surfaces. As we shall see,
the consistency of these restrictions with the final Hamilton
equations will impose the following restrictions on the DO's
$r_{\bar a}(\tau ,\vec \sigma )$

\beq
  \int_{-\infty}^{+\infty}
d\sigma^r\, r_{\bar a}(\tau ,\vec \sigma ) = 0.
 \label{II16}
 \eeq

Therefore Eqs.(\ref{II15}) and (\ref{II16}) are 6 pairs of second
class constraints and we could think to find a Shanmugadhasan
canonical transformation from the DO's $r_{\bar a}(\tau ,\vec
\sigma )$, $\pi_{\bar a}(\tau ,\vec \sigma )$ to a new basis in
which 6 pairs of conjugate variables vanish due to
Eqs.(\ref{II15}),(\ref{II16}) and the physics is concentrated in
the remaining pairs. However, we shall not look for such a
transformation, because it is highly non-trivial due to the fact
that these constraints are defined only on 2-dimensional surfaces.
We shall go on to work with the DO's $r_{\bar a}(\tau ,\vec \sigma
)$, $\pi_{\bar a}(\tau ,\vec \sigma )$ even if this will imply
formal complications.
\bigskip

In conclusion, we have been able to solve {\it all} the
Hamiltonian constraints of tetrad gravity on the linearized WSW
hyper-surfaces of our gauge.

\vfill\eject

\section{The weak ADM Energy, the Lapse and Shift Functions and the 4-metric.}

After the solution of all the constraints, in this Section we
determine the weak ADM energy, the lapse and shift functions and
the 4-metric of our linearized theory. Since we need the weak ADM
energy, namely the Hamiltonian in the rest-frame instant form of
tetrad gravity, see Eq.(\ref{a11}), we need the following results
of Appendix A: Eqs.(\ref{a10}), where there is the expression of
the weak and strong ADM Poincare' generators, and Eqs.(\ref{a12}),
where there is the expression of the weak ADM Poincare' generators
in our gauge in the canonical basis (\ref{I3}).

In Eqs.(\ref{a11}) we put: i)  $\phi =1+{1\over 2}q+O(r^2_{\bar
a})$ with $q(\tau ,\vec \sigma )$ given by the solution
(\ref{II2}) of the linearized Lichnerowicz equation; ii) the
expression (\ref{II14}) for the cotriad momenta. In this way we
get the form ${\hat E}_{ADM}[r_{\bar a}, \pi_{\bar a}]$ of the
weak ADM energy only in terms of the DO's of our completely fixed
gauge.

\subsection{The ADM Energy and the Lapse and Shift Funcions.}

The {\it Hamiltonian linearization} of tetrad gravity in our
completely fixed gauge is defined by {\it approximating the weak
ADM energy} ${\hat E}_{ADM}[r_{\bar a}, \pi_{\bar a}]$ with the
quadratic functional of $r_{\bar a}(\tau ,\vec \sigma )$ and of
$\pi_{\bar a}(\tau ,\vec \sigma )$ contained in it

 \bea
 {\hat E}_{ADM} &=& -\epsilon {\hat
P}^{\tau}_{ADM,R} = {{12\pi G}\over {c^3}} \int d^3\sigma
\sum_{\bar a} \pi^2_{\bar a}(\tau ,\vec \sigma ) +\nonumber \\
 &+& {{24\pi G}\over {c^3}} \sum_{\bar a\bar b} \sum_{uv} \delta^u_{(a)}
 \gamma_{\bar au}\gamma_{\bar bv} \int d^3\sigma_1d^3\sigma_2 {\cal T}^{(o)u}
 _{(a)v}({\vec \sigma}_1 ,{\vec \sigma}_2) \pi_{\bar a}(\tau ,{\vec \sigma}_1)
 \pi_{\bar b}(\tau ,{\vec \sigma}_2)+\nonumber \\
 &+& {{6\pi G}\over {c^3}} \sum_{\bar a\bar b} \sum_{rs} \gamma_{\bar ar}\gamma_{\bar bs}
 \int d^3\sigma d^3\sigma_1 d^3\sigma_2 \Big[  \sum_u {\cal T}^{(o)u}_{(a)r}(\vec \sigma
 ,{\vec \sigma}_1) {\cal T}^{(o)u}_{(a)s}(\vec \sigma ,{\vec \sigma}_2)+\nonumber \\
 &+&\sum_{uv}(\delta^u_{(b)}\delta^v_{(a)}-\delta^u_{(a)}\delta^v_{(b)})
 {\cal T}^{(o)u}_{(a)r}(\vec \sigma ,{\vec \sigma}_1)
 {\cal T}^{(o)v}_{(b)s}(\vec \sigma ,{\vec \sigma}_2)\Big]
 \pi_{\bar a}(\tau ,{\vec \sigma}_1)\, \pi_{\bar b}(\tau ,{\vec \sigma}_2) -\nonumber \\
 &-& {{c^3}\over {16\pi G}} \sum_r \int d^3\sigma
 \Big[ {1\over 6} \Big( \sum_{\bar au} \gamma_{\bar au}
 {{\partial}\over {\partial \sigma^r}} \int d^3\sigma_1
 {{\partial^2_{1u}r_{\bar a}(\tau ,{\vec \sigma}_1)}\over
 {4\pi |\vec \sigma -{\vec \sigma}_1|}} \Big)^2 -\nonumber \\
 &-&{1\over 3}\sum_{\bar a} \Big( \partial_rr_{\bar a}(\tau ,\vec \sigma )\Big)^2+
 {2\over 3} \Big( \sum_{\bar a}\gamma_{\bar ar} \partial_rr_{\bar a}(\tau ,\vec \sigma )
 \Big)^2 -\nonumber \\
 &-&{1\over 3} \sum_{\bar a\bar b} \sum_u \gamma_{\bar ar} \partial_rr_{\bar a}(\tau
 ,\vec \sigma ) \gamma_{\bar bu} {{\partial}\over {\partial \sigma^r}}
 \int d^3\sigma_1 {{\partial^2_{1u}r_{\bar b}(\tau ,{\vec \sigma}_1)}\over
 {4\pi |\vec \sigma -{\vec \sigma}_1|}} \Big] =\nonumber \\
 &&{}\nonumber \\
 &=& {{12\pi G}\over {c^3}} \int d^3\sigma \sum_{\bar a} \pi^2_{\bar
a}(\tau ,\vec \sigma ) +
  {{12\, \pi G}\over {c^3}} \sum_{\bar a\bar b} \sum_{rs} \gamma_{\bar ar}\gamma_{\bar bs}
 \nonumber \\
 &&\int d^3\sigma d^3\sigma_1 d^3\sigma_2 \,  \sum_u {\cal T}^{(o)u}_{(a)r}(\vec \sigma
 ,{\vec \sigma}_1) {\cal T}^{(o)u}_{(a)s}(\vec \sigma ,{\vec
 \sigma}_2)\,  \pi_{\bar a}(\tau ,{\vec \sigma}_1)\,
   \pi_{\bar b}(\tau ,{\vec \sigma}_2) -\nonumber \\
 &-& {{c^3}\over {16\pi G}} \sum_r \int d^3\sigma
 \Big[ {1\over 6} \Big( \sum_{\bar au} \gamma_{\bar au}
 {{\partial}\over {\partial \sigma^r}} \int d^3\sigma_1
 {{\partial^2_{1u}r_{\bar a}(\tau ,{\vec \sigma}_1)}\over
 {4\pi |\vec \sigma -{\vec \sigma}_1|}} \Big)^2 -\nonumber \\
 &-&{1\over 3}\sum_{\bar a} \Big( \partial_rr_{\bar a}(\tau ,\vec \sigma )\Big)^2+
 {2\over 3} \Big( \sum_{\bar a}\gamma_{\bar ar} \partial_rr_{\bar a}(\tau ,\vec \sigma )
 \Big)^2 -\nonumber \\
 &-&{1\over 3} \sum_{\bar a\bar b} \sum_u \gamma_{\bar ar} \partial_rr_{\bar a}(\tau
 ,\vec \sigma ) \gamma_{\bar bu} {{\partial}\over {\partial \sigma^r}}
 \int d^3\sigma_1 {{\partial^2_{1u}r_{\bar b}(\tau ,{\vec \sigma}_1)}\over
 {4\pi |\vec \sigma -{\vec \sigma}_1|}} \Big] + O(r^3_{\bar a}).
 \label{III1}
\eea

\noindent See Eq.(\ref{c13}) of Appendix C for the Fourier
transform of the ADM energy.\bigskip

The determination of the {\it bulk lapse function} $n(\tau ,\vec
\sigma )$ is done by solving the integral equation (\ref{a13})
written in our completely fixed gauge. Therefore we must evaluate
${{\delta {\hat {\cal H}}_R(\tau ,{\vec \sigma}_1)}\over {\delta
\phi (\tau ,\vec \sigma )}}$ and ${{\delta {\hat E}_{ADM}}\over
{\delta \phi (\tau ,\vec \sigma )}}$. From the linearized version
of the super-hamiltonian constraint given before Eq.(\ref{II1}) we
get

\beq {{\delta {\hat {\cal H}}_R(\tau ,{\vec \sigma}_1)}\over
{\delta \phi (\tau ,\vec \sigma )}} =-\epsilon {{c^3}\over {2\pi
G}} \triangle_1 \delta^3(\vec \sigma ,{\vec \sigma}_1) +
O(r^2_{\bar a}),
 \label{III2}
\eeq

\noindent while from Eqs.(\ref{a12}) and (\ref{II1}) we get

 \bea
 {{\delta {\hat E}_{ADM}}\over {\delta \phi (\tau ,\vec \sigma
)}} &=& \epsilon \, {{c^3}\over {4\pi G}} \int d^3\sigma_1 \sum_r
\Big[ 2
\partial_{1r}q(\tau ,{\vec \sigma}_1) - {1\over {\sqrt{3}}} \sum_{\bar a}
\gamma_{\bar ar} \partial_{1r}r_{\bar a}(\tau ,{\vec \sigma}_1)\Big]
\partial_{1r}\delta^3(\vec \sigma ,{\vec \sigma}_1) +O(r^2_{\bar a}) =\nonumber \\
 &=& - \epsilon\, {{c^3}\over {2\pi G}} \Big[ \triangle q(\tau ,\vec \sigma )-
 {1\over {2\sqrt{3}}} \sum_{\bar ar} \gamma_{\bar ar}
 \partial^2_rr_{\bar a}(\tau ,\vec \sigma )\Big] +O(r^2_{\bar a})=\nonumber \\
 &=& 0 + O(r^2_{\bar a}).
 \label{III3}
\eea

\noindent  Then the integral equation (\ref{a13}) for the lapse
function becomes the following partial differential equation

\bea
 -\epsilon {{c^3}\over {2\pi G}} \int d^3\sigma_1 n(\tau
,{\vec \sigma}_1) \triangle_1\delta^3(\vec \sigma ,{\vec
\sigma}_1) &=& -\epsilon {{c^3}\over {2\pi G}} \triangle n(\tau
,\vec \sigma ) = 0+ O(r^2_{\bar a}),\nonumber \\
 &&{}\nonumber \\
 &&\Downarrow \nonumber \\
 n(\tau ,\vec \sigma ) &=& 0 + O(r^2_{\bar a}).
\label{III4}
 \eea

\bigskip

The determination of the {\it bulk shift functions} uses this
result and the linearized form of  Eq.(\ref{a14}) (see
Eq.(\ref{c14}) of Appendix C for the Fourier transform),

\bea
 n_r(\tau ,\vec \sigma ) &=& -\sqrt{3} {{4\pi G}\over {c^3}} \int
d^3\sigma_1 \sum_{wu} (\delta_{wu}\delta_{(a)(b)}
+\delta_{(a)u}\delta_{(b)w} -\delta_{(a)w}\delta_{(b)u})\nonumber \\
 &&\sum_v \int d^3\sigma_2 \Big[ \delta^w_v\delta_{(a)v}\delta^3({\vec \sigma}_1
 ,{\vec \sigma}_2) +{\cal T}^{(o)w}_{(a)v}({\vec \sigma}_1 ,{\vec \sigma}_2)\Big]
 \nonumber \\
 &&\sum_{\bar a} \gamma_{\bar av} \pi_{\bar a}(\tau ,{\vec \sigma}_2)
 G^{(o) ur}_{(b)}({\vec \sigma}_1, \vec \sigma ) + O(r^2_{\bar a}) =\nonumber \\
 &=& {{\partial}\over {\partial \sigma^r}}\, \Big( \sqrt{3} {{2\pi G}\over {c^3}}
 \sum_{\bar av} \gamma_{\bar av}\nonumber \\
 &&\Big[ \sum_{ua,u\not= a}
 [1-2(\delta_{uv}+\delta_{av})]\, \int^{\sigma^u}_{-\infty}
 d\sigma_1^u \int^{\sigma^a}_{-\infty} d\sigma_1^a
 \int^{\infty}_{\sigma_1^u} d\sigma_2^u \int^{\infty}_{\sigma_1^a}
 d\sigma_2^a\nonumber \\
 &&{{\partial^2 \pi_{\bar a}(\tau ,\sigma_2^u \sigma_2^a
 \sigma_2^{k\not= u,a})}\over {(\partial \sigma_2^v)^2}}
 {|}_{\sigma_2^k=\sigma^k}-\nonumber \\
 &-& 2\sum_{u\not= r} [1-2(\delta_{uv}+\delta_{rv})]\,
 \int^{\sigma^r}_{-\infty} d\sigma_1^r \int^{\sigma^u}_{-\infty}
 d\sigma_1^u \int^{\infty}_{\sigma_1^r} d\sigma_2^r
 \int^{\infty}_{\sigma_1^u} d\sigma_2^u\nonumber \\
 && {{\partial^2 \pi_{\bar
 a}(\tau ,\sigma_2^r \sigma_2^u \sigma_2^{k\not= r,u})}\over
 {(\partial \sigma_2^v)^2}} {|}_{\sigma_2^k=\sigma^k}\Big] \Big)+
  O(r^2_{\bar a}).
 \label{III5}
\eea

 In conclusion in our completely fixed gauge the lapse and shift functions
are

\bea
  N(\tau ,\vec \sigma )&=& -\epsilon +n(\tau ,\vec \sigma
 )=-\epsilon + O(r^2_{\bar a}),\nonumber \\
 N_r(\tau ,\vec \sigma )&=& n_r(\tau ,\vec \sigma ) =-\epsilon \,
 {}^4{\hat g}_{\tau r}(\tau ,\vec \sigma ).
 \label{III6}
 \eea

\bigskip

Since the shift functions are of order $O(r_{\bar a})$, we get the
following results :
\bigskip

i) the 4-coordinates associated to our Hamiltonian gauge are {\it
not synchronous}, so that we are using non-time-orthogonal
reference frames and we cannot use Einstein's convention for the
synchronization of clocks \cite{26}; a non-standard definition of
simultaneity of distant time-like observers is needed, consistent
with the Hamiltonian description based on the Cauchy simultaneity
WSW space-like hyper-surfaces $\Sigma_{\tau}$;

ii) there may be coordinate-dependent gravito-magnetic effects;

iii) the {\it velocity of light becomes direction-dependent} (see
Ref.\cite{26}) : if $u^i$ is a unit 3-vector with respect to the
3-metric ${}^3{\tilde \gamma}_{rs}=-\epsilon ({}^3g_{rs}+
{{n_rn_s}\over {\epsilon \, {}^4g_{\tau\tau}}})$, i.e.
${}^3{\tilde \gamma}_{rs}u^ru^s=1$, [see Appendix A of
Ref.\cite{2}, after Eq.(A5)],  the light velocity in direction
$u^i\,\,$  is $w(u^i) = {{ \sqrt{{}^4g_{\tau\tau}}}\over
{{{u^rn_r}\over {\sqrt{{}^4g_{\tau\tau}}}} +1}} = 1 - u^r\, n_r +
O(r^2_{\bar a})$ with $\Big( {{u^rn_r}\over
{\sqrt{{}^4g_{\tau\tau}}}} \Big)^2 = (u^r\, n_r )^2 + O(r^3_{\bar
a}) < 1$.
\bigskip

\subsection{The Linearized 4-Metric.}

After the solution of all the constraints and the determination of
the lapse and shift functions, the {\it 4-metric} of our
linearized space-time with $\Sigma^{(WSW)}_{\tau}$-adapted
coordinates in the 3-orthogonal gauge with $\pi_{\phi}(\tau ,\vec
\sigma )=0$  becomes (we write it in the form of a perturbation of
the Minkowski metric in Cartesian coordinates {\it only to
visualize} the deviations from special relativity of this
background-independent post-Minkowskian space-time in the
4-coordinates associated with our preferred 3-orthogonal gauge)

\bea
{}^4{\hat g}_{AB}(\tau ,\vec \sigma ) &=& {}^4\eta_{AB} +
{}^4h_{AB}(\tau ,\vec \sigma ),\nonumber \\
 &&{}\nonumber \\
 {}^4h_{\tau\tau}(\tau ,\vec \sigma ) &=& 0 + O(r^2_{\bar a}),\nonumber \\
 {}^4h_{\tau r}(\tau ,\vec \sigma ) &=& -\epsilon n_r(\tau ,\vec \sigma )\,
 {\buildrel {def}\over  =}\,
 -\epsilon {2\over {c^2}} A_{GEM,3-0, r}(\tau ,\vec \sigma ) =
 {{\partial}\over {\partial \sigma^r}} \Big( \epsilon \sqrt{3} {{2\pi G}\over {c^3}}
 \sum_{\bar av} \gamma_{\bar av}\nonumber \\
 &&\Big[ \sum_{ua,u \not= a}[1-2(\delta_{uv}+\delta_{av})]\nonumber \\
 &&\int^{\sigma^u}_{-\infty}
 d\sigma_1^u \int^{\sigma^a}_{-\infty} d\sigma_1^a
 \int^{\infty}_{\sigma_1^u} d\sigma_2^u \int^{\infty}_{\sigma_1^a}
 d\sigma_2^a\, {{\partial^2 \pi_{\bar a}(\tau ,\sigma_2^u \sigma_2^a
 \sigma_2^{k\not= u,a})}\over {(\partial \sigma_2^v)^2}}{|}_{\sigma_2^k=\sigma^k} -\nonumber \\
 &-& 2\sum_{u\not= r} [1-2(\delta_{uv}+\delta_{rv})]\nonumber \\
 && \int^{\sigma^r}_{-\infty} d\sigma_1^r \int^{\sigma^u}_{-\infty}
 d\sigma_1^u \int^{\infty}_{\sigma_1^r} d\sigma_2^r
 \int^{\infty}_{\sigma_1^u} d\sigma_2^u\, {{\partial^2 \pi_{\bar
 a}(\tau ,\sigma_2^r \sigma_2^u \sigma^{k\not= r,u})}\over
 {(\partial \sigma_2^v)^2}}{|}_{\sigma_2^k=\sigma^k} \Big] \Big)+\nonumber \\
 &+&O(r^2_{\bar a}),\nonumber \\
 {}^4h_{rs}(\tau ,\vec \sigma ) &=& -\epsilon [{}^3{\hat
 g}_{rs}(\tau ,\vec \sigma )-\delta_{rs}] {\buildrel {def}\over =}
 \delta_{rs}\, k_r(\tau ,\vec \sigma ) =\nonumber \\
 &=&-2\epsilon \Big[ q(\tau ,\vec \sigma ) +
 {2\over {\sqrt{3}}} \sum_{\bar a} \gamma_{\bar ar} r_{\bar a}(\tau ,\vec \sigma )
 \Big] \delta_{rs} + O(r^2_{\bar a}) =\nonumber \\
 &=& -{{2\epsilon}\over {\sqrt{3}}} \sum_{\bar a} \Big[ \gamma_{\bar ar}
 r_{\bar a}(\tau ,\vec \sigma ) -
 {1\over 2} \sum_u\gamma_{\bar au} \int d^3\sigma_1 {{\partial^2_{1u}r_{\bar a}(\tau
 ,{\vec \sigma}_1)}\over {4\pi |\vec \sigma -{\vec \sigma}_1|}} \Big]
 \delta_{rs} + O(r^2_{\bar a}),\nonumber \\
 &&{}\nonumber \\
 &&{}\nonumber \\
 \sqrt{{}^4{\hat g}(\tau ,\vec \sigma )}&=& -\epsilon \sqrt{\hat
 \gamma (\tau ,\vec \sigma )}+ O(r^2_{\bar a}) = - \epsilon\, \phi^6(\tau ,\vec \sigma )
 + O(r^2_{\bar a}) =
 - \epsilon \, [ 1 + 3\, q(\tau ,\vec \sigma )] + O(r^2_{\bar
 a}) =\nonumber \\
 &=& - \epsilon\, \Big[ 1 - {{\sqrt{3}}\over 2}\,  \sum_{u\bar a}
\gamma_{\bar au} \int d^3\sigma_1 {{\partial^2_{1u}r_{\bar a}(\tau
,{\vec \sigma}_1)} \over {4\pi |\vec \sigma -{\vec
\sigma}_1|}}\Big] +O(r^2_{\bar a}),\nonumber \\
 &&{}\nonumber \\
 {}^4{\hat g}^{\tau\tau}(\tau ,\vec \sigma ) &=& \epsilon +O(r^2_{\bar a}),\nonumber \\
 {}^4{\hat g}^{\tau r}(\tau ,\vec \sigma ) &=& -\epsilon \delta^{rs} n_s(\tau ,\vec \sigma ) +
 O(r^2_{\bar a}),\nonumber \\
 {}^4{\hat g}^{rs}(\tau ,\vec \sigma ) &=&-\epsilon \, \Big({}^3{\hat
 g}^{rs} - n^r\, n^s\Big)(\tau ,\vec \sigma )=\nonumber \\
 &=&-\epsilon
 \Big[1-{2\over {\sqrt{3}}}\sum_{\bar a}\Big(\gamma_{\bar ar}r_{\bar a}(\tau ,\vec \sigma
 )- {1\over 2}\sum_u\gamma_{\bar au} \int d^3\sigma_1
 {{\partial^2_{1u} r_{\bar a}(\tau ,{\vec \sigma}_1)}\over {4\pi
 |\vec \sigma -{\vec \sigma}_1|}}\Big)\Big] \delta^{rs} + O(r^2_{\bar
 a}).\nonumber \\
 &&{}
 \label{III7}
\eea

Therefore, the 3-metric on the linearized WSW hyper-surfaces of
our gauge is only conformal to the Euclidean 3-metric, namely the
linearized $\Sigma^{(WSW)}_{\tau}$ of our gauge are {\it
conformally flat}.\hfill\break

The linearized cotriads, triads and adapted cotetrads and tetrads
become [see Eqs.(\ref{a7}) and (\ref{a16})]

\bea
 {}^3{\hat e}_{(a)r}(\tau ,\vec \sigma ) &=& \delta_{(a)r}
\Big[1+{1\over {\sqrt{3}}}\sum_{\bar a}\Big( \gamma_{\bar
ar}r_{\bar a}(\tau ,\vec \sigma )-\nonumber \\
 &-& {1\over 2}\, \sum_{u}\gamma_{\bar au} \int d^3\sigma_1 {{\partial^2_{1u}r_{\bar a}(\tau
,{\vec \sigma}_1)} \over {4\pi |\vec \sigma -{\vec
\sigma}_1|}}\Big) \Big] +O(r^2_{\bar a}),\nonumber \\
 {}^3{\hat e}^r_{(a)}(\tau ,\vec \sigma ) &=& \delta^r_{(a)}
 \Big[1 -{1\over {\sqrt{3}}}\sum_{\bar a}\Big(\gamma_{\bar ar}r_{\bar a}(\tau
,\vec \sigma ) +\nonumber \\
 &+& {1\over 2}\, \sum_{u}\gamma_{\bar au} \int d^3\sigma_1 {{\partial^2_{1u}r_{\bar a}(\tau
,{\vec \sigma}_1)} \over {4\pi |\vec \sigma -{\vec
\sigma}_1|}}\Big) \Big] +O(r^2_{\bar a}),\nonumber \\
 {}^3\hat e(\tau ,\vec \sigma ) &=& \sqrt{\hat \gamma (\tau ,\vec \sigma )}=
 1+3q(\tau ,\vec \sigma ) +O(r^2_{\bar a})=\nonumber \\
 &=& 1- {{\sqrt{3}}\over 2}\, \sum_{\bar au}\gamma_{\bar au}
  \int d^3\sigma_1 {{\partial^2_{1u}r_{\bar a}(\tau
,{\vec \sigma}_1)} \over {4\pi |\vec \sigma -{\vec
\sigma}_1|}}\Big)  +O(r^2_{\bar a}),\nonumber \\
 &&{}\nonumber \\
 {}^3{\hat g}_{rs}(\tau ,\vec \sigma )&=&-\epsilon \, {}^4{\hat
 g}_{rs}(\tau ,\vec \sigma )= {}^3{\hat e}_{(a)r}(\tau ,\vec \sigma )\,
  {}^3{\hat e}_{(a)s}(\tau ,\vec \sigma ) =\nonumber \\
 &=& \Big[1+{2\over {\sqrt{3}}}\sum_{\bar a}\Big(\gamma_{\bar ar}r_{\bar a}(\tau ,\vec \sigma
 )-\nonumber \\
 &-&{1\over 2}\sum_u\gamma_{\bar au} \int d^3\sigma_1
 {{\partial^2_{1u} r_{\bar a}(\tau ,{\vec \sigma}_1)}\over {4\pi
 |\vec \sigma -{\vec \sigma}_1|}}\Big)\Big] \delta^{rs} + O(r^2_{\bar a}),
 \label{III8}
 \eea

\bea
 {}^4_{(\Sigma )}{\check {\tilde E}}^A_{(o)}(\tau ,\vec \sigma ) &=&
 l^A(\tau ,\vec \sigma ) =
 -\epsilon (1; -\delta^{rs}n_s(\tau ,\vec \sigma ))+O(r^2_{\bar a}),\nonumber \\
 {}^4_{(\Sigma )}{\check {\tilde E}}^A_{(a)}(\tau ,\vec \sigma ) &=&
 (0; {}^3{\hat e}^r_{(a)}(\tau ,\vec \sigma ))=\nonumber \\
 &=& (0; \delta^r_{(a)} \Big[1-{1\over {\sqrt{3}}}\sum_{\bar a}
 \Big(\gamma_{\bar ar}r_{\bar a}(\tau ,\vec \sigma )+\nonumber \\
 &+& {1\over {2}} \sum_{u\bar a}
\gamma_{\bar au} \int d^3\sigma_1 {{\partial^2_{1u}r_{\bar a}(\tau
,{\vec \sigma}_1)} \over {4\pi |\vec \sigma -{\vec \sigma}_1|}}
 \Big) \Big]\, ) +O(r^2_{\bar a}),\nonumber \\
 &&{}\nonumber \\
 {}^4_{(\Sigma )}{\check {\tilde E}}^{(o)}_A(\tau ,\vec \sigma )&=&
 l_A(\tau ,\vec \sigma ) =
 -\epsilon (1;0) + O(r_{\bar a}^2),\nonumber \\
 {}^4_{(\Sigma )}{\check {\tilde E}}^{(a)}_A(\tau ,\vec \sigma )&=&
 ( \delta^r_{(a)}n_r(\tau ,\vec \sigma ); {}^3{\hat e}_{(a)r}) =\nonumber \\
  &=& ( \delta^r_{(a)}n_r(\tau ,\vec \sigma ); \,
  \delta_{(a)r}\, \Big[1 + {1\over {\sqrt{3}}}\sum_{\bar a}
 \Big(\gamma_{\bar ar}r_{\bar a}(\tau ,\vec \sigma )-\nonumber \\
 &-& {1\over {2}} \sum_{u\bar a}
\gamma_{\bar au} \int d^3\sigma_1 {{\partial^2_{1u}r_{\bar a}(\tau
,{\vec \sigma}_1)} \over {4\pi |\vec \sigma -{\vec \sigma}_1|}}
 \Big) \Big]\, ) +O(r^2_{\bar a}).
\label{III9}
 \eea

\noindent Since in our gauge we have $\varphi_{(a)}(\tau ,\vec
\sigma ) = 0$, the tetrads and cotetrads ${}^4E^A_{(\alpha )}$ and
${}^4E^{(\alpha )}_A$ of Eqs.(\ref{I1}) coincide with those of
Eqs.(\ref{III9}).\bigskip

See Appendix D for the associated Christoffel symbols, spin
connection, field strength and Riemann tensor of $\Sigma_{\tau}$
and for the 4-tensors of $M^4$.

\subsection{The Energy-Momentum Landau-Lifschitz Pseudo-Tensor.}

Usually the energy of a gravitational wave on the Minkowski
background is evaluated \cite{22,24} as a mean value over various
wave-lengths of the coordinate-dependent Landau-Lifschitz energy
obtained from the Landau-Lifschitz symmetric pseudo-tensor
\cite{27}.

The Landau-Lifschitz pseudo-tensor ${}_{(L)}t^{\mu\nu} =
{}_{(L)}t^{\nu\mu}$, which contains no second derivatives of the
metric and gives meaningful results only in an asymptotically flat
Cartesian coordinate system, was found trying to rewrite the
consequence ${}^4\nabla_{\mu}\, {}^4T^{\mu\nu}\, {\buildrel \circ
\over =}\, 0$ of the Bianchi identities applied to Einstein's
equations in the form $\partial_{\mu}
[{}_{(L)}t^{\mu\nu}+\sqrt{{}^4g}\, {}^4T^{\mu\nu}]\, {\buildrel
\circ \over =}\, 0$. Starting from Einstein's equations
${}^4T^{\mu\nu}\, {\buildrel \circ \over =}\, {{c^3}\over {8\pi
G}}({}^4R ^{\mu\nu}-{1\over 2}{}^4g^{\mu\nu}\, {}^4R)$, one
rewrites them as

\bea
 {}_{(L)}t^{\mu\nu}&+&\sqrt{{}^4g}\,
{}^4T^{\mu\nu}\, {\buildrel \circ \over =}\,
\partial_{\rho} h^{\mu [\nu\rho ]},\qquad with \nonumber \\
 &&{}\nonumber \\
 &&{}\nonumber \\
  h^{\mu [\nu\rho ]} &=& -h^{\mu [\rho\nu
]}={{c^3}\over {16\pi G}} \partial_{\sigma} [{}^4g
({}^4g^{\mu\nu}\, {}^4g^{\rho\sigma}-{}^4g^{\mu\rho}\,
{}^4g^{\nu\sigma})] = {{c^3}\over {16\pi G}}\, \partial_{\sigma}\,
{\cal T}^{\rho\nu\sigma\mu},\nonumber \\
 &&{}\nonumber \\
 &&\partial_{\nu}\partial_{\rho} h^{\mu [\nu\rho ]}=
 \partial_{\nu}\partial_{\rho}\partial_{\sigma}\,
{\cal T}^{\rho\nu\sigma\mu} =0,\nonumber \\
 &&{}\nonumber \\
 {\cal T}^{\alpha\nu\beta\mu}&=&{}^4{\hat g}^{\alpha\beta}\,
{}^4{\hat g} ^{\mu\nu}-{}^4{\hat g}^{\alpha\mu}\, {}^4{\hat
g}^{\beta\nu}=-{\cal T} ^{\nu\alpha\beta\mu}=-{\cal
T}^{\alpha\nu\mu\beta}={\cal T}^{\beta\mu\alpha\nu},\nonumber \\
 &&[{\cal T}^{\alpha\nu\beta\mu}+{\cal T}^{\alpha\beta\mu\nu}+{\cal
T}^{\alpha \mu\nu\beta}=0],\quad {}^4{\hat g}^{\mu\nu} =
\sqrt{{}^4g}\, {}^4g^{\mu\nu}.
 \label{III10}
 \eea

\noindent Then one gets

\bea
 {}_{(L)}t^{\mu\nu} &=&
{{c^3}\over {16\pi G}} [\partial_{\rho}\, {}^4{\hat g}^{\mu\nu}\,
\partial_{\gamma}\, {}^4{\hat g}^{\gamma\rho}-\partial_{\rho}\, {}^4{\hat g}
^{\mu\rho}\partial_{\gamma}\, {}^4{\hat g}^{\nu\gamma}+{1\over
2}{}^4g^{\mu\nu} \, {}^4g_{\rho\sigma}\, \partial_{\gamma}\,
{}^4{\hat g}^{\rho\delta}\,
\partial_{\delta}\, {}^4{\hat g}^{\sigma\gamma}-\nonumber \\
 &-&({}^4g^{\mu\rho}\, {}^4g_{\gamma \delta} \partial_{\sigma}\,
{}^4{\hat g}^{\nu\delta}\, \partial_{\rho}\, {}^4{\hat
g}^{\gamma\sigma}+{}^4g^{\nu\rho}\, {}^4g_{\gamma\delta} \partial
_{\sigma}\, {}^4{\hat g}^{\mu\delta} \, \partial_{\rho}\,
{}^4{\hat g}^{\gamma \sigma})+{}^4g_{\rho\sigma}\,
{}^4g^{\gamma\delta} \partial_{\gamma}\, {}^4{\hat g}^{\mu\rho}
\partial_{\delta}\, {}^4{\hat g}^{\nu\sigma}+\nonumber \\
 &+&{1\over 8} (2\, {}^4g^{\mu\rho}\,
{}^4g^{\nu\sigma}-{}^4g^{\mu\nu}\, {}^4g^{\rho\sigma}) (2\,
{}^4g_{\gamma\delta}\, {}^4g_{\alpha\beta}-{}^4g_{\delta\alpha}\,
{}^4g_{\gamma\beta}) \partial_{\rho}\, {}^4{\hat g}^{\gamma\beta}
\partial _{\sigma}\, {}^4{\hat g}^{\delta\alpha}].
 \label{III11}
 \eea

However, as noted in Ref.\cite{28} and emphasized in
Ref.\cite{29}, if the energy flux carried away to infinity by
gravitational radiation in a given asymptotically flat space-time
is deduced from the Landau-Lifschitz pseudo-tensor, then the
result is {\it reliable only} on an appropriately chosen flat
background metric and then a connection between Bondi energy at
null infinity and ADM energy at spatial infinity can be
established \cite{29}. Since in our approach to Hamiltonian
linearization of Einstein's equation we have no background metric,
the Landau-Lifschitz pseudo-tensor is not a useful quantity.

\medskip

If we write the Landau-Lifschitz in
$\Sigma^{(WSW)}_{\tau}$-adapted coordinates, $ {}_{(L)}t^{AB}(\tau
,\vec \sigma )$ and then we choose our special 3-orthogonal
coordinates, then the Landau-Lifschitz 4-momentum $P^A_{LL} = \int
d^3\sigma\, {}_{(L)}t^{A\tau}(\tau ,\vec \sigma )$ has to be
contrasted with the weak ADM energy ${\hat E}_{ADM}$, namely the
Hamiltonian in the rest-frame instant form of gravity, and with
the vanishing weak ADM momentum ${\hat P}^r_{ADM} \approx 0$
evaluated in those coordinates.

>From an explicit calculation done by using Eqs.(\ref{III7}) it
turns out that $P^A_{LL}$ does not agree with $E_{ADM}$ and ${\hat
P}^r_{ADM}$ [for its expression see Eq.(\ref{V11})]

\bea
 E_{LL} &=& \int d^3\sigma\, {}_{(LL)}t^{\tau \tau}(\tau ,\vec
 \sigma ) =\nonumber \\
 &=& {{c^3}\over {16\pi G}}\, \int d^3\sigma \Big(
-\frac{25}{24}\,(\partial_{\tau}A)^2
-\frac{175}{72}\sum_{i}(\partial_{i}A)^2 -
\frac{5}{12}\,\partial_{\tau}A\sum_{i}\partial_{\tau}f_{i}
+\nonumber \\ && +\frac{1}{4}\sum_{i}(\partial_{\tau}f_{i})^2
-\frac{1}{8}\big(\sum_{i}\partial_{\tau}f_{i}\big)^2
 -\frac{5}{3}\,\partial_{\tau}A\sum_{i}\partial_{i}n_{i} +\nonumber \\
 &-&\frac{5}{12}\sum_{i,j}\partial_{j}A\partial_{j}f_{i}
-\sum_{i}\partial_{\tau}f_{i} \partial_{i}n_{i}
-\frac{5}{3}\sum_{i}\partial_{i}A\partial_{i}f_{i}+\nonumber \\
 &+&\frac{1}{2}\sum_{r,s}(\partial_{s}n_{r})^2
+\frac{1}{2}\sum_{r,s}\partial_{s}n_{r}\partial_{r}n_{s}
-\sum_{r,s}\partial_{s}n_{s}\partial_{r}n_{r}+\nonumber \\
 &-&\frac{1}{2}\sum_{i}(\partial_{i}f_{i})^2
-\frac{1}{8}\sum_{i,j,r}\partial_{i}f_{j}\partial_{i}f_{r}
+\frac{1}{4}\sum_{i,j}(\partial_{j}f_{i})^2 \Big) (\tau ,\vec
\sigma ) +\nonumber \\
 &+& O(r_{\bar a}^3) \not= {\hat E}_{ADM},\nonumber \\
 &&{}\nonumber \\
 &&{}\nonumber \\
 P^r_{LL} &=& \int d^3\sigma\, {}_{(LL)}t^{r \tau}(\tau ,\vec
 \sigma ) =\nonumber \\
 &=& {{c^3}\over {16\pi G}}\, \int d^3\sigma\,
\Big(\frac{115}{36}\,\partial_{\tau}A\partial_{r}A
+\frac{5}{12}\,\partial_{r}A\sum_{i}\partial_{\tau}f_{i}+\frac{2}{3}\,
\partial_{r}A\partial_{\tau}f_{r}+\nonumber\\
 &+&\frac{5}{3}\,\partial_{\tau}A\partial_{r}f_{r}
+\frac{5}{12}\,\partial_{\tau}A\sum_{i}\partial_{r}f_{i}
+\frac{1}{4}\sum_{i,j}\partial_{\tau}f_{i}\partial_{r}f_{j}-\nonumber\\
 &-&\frac{1}{2}\sum_{i}\partial_{\tau}f_{i}\partial_{r}f_{i}
+\partial_{\tau}f_{r}\partial_{r}f_{r}-\frac{7}{3}
\sum_{i}\partial_{i}A\partial_{i}n_{r}+\nonumber\\
 &+&\frac{4}{3}\sum_{i}\partial_{r}A\partial_{i}n_{i}+
 \sum_{i}\partial_{i}A\partial_{r}n_{i}
-\sum_{i}\partial_{i}f_{r}\partial_{i}n_{r}-\nonumber\\
 &-&\sum_{i}\partial_{i}f_{i}\partial_{i}n_{r}+
 \sum_{i}\partial_{r}f_{i}\partial_{i}n_{i}
+\partial_{r}f_{r}\sum_{i}\partial_{i}n_{i}\Big) (\tau ,\vec
\sigma ) +\nonumber \\
 &+& O(r_{\bar a}^3) \not= {\hat P}^r_{ADM},\nonumber \\
 &&{}\nonumber \\
 &&where\nonumber \\
 &&{}\nonumber \\
 A(\tau,\vec \sigma) &=& \sqrt{3} \,\sum_{u \bar a}\gamma_{\bar a
u}\int d^3\sigma_1 \frac{\partial^2_{1u}r_{\bar a}(\tau ,{\vec
\sigma}_1)}{4\pi |\vec \sigma -{\vec \sigma}_1|}\nonumber\\
 &&{}\nonumber\\
  f_{i}(\tau, \vec
\sigma) &=& \frac{2}{\sqrt{3}}\sum_{\bar a}\gamma_{\bar a
i}\,r_{\bar a}(\tau, \vec \sigma).
 \label{III12}
 \eea

Therefore, when in a future paper we will add matter to tetrad
gravity, we will have to devise a method independent from the
Landau-Lifschitz pseudo-tensor to identify the energy of the
matter and its variation due to the emission of gravitational
waves.

\vfill\eject

\section{Comparison with the Lichnerowicz-York Conformal Approach,
with the Standard Linearized Theory in Harmonic Gauges and with
the Post-Newtonian Approximation.}

Since in the literature there are many {\it coordinate-dependent}
definitions of {\it gravito-magnetism}, which are a source of
ambiguities, in this Section we shall review same of them and we
will rephrase them in the language of our linearized
post-Minkowskian space-time. Moreover we will show that our
completely fixed Hamiltonian gauge does not belong to the family
of the harmonic gauges used in the standard background-dependent
linearization of Einstein's equations. Finally we will make some
comments on the post-Newtonian approximation of our space-time.

\subsection{Comparison with the Lichnerowicz-York Conformal
Approach.}

To establish the connection with the Lichnerowicz-York conformal
approach \cite{18,19,16} we need the extrinsic curvature of the
WSW hyper-surfaces of our gauge. From Eq.(\ref{a15}), we get that
the {\it extrinsic curvature and the ADM momentum} of the WSW
hyper-surfaces of our gauge in the linearized theory are (see
Eq.(\ref{c15}) of Appendix C for the Fourier transform)

\bea
 {}^3{\hat K}_{rs}(\tau ,\vec \sigma ) &=& \epsilon {{4\pi
G}\over {c^3}}
\sum_u(\delta_{ru}\delta_{(a)s}+\delta_{su}\delta_{(a)r}-
\delta_{rs}\delta_{(a)u})\, {}^3{\check {\tilde \pi}}^u_{(a)}(\tau
,\vec \sigma ) +  O(r^2_{\bar a})=\nonumber \\
 &=& \epsilon \sqrt{3} {{4\pi G}\over {c^3}} \sum_{\bar a} \Big[ 2 \delta_{rs}
 \gamma_{\bar a s} \pi_{\bar a}(\tau ,\vec \sigma ) +\nonumber \\
 &+&  [1-\delta_{rs}]\sum_w\gamma_{\bar aw} [1-2(\delta_{rw}+\delta_{sw})] \nonumber \\
 &&{{\partial^2}\over {\partial (\sigma^w)^2}}  \int^{\infty}_{\sigma^r} d\sigma_1^r
 \int^{\infty}_{\sigma^s} d\sigma_1^s\,\, \pi_{\bar a}(\tau
 ,\sigma_1^r\sigma_1^s\sigma^{k\not= r,s}) \Big] + O(r^2_{\bar a})=\nonumber \\
 &&{}\nonumber \\
 &=& {}^3{\hat K}^{rs}(\tau ,\vec \sigma ) + O(r^2_{\bar a}),\nonumber \\
 &&{}\nonumber \\
 \partial_s\, {}^3{\hat K}^{rs}(\tau ,\vec \sigma ) &=& 0 +O(r^2_{\bar a}),\nonumber \\
 &&{}\nonumber \\
 {}^3{\hat {\tilde \Pi}}^{rs}(\tau ,\vec \sigma ) &=&{1\over 4}
 \Big[ {}^3{\hat e}^s_{(a)}\, {}^3{\check {\tilde
 \pi}}^r_{(a)}+{}^3{\hat e}^r_{(a)}\, {}^3{\check {\tilde
 \pi}}^s_{(a)}\Big] (\tau ,\vec \sigma )={1\over 4}\Big[
 {}^3{\check {\tilde \pi}}^r_{(s)}+{}^3{\check {\tilde
 \pi}}^s_{(r)}\Big] (\tau ,\vec \sigma ) =\nonumber \\
 &=&{{\sqrt{3}}\over 2}\Big[ \delta_{rs} \sum_{\bar a}\gamma_{\bar
 ar}\pi_{\bar a}(\tau ,\vec \sigma )+\nonumber \\
 &+&{1\over 2} (1-\delta_{rs}) \sum_{\bar aw}\gamma_{\bar
 aw}[1-2(\delta_{rw}+\delta_{sw})]\, \partial_w^2
 \int^{\infty}_{\sigma^r}d\sigma_1^r
 \int^{\infty}_{\sigma^s}d\sigma_1^s\, \pi_{\bar a}(\tau
 ,\sigma_1^r \sigma_1^s \sigma^{k\not= r,s})\Big]+\nonumber \\
 &+&O(r^2_{\bar a})=\nonumber \\
 &=&{{\epsilon c^3}\over {16\pi G}} ({}^3{\hat K}^{ra}-\delta^{rs}
\, {}^3\hat K)+O(r^2_{\bar a})={{\epsilon c^3}\over {16\pi G}} \,
{}^3{\hat K}^{rs} +O(r^2_{\bar a}),\nonumber \\
 &&{}\nonumber \\
 &&\Downarrow \nonumber \\
 &&{}\nonumber \\
 {}^3{\hat {\tilde \Pi}}^{rs}{}_{|r} &=& \partial_r\, {}^3{\hat
 {\tilde \Pi}}^{rs} +O(r^2_{\bar a}) =0 +O(r^2_{\bar a}),
 \label{IV1}
 \eea

\noindent since the trace, proportional to {\it York's time}
${\cal T}=-\epsilon {{64\pi G}\over {3c^3}}\, {}^3K$, turns out to
be

 \beq
 {}^3{\hat K}(\tau ,\vec \sigma )= \delta^{rs}\, {}^3{\hat K}_{rs}(\tau
 ,\vec \sigma ) + O(r^2_{\bar a}) = 0 + O(r^2_{\bar a}).
 \label{IV2}
 \eeq

 This means that {\it the linearized WSW hyper-surfaces are constant mean
 extrinsic curvature (CMC) surfaces and satisfy the maximal slicing condition}.
Moreover, the ADM super-momentum constraints are satisfied.

\bigskip

 Therefore, York's traceless {\it distorsion tensor} ${}^3A_{rs}=
 {}^3K_{rs}-{1\over 3}\, {}^3g_{rs}\, {}^3K = {}^3{\hat K}_{rs} +
 O(r^2_{\bar a})$, see Ref.\cite{16},
 Chapter 4.10 and Appendix C of Ref.\cite{30},
 coincides with the extrinsic curvature in the linearized theory. After a
 conformal rescaling ${}^3g_{rs}=\phi^4 \, {}^3{\check g}_{rs}$
 \footnote{All the quantities evaluated with the rescaled metric having
 $det\, |{}^3{\check g}|=1$ will be denoted with a $\check {}\quad$.} we get
 ${}^3{\check A}^{rs}= {}^3{\check A}^{rs}_{TT}+{}^3{\check A}^{rs}_L
 =\phi^{10}\, {}^3A^{rs} = {}^3{\hat K}^{rs}+ O(r^2_{\bar a})$, where
 ${}^3{\check A}^{rs}_{TT}$ and ${}^3{\check A}^{rs}_L$ are the
 {\it transverse traceless} (TT) and {\it longitudinal} (L) components respectively.
 The longitudinal component satisfies ${}^3{\check A}^{rs}_L{}_{|s} =
 {}^3{\check A}^{rs}{}_{|s}\, {\buildrel {def} \over =}\, ({\check \triangle}_L
 W)^r = {\check \triangle} W^r +{1\over 3} (W^s{}_{|s})^{|r} + {}^3{\check R}^r{}_s
 W^s$, where $W^r(\tau ,\vec \sigma )$ is
 {\it York's gravito-magnetic vector potential}.
 In the linearized theory it satisfies $\triangle \Big( \delta^r_s +
 {{\partial^r\partial_s}\over {3\triangle}}\Big) W^s = \partial_s\, {}^3{\hat K}^{rs} +
 O(r^2_{\bar a})=0 +O(r^2_{\bar a})$, whose solution is

\bea
 W^r(\tau ,\vec \sigma ) &=& 0 + O(r^2_{\bar a}),\nonumber \\
 &&{}\nonumber \\
 &&\Downarrow \nonumber \\
 &&{}\nonumber \\
{}^3{\check A}^{rs}_L &=& (LW)^{rs}\, {\buildrel {def} \over =}\,
W^{s|r}+W^{r|s} -{2\over 3}\, {}^3{\hat g}^{rs} W^u{}_{|u}=
[\delta^r_u\partial^s+\delta^s_u\partial^r-{2\over 3}
\delta^{rs}\partial_u] W^u +O(r^2_{\bar a})= \nonumber \\
 &=& 0 + O(r^2_{\bar a}).
  \label{IV3}
 \eea

Therefore, York's transverse traceless physical degrees of freedom
of momentum type, namely ${}^3{\check A}^{rs}_{TT}={}^3{\check
A}^{rs}-{}^3{\check A}^{rs}_L= {}^3{\hat K}^{rs} + O(r^2_{\bar
a})$, are linear functionals of the $\pi_{\bar a}$'s of our
linearized theory. The coordinate-type physical degrees of freedom
in ${}^3{\check g}_{rs}$ depend upon the $r_{\bar a}$'s
\footnote{In the 3-orthogonal gauge we have ${}^3{\check
g}_{rs}={}^3{\hat g}^{diag}_{rs}$, see Eq.(\ref{a17}).}.

\subsection{Comparison with the Standard Linearized Theory in
Harmonic Gauges and its Associated Gravito-Electro-Magnetic
Analogy.}

Let us remark that our special 3-orthogonal gauge is not a member
of the family of harmonic gauges, because, in coordinates adapted
to our gauge, the condition $\partial_A\, \Big( \sqrt{{}^4g}\,
{}^4g^{AB}\Big) = 0$ becomes

\bea
 &&\Big[\partial_A\, \Big( \sqrt{\hat \gamma}\, {}^4{\hat g}^{\tau
 A}\Big)\Big] (\tau ,\vec \sigma ) = \Big[\partial_{\tau}\,
 \Big( \sqrt{\hat \gamma} - 1\Big) - \sum_s \partial_s\,
 n_s\Big] (\tau ,\vec \sigma )  + O(r^2_{\bar a}) = 0  + O(r^2_{\bar
 a}), \nonumber \\
 &&{}\nonumber \\
 &&\Big[ \partial_A\, \Big( \sqrt{\hat \gamma}\, {}^4{\hat
 g}^{rA}\Big)\Big] (\tau ,\vec \sigma ) = \Big[ \partial_{\tau}\,
 n_r - \partial_r\, \Big(
{2\over {\sqrt{3}}}\sum_{\bar a}[\gamma_{\bar ar}r_{\bar a})(\tau
,\vec \sigma ) +\nonumber \\
 &+& {1\over 2}\sum_u\gamma_{\bar au} \int
d^3\sigma_1 {{\partial^2_{1u} r_{\bar a}(\tau ,{\vec
\sigma}_1)}\over {4\pi |\vec \sigma -{\vec \sigma}_1|}}\Big) \Big]
(\tau ,\vec \sigma ) + O(r^2_{\bar a}) \not= 0 + O(r^2_{\bar a}).
 \label{IV4}
 \eea

Therefore, on the solutions of Einstein's equations \cite{15} our
radiation gauge identifies a 4-coordinate system different from
those compatible with the standard linearized theory coupled to
matter.
\medskip

This theory, in the post-Newtonian approximation, allows to make
the {\it gravito-electro-magnetic} (GEM) re-formulation of
Einstein's theory (see Refs.\cite{16,31}), emphasizing the spin 1
aspects of this spin 2 theory over the Minkowski background in
Cartesian coordinates.

By putting ${}^4g_{\mu\nu} = {}^4\eta_{\mu\nu} + h_{\mu\nu}$ with
$h_{\mu\nu} = {\bar h}_{\mu\nu} + {1\over 2}\, {}^4\eta_{\mu\nu}\,
h$ (${}h = {}^4\eta^{\mu\nu}\, h_{\mu\nu}$, $\sqrt{{}^4g} \,
{}^4g^{\mu\nu} = {}^4\eta^{\mu\nu} - {\bar h}^{\mu\nu}$, ${\bar
h}^{\mu\nu} = {}^4\eta^{\mu\alpha}\, {}^4\eta^{\nu\beta}\,
h_{\alpha\beta}$) Einstein's equations in presence of matter
become $\Box\, {\bar h}_{\mu\nu} = {}^4\eta^{\alpha\beta}\,
\partial_{\alpha}\, \partial_{\beta}\, {\bar h}_{\mu\nu}\, {\buildrel
\circ \over =}\, - {1\over {k\, c}}\, T_{\mu\nu}$ when the {\it
Lorentz gauge condition} $\partial_{\nu}\, {\bar h}^{\mu\nu} =
\partial_{\nu}\, \Big( \sqrt{{}^4g}\, {}^4g^{\mu\nu} \Big) = 0$, viz.
{\it the conditions for harmonic coordinates (which include the
post-Newtonian ones)}, are imposed. Even if in this approximation
there is no back-reaction of the gravitational force on the
sources, it allows to make an analogy with electromagnetism
\cite{23}. ${\bar h}_{\mu\nu}$ replaces the 4-potential $A_{\mu}$
with the coordinate transformation (changing the Christoffel
symbols but not the curvature tensor) ${\bar h}^{'}_{\mu\nu} =
{\bar h}_{\mu\nu} -
\partial_{\mu}\, b_{\nu} - \partial_{\nu}\, b_{\mu} + {}^4\eta_{\mu\nu}
\, \partial_{\alpha}\, b^{\alpha}$ corresponding to the gauge
transformation $A^{'}_{\mu} = A_{\mu} + \partial_{\mu}\, \Lambda$.
Maxwell equations $\Box\, A_{\mu} -
\partial_{\mu}\, \partial_{\alpha}\, A^{\alpha}\, {\buildrel \circ \over
=}\, - {1\over c}\, j_{\mu}$ correspond to the linearized Einstein
equations $\Box\, {\bar h}_{\mu\nu} +\, {}^4\eta_{\mu\nu}\,
\partial_{\alpha}\, \partial_{\beta}\, {\bar h}^{\alpha\beta} -
\partial_{\mu}\, \partial_{\alpha}\, {\bar
h}^{\alpha}{}_{\nu} - \partial_{\nu}\, \partial_{\alpha}\, {\bar
h}^{\alpha}{}_{\mu}\, {\buildrel \circ \over =}\, - {2\over k}\,
T_{\mu\nu}$. The {\it Lorentz gauge} $\partial_{\mu}\, A^{\mu} =
0$ (with residual gauge freedom given by functions $\Lambda$'s
satisfying $\Box\, \Lambda = 0$) with equations $\Box\, A_{\mu}\,
{\buildrel \circ \over =}\, - {1\over c}\, j_{\mu}$ corresponds to
the {\it harmonic gauge (harmonic coordinates)} $\partial_{\nu}\,
{\bar h}^{\mu\nu} = 0$ with $\Box\, {\bar h}_{\mu\nu} = - {2\over
k}\, T_{\mu\nu}$ (with residual freedom in the choice of
coordinates having the $b_{\mu}$'s satisfying $\Box\, b^{\mu} =
0$; in Cartesian coordinates the equations of motion are decoupled
one from the other).
\bigskip

The retarded solution ${\bar h}_{\mu\nu}(x^o, \vec x)\, {\buildrel
\circ \over =}\, {{4\, G}\over {c^4}}\, \int d^3x^{'}\,
{{T_{\mu\nu}(x^o - |\vec x - {\vec x}^{'}|, {\vec x}^{'})}\over
{|\vec x - {\vec x}^{'}|}}$ allows to define:\bigskip

i) the {\it gravito-electric} or Newton potential ${\bar h}_{oo} =
- \epsilon\, {4\over {c^2}}\, \Phi_{GEM}$ determined by $\rho =
T^{oo}/c^2$ as the effective gravitational charge density;

ii) the {\it gravito-magnetic vector potential} (the shift
functions) of Ref.\cite{16,31} ${\bar h}_{oi} = \epsilon\, {2\over
{c^2}}\, A_{GEM\, i}$  determined by $j^i = T^{oi}/c$ as the
effective gravitational current density;

iii) to disregard at the lowest order  the 3-metric ${\bar h}_{ij}
= O(c^{-4})$.

\bigskip

As shown in Ref.\cite{31}, in the harmonic gauge the {\it
gravito-electro-magnetic 4-potential} $(\Phi_{GEM}, {\vec
A}_{GEM})$ {\it satisfies the Lorentz gauge condition}

\beq
 \partial_o\Phi_{GEM}+\vec
\partial \cdot ({1\over 2} {\vec A}_{GEM})=0.
 \label{IV5}
 \eeq

\noindent The factor ${1\over 2}$ derives from the fact that the
effective gravito-magnetic charge ($Q_B=2M_{source}$) is twice the
gravito-electric charge ($Q_E=M_{source}$). The {\it
gravito-electro-magnetic fields} are ${\vec E}_{GEM} =-\vec
\partial \Phi_{GEM}-\partial_o({1\over 2}{\vec A}_{GEM})$, ${\vec
B}_{GEM} = \vec \partial\, \times\, {\vec A}_{GEM}\, {\buildrel
{def}\over =}\, c\, {\vec \Omega}_{GEM}$: in the harmonic gauge
(harmonic coordinates) they satisfy Maxwell equations

\bea
 &&\vec \partial \cdot {\vec E}_{GEM}=4\pi G\rho,\qquad
 \vec \partial \cdot ({1\over 2}{\vec B}_{GEM})=0,\nonumber \\
 && \vec \partial \times
{\vec E}_{GEM}=-\partial_o ({1\over 2}{\vec B}_{GEM}),\qquad \vec
\partial \times ({1\over 2}{\vec B}_{GEM}) =\partial_o{\vec
E}_{GEM}+{{4\pi G}\over c} \vec j,
 \label{IV6}
 \eea

\noindent and one can consider as electro-magnetic-like gauge
transformations the residual coordinate freedom.
\bigskip

In this approximation the GEM fields are determined {\it only by
the matter} and vanish in absence of matter, viz. they are
independent from the physical degrees of freedom (like $r_{\bar
a}$, $\pi_{\bar a}$) of the gravitational field.

If we consider solutions of the homogeneous equations $\Box {\bar
h}_{\mu\nu}\, {\buildrel \circ \over =}\, 0$, $\partial_{\nu}{\bar
h}^{\mu\nu}=0$, they are used to describe {\it gravitational
waves} ${\bar h}_{\mu\nu}=a_{\mu\nu}e^{ik_{\alpha}x^{\alpha}}$
($a_{\mu\nu}=const.$, $k_{\mu}k^{\mu}=0$, $a_{\mu\nu}k^{\nu}=0$)
as the only independent degrees of freedom of the gravitational
field. The residual freedom of coordinate transformations in the
harmonic gauge may be used  to get $a^{'}_{\mu\nu} = a_{\mu\nu} -
c_{\mu}\, k_{\nu} - c_{\nu}\, k_{\mu} + {}^4\eta_{\mu\nu}\,
c_{\alpha}\, c^{\alpha}$   with $a^{'}_{o\nu} = a^{'}_{\mu o} = 0$
and $a^{'}_{\mu}{}^{\mu} = a^{'}_i{}^i = 0$ through the choice
$b^{\mu} = - i\, c^{\mu}\, e^{ik_{\alpha}x^{\alpha}}$ (solution of
$\Box\, b^{\mu} = 0$). In {\it this completely fixed harmonic
gauge (with non-diagonal 3-metric), named TT gauge,} only two
independent degrees of freedom (polarization states) survive: they
are the counterpart of $r_{\bar a}$, $\pi_{\bar a}$ of our
completely fixed 3-orthogonal radiation gauge with $\pi_{\phi}
(\tau ,\vec \sigma ) = 0$. If $k^{\mu} = (0; 0, {{\omega}\over c},
{{\omega}\over c})$, the only non-zero $a^{'}_{\mu\nu}$ of the
transverse gravitational wave are $a^{'}_{xx}$, $a^{'}_{yy} = -
a^{'}_{xx}$, $a^{'}_{xy} = a^{'}_{yx}$, and we get $ds^2 =
\epsilon\, \Big[ (dx^o)^2 - (1 + f_{xx})\, dx^2 - 2\, f_{xy}\,
dxdy - (1 - f_{xx})\, dy^2 - dz^2 \Big]$ with $e_{+} = f_{xx} =
a^{'}_{xx} \, cos\, ({{\omega}\over c}\,(z-x^o) + \varphi )$,
$e_{\times} = f_{xy} = a^{'}_{xy}\, cos\, ({{\omega}\over c}\,
(z-x^o) + \psi )$. In this gauge the lapse and shift functions
$n$, $n_i$ are all {\it zero} and there is {\it no GEM potential
$(\Phi_{GEM}, {\vec A}_{GEM})$ coming from the transverse
gravitational wave}.

\medskip

Let us remark that in the electro-magnetic case on Minkowski
space-time with Cartesian coordinates the radiation gauge $A_o =
\vec \partial \cdot \vec A = 0$, with the transverse fields ${\vec
A}_{\perp}$, ${\vec E}_{\perp}$ as the only physical degrees of
freedom (DO's), is a particular case (name it {\it gauge T}) of
the Lorentz gauge $\partial^{\mu}\, A_{\mu} = 0$, obtainable with
a $\Lambda = - i\, c\, e^{i\, k \cdot x}$ with $k^2=0$ from $\Box
\, \Lambda = 0$. Instead, the 4-coordinate system on the
post-Minkowskian space-times connected with our radiation gauge
for the gravitational waves is different from the 4-coordinate
system which is reinterpreted as the TT harmonic gauge of the
spin-2 theory on the background Minkowski space-time with
Cartesian coordinates.

\bigskip

Instead in our non-harmonic radiation 3-orthogonal gauge with
$\pi_{\phi}(\tau ,\vec \sigma ) = 0$, even in absence of matter
the shift functions do {\it not} vanish and we can define\bigskip

i) a {\it vanishing gravito-electric potential}
$\Phi_{GEM}^{(rad)}(\tau ,\vec \sigma ) = 0 + O(r^2_{\bar a})$;

ii) a {\it gravito-magnetic vector potential} ${\vec
A}^{(rad)}_{GEM}(\tau ,\vec \sigma ) = \{ - \epsilon\, {{c^2}\over
2}\, n_r(\tau ,\vec \sigma )\}$ determined by the shift functions;

iii) a {\it gravito-electric field} ${\vec E}^{(rad)}_{GEM}(\tau
,\vec \sigma ) = - \partial_{\tau}\, [{1\over 2}\, {\vec
A}_{GEM}^{(rad)}(\tau ,\vec \sigma )] = - \epsilon\, {{c^2}\over
4}\, \partial_{\tau}\, n_r(\tau ,\vec \sigma )$ {\it also}
determined by the shift functions.

iv) a {\it gravito-magnetic field} ${\vec B}^{(rad)}_{GEM}(\tau
,\vec \sigma ) = \vec \partial \times {\vec A}_{GEM}^{(rad)}(\tau
,\vec \sigma )\, {\buildrel {def} \over =}\, c\, {\vec
\Omega}_{GEM}^{(rad)}(\tau ,\vec \sigma )$, with ${\vec
\Omega}_{GEM}^{(rad)}(\tau ,\vec \sigma ) = \{ \Omega_{GEM,
r}^{(rad)} = c^{-1}\, \epsilon_{ruv}\, {{\partial n_v}\over
{\partial \sigma^u}}\, \}$ the gravito-magnetic precession angular
velocity (connected with the precessional effects of
Lense-Thirring or dragging of the inertial systems) again
determined by the shift functions.
\bigskip

Since we are not in a harmonic gauge the {\it Lorentz condition is
not satisfied} $\vec \partial \cdot ({1\over 2}\, {\vec
A}_{GEM}^{(rad)})\not= 0$. {\it The analogy with Maxwell equations
is partially lost in our completely fixed (radiation) gauge}:
while the equations deriving from the existence of the potential
$\vec \partial \cdot ({1\over 2}\, {\vec B}_{GEM}^{(rad)})=0$ and
$\vec \partial \times {\vec E}_{GEM}^{(rad)} = -
\partial_{\tau}\, ({1\over 2}\, {\vec B}_{GEM}^{(rad)})$ hold, the other two
are not satisfied: $\vec \partial \cdot {\vec E}_{GEM}^{(rad)}\,
\not= 0$, $\vec \partial \times ({1\over 2}\, {\vec
B}_{GEM}^{(rad)}) \not= \partial_{\tau}\, {\vec E}_{GEM}^{(rad)}$.
This shows how the gravito-electric-magnetic analogy is {\it
coordinate dependent}.

\bigskip

Finally, the coordinate transformation (a passive
4-diffeomorphism) from the $\Sigma^{(WSW)}_{\tau}$-adapted
4-coordinates $\tau , \vec \sigma$ of our completely fixed gauge
to the 4-coordinates $x^{\mu} = x^{\mu} (\tau ,\vec \sigma )$ of
the standard harmonic TT gauge and the relation between the DO's
$r_{\bar a}(\tau ,\vec \sigma )$ and the TT polarizations $e_{+}$,
$e_{\times}$ can be obtained as solution of the following system
of partial differential equations:

\beq
 {}^4g_{\mu\nu}(x) = {{\partial x_{\mu}}\over {\partial
 \sigma_A}}\, {{\partial x_{\nu}}\over {\partial \sigma_B}}\,
 {}^4g_{AB}(\tau ,\vec \sigma ),\qquad
 {}^4g_{\mu\nu} = \left( \begin{array}{cccc} \epsilon & 0 & 0 &
 0\\ 0 & -\epsilon\, e_{+} & - \epsilon\, e_{\times} & 0\\
 0 & - \epsilon\, e_{\times} & + \epsilon\, e_{+} & 0\\
 0 & 0 & 0 & - \epsilon \end{array} \right),
 \label{IV7}
 \eeq

\noindent with ${}^4g_{AB}$ given by Eqs.(\ref{III7}).

If we write $x_{\mu}(\tau ,\vec \sigma ) = f_{\mu}(\tau ,\vec
\sigma ) + g_{\mu}(\tau ,\vec \sigma )$ with $f_{\mu} = O(1)$ and
$g_{\mu} = O(r_{\bar a})$ and if we require the asymptotic
behaviour $f_{\mu}(\tau ,\vec \sigma )\, {\rightarrow}_{|\vec
\sigma | \rightarrow \infty}\, x_{(\infty ) \mu}(0) +
u_{\mu}(P_{ADM})\, \tau + \epsilon_{r\, \mu}(u(P_{ADM}))\,
\sigma^r$ with $u^{\mu}(p) = p^{\mu}/\sqrt{\epsilon\, p^2}$, then
we get that
\bigskip

i) the four equations with ${}^4g_{oo} = \epsilon$, ${}^4g_{oi} =
0$ can be solved to give ${{\partial x_{\mu}}\over {\partial
\tau}}$ in terms of ${{\partial x_{\mu}}\over {\partial
\sigma^r}}$;

ii)  the system of partial differential equations${}^4g_{13} =
{}^4g_{23} = 0$, ${}^4g_{33} = - \epsilon$ and ${}^4g_{11} = -
{}^4g_{22}$ have to be used for the determination of the $\vec
\sigma$-dependence of the four functions $x_{\mu}(\tau ,\vec
\sigma )$;

iii) the use of the solution of point ii) inside point i) allows
to find ${{\partial x_{\mu}}\over {\partial \tau}}$ and then, by
integration, the functions $x_{\mu}(\tau ,\vec \sigma )$;

iv) finally the equations $e_{+} = - \epsilon\, {}^4g_{11}$ and
$e_{\times} = - \epsilon\, {}^4g_{12}$ give the two polarizations
$e_{+}$, $e_{\times}$ in terms of the Dirac observables $r_{\bar
a}(\tau ,\vec \sigma )$.

\subsection{Connection with the Post-Newtonian Approximation.}

The standard post-Newtonian approximation \cite{32} is applied to
the gravitational field created by an isolated compact object,
like the Earth, described by an energy-momentum tensor
$T^{\mu\nu}$ [$T^{oo} = O(c^2)$, $T^{oi} = O(c)$, $T^{ij} =
O(c^o)$]. It is a {\it weak field} (${}^4g = {}^4\eta + {}^4h$)
{\it near zone approximation} giving corrections to Newton
gravity.

\medskip

Following Ref.\cite{33}, if $V = {{G\, M}\over R}$ is the Newton
potential of the compact object (of mass $M$ and radial dimensions
$R_o \leq R$), the post-Newtonian (PN) expansion of the 4-metric
${}^4g_{\mu\nu}$ is a series in the dimensionless parameter $\zeta
\approx {{\sqrt{V}}\over {c^2}} \approx {R\over {c\, T}}$ ($T$ is
a characteristic time of variation of the source). The 1PN
approximation keeps the following terms in ${}^4g_{\mu\nu}$:
${}^4g_{oo} = \epsilon + {{a_{oo}}\over {c^2}} + {{b_{oo}}\over
{c^4}}$, ${}^4g_{oi} = {{a_{oi}}\over {c^3}}$, ${}^4g_{ij} = -
\epsilon \delta_{ij} + {{a_{ij}}\over {c^2}}$.

In Ref.\cite{32} the PN approximation was given in a {\it PN
4-coordinate system} (adiabatic approximation of a
Robertson-Walker line element appropriate to a homogeneous
isotropic cosmological model containing the compact object),
defined as a {\it local, quasi-Cartesian} 4-coordinate system at
rest respect to the universe rest frame and in a {\it standard PN
gauge} where the 4-metric has the form (we use the notation of
$\Sigma^{(WSW)}_{\tau}$ -adapted coordinates) \footnote{In
Refs.\cite{34,33} there are various definitions of 4-coordinate
systems. The {\it standard PN gauge} is said to correspond to the
conditions $\partial_s\, {}^4g_{\tau s} - {1\over 2}\,
\partial_{\tau}\, {}^4g_{ss} = O(c^{-5})$, $\partial_s\,
{}^4g_{rs} - {1\over 2}\, ({}^4g_{ss} - {}^4g_{\tau\tau}) =
O(c^{-4})$. Instead the {\it algebraic spatial isotropy condition}
of Ref.\cite{33}, $-{}^4g_{\bar \tau \bar \tau}\, {}^4{\tilde
g}_{rs}=\delta_{rs}+O(c^{-4})$, contains both the harmonic and the
standard post-Newtonian gauges. Finally a {\it ADM Hamiltonian
gauge}, needed to include 2.5PN gravitational radiation reaction,
is ${}^3K = 0$, $\partial_s\, {}^3g_{rs} - {1\over 3}\,
\partial_r\, {}^3g_{ss} = 0$. }

\bea
 {}^4g_{\tau\tau}  &=&\epsilon e^{-{{2\epsilon}\over {c^2}} V}
 = \epsilon [1-{{2\epsilon}\over {c^2}} V
+O(c^{-4})], \nonumber \\
 &&{}\nonumber \\
 {}^4g_{\tau r} &=& - {{4\epsilon}\over {c^2}}\, V_r +
 O(c^{-4}),\nonumber \\
 &&{}\nonumber \\
 {}^4g_{rs} &=& - \epsilon \, {}^3g_{rs} = - \epsilon \,
 \delta_{rs}\, (1 + {{U}\over {c^2}}) + O(c^{-4}),
 \label{IV8}
 \eea

\noindent with $U = V$ in the standard PN gauge. Here $V$ is the
Newton potential generated by the compact object.

Now our 4-coordinate system, in our completely fixed Hamiltonian
gauge, is by definition in the rest frame of the universe
(rest-frame instant form). In presence of matter, the
linearization condition (as we will show in a future paper on
tetrad gravity plus a perfect fluid, where it will be shown that
both the lapse and shift functions depend on matter so that $ n
\not= 0$)) requires  $T^{AB} = O(r_{\bar a})$ and we have

\bea
 {}^4g_{\tau\tau} &=& \epsilon\, [(-\epsilon + n)^2 - {}^3g_{rs}\,
 n^rn^s] = \epsilon (1 - 2\epsilon\, n[matter]) +O(r^2_{\bar a}),\nonumber
 \\
 &&{}\nonumber \\
 {}^4g_{\tau r} &=& - \epsilon\, {}^3g_{rs}\, n^s = - \epsilon\,
 n_r[r_{\bar a}, matter] + O(r^2_{\bar a}),\nonumber \\
 &&{}\nonumber \\
 {}^4g_{rs} &=& - \epsilon\, {}^3g_{rs} = -\epsilon\,
 \delta_{rs}\, (1 + k_r[r_{\bar a},matter]) + O(r^2_{\bar
 a}),\nonumber \\
 &&{}\nonumber \\
 \phi &=& [det\, {}^3g]^{1/12} = (1 + {U\over {4c^2}}) +
 O(r^2_{\bar a}).
 \label{IV9}
 \eea

\noindent Therefore, in our gauge  the {\it Newton potential is
the lapse function}, $V = c^2\, n[matter]$ (and the equation
determining the lapse tends to the Poisson equation), the {\it
gravito-magnetic potentials are the shift functions}, $V_r =
4c^2\, n_r[r_{\bar a},matter]$, and the {\it Lichnerowicz equation
for the conformal factor $\phi$ amounts to a determination of
$U$}, which may not coincide with the Newton potential like it
happens in the standard PN gauge. By solving the linearized
equations for the DO's (see next Section), $\Box\, r_{\bar a} =
Z_{\bar a}[matter]$, i.e. by considering their Lienard-Wiechert
retarded solution without incoming free radiation, we can recover
a form like the one of Eqs.(\ref{IV8}), which does not depend
explicitly on the DO's $r_{\bar a}$ \footnote{Let us also remark
that strictly speaking to reach Newton gravity one usually
performs a double limit: i) the zero-curvature limit $r_{\bar a},
\pi_{\bar a} \rightarrow 0$, eliminating the genuine degrees of
freedom of the gravitational field, which do not exist in
Newtonian gravity; ii) the $1/c$ expansion. However the zero
curvature limit is not consistent with Einstein's equations with
matter and must be replaced (in a linearized theory) with the
restriction of the DO's $r_{\bar a}$, $\pi_{\bar a}$ to the
Lienard-Wiechert solution. Then the $1/c$ expansion will kill the
curvature for $c \rightarrow \infty$.}.

\bigskip

Let us now consider the relation of our linearized radiation gauge
with Ref.\cite{35}, where the Galileo generally covariant
formulation of Newtonian gravity, depending on 27 fields, was
obtained as a limit $c\, \rightarrow \, \infty$ on the ADM action
of metric gravity.

The final action of Ref.\cite{35} with general Galileo covariance
depends on the 26 fields $A_o$, $\alpha_o$, $A_r$, $\alpha_r$,
${}^3{\check g}_{rs}$, ${\check \gamma}_{rs}$, ${\check
\beta}_{rs}$ (after having put the field $\Theta =1$ by rescaling
the absolute time) . There are 18 first class constraints and 8
pairs of second class ones. It turns out that $\alpha_o$, $A_r$,
$\alpha_r$, three components of ${}^3{\check g}_{rs}$, one
component of the momentum conjugate to ${}^3{\check g}_{rs}$, the
trace ${\check \beta}^T$ and the longitudinal ${\check \beta}^L_r$
parts of ${\check \beta}_{rs}$ in its TT decomposition, and the
longitudinal ${\check \gamma}^L_r$ part of ${\check \gamma}_{rs}$
are Hamiltonian gauge variables, while $A_o$ (the Newton
potential) and the remaining components of ${}^3{\check g}_{rs}$,
${\check \gamma}_{rs}$, ${\check \beta}_{rs}$ are determined,
together with their conjugate momenta, by the second class
constraints. There are no propagating dynamical degrees of
freedom. The gauge variables describe the inertial forces in
arbitrary accelerated non-Galilean reference frames.

To make a comparison with the results of Sections 6 and 7 of
Ref.\cite{35} the starting point is the following parametrization
of the 4-metric of Eq.(\ref{III7}) (we show only the 26 terms
which appear in the Newtonian action)

\begin{eqnarray}
 {}^4g_{\tau\tau}&=&
\epsilon \, [1-{{2A_o}\over {c^2}}+{{2\alpha_o}\over
{c^4}}+O(c^{-6})]= \nonumber \\
 &=& \epsilon\, (1 - 2 \epsilon\, n) + O(r^2_{\bar a}), \nonumber \\
 &&{}\nonumber \\
  {}^4g_{\tau r}&=& \epsilon \, [A_r + {{\alpha_r}\over {c^2}}+O(c^{-4})]=-\epsilon\,
n_r + O(r^2_{\bar a}),\nonumber \\
 &&{}\nonumber \\
  {}^4g_{rs}&=&-\epsilon \,
{}^3g_{rs} =-\epsilon [{}^3{\check g}_{rs}+{{{\check
\gamma}_{rs}}\over {c^2}}+{{{\check \beta}_{rs}}\over
{c^4}}+O(c^{-6})]= \nonumber \\
 &=& -\epsilon \, \delta_{rs}\, (1 +
k_r) + O(r^2_{\bar a}),\nonumber \\
 &&{}\nonumber \\
  &&\Downarrow \nonumber \\
   &&{}\nonumber \\
   n[matter] &=& \epsilon\, [{{A_o}\over {c^2}} + {{\alpha_o}\over
   {c^4}} + O(c^{-6}, r^2_{\bar a})],\nonumber \\\
 n_r[r_{\bar a}, matter] &=& - A_r - {1\over {c^2}} \alpha_r
+O(c^{-4}, r^2_{\bar a}),\nonumber \\
 \delta_{rs}\, (1 + k_r[r_{\bar a}, matter]) &=& {}^3{\check g}_{rs}
 +{{{\check \gamma}_{rs}}\over {c^2}} + {{{\check
 \beta}_{rs}}\over {c^4}} + O(c^{-6}, r^2_{\bar a}).
 \label{IV10}
 \end{eqnarray}

Instead the PN approximation (\ref{IV8}) implies

\beq
 A_o = \epsilon V,\quad A_r=0,\quad \alpha_r = -4 V_r,\quad {}^3{\check
g}_{rs} = \delta_{rs},\quad {\check \gamma}_{rs}= U\, \delta_{rs},
 \label{IV11}
 \eeq

\noindent which are consistent with the  gauge freedom of Ref.
\cite{35} (it is a possible gauge of the general Galileo covariant
description of Newtonian gravity).

\vfill\eject

\section{The Hamilton Equations and their Solution.}

In this Section we shall study the Hamilton equations in the
preferred 3-orthogonal gauge and we will verify that their
solution produces a linearized solution of Einstein's equations.
Since, as said in the Introduction, the admissible WSW
hyper-surfaces must admit a generalized Fourier transform, we
shall use it on our linearized WSW hyper-surfaces (where it
coincides with the ordinary Fourier transform on $R^3$ at the
lowest order) to study the Hamilton equations.

\bigskip

As said in the Introduction, in the rest frame instant form of
dynamics on WSW hyper-surfaces the Dirac Hamiltonian is $H_D =
{\hat E}_{ADM} + \vec \lambda (\tau ) \cdot {\hat {\vec
P}}_{ADM}$. However we can add the gauge fixings ${\hat J}^{\tau
r}_{ADM} \approx 0$ [i.e. the vanishing of the weak ADM boost
generators of Eq.(\ref{a12})] to the three first class constraints
${\hat P}^r_{ADM} \approx 0$ with the consequence $\vec \lambda
(\tau ) = 0$. This means that we have to eliminate the {\it
internal 3-center of mass} of the universe (it is put in the
centroid $x^{\mu}_{(\infty )}(\tau )$ near spatial infinity used
as origin of the 3-coordinates on WSW hyper-surfaces), i.e. 3
pairs of global degrees of freedom among the DO's $r_{\bar a}(\tau
,\vec \sigma )$, $\pi_{\bar a}(\tau ,\vec \sigma )$, reducing them
to canonical variables relative to the internal 3-center of mass.

Like in the case of Klein-Gordon and electro-magnetic fields and
like for every isolated system treated in the rest frame instant
form \cite{5}, we should find the canonical transformation from
the canonical variables $r_{\bar a}(\tau ,\vec \sigma )$,
$\pi_{\bar a}(\tau ,\vec \sigma )$ to a canonical basis containing
the canonical internal 3-center of mass $Q^r_{ADM}$, weakly equal
to the M$\o$ller internal 3-center of energy $R^r_{ADM} = - {\hat
J}^{\tau r}_{ADM}/{\hat E}_{ADM}$, ${\hat P}^r_{ADM} (\approx 0)$
as the conjugate momentum, and the {\it internal relative
variables} $R_{\bar a}$, $\Pi_{\bar a}$. Usually \cite{36,37},
like in the Klein-Gordon case, we should start with a naive
3-center of mass $X^r_{ADM}$, conjugate to ${\hat P}^r_{ADM}$, we
should find the canonical variables $R^{'}_{\bar a}$,
$\Pi^{'}_{\bar a}$ relative to it and then we should use the
Gartenhaus-Schwartz transformation to find \cite{36} the canonical
variables with respect to $Q^r_{ADM}$. Then we should add the
gauge fixings $Q^r_{ADM} \approx R^r_{ADM} = - {\hat J}^{\tau
r}_{ADM}/{\hat E}_{ADM} \approx 0$, which put the internal
3-centers in the centroid $x^{\mu}_{(\infty )}$ origin of the
3-coordinates on the WSW hyper-surfaces and implies $H_D = {\hat
E}_{ADM}$, to the rest-frame condition ${\hat P}^r_{ADM} \approx
0$ and we should go to Dirac brackets with the result $R_{\bar a}
\equiv R^{'}_{\bar a}$, $\Pi_{\bar a} \equiv \Pi^{'}_{\bar a}$.
Presumably also the final relative variables have to satisfy
Eqs.(\ref{c2}).

\bigskip

However, since also the determination of the naive 3-center of
mass $X^r_{ADM}$ of gravity is not trivial, we will fix this final
gauge freedom simply by putting equal to zero the Dirac
multipliers, $\vec \lambda (\tau ) = 0$ as a pre-gauge fixing
condition compatible with ${\hat J}^{\tau r}_{ADM} \approx 0$.

\bigskip

We can now  study the Hamilton equations for the independent
canonical degrees of freedom $r_{\bar a}(\tau ,\vec \sigma )$,
$\pi_{\bar a}(\tau ,\vec \sigma )$ of the gravitational field,
only restricted by Eqs.(\ref{II15}), in our completely fixed
3-orthogonal gauge generated by the Hamiltonian $ {\hat E}_{ADM}$
of Eq.(\ref{III1}). These Hamilton equations replace the equations
$\Box {\bar h}_{\mu\nu}=0$, $\partial_{\nu} {\bar h}^{\mu\nu}=0$
of the standard linearized theory in the harmonic gauge
coordinates.

\subsection{The First Half of Hamilton Equations.}

The first half of the Hamilton equations associated with the
Hamiltonian (\ref{III1}) yield the following expression for the
velocities $\partial_{\tau} r_{\bar a}(\tau ,\vec \sigma )$

\bea
\partial_{\tau}r_{\bar a}(\tau ,\vec \sigma ) &{\buildrel \circ \over =}&
\{ r_{\bar a}(\tau ,\vec \sigma ), {\hat E}_{ADM}  \} = {{24\pi
G}\over {c^3}} \pi_{\bar a}(\tau ,\vec \sigma )+\nonumber
\\
 &+&{{24\pi G}\over {c^3}}
\sum_{u,r,\bar b,(a)} \delta_{(a)u} \Big[ \gamma_{\bar
au}\gamma_{\bar br}\int d^3\sigma_1 {\cal T}^{(o) u}_{(a)r}(\vec
\sigma ,{\vec \sigma}_1)\pi_{\bar b}(\tau ,{\vec
\sigma}_1)+\nonumber \\
 &+& \gamma_{\bar ar}\gamma_{\bar bu} \int d^3\sigma_1 \pi_{\bar b}(\tau ,{\vec \sigma}_1)
 {\cal T}^{(o)u}_{(a)r}({\vec \sigma}_1,\vec \sigma )\Big] +\nonumber \\
 &+& {{12\pi G}\over {c^3}} \sum_{u,r,s,\bar b} \gamma_{\bar ar}\gamma_{\bar bs}
 \Big[ \sum_{(a)}\int d^3\sigma_1 {\cal T}^{(o)u}_{(a)r}(\vec \sigma ,{\vec \sigma}_1)
 \int d^3\sigma_2 {\cal T}^{(o)u}_{(a)s}({\vec \sigma}_1,{\vec \sigma}_2)
 +\nonumber \\
 &+&\sum_{v(a)(b)}(\delta_{(a)v}\delta_{(b)u}-\delta_{(a)u}\delta_{(b)v}) \int d^3\sigma_1
 {\cal T}^{(o)u}_{(a)r}(\vec \sigma ,{\vec \sigma}_1) \int d^3\sigma_2
 {\cal T}^{(o)v}_{(b)s}({\vec \sigma}_1,{\vec \sigma}_2)\Big] \pi_{\bar b}(\tau
 ,{\vec \sigma}_2)+\nonumber \\
 &+& {1\over L} O(r^2_{\bar a}).
\label{V1}
 \eea

By using the expression of ${\cal T}^{(o)u}_{(a)r}$ given in the
last two lines of Eqs.(\ref{II13}) it can be checked that the two
terms linear in the ${\cal T}$'s vanish due to a factor
$\delta_{(a)u}(1-\delta_{(a)u})$. After some calculations it turns
out that the two terms bilinear in the ${\cal T}$'s give the same
result, so that the final expression for the velocities is

\bea
\partial_{\tau}r_{\bar a}(\tau ,\vec \sigma ) &{\buildrel \circ \over =}&
 {{24\pi G}\over {c^3}} \pi_{\bar a}(\tau ,\vec \sigma )+
{{6\pi G}\over {c^3}} \sum_{\bar b rs}\gamma_{\bar
 ar}\gamma_{\bar bs} \sum_{uv,u\not=
  v}[1-2(\delta_{ur}+\delta_{vr})][1-2(\delta_{us}+\delta_{vs})]\nonumber \\
 &&\int^{\sigma^u}_{-\infty}d\sigma_1^u \int^{\sigma_1^u}_{-\infty}
 d\sigma_2^u \int^{\infty}_{\sigma^v} d\sigma_1^v
 \int^{\infty}_{\sigma_1^v} d\sigma_2^v {{\partial^4 \pi_{\bar
 b}(\tau, \sigma_2^u \sigma_2^v \sigma^{k\not= u,v})}\over
 {(\partial \sigma_2^r)^2(\partial \sigma_2^s)^2}}
 +{1\over L} O(r^2_{\bar a}).
 \label{V2}
 \eea

In obtaining this result we have made integrations by parts
justified by the asymptotic vanishing of $\pi_{\bar a}(\tau ,\vec
\sigma )$ and we have used $\partial_x\theta
(x,y)=-\partial_y\theta (x,y)$, $\partial_x\delta
(x,y)=-\partial_y\delta (x,y)$ assumed valid on linearized
conformally flat WSW hyper-surfaces.\bigskip

To invert these equations, to get the momenta in terms of the
velocities, we shall assume  the validity of the Fourier transform
on the linearized WSW CMC-hyper-surfaces. They are conformal to
$R^3$, i.e. they are Euclidean plus corrections of order
$O(r_{\bar a})$, irrelevant when acting on the functions $r_{\bar
a}(\tau ,\vec \sigma )$, $\pi_{\bar a}(\tau ,\vec \sigma )$; for
the scalar product we have ${}^3{\hat g}_{rs} k^rh^s= \vec k\cdot
\vec h +O(r_{\bar a})$. As a consequence functions of order
$O(r_{\bar a})$ are considered as functions over $R^3$ at the
lowest order of approximation.

\bigskip

By using the Fourier transform defined in Appendix C,
Eqs.(\ref{V2}) become Eqs.(\ref{c4})

\beq
\partial_{\tau}{\tilde r}_{\bar a}(\tau ,\vec k) =
\sum_{\bar b} A_{\bar a\bar b}(\vec k)
{\tilde \pi}_{\bar b}(\tau ,\vec k),
 \label{V3}
 \eeq

\noindent with the matrix $A_{\bar a\bar b}(\vec k)$ and its
inverse given in Eqs.(\ref{c5}) and (\ref{c6}), respectively. Let
us remark that, notwithstanding $A_{\bar a\bar b}(\vec k)$
diverges for $k^r \rightarrow 0$, Eqs.(\ref{V3}) are well defined
if ${\tilde \pi}_{\bar a}(\tau ,\vec k)\, {\rightarrow}_{\vec k
\rightarrow 0}\, (k^1 k^2 k^3)^{2 + \epsilon}\, {\tilde f}_{\bar
a}(\vec k)$, $\epsilon > 0$. {\it This condition is stronger of
the requirement} (\ref{c2}), i.e. of the Fourier transform of
Eqs.(\ref{II15}).

\bigskip

As a consequence we get [$\epsilon_{\bar a\bar b}=-\epsilon_{\bar
b\bar a}$, $\epsilon_{\bar 1\bar 2} = 1$]

\bea
 {\tilde \pi}_{\bar a}(\tau ,\vec k) &=& \sum_{\bar b}
A^{-1}_{\bar a\bar b}(\vec k) \partial_{\tau}{\tilde r}_{\bar
a}(\tau ,\vec k),\nonumber \\
 &&{}\nonumber \\
 \pi_{\bar a}(\tau ,\vec \sigma ) &=& \sum_{\bar b} \int d^3\sigma_1
 G_{\bar a\bar b}(\vec \sigma -{\vec \sigma}_1) \partial_{\tau}r_{\bar b}(\tau ,{\vec
 \sigma}_1),\nonumber \\
 &&{}\nonumber \\
 G_{\bar a\bar b}(\vec \sigma -{\vec \sigma}_1) &=& \int {{d^3k}\over {(2\pi )^3}}
 A^{-1}_{\bar a\bar b}(\vec k) e^{i\vec k\cdot (\vec \sigma -{\vec \sigma}_1)}=\nonumber \\
 &=&-{{c^3}\over {8\pi G}} \int {{d^3w_1 d^3w_2 d^3w_3}\over {(4\pi )^3
 |\vec \sigma -{\vec w}_1|\, |{\vec w}_1-{\vec w}_2|\, |{\vec w}_2-{\vec w}_3|}}
 \Big[ \delta_{\bar a\bar b} {{\partial^6 \delta^3({\vec w}_3-{\vec \sigma}_1)}\over
 {(\partial w_3^1)^2 (\partial w_3^2)^2 (\partial w_3^3)^2}}+\nonumber \\
 &+&{1\over 2}\sum_{r,s,t,\bar c, \bar d}
 \epsilon_{\bar a\bar c}\gamma_{\bar c r}\epsilon_{\bar b\bar d}\gamma_{\bar ds}
 (2\delta_{tr}-1)(2\delta_{ts}-1) {{\partial^6 \delta^3({\vec w}_3-{\vec \sigma}_1)}\over
 {(\partial w_3^r)^2 (\partial w_3^s)^2 (\partial w_3^t)^2}}\Big].
\label{V4}
 \eea

The final result of the inversion is

\bea
 \pi_{\bar a}(\tau ,\vec \sigma ) &=& -{{c^3}\over {8\pi G}}
\int {{d^3\sigma_1 d^3\sigma_2 d^3\sigma_3}\over {(4\pi )^3 |\vec
\sigma -{\vec \sigma}_1|\, |{\vec \sigma}_1-{\vec \sigma}_2|\,
|{\vec \sigma}_2- {\vec \sigma}_3|}} \nonumber \\
 &&\Big[ {{\partial^6 \partial_{\tau}r_{\bar a}(\tau ,{\vec \sigma}_3)}\over
 {(\partial \sigma_3^1)^2 (\partial \sigma_3^2)^2 (\partial \sigma_3^3)^2}}+
 \nonumber \\
 &+& {1\over 2} \sum_{r,s,t, \bar b, \bar c, \bar d}
 \epsilon_{\bar a\bar c}\gamma_{\bar cr} \epsilon_{\bar b\bar d}\gamma_{\bar ds}
 (2\delta_{tr}-1)(2\delta_{ts}-1) {{\partial^6
 \partial_{\tau}r_{\bar b}(\tau ,{\vec \sigma}_3)}\over
 {(\partial \sigma_3^r)^2 (\partial \sigma_3^s)^2 (\partial \sigma_3^t)^2}} \Big].
\label{V5}
 \eea

Eqs.(\ref{V5}) satisfies the condition (\ref{II15}) automatically.

\subsection{The Second Half of Hamilton Equations.}

Let us now study the second half of Hamilton equations associated
with the Hamiltonian (\ref{III1}).  After some calculations we get

\bea
\partial_{\tau}\pi_{\bar a}(\tau ,\vec \sigma ) &{\buildrel \circ \over =}&
\{ \pi_{\bar a}(\tau ,\vec \sigma ), {\hat E}_{ADM} \} =\nonumber
\\
 &=& {{c^3}\over {24\pi G}} \sum_r {{\partial^2r_{\bar a}(\tau
 ,\vec \sigma )}\over {\partial (\sigma^r)^2}} - {{c^3}\over {12\pi G}} \sum_{r,\bar b}
 \gamma_{\bar ar}\gamma_{\bar br} {{\partial^2 r_{\bar b}(\tau ,\vec \sigma )}\over
 {(\partial \sigma^r)^2}}-\nonumber \\
 &-& {{c^3}\over {48\pi G}} \sum_{\bar brs}\, \gamma_{\bar ar}\gamma_{\bar
 bs}\, \int {{d^3\sigma_1}\over {4\pi |\vec \sigma -{\vec \sigma}_1|}}
 {{\partial^4r_{\bar b}(\tau ,{\vec \sigma}_1)}\over {(\partial \sigma_1^r)^2
 (\partial \sigma_1^s)^2}} + {{c^3}\over {G L^2}} O(r^2_{\bar a}).
 \label{V6}
 \eea

The Fourier transform of Eqs.(\ref{V6}) is given in Eqs.
(\ref{c7}) and (\ref{c8}), with the remarkable result
$\partial_{\tau}\, {\tilde \pi}_{\bar a}(\tau ,\vec k)\, \cir\, -
|\vec k|^2\, \sum_{\bar b}\, A^{-1}_{\bar a\bar b}(\vec k)\,
{\tilde r}_{\bar a}(\tau ,\vec k)$.

\medskip

This {\it implies} $\int_{-\infty}^{\infty}\, d\sigma^r\, r_{\bar
a}(\tau ,\vec \sigma ) = 0$, i.e. Eq.(\ref{II16}), {\it as the
simplest way to get consistency} between  Eqs. (\ref{V6}) and
(\ref{II15}).
\bigskip

Let us remark that, as shown in Eqs. (\ref{d9}) of Appendix D, our
Hamilton equations imply the {\it satisfaction of the remaining
Einstein equations} ${}^4{\hat R}_{rs}(\tau ,\vec \sigma ) \,
\cir\, 0$.

\bigskip

Eqs. (\ref{V4}) and (\ref{V6}) imply

\bea
\partial_{\tau}^2r_{\bar a}(\tau ,\vec \sigma) &{\buildrel \circ \over =}&
{{24\pi G}\over {c^3}} \partial_{\tau}\pi_{\bar a}(\tau ,\vec
\sigma ) +\nonumber \\
 &+& {{6\pi G}\over {c^3}} \sum_{\bar brs}\gamma_{\bar ar}\gamma_{\bar bs}
 \sum_{uv,u\not= v}[1-2(\delta_{ur}+\delta_{vr})][1-2(\delta_{us}+\delta_{vs})]\nonumber \\
 &&\int_{-\infty}^{\sigma^u} d\sigma_1^u \int_{-\infty}^{\sigma^v} d\sigma_1^v
 \int^{\infty}_{\sigma_1^u} d\sigma_2^u \int^{\infty}_{\sigma_1^v} d\sigma_2^v
 (\partial^2_r \partial^2_s \partial_{\tau}\pi_{\bar b})(\tau ,\sigma_2^u, \sigma_2^v,
 \sigma^{k\not= u,v})+\nonumber \\
 &+& {1\over L} O(r^2_{\bar a}),\nonumber \\
 &&{}\nonumber \\
 &&\Downarrow \nonumber \\
 &&{}\nonumber \\
 \Box r_{\bar a}(\tau ,\vec \sigma ) &=& [\partial_{\tau}^2-\sum_r \partial_r^2]
 r_{\bar a}(\tau ,\vec \sigma ) \, {\buildrel \circ \over =}\nonumber \\
  &{\buildrel \circ \over =}& {1\over 2}\sum_{\bar bru} \gamma_{\bar
  ar}(4\gamma_{\bar br}-\gamma_{\bar bu})\int {{d^3\sigma_1}\over {4\pi |\vec \sigma
-{\vec \sigma}_1|}}
  {{\partial^4 r_{\bar b}(\tau ,{\vec \sigma}_1)}\over {(\partial \sigma_1^r)^2
  (\partial \sigma_1^u)^2}}+\nonumber \\
&+&{1\over 8}\sum_{\bar brs}\sum_{\bar ctw}\sum_{uv}
(-2\delta_{\bar b \bar c}+4\gamma_{\bar bt}\gamma_{\bar
  ct}-\gamma_{\bar bt}\gamma_{\bar cw})\gamma_{\bar ar}\gamma_{\bar bs}
  (1-\delta_{uv})[1-2(\delta_{ur}+\delta_{vr})]\nonumber \\
  && [1-2(\delta_{us}+\delta_{vs})]
\int^{\sigma^u}_{-\infty} d\sigma_1^u \int^{\sigma^v}_{-\infty}
d\sigma_1^v
 \int^{\infty}_{\sigma_1^u} d\sigma_2^u \int^{\infty}_{\sigma_1^v}
 d\sigma_2^v \nonumber \\
 &&\int d^3\sigma_3\,
 {{[(\partial_{3r})^2 (\partial_{3s})^2 (\partial_{3w})^2(\partial_{3t})^2  r_{\bar c}](\tau
 ,{\vec \sigma}_3)}\over {4\pi \sqrt{(\sigma_3^u-\sigma_2^u)^2+(\sigma_3^v-\sigma_2^v)^2+
 (\sigma_3^{k\not=u,v}-\sigma^{k\not= u,v})^2} }}
 +{1\over {L^2}}O(r^2_{\bar a}).
 \label{V7}
 \eea

\bigskip

As shown in Appendix C, Eq.(\ref{c10}), all the terms in the
second member of Eq.(\ref{V7}) cancel, so that we find that the
DO's $r_{\bar a}(\tau ,\vec \sigma )$ satisfy the wave equation.
Actually, as shown in Eq.(\ref{c10}), the Fourier transform of the
equation of motion is

\bea
  &&{\ddot{\tilde r}}_{\bar a}(\tau ,\vec k)+|\vec k|^2{\tilde
r}_{\bar a}(\tau ,\vec k)=0, \nonumber \\
 &&{}\nonumber \\
 &&\Downarrow \nonumber \\
 &&{}\nonumber \\
 &&\Box\, r_{\bar a}(\tau ,\vec \sigma )\, \cir\, 0.
 \label{V8}
 \eea

Therefore, in our radiation gauge we get {\it the wave equation
$\Box\, r_{\bar a}(\tau ,\vec \sigma ) \cir 0$ for the DO's}.
Notwithstanding the presence of gravito-magnetism, which, as said,
should imply the  anisotropy of light propagation, we get an
isotropic propagation of gravitational waves in the radiation
gauge.

\bigskip

The complicated form (\ref{III1}), (\ref{c13}), of the weak ADM
energy shows that {\it in our gauge we do not have the
conventional description of a massless spin two particle over
Minkowski space-time like in the TT harmonic gauge}. This fact is
connected with the lack of a coordinate-independent notion of
gravitational energy density, which has no counterpart in the
theory of massless spin two particles in Minkowski space-time.
Instead we will see in Eqs.(\ref{V11}) and (\ref{V12}) that the
weak ADM 3-momentum and angular momentum do admit a standard
particle interpretation.

\bigskip

The solutions of the Hamilton equations for the DO's $r_{\bar
a}(\tau ,\vec \sigma )$ and $\pi_{\bar a}(\tau ,\vec \sigma )$ are
given in Eq.(\ref{c11}), (\ref{c12})

\bea
 r_{\bar a}(\tau ,\vec{\sigma})&=& \int {{d^3k}\over
{(2\pi )^3}}\, \Big(C_{\bar a}(\vec k)e^{-i\,|\vec
k|\tau}+C^{\ast}_{\bar a}(-\vec k)e^{i\, |\vec k|\tau}\Big)\, e^{
+i\, \vec k\cdot \vec \sigma},\nonumber \\
 &&{}\nonumber \\
 \pi_{\bar a}(\tau ,\vec \sigma ) &=&  \int {{d^3k}\over
{(2\pi )^3}}\,  {\tilde \pi}_{\bar a}(\tau ,\vec k)\, e^{i\, \vec
k \cdot \vec \sigma} =\int {{d^3k}\over {(2\pi )^3}}\,  \sum_{\bar
b}A^{-1}_{\bar a \bar b}\partial_{\tau}{\tilde r}_{\bar b}(\tau
,\vec k)\,  e^{i\, \vec k \cdot \vec \sigma},\nonumber \\
 &=& \int {{d^3k}\over {(2\pi )^3}}\,
 \sum_{\bar b}A^{-1}_{\bar a \bar b}(\vec k)\, \Big[-i|\vec
k|C_{\bar b}(\vec k)e^{-i|\vec k|\tau}+i|\vec k|C_{\bar
b}^{\ast}(-\vec k)e^{i|\vec k|\tau}\Big]\,  e^{i\, \vec k \cdot
\vec \sigma},\nonumber \\
 &&{}
 \label{V9}
 \eea

\noindent with the functions $C_{\bar a}(\vec k)$ {\it vanishing}
as $(k^1\, k^2\, k^3)^{\epsilon}$ for $k^r \rightarrow 0$.

\subsection{Special Solutions for the Background-Independent
Gravitational Waves.}

Special solutions are the following {\it plane waves} whose
3-momentum $\vec h$ cannot lie in any coordinate plane [$ \vec h
\not= \vec 0; (1,0,0); (0,1,0); (0,0,1); (1,1,0); (1,0,1);
(0,1,1)$]

\bea
 r_{\bar a}(\tau ,\vec \sigma ) &=& C_{\bar a}\, e^{i\, (\vec h
 \cdot \vec \sigma - |\vec h|\, \tau )},\quad C_{\bar a} = const.,\nonumber \\
 &&{}\nonumber \\
 &&{\tilde r}_{\bar a}(\tau ,\vec k) = (2\pi )^3\, C_{\bar a}\,
 e^{-i\, |\vec h|\, \tau}\, \delta^3(\vec k - \vec h),\nonumber \\
 &&{}\nonumber \\
 \pi_{\bar a}(\tau ,\vec \sigma ) &=& -i\, |\vec h|\, \sum_{\bar b}\,
 A^{-1}_{\bar a\bar b}(\vec h)\, C_{\bar b}\, e^{i(\vec h\cdot \vec \sigma -
 |\vec h|\, \tau )},\nonumber \\
 &&{}\nonumber \\
 &&{\tilde \pi}_{\bar a}(\tau ,\vec k) = -i\, (2\pi )^3\, |\vec h|\, \sum_{\bar b}\,
 A^{-1}_{\bar a\bar b}(\vec h)\, C_{\bar b}\, e^{-i\, |\vec h|\,
 \tau}\, \delta^3(\vec k - \vec h),\quad {\tilde \pi}_{\bar
 a}(\tau ,\vec 0) =0,\nonumber \\
 &&{}
 \label{V10}
 \eea

\noindent which verify Eqs. (\ref{II15}) and (\ref{II16}). Here
$\pi_{\bar a}$ has been evaluated with Eq.(\ref{V5}) and
Eqs.(\ref{c4}) and (\ref{c7}) are valid also at $\vec k = 0$.
\bigskip

Hoever these solutions do not satisfy the rest frame conditions
${\hat P}^r_{ADM} \approx 0$, which restrict the solutions to
globally outgoing or ingoing wave packets.

\medskip

Before looking for these solutions, we shall give the form of the
weak ADM charges (\ref{a12}) in the radiation gauge with our
solution of the constraints (see Eqs.(\ref{III1}) and (\ref{c13})
for ${\hat E}_{ADM}$). The weak ADM 3-momentum and  spin  of
Eqs.(\ref{a12})  assume the same simple form in terms of Fourier
transformed quantities as for free massless fields due to the fact
that we are in an instant form of dynamics. This is not true for
the ADM boosts, which have a complicated form like the ADM energy.

\bigskip

For the 3-momentum, due to exact cancellations, we get

\bea
 {\hat P}^r_{ADM} &=& - \int d^3\sigma\, \sum_{\bar c}\, \Big[
 \partial_r\, r_{\bar c}(\tau ,\vec \sigma ) -\nonumber \\
 &-& \gamma_{\bar cr}\, \Big(- \sum_{\bar
  a u}\, \gamma_{\bar au}\, \int d^3\sigma_1\, {{\partial^2_{1u}\, r_{\bar a}(\tau
 ,{\vec \sigma}_1)}\over {4\pi\, |\vec \sigma - {\vec \sigma}_1|}}
 -2\, \sum_{\bar a}\, \gamma_{\bar ar}\, \partial_r\, r_{\bar
 a}(\tau ,\vec \sigma )\Big) \Big]\, \pi_{\bar c}(\tau ,\vec
 \sigma ) +\nonumber \\
 &+& \int d^3\sigma d^3\sigma_1\, \sum_{\bar cs}\, \Big(
 \gamma_{\bar cs}\nonumber \\
 &&\Big[ \sum_u\, {1\over {\sqrt{3}}}\, \Big( {1\over {2}}\, \sum_{\bar
 am}\, \gamma_{\bar am}\, \partial_r\, \int d^3\sigma_2\, {{\partial^2_{2m}\, r_{\bar a}(\tau
 ,{\vec \sigma}_2)}\over {4\pi\, |\vec \sigma - {\vec \sigma}_1|}}
+ \, \sum_{\bar a}\, \gamma_{\bar au}\,
\partial_r\, r_{\bar a}(\tau ,\vec \sigma )\Big)\nonumber \\
 &&\delta^u_{(a)}\, {\cal T}^{(o)u}_{(a)s}(\vec \sigma ,{\vec
\sigma}_1) +\nonumber \\
 &+& \sum_{uv}\, \Big( \delta^r_u\, {1\over {\sqrt{3}}}\, \Big( {1\over {2}}\, \sum_{\bar
 am}\, \gamma_{\bar am}\, \partial_v\, \int d^3\sigma_2\, {{\partial^2_{2m}\, r_{\bar a}(\tau
 ,{\vec \sigma}_2)}\over {4\pi\, |\vec \sigma - {\vec \sigma}_1|}}
+ \, \sum_{\bar a}\, \gamma_{\bar au}\,
\partial_v\, r_{\bar a}(\tau ,\vec \sigma )\Big)\, +\nonumber \\
 &+& \delta^r_v\, {1\over {\sqrt{3}}}\, \Big( {1\over {2}}\, \sum_{\bar
 am}\, \gamma_{\bar am}\, \partial_u\, \int d^3\sigma_2\, {{\partial^2_{2m}\, r_{\bar a}(\tau
 ,{\vec \sigma}_2)}\over {4\pi\, |\vec \sigma - {\vec \sigma}_1|}}
+ \, \sum_{\bar a}\, \gamma_{\bar av}\, \partial_u\,
 r_{\bar a}(\tau ,\vec \sigma )\Big)\, \Big)\nonumber \\
 &&\delta^u_{(a)}\, {\cal T}^{(o) v}_{(a)s}(\vec \sigma ,{\vec
 \sigma}_1) \Big] \Big)\, \pi_{\bar c}(\tau ,{\vec \sigma}_1)
 =\nonumber \\
 &&{}\nonumber \\
 &=& - \int d^3\sigma\, \sum_{\bar c}\, \pi_{\bar c}(\tau ,\vec
 \sigma )\, \partial_r\, r_{\bar c}(\tau ,\vec \sigma ) =\nonumber \\
 &=& i \int {{d^3k}\over {(2\pi )^3}}\,  \sum_{\bar c}\, {\tilde \pi}_{\bar c}(\tau
 ,\vec k)\, k^r\, {\tilde r}_{\bar c}(\tau ,-\vec k),
 \label{V11}
 \eea

\noindent where we used ${\cal T}^{(o)u}_{(u)s} = 0$, see
Eq.(\ref{II13}).\bigskip

For the angular momentum, due to similar cancellations, we get

\bea
 {\hat J}^{rs}_{ADM} &=& \int d^3\sigma\, \sum_{\bar c}\,
 \pi_{\bar c}(\tau ,\vec \sigma )\, (\sigma^r\, \partial_s -
 \sigma^s\, \partial_r)\, r_{\bar c}(\tau ,\vec \sigma )
 =\nonumber \\
 &=& - \int {{d^3k}\over {(2\pi )^3}}\,  \sum_{\bar c}\, {\tilde \pi}_{\bar c}(\tau
 ,\vec k)\, \Big( k^r\, {{\partial}\over {\partial k^s}} -  k^s\,
 {{\partial}\over {\partial k^r}}\Big)\, {\tilde r}_{\bar
 c}(\tau , -\vec k).
 \label{V12}
 \eea

\bigskip

Eqs. (\ref{V11}) and (\ref{V12}) reflect the fact that we are in
an instant form of the dynamics. Instead for the boosts we get a
complicated expression like for the energy (see Eqs.(\ref{II13})
for the kernels)

\bea
 {\hat J}^{\tau r}_{ADM} &=& \epsilon\, \int d^3\sigma\,
 \sigma^r\, \Big(\, {{c^3}\over {16\pi\, G}}\, \Big[ {1\over 3}\, \sum_r\,
 \Big( {1\over 2}\, \Big(\sum_{\bar
 am}\, \gamma_{\bar am}\, \partial_r\, \int d^3\sigma_2\, {{\partial^2_{2m}\, r_{\bar a}(\tau
 ,{\vec \sigma}_2)}\over {4\pi\, |\vec \sigma - {\vec
 \sigma}_1|}}\Big)^2-\nonumber \\
 &-& \sum_{\bar a}\, \Big( \partial_r\, r_{\bar a}(\tau ,\vec
 \sigma )\Big)^2 + \sum_{\bar
 am}\, \gamma_{\bar am}\, \partial_r\, \int d^3\sigma_2\, {{\partial^2_{2m}\, r_{\bar a}(\tau
 ,{\vec \sigma}_2)}\over {4\pi\, |\vec \sigma - {\vec
 \sigma}_1|}}\, \sum_{\bar b}\, \gamma_{\bar br}\, \partial_r\,
 r_{\bar b}(\tau ,\vec \sigma ) +\nonumber \\
 &+&2 \Big( \sum_{\bar b}\, \gamma_{\bar br}\, \partial_r\,
 r_{\bar b}(\tau ,\vec \sigma )\Big)^2 \Big) \Big] - {{6\pi \,
 G}\over {c^3}}\, \Big[ 2\, \sum_{\bar a}\, \pi^2_{\bar a}(\tau
 ,\vec \sigma ) +\nonumber \\
 &+&4\, \sum_{\bar bu}\, \gamma_{\bar bu}\, \pi_{\bar b}(\tau
 ,\vec \sigma )\, \int d^3\sigma_1\, \sum_m\, \delta^u_{(a)}\, {\cal
 T}^{(o)u}_{(a)m}(\vec \sigma ,{\vec \sigma}_1)\, \sum_{\bar a}\,
 \gamma_{\bar am}\, \pi_{\bar a}(\tau ,{\vec \sigma}_1) +\nonumber \\
 &+& \int d^3\sigma_1 d^3\sigma_2\, \Big( \sum_{u}\, \sum_m\,
 {\cal T}^{(o)u}_{(a)m}(\vec \sigma ,{\vec \sigma}_1)\, \sum_{\bar
 a}\, \gamma_{\bar am}\, \pi_{\bar a}(\tau ,{\vec
 \sigma}_1)\nonumber \\
 &&\sum_s\, {\cal T}^{(o)u}_{(a)s}(\vec \sigma ,{\vec \sigma}_2)\,
 \sum_{\bar b}\, \gamma_{\bar bs}\, \pi_{\bar b}(\tau ,{\vec
 \sigma}_2)+\nonumber \\
 &+& \sum_{uv}\, (\delta^u_{(b)}\, \delta^v_{(a)} -
 \delta^u_{(a)}\, \delta^v_{(b)})\, \sum_m\,
 {\cal T}^{(o)u}_{(a)m}(\vec \sigma ,{\vec \sigma}_1)\, \sum_{\bar
 a}\, \gamma_{\bar am}\, \pi_{\bar a}(\tau ,{\vec
 \sigma}_1)\nonumber \\
 &&\sum_s\, {\cal T}^{(o)v}_{(b)s}(\vec \sigma ,{\vec \sigma}_2)\,
 \sum_{\bar b}\, \gamma_{\bar bs}\, \pi_{\bar b}(\tau ,{\vec
 \sigma}_2)\Big) \Big] \Big) -\nonumber \\
 &-& {{\epsilon\, c^3}\over {8\pi\, G}}\, \int d^3\sigma\,
 \sum_{uv}\, \delta^r_u\, (\delta_{uv} - 1)\, {1\over 3}\, \Big(
 -{1\over 4}\, \sum_{\bar am}\, \gamma_{\bar am}\, \partial_u\,
 \int d^3\sigma_2\, {{\partial^2_{2m}\, r_{\bar a}(\tau
 ,{\vec \sigma}_2)}\over {4\pi\, |\vec \sigma - {\vec
 \sigma}_1|}} +\nonumber\\
 &+& \sum_{\bar a}\, (\gamma_{\bar av} - \gamma_{\bar
 au})\, \partial_u\, r_{\bar a}(\tau ,\vec \sigma )\Big)\nonumber \\
 &&\Big( -  \sum_{\bar an}\, \gamma_{\bar an}\,
 \int d^3\sigma_2\, {{\partial^2_{2n}\, r_{\bar a}(\tau
 ,{\vec \sigma}_2)}\over {4\pi\, |\vec \sigma - {\vec
 \sigma}_1|}} + 2\, \sum_{\bar b}\, \gamma_{\bar bu}\, r_{\bar
 b}(\tau ,\vec \sigma ) \Big) =\nonumber \\
 &=& \epsilon\int d^3\sigma \sigma^r
\Big(\frac{c^3}{16 \pi G}\sum_{s} \big[2(\partial_{s}q(\tau ,\vec
\sigma ))^2 -\frac{1}{3}\sum_{\bar b}(\partial_{s}r_{\bar b}(\tau
,\vec \sigma ))^2 + \nonumber \\
 &-& \frac{2}{\sqrt{3}}\,\partial_{s}q(\tau ,\vec \sigma )\sum_{\bar
b}\gamma_{\bar b s}\partial_{s}r_{\bar b}(\tau ,\vec \sigma
)+\frac{2}{3}(\sum_{\bar b}\gamma_{\bar b s}\partial_{s}r_{\bar
b})^2(\tau ,\vec \sigma )\big] -\frac{12 \pi
G}{c^3}\big[\sum_{\bar a}\pi_{\bar a}^2(\tau ,\vec \sigma ) +
\nonumber \\
 &+& 4\int d^3{\sigma_1}\int
d^3{\sigma_2}\sum_{u,m,s,(a)}{\cal T}^{(o)u}_{(a)m}(\vec \sigma
,{\vec \sigma}_1)\sum_{\bar b}\gamma_{\bar b m}\pi_{\bar
b}(\tau,{\vec \sigma}_1){\cal T}^{(o)u}_{(a)s}(\vec \sigma ,{\vec
\sigma}_2)\sum_{\bar c}\gamma_{\bar c s}\pi_{\bar c}(\tau,{\vec
\sigma}_2)\big]\Big)+\nonumber\\
 & -& \epsilon \frac{c^3}{8 \pi G}\int d^3\sigma
\sum_{v}(\delta_{rv}-1)\big(2q+\frac{2}{\sqrt{3}}\sum_{\bar
a}\gamma_{\bar a r}r_{\bar
a}\big)\big(\frac{1}{2}\,\partial_{r}q+\frac{1}{\sqrt{3}}\sum_{\bar
b}(\gamma_{\bar b v}-\gamma_{\bar b r})\partial_r r_{\bar
b}\big)(\tau ,\vec \sigma )+\nonumber \\
 &+& O(r_{\bar a}^3).
 \label{V13}
 \eea

\bigskip

The weak ADM 3-momentum of the solutions (\ref{V9}), (\ref{c11}),
has the form

\beq
 {\hat P}^r_{ADM} =
   2\, \sum_{\bar b\bar c}\, \int {{d^3k}\over {(2\pi )^3}}\,
  A^{-1}_{\bar c\bar b}(\vec k)\, |\vec k|\, k_r\, C_{\bar b}(\vec
  k)\, C^*_{\bar c}(\vec k) \approx 0,
 \label{V14}
 \eeq

\noindent so that its vanishing is a condition on the Fourier
coefficients $C_{\bar a}(\vec k)$, which, for instance, {\it
cannot be satisfied by the plane waves} (\ref{V10}).

\bigskip

A {\it class of solutions} of the Hamilton equations  with the
correct vanishing behaviour at spatial infinity, which, besides
Eqs.(\ref{c2}), also satisfies the rest frame condition ${\hat
P}^r_{ADM} = 0$ is obtained  by taking coefficients $C_{\bar
a}(\vec k) =  C_{\bar a}\, {{(k_1 k_2 k_3)^2}\over {|\vec k|}}\,
e^{- {\vec k}^2}$ ($C_{\bar a} = const.$) in Eq.(\ref{V9})

\bea
 r_{\bar a}(\tau ,\vec \sigma ) &=& C_{\bar a}\, \int {{d^3k}\over {(2\pi
 )^3}}\, {{(k_1 k_2 k_3)^2}\over {|\vec k|}}\, e^{- {\vec k}^2}\, \Big[ e^{-i (|\vec
 k|\, \tau - \vec k \cdot \vec \sigma )} +  e^{i (|\vec
 k|\, \tau - \vec k \cdot \vec \sigma )} \Big] =\nonumber \\
 &=& - {{4\, C_{\bar a}}\over {(2\pi )^2}}\, {{\partial^2}\over {\partial^2
 \sigma_1}}\,  {{\partial^2}\over {\partial^2
 \sigma_2}}\,  {{\partial^2}\over {\partial^2
 \sigma_3}}\, {{(1 + |\vec \sigma |^2 - \tau^2)}\over {[1+ (|\vec
 \sigma | + \tau )^2]\, [1+ (|\vec
 \sigma | - \tau )^2]}},\nonumber \\
 &&{}\nonumber \\
 \pi_{\bar a}(\tau ,\vec \sigma ) &=& -i\,  \int {{d^3k}\over {(2\pi
 )^3}}\, {{(k_1 k_2 k_3)^2}\over {|\vec k|}}\, e^{- {\vec
 k}^2}\,\nonumber \\
 &&|\vec k|\, \sum_{\bar b}\, A^{-1}_{\bar a\bar b}(\vec k)\,
 C_{\bar b}\,  \Big[ e^{-i (|\vec
 k|\, \tau - \vec k \cdot \vec \sigma )} -  e^{i (|\vec
 k|\, \tau - \vec k \cdot \vec \sigma )} \Big],\nonumber \\
 &&{}\nonumber \\
 &&{}\nonumber \\
 n_r(\tau ,\vec \sigma ) &=&  i\,\sqrt{3} {{4\pi G}\over {c^3}}
 \sum_{\bar avuc} \gamma_{\bar av}(1-{\delta}_{uc})
 [1-2(\delta_{uv}+\delta_{cv})][1-2(\delta_{ur}+\delta_{cr})]\nonumber  \\
 &&\int {{d^3k}\over {(2\pi^3)}}\, (k_1k_2k_3)^2 e^{-{\vec k}^2}
{{k_r k^2_v}\over{k^2_u k^2_c}}\, \sum_{\bar b}\,  A^{-1}_{\bar
a\bar b}(\vec k)\, C_{\bar b}\, sin\, (|\vec k|\, \tau ) e^{i \vec
k \cdot \vec \sigma},\nonumber \\
 &&{}\nonumber \\
 {\hat E}_{ADM} &=& \sum_{\bar a\bar b}\, \alpha_{\bar a\bar b}\,
 C_{\bar a}C_{\bar b},\nonumber \\
 {\hat P}^r_{ADM} &=& 0,\nonumber \\
 {\hat J}^r_{ADM} &=& {1\over 2}\, \epsilon^{ruv}\, {\hat J}^{uv}_{ADM} =
 \sum_{\bar a\bar b}\, \beta^r_{\bar a\bar b}\, C_{\bar a}C_{\bar
 b}.
 \label{V15}
 \eea

The two constants $C_{\bar 1}$, $C_{\bar 2}$ have to be expressed
in terms of the two boundary constants $M = {\hat E}_{ADM}$, $S =
|{\hat {\vec J}}_{ADM}|$ defining the mass and spin of the
post-Minkowskian Einstein space-time. The solution (\ref{V15})
goes like $|\vec \sigma|^{-8}$ for $|\vec \sigma | \rightarrow
\infty$, i.e. much faster than the behaviour (\ref{I2}). It should
be possible to find solutions saturating Eqs.(\ref{I2}), namely
such that by using Eqs.(\ref{III7}) we should get ${}^3g_{rs}
\rightarrow_{r \rightarrow \infty}\, (1 + {{ {\hat E}_{ADM}}\over
{2\, r}}) \delta_{rs}$, $n_r  = -\epsilon\, {}^4g_{\tau r}
\rightarrow_{r \rightarrow \infty}\, O(r^{-\epsilon})$. By
comparison, in presence of compact matter the expected solution
should have a post-Newtonian behaviour ${}^4g_{\tau\tau}
\rightarrow_{r \rightarrow \infty}\, 1-{{2M}\over r}$,
${}^4g_{\tau r} \rightarrow_{r \rightarrow \infty}\, 4
\epsilon_{ruv} S^u {{\sigma^v}\over {r^3}}$, ${}^4g_{rs}
\rightarrow_{r \rightarrow \infty}\, (1+ {{2M}\over r})
\delta_{rs}$ (see Eq.(19.5) of Ref.\cite{22}).

{\it Therefore, in absence of matter, the rest-frame condition
destroys the transversality property of the TT harmonic gauge
plane waves}.

\vfill\eject

\section{The Time-like Geodesics, the Geodesic Deviation Equation and the
Polarization of Gravitational Waves in the Radiation Gauge.}

\subsection{The Geodesic Equation.}

Let us consider a time-like geodesic with affine parameter $s$,
$\sigma^A(s)=\Big( \tau (s); \sigma^u(s)\Big)$. After the
linearization in the radiation gauge the {\it geodesic equation}
becomes (see Eqs.(\ref{c16}) of Appendix C for Fourier transforms)

\bea
{{d^2 \tau (s)}\over {ds^2}} &=& -{}^4{\hat \Gamma}^{\tau}_{rs}(\sigma
(s)) {{d\sigma^r(s)}\over {ds}} {{d\sigma^s(s)}\over {ds}}+O(r^2_{\bar
a}),\nonumber \\
 &&{}\nonumber \\
 {{d^2 \sigma^u(s)}\over {ds^2}} &=& -\partial_{\tau}n_u(\sigma (s))
 ({{d \tau (a)}\over {ds}})^2-2\, {}^4{\hat \Gamma}^u_{\tau r}(\sigma (s))
 {{d\tau (s)}\over {ds}}{{d\sigma^r(s)}\over {ds}}- {}^3{\hat \Gamma}^u_{rs}(\sigma (s))
 {{d\sigma^r(s)}\over {ds}} {{d\sigma^s(s)}\over {ds}} +\nonumber \\
 &+& O(r^2_{\bar a}),
 \label{VI1}
  \eea

\noindent where Eq.(\ref{d5}) has to be used for the Christoffel
symbols.

\bigskip

If we parametrize the geodesic as $\sigma^A(s)=a^A + b^A s +
f^A(s|r_{\bar a},\pi_{\bar a}]$, where $a^A + b^A\, s$ is the flat
geodesic, we get at the lowest order ${{d^2f^{\tau}(s)}\over
{ds^2}} \approx -b^rb^s\, {}^4{\hat \Gamma}^{\tau}_{rs}(\sigma
(s)) + O(r^2_{\bar a}) \approx -b^rb^s\, {}^4{\hat
\Gamma}^{\tau}_{rs}(a+bs) + O(r^2_{\bar a})$. Therefore the
solution to the first of Eqs.(\ref{VI1}) is

\bea
 \tau(s) &=&a^{\tau}+b^{\tau}s-\int_0^s ds_1 \int_0^{s_1} ds_2
{}^4{\hat \Gamma}^{\tau}_{ru}(a+bs_2)b^r b^u\nonumber\\
 p^{\tau}(s)&=& {{d \tau (s)}\over {ds}} = b^{\tau}-\int_0^s ds_2 {}^4{\hat
\Gamma}^{\tau}_{ru}(a+bs_2)b^r b^u
\qquad\text{with}\qquad{}^4{\hat \Gamma}^{\tau}_{ru}(a+bs_2)\sim
O(r_{\bar a}).
 \label{VI2}
 \eea

Since the tangent $ p^A(s)= {{d \sigma^A (s)}\over {ds}}$ to the
geodesic must satisfy either ${}^4g^{AB}\, p_A\, p_B = \epsilon$
(time-like geodesic with $s$ as proper time) or ${}^4g^{AB}\,
p_A\, p_B = 0$ (null geodesic) at the initial time, the constant
$b^{\tau}$ is determined in terms of the $b^r$'s and of the
initial data. Therefore we will have $b^{\tau} = b_{(o)}^{\tau} +
b^{\tau}_{(1)}$ with $b^{\tau}_{(1)} = O(r_{\bar a})$. Therefore
Eq.(\ref{VI2}) may be rewritten in the form

\bea
 \tau(s) &=&a^{\tau}+[b^{\tau}_{(o)} + b^{\tau}_{(1)}]\, s-\int_0^s ds_1 \int_0^{s_1} ds_2
{}^4{\hat \Gamma}^{\tau}_{ru}(a+b_{(o)}s_2)b^r b^u\nonumber\\
 &&{}\nonumber \\
 p^{\tau}(s)&=& {{d \tau (s)}\over {ds}} = b^{\tau}_{(o)} + b^{\tau}_{(1)}
 -\int_0^s ds_2 {}^4{\hat \Gamma}^{\tau}_{ru}(a+b_{(o)} s_2)\, b^r b^u.
 \label{VI3}
 \eea

The spatial part of Eqs.(\ref{VI1}) becomes

\bea
 {{d^2 f^u(s)}\over {ds^2}} &=& -\partial_{\tau}n_u(a+b_{(o)}s)
 (b^{\tau}_{(o)})^2-2\, {}^4{\hat \Gamma}^u_{\tau r}(a+b_{(o)}s)
 b^{\tau}_{(o)}b^r- {}^3{\hat \Gamma}^u_{rt}(a+b_{(o)}s)
 b^r b^t +O(r^2_{\bar a})\nonumber \\
 &&{}\nonumber \\
 \sigma^u(s)&=&a^u + b^u - \int_0^s ds_1 \int_0^{s_1} ds_2
\Big[(b^{\tau}_{(o)})^2-2\, {}^4{\hat \Gamma}^u_{\tau r}(a+b_{(o)}
s_2)
 b^{\tau}_{(o)}\, b^r- {}^3{\hat \Gamma}^u_{rt}(a+b_{(o)} s_2)
 b^r b^t \Big]\nonumber \\
 &&{}\nonumber \\
 p^u(s) &=& {{d\sigma^u(s)}\over {ds}}=b^u-\int_0^s ds_2
\Big[(b^{\tau}_{(o)})^2-2\, {}^4{\hat \Gamma}^u_{\tau r}(a+b_{(o)}
s_2)
 b^{\tau}_{(o)}\, b^r- {}^3{\hat \Gamma}^u_{rt}(a+b_{(o)} s_2)
 b^r b^t \Big].\nonumber \\
 &&{}
 \label{VI4}
 \eea

In the time-like case at $s = 0$, where $\tau (0) = a^{\tau}$, we
have the condition

\bea
 {}^4g_{AB}p^Ap^B{|}_{s=0}&=&\epsilon \Big[(p^{\tau}(s))^2-2
n_r(s,r_{\bar a}(s))p^r(s)p^{\tau}(s)-[\delta_{rt}-\epsilon
 {}^4h_{rt}(s,r_{\bar a}(s))p^r(s)p^t(s)]\Big]{|}_{s=0} =\nonumber \\
 &=& \epsilon,\nonumber \\
 &&{}\nonumber \\
 1&=&(b^{\tau})^2-2b^{\tau}b^r n_r(a^{\tau},r_{\bar
a}(a^{\tau}))-[\delta_{rt}-\epsilon {}^4h_{rt}(a^{\tau},r_{\bar
a}(a^{\tau}))b^r b^t],\quad at\, s=0,
 \label{VI5}
  \eea

\noindent whose solution is

\bea
 b^{\tau} &=& b^{\tau}_{(o)} + b^{\tau}_{(1)}
 =b^r n_r(a,r_{\bar a}(a))\pm
\sqrt{(b^r n_r(a^{\tau},r_{\bar a}(a^{\tau})))^2+ 1 +
[\delta_{rt}-\epsilon {}^4h_{rt}(a^{\tau},r_{\bar a}(a^{\tau}))]\,
b^r b^t},\nonumber \\
 &&{}\nonumber \\
 b^{\tau}_{(o)} &=& \pm \sqrt{1 + \delta_{rs}b^r b^s},\nonumber \\
 b^{\tau}_{(1)} &=&  n_r(a^{\tau},r_{\bar
a}(a^{\tau}))\, b^r\, \mp {{\epsilon}\over 2}\,
{{{}^4h_{rt}(a^{\tau},r_{\bar a}(a^{\tau}))b^r b^t}\over { \sqrt{1
+ \delta_{rs}b^r b^s}}} + O(r^2_{\bar a}).
 \label{VI6}
 \eea

If we choose $b^r=0$, we can take the solution $f^{\tau}(s)=0$ so
that we get $\tau (s) = a^{\tau} + b^{\tau} s
{\rightarrow}_{b^{\tau}=1}\, s+a^{\tau}$.

\bigskip

With the choice $b^{\tau}=1$, $b^r=0$, $a^{\tau}=0$, we get $\tau
(s)=s$ and ${{d^2f^u(s)}\over {ds^2}} =-\partial_{\tau}n_u(\tau
(s) ,\vec \sigma (s)=\vec a+\vec f(s))=0$ at the lowest order,
with the solution $f^u(s)=-\int^s ds^{'} n_u(s^{'}, a^r)
+O(r^2_{\bar a})$. Therefore, with $b^{\tau}=1$, $b^r=0$,
$a^{\tau}=0$ we get the following expression for the geodesics

\bea \sigma^A(s) &=& \Big( \tau (s)=s;\,\, \sigma^u(s)= a^u-
\int^s_0 dw \, n_u(w, a^v) \Big) + O(r^2_{\bar a}),\nonumber \\
 &&{}\nonumber \\
 {{d\sigma^A(s)}\over {ds}} &=& \Big( 1;\,\, -n_u(s, a^v) \Big)+
 O(r^2_{\bar a}).
 \label{VI7}
 \eea

\noindent This is the {\it trajectory of a massive test particle
in our coordinates}.

\subsection{The Geodesic Deviation Equation.}

If we consider two nearby freely falling particles following two
nearby geodesics (\ref{VI7}) such that $\sigma^A_1(0) = (0; \vec
0)$ and $\sigma^A_2(0) = (0; a^u)$ with an infinitesimal $a^u$, we
get for $\triangle x^A(s) = \sigma^A_2(s) - \sigma^A_1(s)$:
$\triangle x^{\tau}(s) = 0$, $\triangle x^r(s) \approx [
\delta^{ru} - \int_0^s dw\, {{\partial n^r(w, a^v)}\over {\partial
a^u}}]\, a^u$. If $a^u = \delta^{ur}\, a$ is their coordinate
distance, then their {\it proper distance} is $\triangle l =
\sqrt{-\epsilon\, {}^4g_{AB}\, \triangle x^A\, \triangle x^B} =
\triangle x^r\, \sqrt{{}^3g_{rr}} = (1 + {1\over 2} k_r)\,
\triangle x^r + O(r^2_{\bar a}) \approx [1 + {1\over 2} k_r -
\int^s_0 dw\, {{\partial n^r(w, a^v)}\over {\partial a^r}}]\, a +
O(r^2_{\bar a})$ by using Eqs.(\ref{III7}) [in a TT harmonic gauge
one gets $\triangle l \approx [1 + h^{TT}_{rr}]\, a$ if $r$ is one
of the two polarization directions].

\bigskip

As a consequence the connecting vector $\triangle x^A(s)$
satisfies

\beq
 \triangle x^{\tau}(s) =0,\qquad {{d^2 \triangle x^r(s)}\over {ds^2}} = -
 {d\over {d s}}\, \Big({{\partial n^r(s,a^v)}\over {\partial a^u}}\Big)\, a^u.
 \label{VI8}
 \eeq

This is the {\it geodesic deviation equation} along the geodesic
$\sigma^A_1(s)$ of the form (\ref{VI7}): it shows explicitly the
action of the {\it tidal forces} in our 4-coordinate system. More
in general, by using the equation

\bea
 &&{{d\sigma^C(s)}\over {ds}} \, {}^4\nabla_C \triangle x^A(s) = {{d \triangle x^A(s)}
 \over {ds}}+{{d\sigma^C(s)}\over {ds}} \, {}^4{\hat \Gamma}^A_{CE}(s, a^u)
 \triangle x^E(s)+O(r^2_{\bar a}) =\nonumber \\
 &&={{d \triangle x^A(s)} \over {ds}} + {}^4{\hat \Gamma}^A_{\tau E}(s, a^u)
 \triangle x^E(s) +O(r^2_{\bar a}),
 \label{VI9}
 \eea

\noindent the geodesic deviation equation \cite{22,24} takes the
form

\bea
&&{{d\sigma^B(s)}\over {ds}}\, {}^4\nabla_B \Big( {{d\sigma^C(s)}\over
{ds}}\, {}^4\nabla_C\, \triangle x^A(s) \Big) =- {}^4{\hat
R}^A{}_{BCD}(\sigma (s))\, \triangle x^C(s) {{d\sigma^B(s)}\over {ds}}
{{d\sigma^D(s)}\over {ds}},\nonumber \\
 &&{}\nonumber \\
 &&{}\nonumber \\
 &&\Downarrow \nonumber \\
 &&{}\nonumber \\
 &&{{d^2\triangle x^A(s)}\over {ds^2}} + 2\, {}^4{\hat \Gamma}^A_{\tau E}(s, a^u)
 {{d\triangle x^E(s)}\over {ds}}+ {{d\, {}^4{\hat \Gamma}^A_{\tau E}(s, a^u)}\over
 {ds}} \triangle x^E(s) =- {}^4{\hat R}^A{}_{\tau C\tau}(s, a^u)
 \triangle x^C(s),\nonumber \\
 &&{}\nonumber \\
 &&\Downarrow \nonumber \\
 &&{}\nonumber \\
 &&{{d^2 \triangle x^{\tau}(s)}\over {ds^2}} = 0+ O(r^2_{\bar a}),\quad \Rightarrow
 \quad \triangle x^{\tau }(s) = c s + d,\nonumber \\
 &&{}\nonumber \\
 &&{}\nonumber \\
 &&{{d^2 \triangle x^r(s)}\over {ds^2}}+ 2 \Big[ {}^4{\hat \Gamma}^r_{\tau\tau}(s
 ,a^u) c +{}^4{\hat \Gamma}^r_{\tau s}(s, a^u) {{d \triangle x^s(s)}\over {ds}}+
 {{d\, {}^4{\hat \Gamma}^r_{\tau\tau}(s, a^u)}\over {ds}} (c s +d)+\nonumber \\
 &&+ {{d\, {}^4{\hat \Gamma}^r_{\tau s}(s, a^u)}\over {ds}} \triangle x^s(s) =-
 {}^4{\hat R}^r{}_{\tau s\tau}(s, a^u) \triangle x^s(s) + O(r^2_{\bar a}).
 \label{VI10}
  \eea

\bigskip

If we take $\triangle x^{\tau}(s)=0$ as it happens for the
geodesic (\ref{VI7}), i.e. $c = d =0$, the geodesic deviation
equation becomes

\bea
 {{d^2 \triangle x^r(s)}\over {ds^2}} &+& 2\, {}^4{\hat
\Gamma}^r_{\tau s}(s, a^u) {{d \triangle x^s(s)}\over {ds}} +
{{d\, {}^4{\hat \Gamma}^r_{\tau s}(s, a^u)}\over {ds}} \triangle
x^s(s) =\nonumber \\
 &=& - {}^4{\hat R}^r{}_{\tau s\tau}(s, a^u) \triangle x^s(s) +
 O(r^2_{\bar a}).
 \label{VI11}
 \eea

If we write $\triangle x^r(s) = \triangle_{(o)} x^r(s) +
\triangle_{(1)} x^r(s)$, with $ \triangle_{(o)} x^r(s) = (o; a^u)$
the flat deviation and $ \triangle_{(1)} x^r(s) = \Big( 0; -
\int^s_0 dw\, [n^u(w,a^v) - n^u(w,\vec 0)]\Big) \approx \Big( 0; -
\int^s_0 dw\, {{\partial n^u(w,a^v)}\over {\partial
a^v}}{|}_{a^v=0}\, a^v\Big) = O(r_{\bar a})$, Eq.(\ref{VI11})
becomes

\bea
 &&{{d^2}\over {ds^2}}\, \Big( \triangle_{(o)} x^r(s) +
\triangle_{(1)} x^r(s)\Big) = {{d^2\, \triangle_{(1)} x^r(s)}\over
{ds^2}} =\nonumber \\
 &&= - 2\, {}^4{\hat \Gamma}^r_{\tau
s}(s, a^u) {{d \triangle_{(o)} x^s(s)}\over {ds}} - {{d\,
{}^4{\hat \Gamma}^r_{\tau s}(s, a^u)}\over {ds}} \triangle_{(o)}
x^s(s) -\nonumber \\
 &&- {}^4{\hat R}^r{}_{\tau s\tau}(s, a^u) \triangle_{(o)} x^s(s) +
 O(r^2_{\bar a}).
 \label{VI12}
 \eea

\noindent The use of Eqs.(\ref{d5}) and (\ref{d7}) allows to check
that Eq.(\ref{VI12}) coincides with Eq.(\ref{VI8}).

\subsection{The Tidal Forces Generated by the DO's on Test Particles.}

Eq.(\ref{VI8}) can be used to see the effect of the tidal forces
generated by the two DO's $r_{\bar a}(\tau ,\vec \sigma )$ of our
radiation gauge, replacing the two polarizations of the TT
harmonic gauge, on a sphere of test particles surrounding a test
particle sitting in the origin $(\tau = 0; \vec \sigma = \vec 0)$
at $s=0$ and described by the geodesic $\sigma_o^A(s) = (s;
\sigma^u_o(s) = - \int^s_0 dw\, n^u(w,\vec 0)\, )$. If we define
$a^u = a\, \eta^u$ with ${}^3g_{uv}\, \eta^u\, \eta^v = 1 +
O(r_{\bar a})$ and $a << 1$ (the radius of the sphere), the test
particles on the sphere at $s=0$ will follow the geodesics
$\sigma^A(s) = (s; \sigma^u(s) = a\, \eta^u - \int^s_0 dw\,
n^u(s,a\, \eta^v)\, )$. The connecting vectors $\triangle x^A(s) =
\triangle_{(o)} x^A(s) + \triangle_{(1)} x^A(s) = \sigma^A(s) -
\sigma^A_o(s)$ satisfy Eq.(\ref{VI8}), namely

\beq
 \triangle x^{\tau}(s) =0,\qquad {{d^2 \triangle_{(1)}
 x^r(s)}\over {ds^2}} = - {d\over {d s}}\, \Big({{\partial n^r(s,a^v)}\over {\partial
 a^v}}{|}_{a^v=0}\Big)\, a\, \eta^v + O(r^2_{\bar a}).
 \label{VI13}
 \eeq

We have to solve these equations with the shift functions
(\ref{III5}) evaluated on the solutions (\ref{V15}), which are not
transverse plane waves due to the rest frame conditions (this
forces us to consider a sphere of test particles).

\bigskip

The quasi-Shanmugadhasan canonical basis (\ref{I2}) adapted to our
gauge is not unique, because there is still the freedom associated
with the numerical constants $\gamma_{\bar ar}$ (their variation
amounts to a redefinition of the DO's). As a consequence of the
conditions stated after Eqs.(\ref{I5}), the $\gamma_{\bar ar}$'s
may be written in term of an angle $\psi$ with $0 \leq \psi \leq
2\pi$

\bigskip

\bea
 \gamma_{\bar 1r} &=& \Big( {{cos\, \psi}\over {\sqrt{6}}} -
{{sin\, \psi}\over {\sqrt{2}}},\, {{cos\, \psi}\over {\sqrt{6}}} +
{{sin\, \psi}\over {\sqrt{2}}},\, - \sqrt{ {2\over 3}}\, cos\,
\psi \Big),\nonumber \\
 \gamma_{\bar 2r} &=& \Big( - {{cos\, \psi}\over {\sqrt{2}}} -
{{sin\, \psi}\over {\sqrt{6}}},\, {{cos\, \psi}\over {\sqrt{2}}} -
{{sin\, \psi}\over {\sqrt{6}}},\, \sqrt{ {2\over 3}}\, cos\, \psi
\Big).
 \label{VI14}
 \eea

\bigskip

By defining the DO's, i.e.our polarizations, with the convention
$\psi = 0$, we get $\gamma_{\bar 1r} = {1\over {\sqrt{6}}}\,
(1,1,-2)$, $\gamma_{\bar 2r} = {1\over {\sqrt{2}}}\, (-1, 1, 0)$.
Then Eqs.(\ref{III7}) imply

\bigskip

\bea
 -\epsilon\,&& {}^3{\hat g}^{(\bar 1)}_{rs}(\tau ,\vec \sigma ) =
 \delta_{rs} + {}^4h_{rs}^{(\bar 1)}(\tau ,\vec \sigma ) {\buildrel {def}\over =}
 \delta_{rs}\, [1 + k^{(\bar 1)}_r(\tau ,\vec \sigma )] + O(r^2_{\bar a}),\nonumber \\
 &&{}\nonumber \\
 &&k^{\bar 1}_1(\tau ,\vec \sigma ) = k^{\bar 1}_2(\tau ,\vec \sigma
) = -\epsilon\, {{\sqrt{2}}\over 3}\, [ r_{\bar 1}(\tau ,\vec
\sigma ) - {1\over 2} \int d^3\sigma_1\, {{(\partial^2_{1,1} +
\partial^2_{1,2} + \partial^2_{1,3})\, r_{\bar 1}(\tau
 ,{\vec \sigma}_1)}\over {4\pi |\vec \sigma -{\vec
 \sigma}_1|}}],\nonumber \\
 &&k^{\bar 1}_3(\tau ,\vec \sigma ) = -\epsilon\, {{\sqrt{2}}\over
3}\, [ -2 r_{\bar 1}(\tau ,\vec \sigma ) - {1\over 2} \int
d^3\sigma_1\, {{(\partial^2_{1,1} + \partial^2_{1,2} +
\partial^2_{1,3})\, r_{\bar 1}(\tau
 ,{\vec \sigma}_1)}\over {4\pi |\vec \sigma -{\vec
 \sigma}_1|}}],\nonumber \\
 &&{}\nonumber \\
 &&{}\nonumber \\
 -\epsilon\,&& {}^3{\hat g}^{(\bar 2)}_{rs}(\tau ,\vec \sigma ) =
 \delta_{rs} + {}^4h_{rs}^{(\bar 2)}(\tau ,\vec \sigma ) {\buildrel {def}\over =}
 \delta_{rs}\, [1 + k^{(\bar 2)}_r(\tau ,\vec \sigma )] + O(r^2_{\bar
 a}),\nonumber \\
 &&{}\nonumber \\
 &&k^{\bar 2}_1(\tau ,\vec \sigma ) = -\epsilon\, \sqrt{{2\over
3}}\, [ - r_{\bar 2}(\tau ,\vec \sigma ) + {1\over 2} \int
d^3\sigma_1\, {{(\partial^2_{1,1} - \partial^2_{1,2} )\, r_{\bar
1}(\tau ,{\vec \sigma}_1)}\over {4\pi |\vec \sigma -{\vec
\sigma}_1|}}],\nonumber \\
 &&k^{\bar 2}_2(\tau ,\vec \sigma ) = -\epsilon\, \sqrt{{2\over
3}}\, [  r_{\bar 2}(\tau ,\vec \sigma ) + {1\over 2} \int
d^3\sigma_1\, {{(\partial^2_{1,1} - \partial^2_{1,2} )\, r_{\bar
1}(\tau ,{\vec \sigma}_1)}\over {4\pi |\vec \sigma -{\vec
\sigma}_1|}}],\nonumber \\
 &&k^{\bar 2}_3(\tau ,\vec \sigma ) = -\epsilon\, \sqrt{{2\over 3}}\,
[ 0 + {1\over 2} \int d^3\sigma_1\, {{(\partial^2_{1,1} -
\partial^2_{1,2} )\, r_{\bar 1}(\tau ,{\vec \sigma}_1)}\over {4\pi
|\vec \sigma -{\vec \sigma}_1|}}].
 \label{VI15}
 \eea

\bigskip

\begin{figure}
\vspace{1mm} \centerline{\epsfig{figure=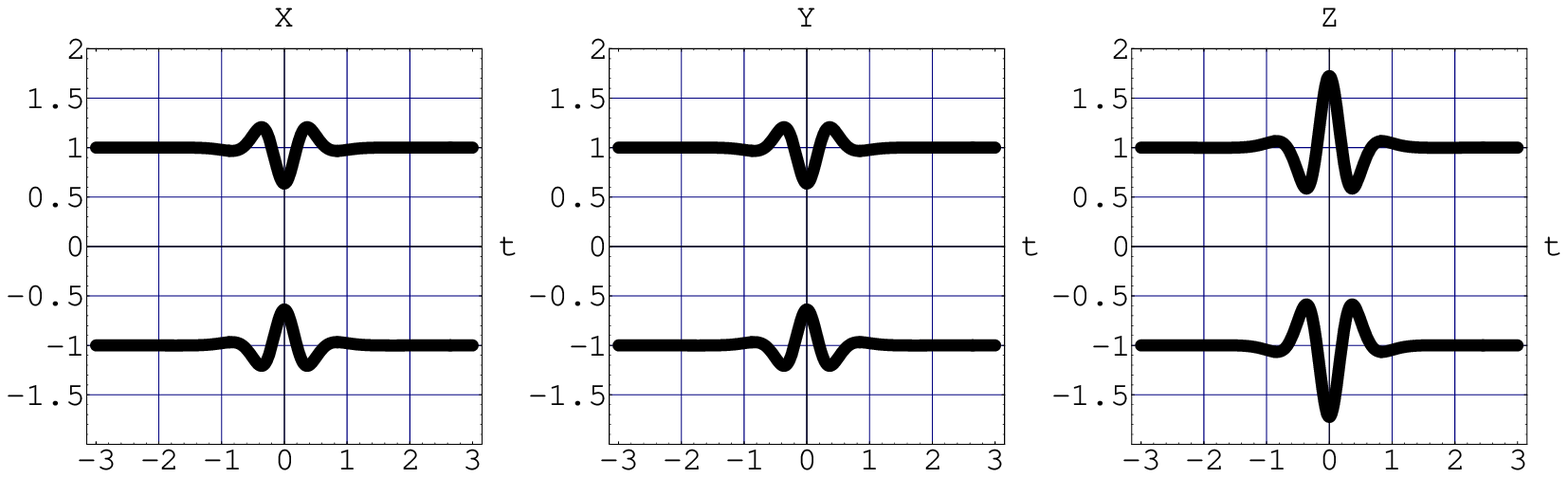,width=15cm}}
\vspace{5mm} \centerline{\epsfig{figure=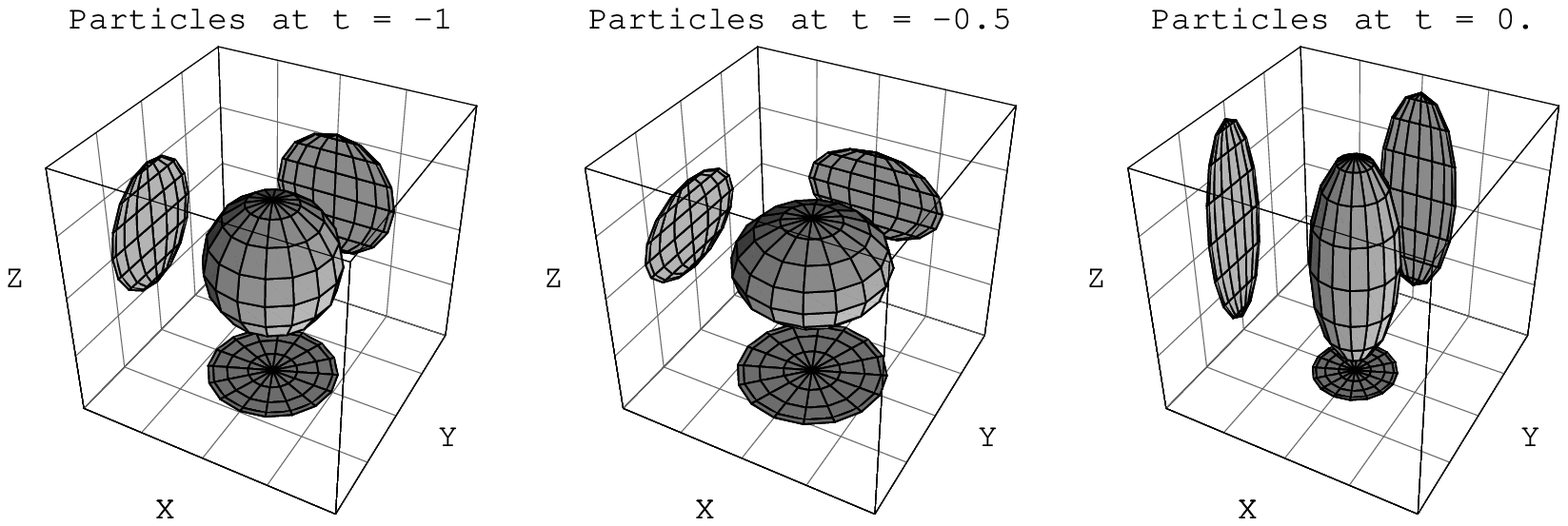,width=15cm}}
\vspace{5mm} \caption{Deformation of a sphere of particle at rest
induced by the passage of the gravitational wave packet of Eq.
(\ref{V15}) for $C_{\bar 1} \not= 0 $, $C_{\bar 2} = 0$.}
\label{fig:R1}
\end{figure}

\begin{figure}
\vspace{1mm} \centerline{\epsfig{figure=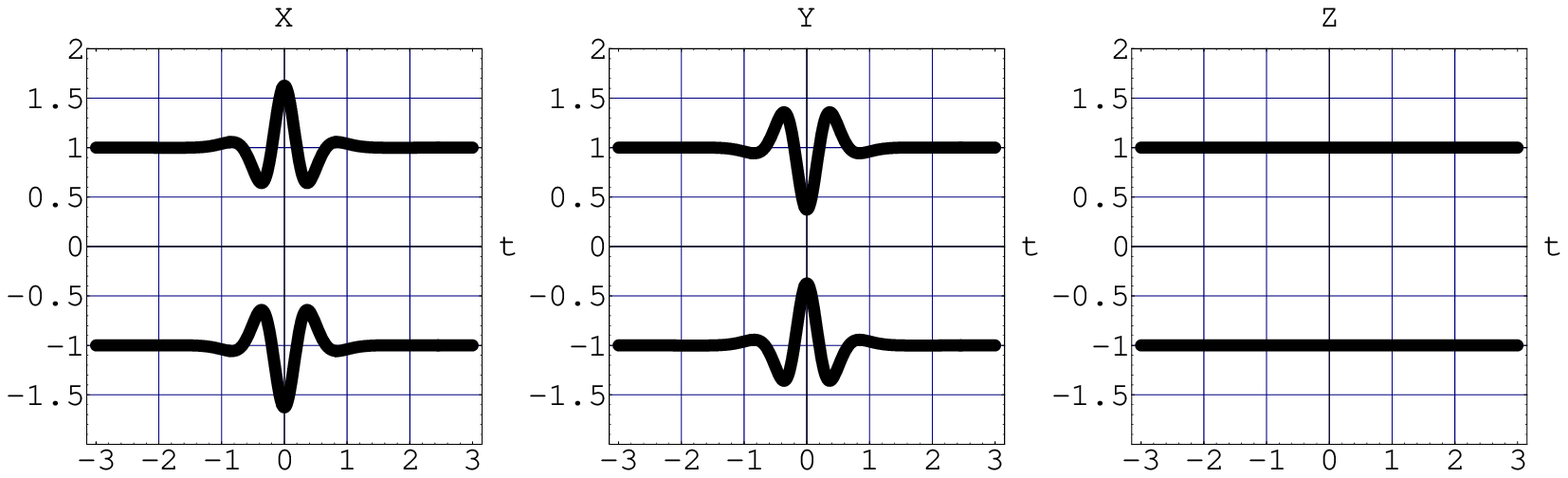,width=15cm}}
\vspace{5mm} \centerline{\epsfig{figure=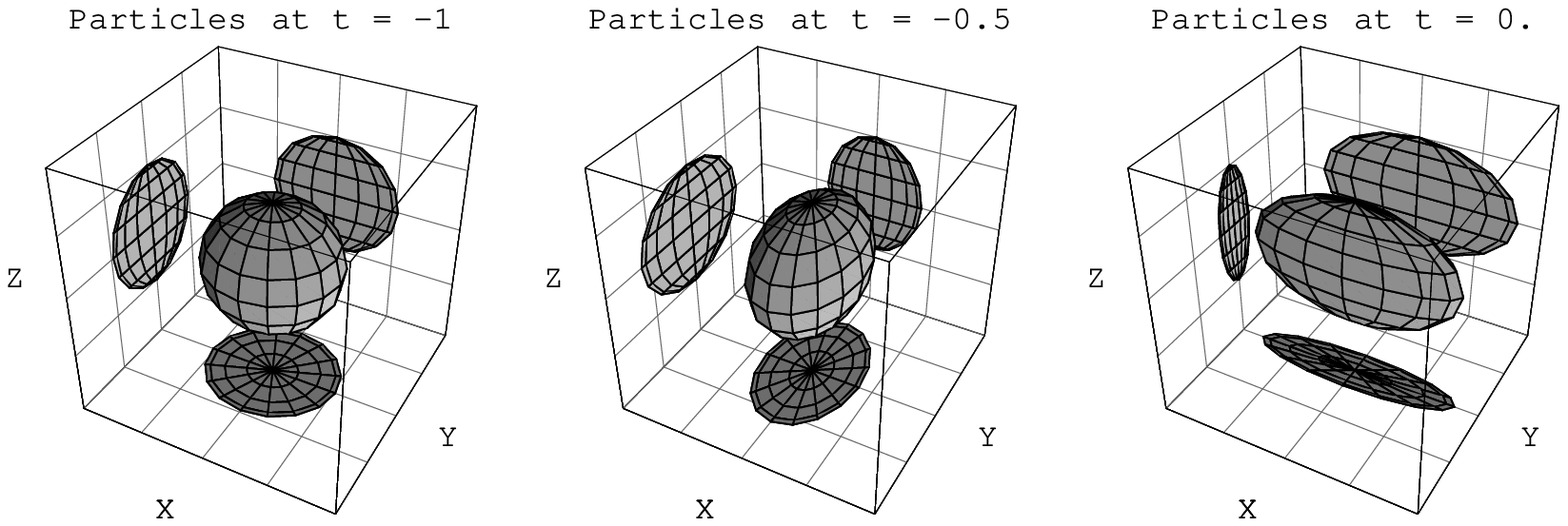,width=15cm}}
\vspace{5mm} \caption{Deformation of a sphere of particle at rest
induced by the passage of the gravitational wave packet of Eq.
(\ref{V15}) for $C_{\bar 1} = 0 $, $C_{\bar 2} \not= 0$.}
\label{fig:R2}
\end{figure}

If we  consider our (non transverse) solution given by the wave
packet (\ref{V15}) and we solve Eqs.(\ref{VI13}) numerically for a
sphere of particle at rest around the origin of the 3-coordinates
on a WSW hyper-surface, we obtain the two 3-dimensional
deformation patterns replacing the usual 2-dimensional ones for
the polarization in the TT harmonic gauge:

i) in figure \ref{fig:R1} there is the deformation pattern for the
case $C_{\bar 1} \not= 0$, $C_{\bar 2} = 0$, namely for $r_{\bar
1}(\tau ,\vec \sigma ) \not= 0$, $r_{\bar 2}(\tau ,\vec \sigma ) =
0$;

ii) in figure \ref{fig:R2} there is the deformation pattern for
the case $C_{\bar 1} = 0$, $C_{\bar 2} \not= 0$, namely for
$r_{\bar 1}(\tau ,\vec \sigma ) = 0$, $r_{\bar 2}(\tau ,\vec
\sigma ) \not= 0$.

\noindent In the two figures are reported the snapshots at three
different times ($t = -1, -0.5, 0$) of the sphere of particles
originally at rest (bottom) and the time evolution (from $t=-3$ to
$t=3$) of the six particles at the intersection of the three axes
and the sphere of particle (top), whose initial 3-coordinates are
$(1,0,0)$ and $(-1,0,0)$ on the $x$-axis, $(0,1,0)$ and $(0,-1,0)$
on the $y$-axis,  $(0,0,1)$ and $(0,0,-1)$ on the $z$-axis,
respectively. Since a particle on a $x$ axis will remain on the
same axes during the evolution, only the $x$ coordinates are
reported for the two particles on $x$ axes. The same
representation rule has been applied for the particles lying on
the other axes.

\subsection{A Relativistic Harmonic Oscillator as a Resonant Detector.}

Let us remark that in a TT harmonic gauge the basic idealization
of a test resonant detector on the Earth for gravitational waves
is a non-relativistic damped harmonic oscillator \cite{22,24}.
Given two equal masses $m$ located along the $r$ axis at the
positions $r_1(t)$ and $r_2(t)$, connected by a massless spring
with spring constant $k$, damping constant $\nu$ and unstretched
length $l_o$, one considers the proper extension $l(t) =
\int^{r_2(t)}_{r_1(t)} dt\, \sqrt{1 + h^{TT}_{rr}(t)} \approx [ 1
+ {1\over 2}\, h^{TT}_{rr}(t)]\, l_o$ in the metric of the
gravitational wave and defines $\xi = l - l_o \approx r_2 - r_1 -
l_o + {1\over 2}\, h^{TT}_{rr}(r_2-r_1)$. This leads to modify the
equation of motion ${{d^2 \xi}\over {dt^2}} + 2 \gamma\, {{d
\xi}\over {dt}} + \omega^2_o\, \xi = 0$ ($\omega_o^2 = 2k/m$,
$\gamma = \nu / m$), valid in absence of gravitational radiation,
to

 \beq
 {{d^2 \xi}\over {dt^2}} + 2 \gamma\, {{d \xi}\over {dt}} +
\omega^2_o\, \xi = {1\over 2}\, l_o\, {{d^2 h^{TT}_{rr}}\over
{dt^2}}.
 \label{VI16}
 \eeq

\noindent This same equation can be obtained as a consequence of
the tidal force at the lowest order from the geodesic deviation
equation \cite{24}, by considering the center of mass of the
detector as moving along a geodesic and by identifying $\xi$ with
the spatial part of the connecting vector $\triangle x^A(s)$ with
$\triangle x^{\tau}(s) =0$ ($\xi$ is twice the connecting vector
from the center of mass to one of the masses). Note that due to
the additional force terms the nearby world-line is no more a
geodesic.

\bigskip

Let us try to generalize this resonant detector to {\it special
relativity} and to see the effect on it of our post-Minkowskian
gravitational wave. As shown in Ref.\cite{36}, there is a
description of the relativistic 2-body problem on arbitrary
space-like hyper-surfaces $\Sigma_{\tau}$, leaves of a 3+1
splitting of Minkowski space-time. If $z^{\mu}(\tau ,\vec \sigma
)$ is the embedding of $\Sigma_{\tau}$ into Minkowski space-time,
the two positive energy scalar particles are described by
3-coordinates $\eta^r_i(\tau )$, $i=1,2$, on $\Sigma_{\tau}$, such
that $x^{\mu}_i(\tau ) = z^{\mu}(\tau , {\vec \eta}_i(\tau ))$ and
by the conjugate momenta $\kappa_{ir}(\tau )$. When the
hyper-surfaces are the Wigner hyper-planes, orthogonal to the
total 4-momentum of the 2-body system, we know how to replace the
canonical coordinates ${\vec \eta}_i$, ${\vec \kappa}_i$ with: i)
the canonical internal relativistic center of mass ${\vec q}_{+}$
and its conjugate momentum ${\vec \kappa}_{+} \approx 0$ (it is
vanishing since we are in the rest-frame instant form of
dynamics); ii) relative coordinates $\vec \rho$ and $\vec \pi$.
Even if we do not know explicitly how to express the old basis
${\vec \eta}_i$, ${\vec \kappa}_i$ in terms of the new basis due
to the non-linearity of the canonical transformation, we are able
to evaluate the Hamiltonian for the relative motion in the rest
frame. In the free case it is $H = \sqrt{m_1^2 + 2{\vec \pi}^2} +
\sqrt{m_2^2 + 2 {\vec \pi}^2}$. We can add action-at-a-distance
interactions either inside the square roots (DrozVincent - Komar -
Todorov  (DVKT) models \cite{38}),  $H = \sqrt{m_1^2 + V({\vec
\rho}^2) + 2{\vec \pi}^2} + \sqrt{m_2^2 + V({\vec \rho}^2) + 2
{\vec \pi}^2}$, or outside them (like the Coulomb potential for
charged particles \cite{39}),  $H = \sqrt{m_1^2 + 2{\vec \pi}^2} +
\sqrt{m_2^2 + 2 {\vec \pi}^2} + U({\vec \rho}^2)$. Due to the
square roots we cannot find explicitly  the Lagrangian and the
Euler-Lagrange equations for the relative motion in the rest
frame, except in the special equal mass case. For $m_1 = m_2 = m$
we get \cite{36} $L = - m\, \sqrt{4 - {\dot {\vec \rho}}^2}$ and
an interaction of the DVKT type is introduced by the replacement
$m \mapsto M = \sqrt{m^2 + V({\vec \rho}^2)}$. The Euler-Lagrange
equations are

\bea
 &&{d\over {d\tau}}\, {{M\, {\dot {\vec \rho}}}\over {\sqrt{4 - {\dot {\vec \rho}}^2} }}
 \cir {{\sqrt{4 - {\dot {\vec \rho}}^2}}\over M}\, V^{'}({\vec
 \rho}^2)\, \vec \rho,\qquad V^{'}({\vec \rho}^2) = {{d V({\vec
 \rho}^2)}\over {d\, {\vec \rho}^2}},\nonumber \\
 &&{}\nonumber \\
 &&or\nonumber \\
 &&{}\nonumber \\
 &&{\ddot {\vec \rho}} \cir {{4-{\dot {\vec \rho}}^2}\over {m^2 +
 V({\vec \rho}^2)}}\, V^{'}({\vec \rho}^2)\, [\vec \rho - {1\over
 2}\, \vec \rho \cdot {\dot {\vec \rho}}\, {\dot {\vec \rho}}].
 \label{VI17}
 \eea

\noindent For $V = - {m\over 2} k\, {\vec \rho}^2$ (relativistic
harmonic oscillator), the non-relativistic limit of these
equations becomes ${\ddot {\vec \rho}} + \omega_o^2\, \vec \rho =
0$. A damping term $- \nu\, {\dot {\vec \rho}}$ may be added by
hand.

\bigskip

Like in the non-relativistic case, we can adapt this relativistic
model, in the case of a test 2-body system, to curved space-time:
i) the relativistic center of mass moves along a geodesic
$\Gamma$, with 4-coordinates $\sigma^A(s)$; ii) the relative
variable $\vec \rho$ is the connecting vector in $\Sigma_{\tau}$
to a nearby (non-geodesic) world-line $\Gamma_1$. However, since
$\Sigma_{\tau}$ is a Riemannian 3-manifold, the flat vector $\vec
\rho$ has to be reinterpreted as the field of tangent vectors to
the 3-geodesic joining $\Gamma$ to $\Gamma_1$ on each
$\Sigma_{\tau}$ \cite{40}: if $\theta^r(s,\zeta )$ is a 3-geodesic
with affine parameter $\zeta$ \footnote{ ${{\partial^2
\theta^r(s,\zeta )}\over {\partial \zeta^2}} + {}^3{\hat
\Gamma}^r_{uv}(\theta (s,\zeta))\, {{\partial \theta^u(s,\zeta
)}\over {\partial \zeta}}\, {{\partial \theta^v(s,\zeta )}\over
{\partial \zeta}} = 0$ with the 3-Christoffel symbol of
Eqs.(\ref{d1}).} such that $\Gamma$ is $\theta^r(s,0)$ and
$\Gamma_1$ is $\theta^r(s,1)$, then we replace $V({\vec
\rho}^2(\tau ))$ with $V({\cal R}^2(s))$, where ${\cal R}^2(s) =
\int_0^1 d\zeta\, {{\partial \theta^r(s,\zeta )}\over {\partial
\zeta}}\, {}^3g_{rs}(\tau , \theta(s,\zeta ))\, {{\partial
\theta^s(s,\zeta )}\over {\partial \zeta}}$ (it is independent
from $\zeta$ because the tangent vector ${{\partial
\theta^r(s,\zeta )}\over {\partial \zeta}}$ is parallel
transported along the 3-geodesic). Analogously we make the
replacements $\vec \rho \cdot {\dot {\vec \rho}} \mapsto {\cal
R}_1(s) = \int_0^1 d\zeta\, {{\partial \theta^r(s,\zeta )}\over
{\partial \zeta}}\, {}^3g_{rs}(\tau ,\theta (s,\zeta ))\,
{{\partial^2 \theta^s(s,\zeta )}\over {\partial s\, \partial
\zeta}}$ and ${\dot {\vec \rho}}^2 \mapsto {\cal R}_2(s) =
\int_0^1 d\zeta\, {{\partial^2 \theta^r(s,\zeta )}\over {\partial
s\, \partial \zeta}}\, {}^3g_{rs}(\tau ,\theta (s,\zeta ))\,
{{\partial^2 \theta^s(s,\zeta )}\over {\partial s\,
\partial \zeta}}$.

By identifying $\rho^r(\tau ) \mapsto \triangle \sigma^r(s)$,
${\dot \rho}^r(\tau ) \mapsto {{d \triangle \sigma^r(s)}\over
{ds}}$, ${\ddot {\vec \rho}}^r \mapsto {{d^2 \triangle
\sigma^r(s)}\over {ds^2}}$, and by adding the force terms (damping
included) of Eqs.(\ref{VI17}) in the geodesic deviation equation
(\ref{VI9}) we get

\bea
 {{d^2 \triangle x^r(s)}\over {ds^2}} &+& \Big[ 2\, {}^4{\hat
\Gamma}^r_{\tau s}(s) + {{4-{\cal R}_2(s)}\over {m^2 + V({\cal
R}^2(s))}}\, V^{'}({\cal R}^2(s))\, (\nu + {1\over 2}\, {\cal
 R}_1(s)) \Big]\, {{d \triangle x^s(s)}\over {ds}} +\nonumber \\
 &+&\Big[  {{d\, {}^4{\hat \Gamma}^r_{\tau s}(s)}\over
{ds}} + {}^4{\hat R}^r{}_{\tau s\tau}(s) - {{4-{\cal R}_2(s)}\over
{m^2 + V({\cal R}^2(s))}}\, V^{'}({\cal R}^2(s)) \Big]\, \triangle
x^s(s) = 0.
 \label{VI18}
 \eea

\noindent {\it This is the post-Minkowskian counterpart in our
coordinates of the non-relativistic equation (\ref{VI16}) of the
TT harmonic gauge}.

For very small $|\vec \rho |$ we can make the approximations
${\cal R}^2 \approx  {{\partial \theta^r(s,\zeta )}\over {\partial
\zeta}}{|}_{\zeta =0}\, {}^3g_{rs}(\sigma^A(s))\, {{\partial
\theta^s(s,\zeta )}\over {\partial \zeta}}{|}_{\zeta =0}$, ${\cal
R}_1 \approx {{\partial \theta^r(s,\zeta )}\over {\partial
\zeta}}{|}_{\zeta =0}\, {}^3g_{rs}(\sigma^A(s))\, {{\partial^2
\theta^s(s,\zeta )}\over {\partial s\, \partial \zeta}}{|}_{\zeta
=0}$, ${\cal R}_2 \approx {{\partial^2 \theta^r(s,\zeta )}\over
{\partial s\, \partial \zeta}}{|}_{\zeta =0} \,
{}^3g_{rs}(\sigma^A(s))\, {{\partial^2 \theta^s(s,\zeta )}\over
{\partial s\, \partial \zeta}}{|}_{\zeta =0}$: this is a
pole-dipole approximation of the test 2-body problem.

\vfill\eject

\section{The Embedding of the WSW Hyper-surfaces, the Associated Congruence of Time-Like
Observers and the Null Geodesics.}

\subsection{The Post-Minkowskian WSW Hyper-surfaces.}

The WSW triads  ${}^3e^{(WSW)\, r}_{(a)}(\tau ,\vec \sigma )\,
{\rightarrow}_{|\vec \sigma | \rightarrow \infty}\,
\delta^r_{(a)}$, parallel transported from spatial infinity on the
linearized WSW CMC hyper-surface, and the associated cotriads
${}^3e^{(WSW)}_{(a)r}(\tau ,\vec \sigma )$
\footnote{${}^3e^{(WSW)\, r}_{(a)}={}^3{\hat g}^{rs}\,
{}^3e^{(WSW)}_{(a)s}= -\epsilon \delta^{rs}\,
{}^3e^{(WSW)}_{(a)s}+2\epsilon \delta^r_{(a)}[q+{2\over
{\sqrt{3}}} \sum_{\bar a}\gamma_{\bar ar}r_{\bar a}]+O(r^2_{\bar
a})$.} are the solution of Eqs.(\ref{a18}), whose linearized form
is

\bea
 &&\partial_r \Big( \sqrt{\hat \gamma}\, {}^3e^{(WSW)
r}_{(a)}\Big)(\tau ,\vec \sigma ) =\partial_r\, {}^3e^{(WSW)
r}_{(a)}(\tau ,\vec \sigma )+3\delta^r_{(a)} \partial_rq(\tau
,\vec \sigma ) +O(r^2_{\bar a})=0,\nonumber \\
 &&{}\nonumber \\
 &&\partial_1\, {}^3e^{(WSW)}_{(2) 3}(\tau ,\vec \sigma ) +
 \partial_3\, {}^3e^{(WSW)}_{(1)2}(\tau ,\vec \sigma )  +
 \partial_2\, {}^3e^{(WSW)}_{(3)1}(\tau ,\vec \sigma ) =\nonumber \\
 &&=\Big[ {}^3{\hat \Gamma}^2_{13} + {}^3{\hat \Gamma}^1_{32} + {}^3{\hat \Gamma}^3_{21}\Big]
 (\tau ,\vec \sigma ) + O(r^2_{\bar a}) =0 +  O(r^2_{\bar a}),
 \label{VII1}
 \eea

\noindent where Eqs.(\ref{d1}) have been used.

\medskip

These equations imply

\bea
 {}^3e^{(WSW)\, r}_{(a)}(\tau ,\vec \sigma ) &=&
 f^r_{(a)\perp}(\tau ,\vec \sigma ) + {{\sqrt{3}}\over 2}\,
 \delta^r_{(a)}\,
 \sum_{\bar au}\, \gamma_{\bar au}\, \int d^3\sigma_1\,
 {{\partial^2_{1u}\, r_{\bar a}(\tau ,{\vec \sigma}_1)}\over
 {4\pi\, |\vec \sigma - {\vec \sigma}_1|}},\nonumber \\
 &&{}\nonumber \\
 &&\partial_r\, f^r_{(a)\perp}(\tau ,\vec \sigma ) = 0,\qquad
f^r_{(a)\perp}(\tau ,\vec \sigma )\, {\rightarrow}_{|\vec \sigma|
\rightarrow \infty}\, \delta^r_{(a)},\nonumber \\
 &&{}\nonumber \\
 &&\partial_1\, f^3_{(2)\perp} + \partial_3\, f^2_{(1)\perp} +
 \partial_2\, f^1_{(3)\perp} = 0 + O(r^2_{\bar a}),\nonumber \\
 &&{}\nonumber \\
 &&\Rightarrow\quad f^r_{(a)\perp} = \delta^r_{(a)} + (\delta^{rs}
 + {{\partial^r\, \partial^s}\over {\triangle}})\, g_{(a)s},
 \label{VII2}
 \eea

\noindent with $g_{(a)s}$ arbitrary. The simplest solution is to
take $g_{(a)s} = 0$, so that the WSW triads  are

\beq
 {}^3e^{(WSW)r}_{(a)} = \delta^r_{(a)}\, \Big[ 1 +  {{\sqrt{3}}\over 2}\,
 \sum_{\bar au}\, \gamma_{\bar au}\, \int d^3\sigma_1\,
 {{\partial^2_{1u}\, r_{\bar a}(\tau ,{\vec \sigma}_1)}\over
 {4\pi\, |\vec \sigma - {\vec \sigma}_1|}}\Big].
 \label{VII3}
 \eeq

Eqs.(\ref{a19}), allow to find the associated WSW adapted tetrads
(preferred ADM Eulerian observers or  {\it asymptotic fixed stars}
giving a {\it local compass of inertia} to be compared with local
(Fermi-Walker transported or not) gyroscopes)

\bea
 &&{}^4_{(\Sigma )}{\check {\tilde E}}^{(WSW) A}_{(o)}(\tau
,\vec \sigma ) =-\epsilon \Big( 1;\, -n_r(\tau ,\vec \sigma
)\Big),\nonumber \\ &&{}^4_{(\Sigma )}{\check {\tilde E}}^{(WSW)
A}_{(a)}(\tau ,\vec \sigma ) = \Big( 0;\, {}^3e^{(WSW)\,
r}_{(a)}(\tau ,\vec \sigma )\Big) .
 \label{VII4}
 \eea

\medskip

Once the WSW triads are known, Eqs.(\ref{a20}) and (\ref{a21})
give the embedding of the linearized WSW CMC hyper-surfaces into
the linearized space-time

\bea z^{\mu}_{(WSW)}(\tau ,\vec \sigma ) &=& x^{\mu}_{(\infty
)}(0) + b^{\mu}_A(\tau ,\vec \sigma ) F^A(\tau ,\vec \sigma
),\nonumber \\
 &&{}\nonumber \\
 &&F^{\tau}(\tau ,\vec \sigma ) = -\epsilon \tau ,\nonumber \\
 &&F^s(\tau ,\vec \sigma ) ={}^3e^{(WSW)\, s}_{(a)}(\tau ,\vec
\sigma ) \delta_{(a)r} \sigma^r +\epsilon \tau n_s(\tau ,\vec
\sigma ),
 \label{VII5}
  \eea

\noindent with the transition coefficients $b^{\mu}_A={{\partial
z^{\mu}_{(WSW)}}\over {\partial \sigma^A}}$ solution of the linear
partial differential equations

\beq
 F^B(\tau ,\vec \sigma) {{\partial b^{\mu}_B(\tau ,\vec \sigma
)}\over {\partial \sigma^A}} = \Big( \delta^B_A - {{\partial
F^B(\tau ,\vec \sigma )}\over {\partial \sigma^A}}\Big)
b^{\mu}_B(\tau ,\vec \sigma ).
 \label{VII6}
  \eeq

In adapted coordinates Eqs.(\ref{VII5}) of the embedding of
linearized WSW CMC hyper-surfaces $\Sigma_{\tau}$ become

\beq
 z^A_{(WSW)}(\tau ,\vec \sigma ) = x^A_{(\infty )}(0) + F^A(\tau
 ,\vec \sigma ),
 \label{VII7}
 \eeq

\noindent with $z^{\tau}_{(WSW)}(\tau ,\vec \sigma ) =
x^{\tau}_{(\infty )}(0) - \epsilon\, \tau$. The time-like
evolution vector is $z^{\mu}_{(WSW)\tau}(\tau ,\vec \sigma ) =
\Big( -\epsilon; [\epsilon\, n_r + \epsilon\, \tau\,
\partial_{\tau}\, n_r + \partial_{\tau}\, {}^3e_{(a)}^{(WSW) r}\,
\sigma^a](\tau ,\vec \sigma ) \Big)$.

\subsection{Congruences of Timelike Observers.}

In Eqs.(\ref{III9}) we have the expression of the
$\Sigma^{(WSW)}_{\tau}$-adapted tetrads and cotetrads. In
particular there is the expression for the contra- and co-variant
normals, $l^A(\tau ,\vec \sigma )$ and $l_A(\tau ,\vec \sigma )$,
to $\Sigma^{(WSW)}_{\tau}$. The associated 4-velocity field
defines a (non-rotating, surface forming) congruence of time-like
observers orthogonal to $\Sigma^{(WSW)}_{\tau}$, using the
parameter $\tau$, labeling the leaves of the foliation, as
evolution parameter \footnote{This is the {\it hyper-surface point
of view} according to Ref.\cite{41}. Instead, the {\it threading
point of view} is a description involving only a rotating
congruence of observers (like the one which can be built with the
4-velocity field associated to $z^A_{(WSW)\tau}(\tau ,\vec \sigma
){}{}$) : since this congruence, being rotating, is not
surface-forming (non-zero vorticity), in each point we can only
divide the tangent space in the direction parallel to the
4-velocity and the orthogonal complement (the local rest frame).
On the other hand, the {\it slicing point of view}, originally
adopted in ADM canonical gravity, uses two congruences: the
non-rotating one with the normals to $\Sigma_{\tau}$ as 4-velocity
fields and a second (rotating, non-surface-forming) congruence of
observers, whose 4-velocity field is the field of time-like unit
vectors determined by the $\tau$ derivative of the embeddings
identifying the leaves $\Sigma_{\tau}$ (the so-called evolution
vector field). Furthermore, it uses the affine parameter
describing the world-lines of this second family of observers as
Hamiltonian evolution parameter.}. Let us remark that $\tau$ {\it
is not in general the proper time of any observer of the
congruence}.

As for any congruence, we have the decomposition ($P_{AB} =
{}^4g_{AB} - l_A\, l_B$)

\bea
 {}^4\nabla_{A}\, l_{B} &=&  l_{A}\, a_{B} + {1\over
3}\, \Theta \, P_{AB} + \sigma_{AB} + \omega_{AB},\nonumber \\
 &&a^{A} = l^{B}\, {}^4\nabla_{B}\, l^{A},\qquad
 4-acceleration,\nonumber \\
 &&\Theta = {}^4\nabla_{A}\, l^{A},\quad scalar\, (volume)\, rate\,
 of\, expansion,\nonumber \\
 &&\sigma_{AB} = {1\over 2}\, (a_{A}\, l_{B} + a_{B}\,
 l_{A}) + {1\over 2}\, ({}^4\nabla_{A}\,
l_{B} + {}^4\nabla_{B}\, l_{A}) - {1\over 3}\, \Theta\,
P_{AB},\nonumber \\
 &&{}{}{}{} rate\, of\, shear\, tensor\, (with\, magnitude\,
\sigma^2={1\over 2} \sigma_{AB}\sigma^{AB}),\nonumber \\
 &&\omega_{AB} = - \omega_{BA} = \epsilon_{ABCD}\,
 \omega^{C}\, l^{D} = {1\over 2}\, (a_{A}\, l_{B} - a_{B}\,
 l_{A}) + {1\over 2}\, ({}^4\nabla_{A}\, l_{B} - {}^4\nabla_{B}\,
  l_{A}) = 0,\nonumber \\
 && twist\, or\, vorticity\, tensor,\qquad \omega^{A} = {1\over
2}\, \epsilon^{ABCD}\, \omega_{BC}\, l_{D} = 0,\quad vorticity\,
vector.
 \label{VII8}
 \eea

\noindent $\Theta$ is the {\it expansion} (it measures the average
expansion of the infinitesimally nearby world-lines surrounding a
given world-line in the congruence), $\sigma _{AB}$ the {\it
shear} (it measures how an initial sphere in the tangent space to
the given world-line, which is Lie transported along $l^{A}$, will
distort toward an ellipsoid with principal axes given by the
eigenvectors of $\sigma^{A}{}_{B}$ with rate given by the
eigenvalues of $\sigma^{A}{}_{B}$) and $\omega_{AB}$ the {\it
twist or vorticity} (it measures the rotation of the
infinitesimally nearby world-lines surrounding the given one);
$\sigma_{AB}$ and $\omega _{AB}$ are purely spatial ($\sigma_{AB}
l^{B} = \omega_{AB} l^{B} = 0$). Due to the Frobenius theorem, the
congruence is (locally) hyper-surface orthogonal if and only if
$\omega_{AB}=0$. The equation ${1\over l}\, l^{A}\,
\partial_{A}\, l = {1\over 3}\, \Theta$ defines a representative
length $l$ along the world-line of $l^{A}$, describing the volume
expansion (contraction) behaviour of the congruence completely.

\medskip

The linearized acceleration of the observers vanishes (i.e. at the
lowest level we get inertial observers) so that there is no
gravito-electric force on test particles in the sense of
Ref.\cite{41} (geodesic in local rest frame)

\beq
 a^A = l^B\, \partial_B\, l^A + {}^4\Gamma^A_{BC}\, l^Bl^C = ( 0;
-\partial_{\tau} \, n_r + {}^4\Gamma^r_{\tau\tau}) + O(r^2_{\bar
a}) = 0 + O(r^2_{\bar a}).
 \label{VII9}
 \eeq

\bigskip

By using Eqs.(\ref{d5})  we get ${}^4\nabla_A\, l_B = -
{}^4\Gamma^C_{AB}\, l_C = \epsilon\, {}^4\Gamma^{\tau}_{AB} +
O(r^2_{\bar a})$. The expansion of the congruence is

\beq
 \Theta = {}^4\nabla_A\, l^A = \partial_A\, l^A +
{}^4\Gamma^A_{AB}\, l^B = \epsilon \vec \partial \cdot \vec n -
\epsilon\, \vec \partial \cdot \vec n + O(r^2_{\bar a}) = 0 +
O(r^2_{\bar a}).
 \label{VII10}
 \eeq

The shear is not zero so that there is gravito-magnetism on test
particles in the sense of Ref.\cite{41} (geodesic in local rest
frame)

\beq
 \sigma_{AB} = {1\over 2} \, (a_A\, l_B + a_B\, l_A) +{1\over 2}\,
({}^4\nabla_A\, l_B + {}^4\nabla_B\, l_A) = \epsilon\,
{}^4\Gamma^{\tau}_{AB} + O(r^2_{\bar a}),
 \label{VII11}
 \eeq

\noindent while the vorticity vanishes (the congruence is surface
forming)

\beq
 \omega_{AB} = {1\over 2} \, (a_A\, l_B - a_B\, l_A) +{1\over 2}\,
({}^4\nabla_A\, l_B - {}^4\nabla_B\, l_A) = 0 + O(r^2_{\bar a}).
 \label{VII12}
 \eeq

\bigskip

If $x^A_{{\vec \sigma}_o}(\tau )$ is the time-like world-line, in
adapted coordinates,  of the observer crossing the leave
$\Sigma_{\tau_o}$ at ${\vec \sigma}_o$, we have

\bea
 x^A_{{\vec \sigma}_o}(\tau ) &=& \Big( \tau ; {\vec \rho}_{{\vec
 \sigma}_o}(\tau )\Big) =\nonumber \\
 &=& \Big( \tau_o - \epsilon\, [z^{\tau}_{(WSW)}(\tau ,  {\vec \rho}_{{\vec
 \sigma}_o}(\tau )\, ) - z^{\tau}_{(WSW)}(\tau_o, {\vec
 \sigma}_o)]; \nonumber \\
 &&\sigma^r_o + z^r_{(WSW)}(\tau ,  {\vec \rho}_{{\vec
 \sigma}_o}(\tau )\, ) - z^r_{(WSW)}(\tau_o, {\vec
 \sigma}_o)\Big),\nonumber \\
 &&{}\nonumber \\
 x^A_{{\vec \sigma}_o}(\tau_o) &=& \Big( \tau_o; {\vec
 \sigma}_o\Big),\qquad {\vec
 \rho}_{{\vec \sigma}_o}(\tau_o) = {\vec \sigma}_o.
 \label{VII13}
 \eea

\noindent The effective trajectory $ {\vec \rho}_{{\vec
 \sigma}_o}(\tau )$ is determined by solving the equations $ {\vec \rho}_{{\vec
 \sigma}_o}(\tau ) = {\vec \sigma}_o + {\vec z}_{(WSW)}(\tau ,  {\vec \rho}_{{\vec
 \sigma}_o}(\tau )\, ) - {\vec z}_{(WSW)}(\tau_o, {\vec
 \sigma}_o)$ with ${\vec z}_{(WSW)}$ given by Eq.(\ref{VII7}). The
 4-velocity ${\dot x}^A_{{\vec \sigma}_o}(\tau ) = {{d x^A_{{\vec
 \sigma}_o}(\tau )}\over {d \tau}}$ satisfies

\beq
 l^A_{{\vec \sigma}_o}(\tau ) = l^A(\tau , {\vec
 \rho}_{{\vec \sigma}_o}(\tau )) = {{{\dot x}^A_{{\vec
 \sigma}_o}(\tau )}\over {\sqrt{{}^4g_{BC}(x_{{\vec \sigma}_o}(\tau ))\,
 {\dot x}^B_{{\vec \sigma}_o}(\tau )\, {\dot x}^C_{{\vec \sigma}_o}(\tau
 )} }},
 \label{VII14}
 \eeq

\noindent with the observer acceleration given by $a^A_{{\vec
\sigma}_o}(\tau ) = {{d l^A_{{\vec
 \sigma}_o}(\tau )}\over {d \tau}} = 0 + O(r^2_{\bar a})$, $a^A_{{\vec
 \sigma}_o}(\tau )\, l_{{\vec \sigma}_o\, A}(\tau ) = 0$.

Since ${\dot x}^{\tau}_{{\vec \sigma}_o}(\tau ) = \tau$, ${\dot
x}^r_{{\vec \sigma}_o}(\tau ) = O(r_{\bar a})$, we get
${}^4g_{AB}\, {\dot x}^A_{{\vec \sigma}_o}\, {\dot x}^B_{{\vec
\sigma}_o} = \epsilon + O(r^2_{\bar a})$, namely that to this
order $\tau$ is the proper time of any observer of the congruence.

\medskip

Yet, the ADM canonical formalism gives us an additional
information. Actually, on each space-like hyper-surface
$\Sigma^{(WSW)}_{\tau}$ of the foliation, there is a privileged
contravariant space-like direction identified by the lapse and
shift gauge variables

\bea
 {\cal N}^A(\tau ,\vec \sigma ) &=& {1\over {|\vec N(\tau
,\vec \sigma )|}}\, \Big( 0; n^r(\tau ,\vec \sigma )\Big) = \Big(
0; n_r(\tau ,\vec \sigma ) \Big) + O(r^2_{\bar a}),\nonumber \\
 {\cal N}_A(\tau ,\vec \sigma ) &=& |\vec N(\tau ,\vec \sigma
 )|\, \Big( 1; {{N_r(\tau ,\vec \sigma )}\over {|\vec N(\tau ,\vec
 \sigma )|^2}}\Big) = \Big( |\vec n(\tau ,\vec \sigma )|;
 n_r(\tau ,\vec \sigma )\Big) + O(r^2_{\bar a}),\nonumber \\
 &&{}\nonumber \\
 &&{\cal N}^A(\tau ,\vec \sigma )\, l_A(\tau ,\vec \sigma
 ) = 0,\qquad {\cal N}^{\mu}(\tau ,\vec \sigma )\, {\cal N}_{\mu}(\tau ,\vec \sigma )
 = - \epsilon,\nonumber \\
 && |\vec N(\tau ,\vec \sigma )| = \sqrt{({}^3g_{rs}\, N^r\, N^s)(\tau ,\vec \sigma
 )} = |\vec n(\tau ,\vec \sigma )| + O(r^2_{\bar a}).
 \label{VII15}
 \eea

If 4-coordinates exist, corresponding to an on-shell complete
Hamiltonian gauge fixing, such that the vector field identified by
${\cal N}^A(\tau ,\vec \sigma )$ on each $\Sigma^{(WSW)}_{\tau}$
is surface-forming (zero vorticity \footnote{This requires that
${\cal N}_A\, d\sigma^A$ is  a closed 1-form, namely that we have
$\partial_{\tau}\, n_r = \partial_r\, |\vec n| + O(r^2_{\bar a})$
and $\partial_r\, n_s = \partial_s\, n_r + O(r^2_{\bar a}) $. In
turn, this requires $n_r = \partial_r\, f + O(r^2_{\bar a})$ with
$\partial_{\tau}\, f = |\vec n| + const.$ Our gauge has a
non-surface forming ${\cal N}^A$.}), then each $\Sigma_{\tau}$ can
be foliated with 2-surfaces, and the 3+1 splitting of space-time
becomes a (2+1)+1 splitting corresponding to the 2+2 splittings
studied by Stachel and d'Inverno \cite{42}.

We have therefore a natural candidate for one of the three spatial
triads of each observer: $E^A_{{\vec \sigma}_o\, ({\cal N})}(\tau
) = {\cal N}^A_{{\vec \sigma}_o}(\tau ) = {\cal N}^A(\tau ,{\vec
\rho}_{{\vec \sigma}_o}(\tau ))$. By means of $l^A_{{\vec
\sigma}_o}(\tau ) = l^A(\tau ,{\vec \rho}_{{\vec \sigma}_o}(\tau
))$ and ${\cal N}^A_{{\vec \sigma}_o}(\tau )$, we can construct
two null vectors at each space-time point

\bea
 &&{\cal K}^A_{{\vec \sigma}_o}(\tau ) = \sqrt{{{|\vec n|}\over
 2}}\, \Big( l^A_{{\vec \sigma}_o}(\tau ) + {\cal N}^A_{{\vec
\sigma}_o}(\tau ) \Big) = \sqrt{{{1}\over
 2}}\, \Big( l^A_{{\vec \sigma}_o}(\tau ) + {\cal N}^A_{{\vec
\sigma}_o}(\tau ) \Big) + O(r^2_{\bar a}),\nonumber \\
 &&{\cal L}^A_{{\vec \sigma}_o}(\tau ) = {1\over {\sqrt{2\, |\vec
 n|}}}\, \Big( l^A_{{\vec \sigma}_o}(\tau ) - {\cal N}^A_{{\vec
\sigma}_o}(\tau ) \Big) = {1\over {\sqrt{2}}}\, \Big( l^A_{{\vec
\sigma}_o}(\tau ) - {\cal N}^A_{{\vec \sigma}_o}(\tau ) \Big) +
O(r^2_{\bar a}).
 \label{VII16}
 \eea

\noindent and then we can arrive at a {\it null tetrad} of the
type used in the Newman-Penrose formalism \cite{43}. The last two
axes of the spatial triad can be chosen as two space-like circular
complex polarization vectors $E^{\mu}_{{\vec \sigma}_o\, (\pm
)}(\tau )$, like in electromagnetism. They are built starting from
the transverse helicity polarization vectors $E^{\mu}_{{\vec
\sigma}_o\, (1,2)}(\tau )$, which are the first and second columns
of the standard Wigner helicity boost generating ${\cal
K}^{\mu}_{{\vec \sigma}_o}(\tau )$ from the reference vector
${\buildrel \circ \over {\cal K}}{}^{\mu}_{{\vec \sigma}_o}(\tau )
= |\vec N|\, \Big( 1; 001\Big)$.

Let us call $E^{(ADM) \mu}_{{\vec \sigma}_o\, (\alpha )}(\tau )$
the ADM tetrad formed by $l^{\mu}_{{\vec \sigma}_o}(\tau )$,
${\cal N}^{\mu}_{{\vec \sigma}_o}(\tau )$, $E^{\mu}_{{\vec
\sigma}_o\, (1,2)}(\tau )$. This tetrad will not be in general
Fermi-Walker transported along the world-line $x^{\mu}_{{\vec
\sigma}_o}(\tau )$ of the observer.

\medskip

Another possible (but only on-shell) choice of the spatial triad
together with the unit normal to $\Sigma^{(WSW)}_{\tau}$ is the
{\it local WSW (on-shell) compass of inertia} defined in
Eqs.(\ref{VII4}). This local compass corresponds to the {\it
standard of non rotation with respect to the fixed stars} and its
$\tau$-evolution, dictated by Einstein's equations, does not
correspond to the FW transport, which is defined independently
from them using only local geometrical and group-theoretical
concepts. The WSW local compass of inertia corresponds to pointing
to the fixed stars with a telescope and is needed in a satellite
like Gravity Probe B to detect the frame dragging (or
gravito-magnetic Lense-Thirring effect) of the inertial frames by
means of the rotation with respect to it of a FW transported
gyroscope.

\medskip

Given the 4-velocity $l^A_{{\vec \sigma}_o}(\tau ) = E^A_{{\vec
\sigma}_o}(\tau )$ of the observer, the observer spatial triads
$E^A_{{\vec \sigma}_o\, (a)}(\tau )$, $a = 1,2,3$, have to be
chosen in a conventional way, namely by means of a conventional
assignment of an origin for the local measurements of rotations.
Usually, the choice corresponds to Fermi-Walker (FW) transported
({\it gyroscope-type transport, non-rotating observer}) tetrads
$E^{(FW)\, A}_{{\vec \sigma}_o\, (\alpha )}(\tau )$, such that (we
show also the implication of the linearization)

\begin{eqnarray}
{D\over {D\tau}}\, E^{(FW)\, A}_{{\vec \sigma}_o\, (a)}(\tau ) &=&
\Omega^{(FW)}_{{\vec \sigma}_o}{}^A{}_B(\tau )\, E^{(FW)\,
B}_{{\vec \sigma}_o\, (a)}(\tau ) = l^A_{{\vec \sigma}_o}(\tau )\,
a_{{\vec \sigma}_o\, B}(\tau )\, E^{(FW)\, B}_{{\vec \sigma}_o\,
(a)}(\tau ) = 0 + O(r^2_{\bar a}),\nonumber \\
 &&\Omega^{(FW)}_{{\vec \sigma}_o}{}^{AB}(\tau ) =
 a^{A}_{{\vec \sigma}_o}(\tau )\, l^{B}_{{\vec \sigma}_o}(\tau
 ) - a^{B}_{{\vec \sigma}_o}(\tau )\, l^{A}_{{\vec \sigma}_o}(\tau
 ).
 \label{VII17}
 \end{eqnarray}

\noindent The triad $E^{(FW)\, A}_{{\vec \sigma}_o\, (a)}(\tau )$
is the correct relativistic generalization of global Galilean
non-rotating frames. Naturally any other choice of the triads (Lie
transport, co-rotating-FW transport,...) is possible. A generic
triad $E^{A}_{{\vec \sigma}_o\, (a)}(\tau )$ will satisfy ${D\over
{D\tau}}\, E^{A}_{{\vec \sigma}_o\, (a)}(\tau ) = \Omega_{{\vec
\sigma}_o}{}^{A}{}_{B}(\tau )\, E^{B}_{{\vec \sigma}_o\, (a)}(\tau
)$ with $\Omega^{AB}_{{\vec \sigma}_o} = \Omega^{(FW)\, AB}_{{\vec
\sigma}_o} + \Omega^{(SR)\, AB}_{{\vec \sigma}_o}$ with the
spatial rotation part $\Omega^{(SR)\, AB}_{{\vec \sigma}_o} =
\epsilon^{ABCD}\, l_{{\vec \sigma}_o\, C}\, J_{{\vec \sigma}_o\,
D}$, $J^A_{{\vec \sigma}_o}\, l_{{\vec \sigma}_o\, A} = 0$,
producing a rotation of the gyroscope in the local space-like
2-plane orthogonal to $l^A_{{\vec \sigma}_o}$ and $J^A_{{\vec
\sigma}_o}$.

\medskip

In the linearized theory on the WSW hyper-surfaces of our gauge FW
transport implies no $\tau$-dependence for the triads $E^{(FW)\,
A}_{{\vec \sigma}_o\, (a)}(\tau )$. Neither the ADM tetrads nor
the WSW tetrads (\ref{VII4}) are FW transported.

\bigskip

See Ref.\cite{41} for the description of a geodesics $y^{\mu}(\tau
)$, the world-line of a scalar test particle, from the point of
view of those observers $\gamma_{{\vec \sigma}_o, y(\tau )}$ in
the congruence which intersect it, namely such that at $\tau$ it
holds $x^{\mu}_{{\vec \sigma}_o, y(\tau )}(\tau ) = y^{\mu}(\tau
)$. The family of these observers is called a {\it relative
observer world 2-sheet} in Ref.\cite{41}.

\subsection{The Null Geodesics, the Deformed Light-Cone and the
Eikonal.}

The solution  of Eqs.(\ref{VI1}) for {\it null geodesics} has
still the form

\bea
 \tau(s) &=&a^{\tau}+[b^{\tau}_{(o)} + b^{\tau}_{(1)}]\, s-\int_0^s ds_1 \int_0^{s_1} ds_2
{}^4{\hat \Gamma}^{\tau}_{ru}(a+b_{(o)}s_2)b^r b^u\nonumber\\
 &&{}\nonumber \\
 \sigma^u(s)&=&a^u + b^u - \int_0^s ds_1 \int_0^{s_1} ds_2
\Big[(b^{\tau}_{(o)})^2-2\, {}^4{\hat \Gamma}^u_{\tau r}(a+b_{(o)}
s_2)
 b^{\tau}_{(o)}\, b^r- {}^3{\hat \Gamma}^u_{rt}(a+b_{(o)} s_2)
 b^r b^t \Big]\nonumber \\
 &&{}\nonumber \\
 &&with\nonumber \\
 &&{}\nonumber \\
 p^{\tau}(s)&=& {{d \tau (s)}\over {ds}} = b^{\tau}_{(o)} + b^{\tau}_{(1)}
 -\int_0^s ds_2 {}^4{\hat \Gamma}^{\tau}_{ru}(a+b_{(o)} s_2)\, b^r
 b^u,\nonumber \\
 &&{}\nonumber \\
  p^u(s) &=& {{d\sigma^u(s)}\over {ds}}=b^u-\int_0^s ds_2
\Big[(b^{\tau}_{(o)})^2-2\, {}^4{\hat \Gamma}^u_{\tau r}(a+b_{(o)}
s_2) b^{\tau}_{(o)}\, b^r- {}^3{\hat \Gamma}^u_{rt}(a+b_{(o)} s_2)
 b^r b^t \Big],
 \label{VII18}
 \eea

\noindent but now at $s=0$ we have

\bea
 {}^4g_{AB}p^Ap^B{|}_{s=0}&=&\epsilon \Big[(p^{\tau}(s))^2-2
n_r(s,r_{\bar a}(s))p^r(s)p^{\tau}(s)-[\delta_{rt}-\epsilon
 {}^4h_{rt}(s,r_{\bar a}(s))p^r(s)p^t(s)]\Big]{|}_{s=0} =\nonumber \\
 &=& 0,\nonumber \\
 &&{}\nonumber \\
 0&=&(b^{\tau})^2-2b^{\tau}b^r n_r(a^{\tau},r_{\bar
a}(a^{\tau}))-[\delta_{rt}-\epsilon {}^4h_{rt}(a^{\tau},r_{\bar
a}(a^{\tau}))]\, b^r b^t,\quad at\, s=0,
 \label{VII19}
  \eea

\noindent whose solution is

\bea
 b^{\tau} &=& b^{\tau}_{(o)} + b^{\tau}_{(1)} =
 =b^r n_r(a^{\tau},r_{\bar a}(a^{\tau}))\pm
\sqrt{(b^r n_r(a^{\tau},r_{\bar
a}(a^{\tau})))^2+[\delta_{rt}-\epsilon {}^4h_{rt}(a^{\tau},r_{\bar
a}(a^{\tau}))]\, b^r b^t},\nonumber \\
 &&{}\nonumber \\
 b^{\tau}_{(o)} &=& \pm \sqrt{\delta_{rs}b^r b^s},\nonumber \\
 b^{\tau}_{(1)} &=&  n_r(a^{\tau},r_{\bar
a}(a^{\tau}))\, b^r\, \mp {{\epsilon}\over 2}\,
{{{}^4h_{rt}(a^{\tau},r_{\bar a}(a^{\tau}))b^r b^t}\over {
\sqrt{\delta_{rs}b^r b^s}}} + O(r^2_{\bar a}).
 \label{VII20}
 \eea

Therefore, if we consider the family of null geodesics emanating
from a fixed point of space-time and parametrized by $\{ b^r \}$,
we obtain the null surface describing the {\it deformed
light-cone} through that point in our post-Minkowskian Einstein
space-time.

\bigskip

Let us now consider the {\it eikonal equation} \cite{7}
${}^4g^{AB}(\tau ,\vec \sigma )\, \partial_A\, U(\tau ,\vec \sigma
)\, \partial_B\, U(\tau ,\vec \sigma ) = 0$, whose solution, the
so-called {\it optical function} $U$, is used to find the null
hyper-surfaces $U(\tau ,\vec \sigma ) = 0$ tangent to the deformed
light-cones, generalizing the planes $x^{\pm} = {1\over
{\sqrt{2}}}\, (x^o \pm x^3) = 0$ tangent to the light-cone in
special relativity, where they are used for the front (or null)
form of dynamics in light-cone coordinates.

In Minkowski space-time with Cartesian coordinates the eikonal
equation is $(\partial_{\tau}\, U_M(\tau ,\vec \sigma ))^2 - (\vec
\partial\, U_M(\tau ,\vec \sigma ))^2 = 0$, namely
$\partial_{\tau}\, U_M = \alpha\, |\vec \partial \, U_M|$ with
$\alpha = \pm$. The solutions are $U_M(\tau ,\vec \sigma ) =
f(\tau \pm \sum_r\, A_r\, \sigma^r)$ with $\sum_r\, A^2_r = 1$.

\medskip

After the linearization in our gauge the eikonal equation becomes
(${\hat \gamma}^{rs} = \gamma_r\, \delta^{rs}$)

\beq
 (\partial_{\tau}\, U(\tau ,\vec \sigma ))^2 - 2\,
 \partial_{\tau}\, U(\tau ,\vec \sigma )\, \sum_r\, n_r(\tau ,\vec
 \sigma )\, \partial_r\, U(\tau ,\vec \sigma ) - \sum_r\,
 \gamma_r(\tau ,\vec \sigma )\, (\partial_r\, U(\tau ,\vec \sigma
 ))^2 = 0.
 \label{VII21}
 \eeq

Let us put $U = U_M + V$ with $V = O(r_{\bar a})$. By disregarding
terms of order $V^2$ the eikonal equation becomes the following
quasi-linear partial differential equation for $V$

\bea
 &&2\, \Big[ \alpha |\vec \partial\, U_M| - \sum_r\, n_r\,
 \partial_r\, U_M\Big]\, \partial_{\tau}\, V - 2\, \sum_r\, \Big[
 \alpha\, |\vec \partial\, U_M|\, n_r + \gamma_r\, \partial_r\,
 U_m\Big]\, \partial_r V =\nonumber \\
 &&= \sum_r\, (\gamma_r - 1)\, (\partial_r\, U_M)^2 + 2\, \alpha\,
 |\vec \partial\, U_M|\, \sum_r\, n_r\, \partial_r\, U_M.
 \label{VII22}
 \eea

If we write this equation in the form $a\, \partial_{\tau}\, V +
\sum_r\, a_r\, \partial_r\, V = F$, then a solution $U = U_M + V$
of the linearized eikonal equation can be obtained with the method
of characteristics \cite{25} if it is possible to find an explicit
solution of the system

\beq
 {{dV}\over F} = {{d\tau}\over a} = {{d\sigma^r}\over {a_r}}.
 \label{VII23}
 \eeq

\vfill\eject

\section{Conclusions.}

We have defined a background-independent Hamiltonian linearization
of vacuum canonical tetrad gravity in a completely fixed
3-orthogonal gauge in the framework of its rest-frame instant
form, where the evolution is governed by the weak ADM energy. In
this {\it non-harmonic} gauge every quantity is expressed in terms
of the DO's $r_{\bar a}(\tau ,\vec \sigma )$, $\pi_{\bar a}(\tau
,\vec \sigma ) $, $\bar a =1,2$, parametrizing the independent
degrees of freedom of the gravitational field. The method is based
on the linearization of the Lichnerowicz equation for the
conformal factor of the 3-metric and on the restriction of the
weak ADM energy to its part quadratic in the DO's.

\medskip

As a consequence we succeed for the first time in solving all the
constraints of tetrad gravity (super-hamiltonian constraint
included) and to find a solution of the linearized Einstein
equations in the uniquely defined 4-coordinate system induced by
the chosen gauge, which corresponds to a {\it post-Minkowskian
Einstein space-time} of the Christodoulou-Klainermann type. The
DO's $r_{\bar a}(\tau ,\vec \sigma )$ turn out to satisfy the
massless wave equation even if we are not in a harmonic gauge and
we get non-zero shift functions (namely our 4-coordinates are
non-synchronous). Besides re-opening the Hamiltonian approach to
gravity, we can show explicitly the role played by the
deterministically predictable DO's of the gravitational field in
deforming the structures of the flat Minkowski space-time without
having used it as a background for the propagation of a massless
spin 2 field as it happens in the standard treatment of
gravitational waves in the harmonic gauges. In particular the two
configuration DO's $r_{\bar a}(\tau ,\vec \sigma )$, $\bar a
=1,2$, of our gauge, where the 3-metric is diagonal, replace the
two polarizations of the TT harmonic gauge (with its non-diagonal
3-metric) and, through the geodesic deviation equation, they
induce two well defined patterns of deformation on a sphere of
test particles. Even if we were able to eliminate the background
with our Hamiltonian linearization, our results are still
coordinate (i.e. gauge)-dependent like in all existing treatments
of gravitational waves: indeed our DO's $r_{\bar a}$, $\pi_{\bar
a}$ are coordinate-dependent non-tensorial quantities. To get a
coordinate-independent description of them, we have to verify the
main conjecture of Ref.\cite{15}. It states that by means of a
Hamiltonian re-formulation of the Newman-Penrose formalism
\cite{43} it should be possible to find a Shanmugadhsan canonical
basis in which the DO's are also (coordinate-independent) Bergmann
observables and also the gauge variables are
coordinate-independent.

We have also made some comments on the coordinate dependence of
the gravito-magnetic effects: for instance in our non-harmonic
gauge the gravito-electric-magnetic analogy does not hold. Again
the verification of the main conjecture of Ref.\cite{15} would
allow a coordinate-independent description of gravito-magnetism.

\medskip

These results about background-independent gravitational waves in
post-Minkowskian space-times are welcome because {\it they open
the possibility, after the introduction of matter, to study the
emission of gravitational waves from relativistic sources without
any kind of post-Newtonian approximation}. For instance this is
the case for the relativistic motion (but still in the weak field
regime) of the binaries before the beginning of the final inspiral
phase: it is known that in this phase the post-Newtonian
approximation does not work and that , till now, only numerical
gravity may help. In a future paper \cite{44} we will add a
relativistic perfect fluid, described by suitable Lagrangian
\cite{45} or Eulerian \cite{46} variables, to tetrad gravity, we
will define a Hamiltonian linearization of the system in the
completely fixed 3-orthogonal gauge, we will find the Hamilton
equations for the DO's both of the gravitational field and of the
fluid, we will find the relativistic version of the Newton and
gravito-magnetic action-at-a-distance potentials acting inside the
fluid present in the weak ADM energy and finally, by using a
multipolar expansion, we will find the relativistic counterpart of
the post-Newtonian quadrupole emission formula.

Let us remark that till now we have a treatment of the generation
of gravitational waves from a compact localized source of size $R$
and mean internal velocity $v$ only \cite{22} for {\it nearly
Newtonian slow motion} sources for which $v << c$,
${{\lambda}\over {2\pi}} >> R$: outgoing gravitational waves are
observed in the {\it radiation zone} (far field, $r >>
{{\lambda}\over {2\pi}}$), while deep in the {\it near zone} ($R <
r << {{\lambda}\over {2\pi}}$), for example $r \leq 1000 \,
{{\lambda}\over {2\pi}}$, vacuum Newtonian gravitation theory is
valid. On the contrary with our approach in suitable 4-coordinates
we are going to obtain a {\it weak field approximation but with
fast relativistic motion in the source} subject to the restriction
that the total invariant mass and the mass currents are compatible
with the weak field requirement. This is enough to get
relativistic results conceptually equivalent to the re-summation
of the post-Newtonian expansion.

Moreover we will have to explore whether our Hamiltonian approach
is suitable for doing  Hamiltonian numerical gravity on the full
non-linearized theory.

\vfill\eject

\appendix

\section{Results on tetrad gravity.}

In this Appendix we reproduce those results of Ref.\cite{3}, which
are needed in this paper.

The Lichnerowicz equation in the 3-orthogonal gauge
$\pi_{\phi}(\tau ,\vec \sigma ) \approx 0$ given in Eq.(193) of
Ref.\cite{3} is (see Eq.(\ref{a4}) for the definition of the
${\cal T}^u_{(a)r}$'s)

\begin{eqnarray}
(-{\tilde \triangle}[r_{\bar a}] &+&{1\over 8}{}^3{\tilde
R}[r_{\bar a}])(\tau ,\vec \sigma ) \phi (\tau ,\vec \sigma )={{12
\pi^2 G^2}\over {c^6}} \Big[ 2(\phi^{-7}  \sum_{\bar a}
\pi^2_{\bar a})(\tau ,\vec \sigma )+\nonumber \\ &+&4\Big(
\phi^{-5} \sum_ue^{{1\over {\sqrt{3}}}\sum_{\bar a}\gamma_{\bar
au}r_{\bar a}} \sum_{\bar b}\gamma_{\bar bu}\pi_{\bar b}\Big)(\tau
,\vec \sigma )\times \nonumber \\ &&\int d^3\sigma_1 \sum_r
\delta^u_{(a)} {\cal T}^u_{(a)r}(\vec \sigma ,{\vec \sigma}_1,\tau
) \Big( \phi^{-2} e^{-{1\over {\sqrt{3}}}\sum_{\bar a}
\gamma_{\bar ar}r_{\bar a}}\sum_{\bar b}\gamma_{\bar br} \pi_{\bar
b}\Big) (\tau ,{\vec \sigma}_1)+\nonumber \\ &+&\phi^{-3}(\tau
,\vec \sigma ) \int d^3\sigma_1d^3\sigma_2 \Big( \sum_u e^{{2\over
{\sqrt{3}}}\sum_{\bar a} \gamma_{\bar au}r_{\bar a}(\tau ,\vec
\sigma )} \times \nonumber \\ &&\sum_r{\cal T}^u_{(a)r}(\vec
\sigma ,{\vec \sigma}_1,\tau ) \Big( \phi^{-2} e^{-{1\over
{\sqrt{3}}}\sum_{\bar a} \gamma_{\bar ar}r_{\bar a}}\sum_{\bar
b}\gamma_{\bar br} \pi_{\bar b}\Big) (\tau ,{\vec \sigma}_1)\times
\nonumber \\ &&\sum_s {\cal T}^u_{(a)s}(\vec \sigma ,{\vec
\sigma}_2,\tau ) \Big( \phi^{-2} e^{-{1\over {\sqrt{3}}}\sum_{\bar
a} \gamma_{\bar as}r_{\bar a}}\sum_{\bar c}\gamma_{\bar cs}
\pi_{\bar c}\Big) (\tau ,{\vec \sigma}_2)+\nonumber \\
&+&\sum_{uv} e^{{1\over {\sqrt{3}}}\sum_{\bar a}(\gamma_{\bar
au}+\gamma_{\bar av})r_{\bar a}(\tau ,\vec \sigma )}
(\delta^u_{(b)}\delta^v_{(a)}-\delta^u_{(a)} \delta^v_{(b)})\times
\nonumber \\ &&\sum_r {\cal T}^u_{(a)r}(\vec \sigma ,{\vec
\sigma}_1,\tau ) \Big( \phi^{-2} e^{-{1\over {\sqrt{3}}}\sum_{\bar
a} \gamma_{\bar ar}r_{\bar a}}\sum_{\bar b}\gamma_{\bar br}
\pi_{\bar b}\Big) (\tau ,{\vec \sigma}_1)\nonumber \\ &&\sum_s
{\cal T}^v_{(b)s}(\vec \sigma ,{\vec \sigma}_2,\tau ) \Big(
\phi^{-2} e^{-{1\over {\sqrt{3}}}\sum_{\bar a} \gamma_{\bar
as}r_{\bar a}}\sum_{\bar c}\gamma_{\bar cs} \pi_{\bar c}\Big)
(\tau ,{\vec \sigma}_2)\, \Big)\, \Big].
 \label{a1}
 \end{eqnarray}

This equation for the conformal factor $\phi (\tau ,\vec \sigma )$
of the 3-metric is implied by the super-hamiltonian constraint,
Eq.(191) of Ref.\cite{3}, which is

\bea
 {\tilde {\cal H}}_R(\tau ,\vec \sigma )\,
 {|}_{\pi_{\phi}=0} &=& \epsilon \phi (\tau ,\vec \sigma )\, \Big[ \, {{c^3}\over
{16\pi G}} (-8{\tilde \triangle}[r_{\bar a}] +{}^3{\tilde
R}[r_{\bar a}])\phi -\nonumber \\
 &-&{{6\pi G}\over {c^3}}\Big[ \Big( 2\phi^{-7}  \sum_{\bar a}
 \pi^2_{\bar a}\Big)(\tau ,\vec \sigma )+\nonumber \\
 &+&4\Big( \phi^{-5}
\sum_ue^{{1\over {\sqrt{3}}}\sum_{\bar a}\gamma_{\bar au}r_{\bar
a}} \sum_{\bar b}\gamma_{\bar bu}\pi_{\bar b}\Big)(\tau ,\vec
\sigma )\times \nonumber \\
 &&\int d^3\sigma_1 \sum_r
\delta^u_{(a)} {\cal T}^u_{(a)r}(\vec \sigma ,{\vec \sigma}_1,\tau
) \Big( \phi^{-2} e^{-{1\over {\sqrt{3}}}\sum_{\bar a}
\gamma_{\bar ar}r_{\bar a}}\sum_{\bar b}\gamma_{\bar br} \pi_{\bar
b}\Big) (\tau ,{\vec \sigma}_1)+\nonumber \\
 &+&\phi^{-3}(\tau ,\vec \sigma )
\int d^3\sigma_1d^3\sigma_2 \Big( \sum_u e^{{2\over
{\sqrt{3}}}\sum_{\bar a} \gamma_{\bar au}r_{\bar a}(\tau ,\vec
\sigma )} \times \nonumber \\ &&\sum_r{\cal T}^u_{(a)r}(\vec
\sigma ,{\vec \sigma}_1,\tau ) \Big( \phi^{-2} e^{-{1\over
{\sqrt{3}}}\sum_{\bar a} \gamma_{\bar ar}r_{\bar a}}\sum_{\bar
b}\gamma_{\bar br} \pi_{\bar b}\Big) (\tau ,{\vec \sigma}_1)\times
\nonumber \\ &&\sum_s {\cal T}^u_{(a)s}(\vec \sigma ,{\vec
\sigma}_2,\tau ) \Big( \phi^{-2} e^{-{1\over {\sqrt{3}}}\sum_{\bar
a} \gamma_{\bar as}r_{\bar a}}\sum_{\bar c}\gamma_{\bar cs}
\pi_{\bar c}\Big) (\tau ,{\vec \sigma}_2)+\nonumber \\
&+&\sum_{uv} e^{{1\over {\sqrt{3}}}\sum_{\bar a}(\gamma_{\bar
au}+\gamma_{\bar av})r_{\bar a}(\tau ,\vec \sigma )}
(\delta^u_{(b)}\delta^v_{(a)}-\delta^u_{(a)} \delta^v_{(b)})\times
\nonumber \\ &&\sum_r {\cal T}^u_{(a)r}(\vec \sigma ,{\vec
\sigma}_1,\tau ) \Big( \phi^{-2} e^{-{1\over {\sqrt{3}}}\sum_{\bar
a} \gamma_{\bar ar}r_{\bar a}}\sum_{\bar b}\gamma_{\bar br}
\pi_{\bar b}\Big) (\tau ,{\vec \sigma}_1)\nonumber \\ &&\sum_s
{\cal T}^v_{(b)s}(\vec \sigma ,{\vec \sigma}_2,\tau ) \Big(
\phi^{-2} e^{-{1\over {\sqrt{3}}}\sum_{\bar a} \gamma_{\bar
as}r_{\bar a}}\sum_{\bar c}\gamma_{\bar cs} \pi_{\bar c}\Big)
(\tau ,{\vec \sigma}_2)\, \Big)\, \Big]\, \Big] \approx 0.
 \label{a2}
 \end{eqnarray}

The scalar curvature and the Laplace-Beltrami operator appearing
in these equations, Eq.(190) of Ref.\cite{3}, are

\begin{eqnarray}
{}^3\hat R [\phi ,r_{\bar a}]&=& -\sum_{uv}\{ (2\partial_vln\,
\phi +{1\over {\sqrt{3}}}\sum_{\bar a}\gamma_{\bar au}
\partial_vr_{\bar a})(4\partial_vln\, \phi -{1\over {\sqrt{3}}}\sum_{\bar b}\gamma
_{\bar bu}\partial_vr_{\bar b})+\nonumber \\
&&+\phi^{-4}e^{{2\over {\sqrt{3}}}\sum_{\bar c}\gamma_{\bar
cv}r_{\bar c}} [2\partial^2_vln\, \phi +{1\over
{\sqrt{3}}}\sum_{\bar a}\gamma_{\bar au}\partial_v^2r_{\bar
a}+\nonumber \\
 &&+{2\over {\sqrt{3}}}(2\partial_vln\, \phi +{1\over {\sqrt{3}}}\sum_{\bar a}\gamma
_{\bar au}\partial_vr_{\bar a})\sum_{\bar b}(\gamma_{\bar
bu}-\gamma_{\bar bv})
\partial_vr_{\bar b}-\nonumber \\
&&-(2\partial_vln\, \phi +{1\over {\sqrt{3}}}\sum_{\bar
a}\gamma_{\bar av}\partial_vr_{\bar a})(2\partial_vln\, \phi
+{1\over {\sqrt{3}}}\sum_{\bar b}\gamma_{\bar bu}
\partial_vr_{\bar b}] \} +\nonumber \\
&&+\phi^{-4}\sum_ue^{{2\over {\sqrt{3}}}\sum_{\bar c}\gamma_{\bar
cu}r_{\bar c}} [-2\partial^2_uln\, \phi +{2\over
{\sqrt{3}}}\sum_{\bar a}\gamma_{\bar au} \partial^2_ur_{\bar
a}+\nonumber \\
 &&+(2\partial_uln\, \phi +{1\over
{\sqrt{3}}}\sum_{\bar a}\gamma_{\bar au}\partial_ur_{\bar
a})(2\partial_uln\, \phi -{2\over {\sqrt{3}}}\sum_{\bar
b}\gamma_{\bar bu} \partial_ur_{\bar b})],\nonumber \\
 &&{}\nonumber \\
  \tilde \triangle [r_{\bar a}]&=& \partial_r
[{}^3{\tilde g}^{rs}\, \partial_s] ={}^3{\tilde g}^{rs}\,
{}^3{\tilde \nabla}_r\, {}^3{\tilde \nabla}_s=  \sum_re^{-{2\over
{\sqrt{3}}}\sum_{\bar a}\gamma_{\bar ar} r_{\bar a}}
[\partial_r^2-{2\over {\sqrt{3}}} \sum_{\bar b}\gamma_{\bar br}
\partial_rr_{\bar b} \partial_r].
 \label{a3}
 \end{eqnarray}

If ${\tilde \pi}^{\vec \alpha}_{(a)}(\tau ,\vec \sigma ) \approx
0$ and ${\tilde \pi}^{\vec \xi}_r(\tau ,\vec \sigma ) \approx 0$
are the Abelianized momenta corresponding to the rotation and
super-momentum constraints, the connection of the old cotriad
momenta to the final ones, see Eq.(\ref{I4}), after the
quasi-Shanmugadhasan canonical transformation (\ref{I3}), i.e.
Eq.(185) of Ref.\cite{3}, is

\begin{eqnarray}
{}^3{\hat {\tilde \pi}}^r_{(a)}(\tau ,\vec \sigma )&=& {}^3{\tilde
\pi}^r_{(a)}(\tau ,\vec \sigma ){|}_{\alpha_{(a)}=
 0, \xi^r=\sigma^r, {\tilde
 \pi}^{\vec \alpha}_{(a)}= {\tilde \pi}^{\vec \xi}_r=0}= \nonumber \\
 &=& \sum_s \int d^3\sigma_1 {\cal K}^r_{(a)s}(\vec \sigma ,{\vec \sigma}_1;\tau )
 \, {\tilde \Pi}^s(\tau ,{\vec \sigma}_1) =\nonumber \\
 &=&\sum_s \int
d^3\sigma_1 {\cal K}^r_{(a)s}(\vec \sigma ,{\vec \sigma}_1;\tau )
(\phi^{-2}e^{-{1\over {\sqrt{3}}}\sum_{\bar a}\gamma_{\bar
as}r_{\bar a}}) (\tau ,{\vec \sigma}_1) \nonumber \\ &&\Big[
{{\pi_{\phi}}\over {6\, \phi}} +\sqrt{3} \sum_{\bar b}
\gamma_{\bar bs} \pi_{\bar b}\Big] (\tau ,{\vec
\sigma}_1),\nonumber \\
 \rightarrow_{\pi_{\phi} \rightarrow 0}&&
\sqrt{3} \sum_{s,\bar b} \gamma_{\bar bs}\int d^3\sigma_1 {\cal
K}^r_{(a)s}(\vec \sigma ,{\vec \sigma}_1;\tau |\phi ,r_{\bar a}]
(\phi^{-2}e^{-{1\over {\sqrt{3}}}\sum_{\bar a}\gamma_{\bar
as}r_{\bar a}} \pi_{\bar b})(\tau ,{\vec \sigma}_1),\nonumber \\
 &&{}\nonumber \\
 &&\text{with the kernel} \nonumber \\
 &&{}\nonumber \\
 {\cal K}^r_{(a)s}(\vec \sigma ,{\vec \sigma}_1,\tau ) &=&
 {\tilde {\cal K}}^r_{(a)s}(\vec \sigma ,{\vec \sigma}_1,\tau
|\phi ,r_{\bar a}]\, {\buildrel {def} \over =}\,
\delta^r_{(a)}\delta^r_s\delta^3(\vec \sigma ,{\vec
\sigma}_1)+{\cal T}^r _{(a)s}(\vec \sigma ,{\vec \sigma}_1,\tau
),\nonumber \\
 &&{}\nonumber \\
  {\cal T}^r_{(a)s}(\vec \sigma ,{\vec \sigma}_1,\tau ) &=&
 {\tilde {\cal T}}^r_{(a)s}(\vec \sigma ,{\vec \sigma}_1,\tau |\phi ,r_{\bar a}]=\nonumber \\
 &=&-Q_s(\tau ,{\vec \sigma}_1){{\partial G^{rs}_{(a)}(\vec \sigma ,{\vec \sigma}_1;\tau |
 \phi ,r_{\bar a}]}\over {\partial \sigma_1^s}}-\sum_v {{\partial Q_s(\tau
 ,{\vec \sigma}_1)}\over {\partial \sigma_1^v}} G^{rv}_{(a)}(\vec \sigma
 ,{\vec \sigma}_1;\tau |\phi ,r_{\bar a}], \nonumber \\
 &&
 \label{a4}
 \eea

\noindent The canonicity of the transformation (\ref{I3}) implies
that the kernels ${\cal T}^r_{(o)s}$'s are determined by the
kernels $G^{rs}_{(a)}$'s.

Eqs.(156) of Ref.\cite{3}, giving the partial differential
equations for the kernels $G^{ru}_{(a)}$'s, are

\bea
 &&\epsilon_{(a)(r_1)(b)} \delta^s_{r_1} {{\delta^3({\vec
\sigma}_1,{\vec \sigma})}\over {Q_{r_1}(\tau ,\vec \sigma )}}
-\sum_u \epsilon_{(u)(r_1)(b)} {{Q_u(\tau ,\vec \sigma )}\over
{Q_{r_1}(\tau ,\vec \sigma )}} {{\partial G^{su}_{(a)}({\vec
\sigma}_1 ,{\vec \sigma};\tau )}\over {\partial
\sigma^{r_1}}}=\nonumber \\
 =&& \epsilon_{(a)(r_2)(b)} \delta^s_{r_2} {{\delta^3({\vec \sigma}_1,{\vec
\sigma})}\over {Q_{r_2}(\tau ,\vec \sigma )}} -\sum_u
\epsilon_{(u)(r_2)(b)} {{Q_u(\tau ,\vec \sigma )}\over
{Q_{r_2}(\tau ,\vec \sigma )}} {{\partial G^{su}_{(a)}({\vec
\sigma}_1
 ,{\vec \sigma};\tau )}\over {\partial \sigma^{r_2}}},\nonumber \\
 &&{}\nonumber \\
 &&\Downarrow \nonumber \\
 &&{}\nonumber \\
 &&{1\over {Q_{r_1}^2(\tau ,\vec \sigma )}}\, {{\partial G^{sr_2}_{(a)}
 ({\vec \sigma}_1 ,{\vec \sigma};\tau )}\over {\partial \sigma^{r_1}}}+
 {1\over {Q_{r_2}^2(\tau ,\vec \sigma )}}\, {{\partial G^{sr_1}_{(a)}
 ({\vec \sigma}_1 ,{\vec \sigma};\tau )}\over {\partial \sigma^{r_2}}}
   =\nonumber \\
   =&& \Big[ {{\delta_{(a)r_1}\delta^s_{r_2}}\over {Q_{r_1}(\tau ,\vec \sigma )
   Q^2_{r_2}(\tau ,\vec \sigma )}}+
 {{\delta_{(a)r_2}\delta^s_{r_1}}\over {Q^2_{r_1}(\tau ,\vec \sigma )
   Q_{r_2}(\tau ,\vec \sigma )}}\Big]   \delta^3({\vec \sigma}_1,{\vec \sigma}),
   \nonumber \\
   &&{}\nonumber \\
 &&\Downarrow \nonumber \\
 &&{}\nonumber \\
 1)&&\,\, s=a\quad \text{ homogeneous equations}:\nonumber \\
 &&{}\nonumber \\
 &&{1\over {Q_1^2(\tau ,\vec \sigma )}}\, {{\partial G^{a2}_{(a)}
 ({\vec \sigma}_1 ,{\vec \sigma};\tau )}\over {\partial \sigma^1}} +
 {1\over {Q_2^2(\tau ,\vec \sigma )}}\, {{\partial G^{a1}_{(a)}
 ({\vec \sigma}_1 ,{\vec \sigma};\tau )}\over {\partial \sigma^2}} =\nonumber \\
 &&= {1\over {Q_2^2(\tau ,\vec \sigma )}}\, {{\partial G^{a3}_{(a)}
 ({\vec \sigma}_1 ,{\vec \sigma};\tau )}\over {\partial \sigma^2}} +
 {1\over {Q_3^2(\tau ,\vec \sigma )}}\, {{\partial G^{a2}_{(a)}
 ({\vec \sigma}_1 ,{\vec \sigma};\tau )}\over {\partial \sigma^3}} =\nonumber \\
 &&= {1\over {Q_3^2(\tau ,\vec \sigma )}}\, {{\partial G^{a1}_{(a)}
 ({\vec \sigma}_1 ,{\vec \sigma};\tau )}\over {\partial \sigma^3}} +
 {1\over {Q_1^2(\tau ,\vec \sigma )}}\, {{\partial G^{a3}_{(a)}
 ({\vec \sigma}_1 ,{\vec \sigma};\tau )}\over {\partial \sigma^1}} = 0,
 \qquad a=1,2,3;\nonumber \\
 &&{}\nonumber \\
 2)&&\,\, s\not= a\,\, [s\not= r,\, r\not= a]\quad \text{in-homogeneous equations}:\nonumber \\
 &&{}\nonumber \\
 &&{1\over {Q_s^2(\tau ,\vec \sigma )}}\, {{\partial G^{sr}_{(a)}
 ({\vec \sigma}_1 ,{\vec \sigma};\tau )}\over {\partial \sigma^s}} +
 {1\over {Q_r^2(\tau ,\vec \sigma )}}\, {{\partial G^{ss}_{(a)}
 ({\vec \sigma}_1 ,{\vec \sigma};\tau )}\over {\partial \sigma^r}} =\nonumber \\
  &&={1\over {Q_r^2(\tau ,\vec \sigma )}}\, {{\partial G^{sa}_{(a)}
 ({\vec \sigma}_1 ,{\vec \sigma};\tau )}\over {\partial \sigma^r}} +
 {1\over {Q_a^2(\tau ,\vec \sigma )}}\, {{\partial G^{sr}_{(a)}
 ({\vec \sigma}_1 ,{\vec \sigma};\tau )}\over {\partial \sigma^a}} =0,\nonumber \\
 &&{1\over {Q_a^2(\tau ,\vec \sigma )}}\, {{\partial G^{ss}_{(a)}
 ({\vec \sigma}_1 ,{\vec \sigma};\tau )}\over {\partial \sigma^a}} +
 {1\over {Q_s^2(\tau ,\vec \sigma )}}\, {{\partial G^{sa}_{(a)}
 ({\vec \sigma}_1 ,{\vec \sigma};\tau )}\over {\partial \sigma^s}}
 = {{\delta^3({\vec \sigma}_1 ;{\vec \sigma})}\over {Q_a(\tau ,\vec \sigma )
 Q_s^2(\tau ,\vec \sigma )}}.
 \label{a5}
 \eea

Eqs.(159) of Ref.\cite{3} for the kernels ${\cal K}^r_{(a)u}$'s,
which ensure the simultaneous satisfaction of the rotation and
super-momentum constraints, are [$Q_u = \phi^2\, e^{{1\over
{\sqrt{3}}}\, \sum_{\bar a}\, \gamma_{\bar au}\, r_{\bar a}}$]

\bea
 &&\sum_r Q_r(\tau ,\vec \sigma ) \Big[ \delta_{(a)r} {\cal
K}^r_{(b)u} - \delta_{(b)r} {\cal K}^r_{(a)u} \Big] (\vec \sigma
,{\vec \sigma}_1;\tau ) =\nonumber \\
 &&= Q_u(\tau ,{\vec \sigma}_1 ) \Big[ Q_a(\tau ,\vec \sigma ){{\partial G^{au}_{(b)}
(\vec \sigma ,{\vec \sigma}_1;\tau )}\over {\partial
\sigma^u_1}}-Q_b(\tau ,\vec \sigma ){{\partial G^{bu}_{(a)} (\vec
\sigma ,{\vec \sigma}_1;\tau )}\over {\partial \sigma^u_1}} \Big]
+\nonumber \\
 &&+\sum_v {{\partial Q_u(\tau ,{\vec \sigma}_1 )}\over {\partial \sigma^v_1}}
\Big[ Q_a(\tau ,\vec \sigma ) G^{av}_{(b)}(\vec \sigma ,{\vec
\sigma}_1;\tau )- Q_b(\tau ,\vec \sigma ) G^{bv}_{(a)}(\vec \sigma
,{\vec \sigma}_1;\tau )\Big] = \nonumber \\
 &&= 0,\qquad a\not= b,\nonumber \\
 &&{}\nonumber \\
 &&{\hat D}^{(\hat \omega )}_{(a)(b)r}(\tau ,\vec \sigma ) {\cal K}^r_{(b)u}
 (\vec \sigma ,{\vec \sigma}_1;\tau ) =\nonumber \\
 &&=\Big( \delta_{(a)(b)}\partial_r + \epsilon_{(a)(b)(c)}\, {}^3{\hat \omega}_{r(c)}
 (\tau ,\vec \sigma )\Big) {\cal K}^r_{(b)u}
 (\vec \sigma ,{\vec \sigma}_1;\tau ) =\nonumber \\
 &&= \Big( \delta_{(a)(b)} \partial_r + \sum_u(\delta_{(a)r}\delta_{(b)u}-
 \delta_{(a)u}\delta_{(b)r}) {{\partial_uQ_r(\tau ,\vec \sigma )}\over
 {Q_u(\tau ,\vec \sigma )}}\Big)
\Big[ \delta_{(b)}^r\delta_{(b)u} \delta^3(\vec \sigma ,{\vec
\sigma}_1) -\nonumber \\
 &-& Q_u(\tau ,{\vec \sigma}_1 ) {{\partial G^{ru}_{(b)}(\vec \sigma ,{\vec \sigma}_1;\tau )}
 \over {\partial \sigma^u_1}} -\sum_v {{\partial Q_u(\tau ,{\vec \sigma}_1 )}\over
 {\partial \sigma^v_1}}\, G^{rv}_{(b)}(\vec \sigma ,{\vec \sigma}_1;\tau ) \Big]
 = 0,\nonumber \\
 &&{}\nonumber \\
 &&{}^3{\hat \omega}_{r(a)}(\tau ,\vec \sigma ) =\sum_u
 \epsilon_{(a)(m)(n)}\delta_{(m)r}\delta_{(n)u}{{\partial_uQ_r(\tau ,\vec \sigma )}\over
 {Q_u(\tau ,\vec \sigma )}}
 \label{a6}
  \eea

The cotriads, the triads, the 3-metric, the Christoffel symbols
and  the spin connection on the WSW hyper-surfaces of our gauge,
given in Eq.(184) of Ref.\cite{3}, have the following expressions
in the canonical basis (\ref{I3})

\begin{eqnarray}
{}^3{\hat e}_{(a)r}&=&
 \delta_{(a)r} \phi^2 e^{{1\over {\sqrt{3}}}\sum_{\bar a}
\gamma_{\bar ar}r_{\bar a}},\nonumber \\
 &&{}\nonumber \\
 {}^3{\hat e}^r_{(a)}&=&
 \delta^r_{(a)} \phi^{-2} e^{-{1\over {\sqrt{3}}}\sum_{\bar a}
\gamma_{\bar ar}r_{\bar a}}, \nonumber \\
 &&{}\nonumber \\
 {}^3{\hat g}_{rs}&=&
 \delta_{rs} \phi^4 e^{{2\over {\sqrt{3}}}\sum_{\bar
a}\gamma_{\bar ar}r_{\bar a}},\nonumber \\
 {}^3{\hat g}^{rs}&=& \delta^{rs} \phi^{-4} e^{- {2\over {\sqrt{3}}}\sum_{\bar
a}\gamma_{\bar ar}r_{\bar a}}, \nonumber \\
 &&{}\nonumber \\
  {}^3\hat e &=& \sqrt{\hat \gamma}= \phi^6,\nonumber \\
 &&{}\nonumber \\
 {}^3{\hat \Gamma}^r_{uv}&=& -\delta_{uv}
 \sum_s\delta^r_s e^{{2\over {\sqrt{3}}}\sum_{\bar
a}(\gamma _{\bar au}-\gamma_{\bar as})r_{\bar a}} \Big[
2\partial_sln\, \phi +{1\over {\sqrt{3}}} \sum_{\bar
 b}\gamma_{\bar bu}\partial_sr_{\bar b}\Big] +\nonumber \\
 &&+\delta^r_u\Big[ 2\partial_vln\, \phi +{1\over
{\sqrt{3}}}\sum_{\bar a}\gamma_{\bar au}
\partial_vr_{\bar a}\Big] +\delta^r_v\Big[ 2\partial_uln\, \phi +{1\over {\sqrt{3}}}
 \sum_{\bar a}\gamma_{\bar av}\partial_ur_{\bar a}\Big], \nonumber \\
 &&{}\nonumber \\
 &&\sum_u\, {}^3{\hat \Gamma}^u_{uv} = 6 \partial_v ln\, \phi, \nonumber \\
 &&{}\nonumber \\
 {}^3{\hat \omega}_{r(a)} &=&
  \epsilon_{(a)(b)(c)} \,
 \delta_{(b)r}\delta_{(c)u} \,  e^{{1\over {\sqrt{3}}}\sum
_{\bar a}(\gamma_{\bar ar}-\gamma_{\bar au})r_{\bar a}}\Big[
2\partial_uln\, \phi +{1 \over {\sqrt{3}}}\sum_{\bar
b}\gamma_{\bar br} \partial_ur_{\bar b}\Big]\nonumber \\
 &&\rightarrow_{r_{\bar a}\rightarrow 0}\, \epsilon_{(a)(b)(c)} \,
\delta_{(b)r}\delta_{(c)u}\, 2\partial_uln\, \phi .
 \label{a7}
\end{eqnarray}

Eqs.(102) and (105) of Ref.\cite{3} define the Green function of
the covariant divergence

\begin{equation}
{\hat D}^{(\omega )}_{(a)(b)r}(\tau ,\vec \sigma )\,
\zeta^{(\omega )r}_{(b)(c)} (\vec \sigma ,{\vec \sigma }^{'};\tau
)=-\delta_{(a)(c)} \delta^3(\vec \sigma , {\vec \sigma }^{'}).
 \label{a8}
 \end{equation}

\begin{eqnarray}
 &&\zeta^{(\omega )r}_{(a)(b)}(\vec \sigma ,{\vec \sigma }^{'};\tau
)=d^r_{\gamma _{pp^{'}}}(\vec \sigma ,{\vec \sigma }^{'}) \Big(
P_{\gamma_{pp^{'}}}\, e^{\int^{\vec \sigma }_{{\vec \sigma }^{'}}
d\sigma_1^s\, {\hat R}^{(c)}\, {}^3\omega_{s(c)}(\tau ,{\vec
\sigma }_1)} \Big)_{(a)(b)} \nonumber \\
 \rightarrow_{\phi \rightarrow 1,r_{\bar a} \rightarrow 0}&&
 \zeta^{(o)r}_{(a)(b)}(\vec \sigma ,{\vec \sigma}_1)
 =-{{\sigma^r-\sigma_1^r}\over {4\pi |\vec
 \sigma -{\vec \sigma}_1|^3}} \delta_{(a)(b)}=-\delta_{(a)(b)}
 {{\partial}\over {\partial \sigma_1^r}}{1\over {4\pi |\vec \sigma
 -{\vec \sigma}_1|}} =\nonumber \\
 &=& \delta_{(a)(b)}\, c^r(\vec \sigma -{\vec \sigma}_1),
 \label{a9}
 \end{eqnarray}

\noindent since the linearization reduces $d^r_{\gamma}(\vec
\sigma ,{\vec \sigma}_1) $ to the Green function of the ordinary
divergence $c^r(\vec \sigma - {\vec \sigma}_1)$.

The weak and strong ADM Poincare' generators of Eqs.(25) of
Ref.\cite{3} are

\begin{eqnarray}
{\hat P}^{\tau}_{ADM}&=& \int d^3\sigma \epsilon [ {{c^3}\over
{16\pi G}}\sqrt{\gamma}\,\, {}^3g^{\check r\check
s}({}^3\Gamma^{\check u} _{\check r\check v}\, {}^3\Gamma^{\check
v}_{\check s\check u}-{}^3\Gamma^{\check u}_{\check r\check s}\,
{}^3\Gamma^{\check v} _{\check v\check u})-\nonumber \\
 &-&{{8\pi G}\over {c^3\, \sqrt{\gamma} }} {}^3G_{\check r\check s\check u\check v}\,
{}^3{\tilde \Pi}^{\check r\check s}\, {}^3{\tilde \Pi}^{\check
u\check v}](\tau ,\vec \sigma ),\nonumber \\
 {\hat P}^{\check r}_{ADM}&=&- 2\int d^3\sigma \, {}^3\Gamma ^{\check r}_{\check
s\check u}(\tau ,\vec \sigma )\, {}^3{\tilde \Pi}^{\check s\check
u}(\tau ,\vec \sigma ),\nonumber \\
 {\hat J}^{\tau \check r}_{ADM}&=&-{\hat J}^{\check r\tau}_{ADM}= \int d^3\sigma
\epsilon \{ \sigma^{\check r}\nonumber \\
 &&[ {{c^3}\over {16\pi G}} \sqrt{\gamma}\,\,
{}^3g^{\check n\check s}({}^3\Gamma^{\check u}_{\check n\check
v}\, {}^3\Gamma^{\check v}_{\check s\check u}-{}^3\Gamma^{\check
u} _{\check n\check s}\, {}^3\Gamma^{\check v}_{\check v\check
u})-{{8\pi G}\over {c^3\, \sqrt{\gamma}}} {}^3G_{\check n\check
s\check u\check v}\, {}^3{\tilde \Pi}^{\check n\check s}\,
{}^3{\tilde \Pi}^{\check u\check v}]+\nonumber \\
 &+&  {{c^3}\over {16\pi G}}
\delta^{\check r}_{\check u}({}^3g_{\check v\check s}-\delta
_{\check v\check s})
\partial_{\check n}[\sqrt{\gamma}({}^3g^{\check n\check s} \,
{}^3g^{\check u\check v}-{}^3g^{\check n\check u}\, {}^3g^{\check
s\check v})] \} (\tau ,\vec \sigma ),\nonumber \\
 {\hat J}^{\check
r\check s}_{ADM}&=& \int d^3\sigma  [(\sigma^{\check r}\,
{}^3\Gamma^{\check s} _{\check u\check v}-\sigma^{\check s}\,
{}^3\Gamma^{\check r}_{\check u\check v})\, {}^3{\tilde
\Pi}^{\check u\check v}](\tau ,\vec \sigma ),\nonumber \\
 &&{}\nonumber \\
P^{\tau}_{ADM}&=&{\hat P}^{\tau}_{ADM}+\int d^3\sigma {\tilde
{\cal H}}(\tau ,\vec \sigma )\approx {\hat
P}^{\tau}_{ADM},\nonumber \\ P^{\check r}_{ADM}&=&{\hat P}^{\check
r}_{ADM}+\int d^3\sigma \, {}^3{\tilde {\cal H}}^{\check r}(\tau
,\vec \sigma )\approx {\hat
 P}^{\check r}_{ADM},\nonumber \\
 J^{\tau \check r}_{ADM}&=&{\hat J}^{\tau \check r}_{ADM}+{1\over 2}
\int d^3\sigma \sigma^{\check r}\, {\tilde {\cal H}}(\tau ,\vec
\sigma ) \approx {\hat J}^{\tau \check r}_{ADM},\nonumber \\
J^{\check r\check s}_{ADM}&=&{\hat J}^{\check r\check
s}_{ADM}+\int d^3\sigma [\sigma^{\check s}\, {}^3{\tilde {\cal
H}}^{\check r}(\tau ,\vec \sigma )- \sigma^{\check r}\,
{}^3{\tilde {\cal H}}^{\check s}(\tau ,\vec \sigma )] \approx
{\hat J}^{\check r\check s}_{ADM}.
 \label{a10}
 \end{eqnarray}

Eq.(197) of Ref.\cite{3} identifies the Hamiltonian of the
rest-frame instant form of tetrad gravity with the weak ADM energy

\beq
 {\hat H}^{(WSW){'}}_{(D)ADM}= -\epsilon {\hat
P}^{\tau}_{ADM,R} = {\hat E}_{ADM}.
 \label{a11}
  \eeq

The expression of the weak ADM Poincare' generators after the
quasi-Shanmugadhasan transformation (\ref{I3}) and the restriction
to the $\pi_{\phi}(\tau ,\vec \sigma )=0$ 3-orthogonal gauge,
given in Eqs.(227) of Appendix B of Ref.\cite{3}, is

\begin{eqnarray}
{\hat P}^{\tau}_{ADM,R}&=&\epsilon \int d^3\sigma \Big(
{{c^3}\over {16\pi G}} \Big[ \phi^2 \sum_r e^{-{2\over
{\sqrt{3}}}\sum_{\bar a}\gamma_{\bar ar}r_{\bar a}} \times
\nonumber \\
 &&\Big( 8(\partial_rln\, \phi )^2 -{1\over 3}\sum_{\bar b}(\partial_rr_{\bar b})^2-\nonumber \\
 &-&{4\over {\sqrt{3}}} \partial_rln\, \phi \sum_{\bar b}\gamma_{\bar br}\partial_rr_{\bar b}+
 {2\over 3}(\sum_{\bar b}\gamma_{\bar br}\partial_rr_{\bar b})^2
\Big) \Big] (\tau ,\vec \sigma )-\nonumber \\
 &&{}\nonumber \\
 &-&{{6\pi G}\over {c^3}} \phi^{-2}(\tau ,\vec \sigma )\Big[
 2(\phi^{-4}\sum_{\bar a} \pi^2_{\bar a})(\tau ,\vec \sigma )+\nonumber \\
 &+&4(\phi^{-2}\sum_ue^{{1\over {\sqrt{3}}}\sum_{\bar a}\gamma_{\bar
au}r_{\bar a}} \sum_{\bar b}\gamma_{\bar bu}\pi_{\bar b})(\tau
,\vec \sigma )\times \nonumber \\ &&\int d^3\sigma_1 \sum_r
\delta^u_{(a)} {\tilde {\cal T}}^u_{(a)r}(\vec \sigma ,{\vec
\sigma}_1,\tau |\phi ,r_{\bar a}] \Big( \phi^{-2} e^{-{1\over
{\sqrt{3}}}\sum_{\bar a} \gamma_{\bar ar}r_{\bar a}}\sum_{\bar
b}\gamma_{\bar br} \pi_{\bar b}\Big) (\tau ,{\vec
\sigma}_1)+\nonumber \\
 &&{}\nonumber \\
 &+&\int d^3\sigma_1d^3\sigma_2 \Big( \sum_u e^{{2\over {\sqrt{3}}}\sum_{\bar a}
\gamma_{\bar au}r_{\bar a}(\tau ,\vec \sigma )} \times \nonumber
\\ &&\sum_r{\tilde {\cal T}}^u_{(a)r}(\vec \sigma ,{\vec
\sigma}_1,\tau |\phi ,r_{\bar a}] \Big( \phi^{-2} e^{-{1\over
{\sqrt{3}}}\sum_{\bar a} \gamma_{\bar ar}r_{\bar a}}\sum_{\bar
b}\gamma_{\bar br} \pi_{\bar b}\Big) (\tau ,{\vec \sigma}_1)\times
\nonumber \\ &&\sum_s {\tilde {\cal T}}^u_{(a)s}(\vec \sigma
,{\vec \sigma}_2,\tau |\phi ,r_{\bar a}] \Big( \phi^{-2}
e^{-{1\over {\sqrt{3}}}\sum_{\bar a} \gamma_{\bar as}r_{\bar
a}}\sum_{\bar c}\gamma_{\bar cs} \pi_{\bar c}\Big) (\tau ,{\vec
\sigma}_2)+\nonumber \\ &+&\sum_{uv} e^{{1\over
{\sqrt{3}}}\sum_{\bar a}(\gamma_{\bar au}+\gamma_{\bar av})r_{\bar
a}(\tau ,\vec \sigma )}
(\delta^u_{(b)}\delta^v_{(a)}-\delta^u_{(a)} \delta^v_{(b)})\times
\nonumber \\ &&\sum_r {\tilde {\cal T}}^u_{(a)r}(\vec \sigma
,{\vec \sigma}_1,\tau |\phi ,r_{\bar a}] \Big( \phi^{-2}
e^{-{1\over {\sqrt{3}}}\sum_{\bar a} \gamma_{\bar ar}r_{\bar
a}}\sum_{\bar b}\gamma_{\bar br} \pi_{\bar b}\Big) (\tau ,{\vec
\sigma}_1)\nonumber \\ &&\sum_s {\tilde {\cal T}}^v_{(b)s}(\vec
\sigma ,{\vec \sigma}_2,\tau |\phi ,r_{\bar a}] \Big( \phi^{-2}
e^{-{1\over {\sqrt{3}}}\sum_{\bar a} \gamma_{\bar as}r_{\bar
a}}\sum_{\bar c}\gamma_{\bar cs} \pi_{\bar c}\Big) (\tau ,{\vec
\sigma}_2)\, \Big)\, \Big]\, \Big) ,\nonumber \\
 &&{}\nonumber \\
 {\hat P}^r_{ADM,R}&=&
-\int d^3\sigma \phi^{-2}(\tau ,\vec \sigma ) \Big( \phi^{-2}(\tau
,\vec \sigma ) \nonumber \\
 &&\Big[ e^{-{2\over {\sqrt{3}}}\sum_{\bar a}\gamma_{\bar ar}r_{\bar a}}
\sum_{\bar b}\partial_rr_{\bar b}\pi_{\bar b}+\nonumber \\
 &+&2\sqrt{3}  e^{-{2\over {\sqrt{3}}}\sum_{\bar a}\gamma_{\bar ar}r_{\bar a}}
   (2\partial_r ln\, \phi +{1\over {\sqrt{3}}}\sum_{\bar b}\gamma_{\bar
br}\partial_r r_{\bar b})\nonumber \\
 &&\sum_{\bar c}\gamma_{\bar cr}\pi_{\bar c}
\Big] (\tau ,\vec \sigma )+\nonumber \\
 &&{}\nonumber \\
 &+& \sqrt{3} \int d^3\sigma_1 \sum_s
 \Big[ -\sum_u\Big( e^{{1\over {\sqrt{3}}}\sum_{\bar a}(\gamma
 _{\bar au}-2\gamma_{\bar ar})r_{\bar a}}(2\partial_rln\, \phi +{1\over {\sqrt{3}}}
 \sum_{\bar b}\gamma_{\bar bu}\partial_rr_{\bar b})\Big)(\tau ,\vec \sigma )\nonumber \\
 &&\delta^u_{(a)}{\tilde {\cal T}}^u_{(a)s}(\vec \sigma ,{\vec \sigma}_1,\tau |\phi ,r_{\bar a}]
 \nonumber \\
 &+&\sum_{uv}\Big(  e^{-{1\over {\sqrt{3}}}\sum_{\bar a}\gamma_{\bar au}r_{\bar
a}} \Big( \delta^r_u(2\partial_vln\, \phi +{1\over
{\sqrt{3}}}\sum_{\bar b}\gamma_{\bar bu}\partial_vr_{\bar
b})+\nonumber \\
 &+&\delta^r_v(2\partial_uln\, \phi +{1\over {\sqrt{3}}}
 \sum_{\bar b}\gamma_{\bar bv}\partial_ur_{\bar b})\Big)\Big) (\tau ,\vec \sigma )
 \nonumber \\
 &&\delta^u_{(a)}{\tilde {\cal T}}^v_{(a)s}
(\vec \sigma ,{\vec \sigma}_1,\tau |\phi ,r_{\bar a}]
\Big]\nonumber \\
 &&\Big( \phi^{-2}e^{-{1\over {\sqrt{3}}}\sum_{\bar a}\gamma_{\bar as}r_{\bar a}}
\sum_{\bar c}\gamma_{\bar cs}\pi_{\bar c}\Big) (\tau ,{\vec
\sigma}_1)\Big),\nonumber \\
 &&{}\nonumber \\
 {\hat J}^{rs}_{ADM,R}&=&\int d^3\sigma \phi^{-2}(\tau ,\vec \sigma )
\Big( \phi^{-2}(\tau ,\vec \sigma ) \nonumber \\
 &&\Big[ e^{-{2\over {\sqrt{3}}}\sum_{\bar a}\gamma_{\bar ar}r_{\bar a}}
\sum_{\bar b}(\sigma^r \partial_s - \sigma^s \partial_r) r_{\bar
 b}\pi_{\bar b})+\nonumber \\
 &+&2\sqrt{3}\sum_u e^{-{2\over {\sqrt{3}}}\sum_{\bar a}\gamma_{\bar au}r_{\bar a}}
  (\sigma^r \delta^s_u-\sigma^s\delta^r_u)
  (2\partial_uln\, \phi +{1\over {\sqrt{3}}}\sum_{\bar b}\gamma_{\bar
bu}\partial_ur_{\bar b})
 \sum_{\bar c}\gamma_{\bar cu}\pi_{\bar c}
\Big] (\tau ,\vec \sigma )+\nonumber \\
 &&{}\nonumber \\
 &+& \sqrt{3}\int d^3\sigma_1 \sum_w \Big[ \Big(-\sigma^r
 \sum_u\Big( e^{{1\over {\sqrt{3}}}\sum_{\bar a}(\gamma
 _{\bar au}-2\gamma_{\bar as})r_{\bar a}}(2\partial_sln\, \phi +{1\over {\sqrt{3}}}
 \sum_{\bar b}\gamma_{\bar bu}\partial_sr_{\bar b})\Big)(\tau ,\vec \sigma )-\nonumber \\
 &-&\sigma^s
 \sum_u\Big( e^{{1\over {\sqrt{3}}}\sum_{\bar a}(\gamma
 _{\bar au}-2\gamma_{\bar ar})r_{\bar a}}(2\partial_rln\, \phi +{1\over {\sqrt{3}}}
 \sum_{\bar b}\gamma_{\bar bu}\partial_rr_{\bar b})\Big)(\tau ,\vec \sigma )\Big)\nonumber \\
 &&\delta^u_{(a)}{\tilde {\cal T}}^u_{(a)w}(\vec \sigma ,{\vec \sigma}_1,\tau |\phi ,r_{\bar a}]
 +\nonumber \\
 &+&\sum_{uv}\Big(  e^{-{1\over {\sqrt{3}}}\sum_{\bar a}\gamma_{\bar au}r_{\bar
a}} \Big( (\sigma^r\delta^s_u-\sigma^s\delta^r_u)(2\partial_vln\,
\phi +{1\over {\sqrt{3}}}\sum_{\bar b}\gamma_{\bar
bu}\partial_vr_{\bar b})+\nonumber \\
 &+&(\sigma^r\delta^s_v-\sigma^s\delta^r_v)(2\partial_uln\, \phi +{1\over {\sqrt{3}}}
 \sum_{\bar b}\gamma_{\bar bv}\partial_ur_{\bar b})\Big)\Big) (\tau ,\vec \sigma )
 \nonumber \\
 &&\delta^u_{(a)}{\tilde {\cal T}}^v_{(a)w}
(\vec \sigma ,{\vec \sigma}_1,\tau |\phi ,r_{\bar a}]
\Big]\nonumber \\
 &&\Big( \phi^{-2}e^{-{1\over {\sqrt{3}}}\sum_{\bar a}\gamma_{\bar as}r_{\bar a}}
\sum_{\bar c}\gamma_{\bar cs}\pi_{\bar c}\Big) (\tau ,{\vec
\sigma}_1)\Big),\nonumber \\
 &&{}\nonumber \\
 {\hat J}^{\tau r}_{ADM,R}&=&\epsilon \int d^3\sigma   \sigma^r
\Big(  {{c^3}\over {16\pi G}} \Big[ \phi^2 \sum_r e^{-{2\over
{\sqrt{3}}}\sum_{\bar a}\gamma_{\bar ar}r_{\bar a}} \times
\nonumber \\
 &&\Big( 8(\partial_rln\, \phi )^2 -{1\over 3}\sum_{\bar b}(\partial_rr_{\bar b})^2-\nonumber \\
 &-&{4\over {\sqrt{3}}} \partial_r\phi \sum_{\bar b}\gamma_{\bar br}\partial_rr_{\bar b}+
 {2\over 3}(\sum_{\bar b}\gamma_{\bar br}\partial_rr_{\bar b})^2
\Big) \Big] (\tau ,\vec \sigma )-\nonumber \\
 &&{}\nonumber \\
 &-&{{6\pi G}\over {c^3}} \phi^{-2}(\tau ,\vec \sigma )
   \Big\{  2(\phi^{-4}\sum_{\bar a} \pi^2_{\bar a})(\tau ,\vec \sigma )+\nonumber
\\
 & & + 4 \Big[ \phi^{-2}\sum_ue^{{1\over {\sqrt{3}}}\sum_{\bar a}
              \gamma_{\bar au}r_{\bar a}} \sum_{\bar b}
              \gamma_{\bar bu}\pi_{\bar b})(\tau ,\vec \sigma )\times \nonumber \\
 & & ~~  \int d^3\sigma_1 \sum_m \delta^u_{(a)} {\tilde {\cal T}}^u_{(a)m}
         (\vec\sigma ,{\vec\sigma}_1,\tau |\phi ,r_{\bar a}] \Big( \phi^{-2}
          e^{-{1\over {\sqrt{3}}}\sum_{\bar a}
          \gamma_{\bar am}r_{\bar a}}\sum_{\bar b}\gamma_{\bar bm}
          \pi_{\bar b}\Big) (\tau ,{\vec \sigma}_1)+\nonumber \\
 &&{}\nonumber \\
 && + \int d^3\sigma_1d^3\sigma_2 \Big( \sum_u e^{{2\over {\sqrt{3}}}\sum_{\bar a}
      \gamma_{\bar au}r_{\bar a}(\tau ,\vec \sigma )} \times \nonumber \\
 && ~~ \sum_m{\tilde {\cal T}}^u_{(a)m}(\vec \sigma ,{\vec\sigma}_1,\tau
       |\phi ,r_{\bar a}] \Big( \phi^{-2} e^{-{1\over {\sqrt{3}}}\sum_{\bar a}
        \gamma_{\bar am}r_{\bar a}}\sum_{\bar b}\gamma_{\bar bm}
        \pi_{\bar b}\Big) (\tau ,{\vec \sigma}_1)\times \nonumber \\
 && ~~ \sum_s {\tilde {\cal T}}^u_{(a)s}
             (\vec \sigma ,{\vec \sigma}_2,\tau |\phi ,r_{\bar a}]
          \Big( \phi^{-2} e^{-{1\over {\sqrt{3}}}\sum_{\bar a}
          \gamma_{\bar as}r_{\bar a}}\sum_{\bar c}\gamma_{\bar cs}
         \pi_{\bar c}\Big) (\tau ,{\vec \sigma}_2)+\nonumber \\
 && + \sum_{uv} e^{{1\over {\sqrt{3}}}\sum_{\bar a}(\gamma_{\bar au}+\gamma_{\bar
av})r_{\bar a}(\tau ,\vec \sigma )}
(\delta^u_{(b)}\delta^v_{(a)}-\delta^u_{(a)} \delta^v_{(b)})\times
\nonumber \\
 && ~~ \sum_m {\tilde {\cal T}}^u_{(a)m}(\vec \sigma
,{\vec \sigma}_1,\tau |\phi ,r_{\bar a}] \Big( \phi^{-2}
e^{-{1\over {\sqrt{3}}}\sum_{\bar a} \gamma_{\bar am}r_{\bar
a}}\sum_{\bar b}\gamma_{\bar bm} \pi_{\bar b}\Big) (\tau ,{\vec
\sigma}_1)\nonumber \\
 &&~~ \sum_s {\tilde {\cal T}}^v_{(b)s}
     (\vec \sigma ,{\vec \sigma}_2,\tau |\phi ,r_{\bar a}] \Big( \phi^{-2}
e^{-{1\over {\sqrt{3}}}\sum_{\bar a} \gamma_{\bar as}r_{\bar
a}}\sum_{\bar c}\gamma_{\bar cs} \pi_{\bar c}\Big) (\tau ,{\vec
\sigma}_2)\,
  \Big)\, \Big]\, \Big\} -\nonumber \\
 &&{}\nonumber \\
 &-&{{\epsilon c^3}\over {8\pi G}} \int d^3\sigma \Big[ \phi^{-2}
\sum_{uv}\delta^r_u(\delta_{uv}-1) (\phi^4e^{{2\over
{\sqrt{3}}}\sum_{\bar a} \gamma_{\bar au}r_{\bar a}}-1)\nonumber
\\
 &&e^{-{2\over {\sqrt{3}}}\sum_{\bar a}(\gamma_{\bar av}-\gamma_{\bar au})r_{\bar a}}
 (\partial_uln\, \phi +{1\over {\sqrt{3}}}\sum_{\bar b}(\gamma_{\bar bv}-
 \gamma_{\bar bu})\partial_ur_{\bar b})\Big] (\tau ,\vec \sigma ).
 \label{a12}
 \end{eqnarray}

Eqs.(194) of Ref.\cite{3} give the equation for the determination
of the lapse function $n(\tau ,\vec \sigma )$ as a consequence of
the time constancy of the gauge fixing $\rho (\tau ,\vec \sigma )
= {{\pi_{\phi}(\tau ,\vec \sigma )}\over {2\, \phi(\tau ,\vec
\sigma )}} \approx 0$

\begin{eqnarray}
\partial_{\tau} \rho (\tau ,\vec \sigma )\, &{\buildrel \circ \over =}\,&
\{ \rho (\tau ,\vec \sigma ), {\hat H}_{(D)ADM,R} \} = \int
d^3\sigma_1 n(\tau ,{\vec \sigma}_1) \{ \rho (\tau ,\vec \sigma ),
{\hat {\cal H}}_R(\tau ,{\vec \sigma}_1) \} +\nonumber \\ &+&
{\tilde \lambda}_{\tau}(\tau ) \{ \rho (\tau ,\vec \sigma ),{\hat
P}^{\tau} _{ADM,R} \} + {\tilde \lambda}_r(\tau ) \{ \rho (\tau
,\vec \sigma ), {\hat P}^r _{ADM,R} \} \approx \nonumber \\
&\approx& -{1\over 2}\phi (\tau ,\vec \sigma ) \Big[ \int
d^3\sigma_1 n(\tau ,{\vec \sigma}_1) {{\delta {\hat {\cal
H}}_R(\tau ,{\vec \sigma}_1)}\over {\delta \phi (\tau ,\vec \sigma
)}}+\nonumber \\ &+& {\tilde \lambda}_{\tau} {{\delta {\hat
P}^{\tau}_{ADM,R}}\over {\delta \phi (\tau ,\vec \sigma
)}}+{\tilde \lambda}_r(\tau ) {{\delta {\hat P}^r_{ADM,R}}\over
{\delta \phi (\tau ,\vec \sigma )}}  \Big]\approx 0,\nonumber \\
 &&{}\nonumber \\
\Rightarrow&& \quad n(\tau ,\vec \sigma ) - \hat n(\tau ,\vec
\sigma |r_{\bar a} ,\pi_{\bar a},{\tilde \lambda}_A] \approx
0,\nonumber \\ &&{}\nonumber \\
\partial_{\tau}&& \Big[ n(\tau ,\vec \sigma ) - \hat n(\tau ,\vec \sigma
|r_{\bar a},\pi_{\bar a},{\tilde \lambda}_A] \Big] =\nonumber \\
&=&\lambda_n(\tau ,\vec \sigma )- \{ \hat n(\tau ,\vec \sigma
|r_{\bar a},\pi_{\bar a},{\tilde \lambda}_A] , {\hat
H}^{(WSW)}_{(D)ADM,R} \} \approx 0,\nonumber \\ &&{}\nonumber \\
\Rightarrow&& \quad \lambda_n(\tau ,\vec \sigma )\quad
\text{determined};\nonumber \\
 &&{}\nonumber \\
 && \text{the rest-frame instant form expression of this
 equation is}\nonumber \\
 &&{}\nonumber \\
 \int d^3\sigma_1 && n(\tau ,{\vec \sigma}_1){{\delta {\hat {\cal H}}_R(\tau ,{\vec
\sigma}_1)}\over {\delta \phi (\tau ,\vec \sigma )}} =-\epsilon
{{\delta {\hat P}^{\tau}_{ADM,R}}\over {\delta \phi (\tau ,\vec
\sigma )}}.
 \label{a13}
 \end{eqnarray}

For $\pi_{\phi}(\tau ,\vec \sigma ) = 0$ the shift functions
$n_r(\tau ,\vec \sigma )$ of our gauge, as functions of the lapse
function $n(\tau ,\vec \sigma )$, have the form given in Eq.(187)
of Ref.\cite{3}, i.e.

\begin{eqnarray}
 n_r(\tau ,\vec \sigma ) &\approx&
 -{{\epsilon \, 4\sqrt{3} \pi G}\over {c^3}}
  \Big[ \phi^2 e^{{1\over {\sqrt{3}}}\sum_{\bar a}\gamma_{\bar ar}r_{\bar a}}
\Big] (\tau ,\vec \sigma ) \int d^3\sigma_1 [\epsilon - n(\tau
,{\vec \sigma}_1)] \phi^{-2}(\tau ,{\vec \sigma}_1)\nonumber \\
 &&\sum_{wu}e^{{1\over {\sqrt{3}}}\sum_{\bar a}(\gamma_{\bar aw}+\gamma_{\bar au})
 r_{\bar a}(\tau ,{\vec \sigma}_1)} \, \Big( \delta_{wu}\delta_{(b)(d)}+\delta_{(b)u}
 \delta_{(d)w}-\delta_{(b)w}\delta_{(d)u}\Big)\nonumber \\
 &&\sum_v \int d^3\sigma_2 {\cal K}^w_{(b)v}({\vec \sigma}_1, {\vec \sigma}_2;\tau )
 \Big[ \phi^{-2} e^{-{1\over {\sqrt{3}}}\sum_{\bar a}
 \gamma_{\bar au}r_{\bar a}}\Big] (\tau ,{\vec \sigma}_2 )\nonumber \\
 &&\sum_{\bar b}\gamma_{\bar bv}\pi_{\bar b}(\tau ,{\vec \sigma}_2) \, G^{ur}_{(d)}({\vec
 \sigma}_1, \vec \sigma ;\tau ).
 \label{a14}
 \end{eqnarray}

For $\pi_{\phi} = 0$ the extrinsic curvature of our WSW
hyper-surfaces, the ADM metric momentum and the DeWitt
supermetric, given by  Eqs.(186) of Ref.\cite{3}, are

\bea
 {}^3{\hat K}_{rs}(\tau ,\vec \sigma ) &=&{{\epsilon \, 4\pi G}\over {c^3}} [e^{
{1\over {\sqrt{3}}}\sum_{\bar c} (\gamma_{\bar cr}+\gamma_{\bar
cs})r_{\bar c} } \sum_u(\delta_{ru}\delta_{(a)s}
+\delta_{su}\delta_{(a)r}-\delta_{rs}\delta_{(a)u})\nonumber \\
 &&e^{{1\over {\sqrt{3}}} \sum_{\bar c}\gamma_{\bar cu}r_{\bar c}}\,
{}^3{\hat {\tilde \pi}}^u_{(a)}](\tau ,\vec \sigma ),\nonumber \\
 &&{}\nonumber \\
{}^3{\hat K}(\tau ,\vec \sigma ) &=& -{{\epsilon \, 4\pi G}\over
{c^3}} [\phi^{-4} \sum_u \delta_{(a)u} e^{ {1\over
{\sqrt{3}}}\sum_{\bar c}\gamma_{\bar cu}r_{\bar c}}\, {}^3{\hat
{\tilde \pi}} ^u_{(a)}](\tau ,\vec \sigma )=\nonumber
\\
 &=&  -{{\epsilon \, 4\sqrt{3} \pi G
}\over {c^3}} \phi^{-4}(\tau ,\vec \sigma )  \sum_u \delta_{(a)u}
e^{ {1\over {\sqrt{3}}}\sum_{\bar c}\gamma_{\bar cu}r_{\bar
c}(\tau ,\vec \sigma )}\nonumber \\
 &&\sum_s\sum_{\bar b}
\gamma_{\bar bs}\int d^3\sigma_1 {\cal K}^r_{(a)s}(\vec \sigma
,{\vec \sigma}_1;\tau |\phi ,r_{\bar a}] (\phi^{-2}e^{-{1\over
{\sqrt{3}}}\sum_{\bar a}\gamma_{\bar as}r_{\bar a}} \pi_{\bar
b})(\tau ,{\vec \sigma}_1) ,\nonumber \\
 &&{}\nonumber \\
 &&{}\nonumber \\
 {}^3{\hat {\tilde \Pi}}^{rs}(\tau ,\vec \sigma )&=&{1\over
4}[{}^3{\hat e}^r _{(a)}\, {}^3{\hat {\tilde \pi}}^s_{(a)}
+{}^3{\hat e}^s_{(a)}\, {}^3{\hat {\tilde \pi}}^r_{(a)}](\tau
,\vec \sigma )=\nonumber \\
 &=&{1\over 4}\phi^{-2}(\tau ,\vec \sigma )[e^{-{1\over {\sqrt{3}}}\sum_{\bar a}
\gamma_{\bar ar}r_{\bar a}} \delta^r_{(a)}\, {}^3{\hat {\tilde
\pi}}^s_{(a)}+ \nonumber \\
 &+&e^{-{1\over {\sqrt{3}}}\sum_{\bar a}\gamma_{\bar as}r_{\bar a}}
\delta^s_{(a)}\, {}^3{\hat {\tilde \pi}}^r_{(a)}](\tau ,\vec
\sigma )=\nonumber \\
 &=&{{\sqrt{3}}\over 4}\phi^{-2}(\tau ,\vec \sigma )\Big[ e^{-{1\over {\sqrt{3}}}\sum_{\bar a}
\gamma_{\bar ar}r_{\bar a}} \delta^r_{(a)}\nonumber \\
 && \sum_u \int d^3\sigma_1 {\cal K}^r_{(a)u}(\vec \sigma
 ,{\vec \sigma}_1,\tau |\phi ,r_{\bar a},\tilde \Pi ] +\nonumber \\
&+&e^{-{1\over {\sqrt{3}}}\sum_{\bar a}\gamma_{\bar as}r_{\bar a}}
\delta^s_{(a)}
 \sum_u \int d^3\sigma_1 {\cal K}^s_{(a)u}(\vec \sigma
 ,{\vec \sigma}_1,\tau |\phi ,r_{\bar a},\tilde \Pi ] \Big] \nonumber \\
 &&(\phi^{-2}e^{-{1\over {\sqrt{3}}}\sum_{\bar a}\gamma_{\bar au}r_{\bar
a}}) (\tau ,{\vec \sigma}_1)\, \sum_{\bar b}\gamma_{\bar bu}
\pi_{\bar b}(\tau ,{\vec \sigma}_1),\nonumber \\
 &&{}\nonumber \\
 {}^3{\hat G}_{rsuv}(\tau ,\vec \sigma
) &=& [{}^3{\hat g}_{ru}\, {}^3{\hat g} _{sv}+{}^3{\hat g}_{rv}\,
{}^3{\hat g}_{su} -{}^3{\hat g}_{rs}\, {}^3{\hat g} _{uv}](\tau
,\vec \sigma )=\nonumber \\
 &=&\phi^8(\tau ,\vec \sigma
)[e^{{2\over {\sqrt{3}}}\sum_{\bar a}(\gamma_{\bar
ar}+\gamma_{\bar as})r_{\bar a}}
(\delta_{ru}\delta_{sv}+\delta_{rv}\delta _{su})-\nonumber \\
 &-&e^{{2\over {\sqrt{3}}}\sum_{\bar a}(\gamma_{\bar ar}+\gamma_{\bar au})
r_{\bar a}} \delta_{rs}\delta_{uv}](\tau ,\vec \sigma ).
 \label{a15}
 \end{eqnarray}

The tetrads and cotetrads adapted to the WSW hyper-surfaces of our
gauge, given in Eq.(1) of Ref.\cite{3}, are

\begin{eqnarray}
{}^4_{(\Sigma )}{\check E}^{\mu}_{(\alpha )}&=&\lbrace
{}^4_{(\Sigma )}{\check E}^{\mu}_{(o)}=l^{\mu}= {\hat
b}^{\mu}_l={1\over N}(b^{\mu} _{\tau}-N^rb^{\mu}_r); \,\,
{}^4_{(\Sigma )}{\check E}^{\mu}_{(a)}={}^3e^s_{(a)} b^{\mu}_s
\rbrace ,\nonumber \\ {}^4_{(\Sigma )}{\check E}_{\mu}^{(\alpha
)}&=&\lbrace {}^4_{(\Sigma )}{\check E}_{\mu}^{(o)}=\epsilon
l_{\mu}= {\hat b}^l_{\mu}= N b^{\tau} _{\mu};\,\, {}^4_{(\Sigma
)}{\check E}_{\mu}^{(a)}={}^3e_s^{(a)} {\hat b}^s_{\mu}\rbrace
,\nonumber \\ &&{}\nonumber \\ {}^4_{(\Sigma )}{\check
E}^{\mu}_{(\alpha )}&& {}^4g_{\mu\nu}\,\,\, {}^4_{(\Sigma
)}{\check E}^{\nu}_{(\beta )} = {}^4\eta_{(\alpha )(\beta )},
\nonumber \\
 &&{}\nonumber \\
{}^4_{(\Sigma )}{\check {\tilde E}}^A_{(\alpha )}&=&{}^4_{(\Sigma
)}{\check E} ^{\mu}_{(\alpha )}\, b^A_{\mu},\quad \Rightarrow
{}^4_{(\Sigma )}{\check {\tilde E}}^A_{(o)}=\epsilon l^A,\nonumber
\\
 &&{}\nonumber \\
&&{}^4_{(\Sigma )}{\check {\tilde E}}^{\tau}_{(o)}={1\over
N},\quad\quad {}^4_{(\Sigma )}{\check {\tilde
E}}^{\tau}_{(a)}=0,\nonumber \\ &&{}^4_{(\Sigma )}{\check {\tilde
E}}^r_{(o)}=-{{N^r}\over N},\quad\quad {}^4_{(\Sigma )}{\check
{\tilde E}}^r_{(a)}={}^3e^r_{(a)};\nonumber \\ {}^4_{(\Sigma
)}{\check {\tilde E}}_A^{(\alpha )}&=&{}^4_{(\Sigma )}{\check E}
_{\mu}^{(\alpha )}\, b^{\mu}_A,\quad \Rightarrow {}^4_{(\Sigma
)}{\check {\tilde E}}_A^{(o)} = l_A,\nonumber \\
 &&{}\nonumber \\
 &&{}^4_{(\Sigma )}{\check {\tilde E}}^{(o)}_{\tau}=N,\quad\quad
{}^4_{(\Sigma )}{\check {\tilde E}}^{(a)}_{\tau}=N^r\,
{}^3e^{(a)}_r=N^{(a)}, \nonumber \\
 &&{}^4_{(\Sigma )}{\check {\tilde E}}^{(o)}_r=0,\quad\quad
{}^4_{(\Sigma )}{\check {\tilde
E}}^{(a)}_r={}^3e^{(a)}_r,\nonumber \\
 &&{}\nonumber \\
 &&{}^4_{(\Sigma )}{\check E}^A_{(\alpha )}\,
{}^4g_{AB}\, {}^4_{(\Sigma )}{\check E}^B_{(\beta
)}={}^4\eta_{(\alpha )(\beta )}.
 \label{a16}
 \end{eqnarray}

The 3-metric on our WSW hyper-surfaces and the line element of our
space-time, given by Eqs.(183) of Ref.\cite{3}, are

\begin{eqnarray}
{}^3{\hat g}_{rs}&=&e^{2q}\, \left( \begin{array}{ccc} e^{ {2\over
{\sqrt{3}}} \sum_{\bar a}\gamma_{\bar a1}r_{\bar a}}&0&0\\ 0& e^{
{2\over {\sqrt{3}}} \sum_{\bar a}\gamma_{\bar a2}r_{\bar a}}&0\\
0&0& e^{ {2\over {\sqrt{3}}} \sum_{\bar a}\gamma_{\bar a3}r_{\bar
a}} \end{array} \right) = \phi^4\,\, {}^3{\hat g}^{diag}_{rs},
\nonumber \\
 &&{}\nonumber \\
  \hat \gamma &=&{}^3\hat g= {}^3{\hat
e}^2=e^{6q}= \phi^{12},\quad\quad det\, |{\hat g}^{diag}_{rs}|
=1,\nonumber \\
 &&{}\nonumber \\
  d{\hat s}^2&=&
 \epsilon \Big( [-\epsilon +n]^2(d\tau )^2-\nonumber \\
 &&-\delta_{uv}[\phi^2 e^{ {2\over {\sqrt{3}}}\sum_{\bar
a} \gamma_{\bar au} r_{\bar a}} d\sigma^u+\phi^{-2} e^{- {2\over
{\sqrt{3}}}\sum_{\bar a} \gamma_{\bar au} r_{\bar a}}\, n_u\,
d\tau ]\nonumber \\
 &&[\phi^2 e^{ {2\over {\sqrt{3}}}\sum_{\bar a}
\gamma_{\bar av} r_{\bar a}} d\sigma^v+\phi^{-2} e^{- {2\over
{\sqrt{3}}}\sum_{\bar a} \gamma_{\bar av} r_{\bar a}}\, n_v\,
d\tau ] \Big),\nonumber \\
 &&{}\nonumber \\
 q&=& 2 ln\, \phi = {1\over 6}\,
ln\, {}^3\hat g,\qquad r_{\bar a}={{\sqrt{3}}\over 2} \sum_r
\gamma_{\bar ar}\, ln\, {{{}^3{\hat g} _{rr}}\over {{}^3\hat g}}.
 \label{a17}
 \end{eqnarray}

The Frauendiener equations for the WSW triads, given in Eq.(200)
of Ref.\cite{3}, are

\begin{eqnarray}
&&{}^3\nabla_r\, {}^3e^{(WSW) r}_{(1)}={}^3\nabla_r\, {}^3e^{(WSW)
r}_{(2)}=0,\nonumber \\
 &&{}^3\nabla_r\, {}^3e^{(WSW) r}_{(3)}=-\alpha
{}^3K,\nonumber \\
 &&{}^3e^{(WSW) r}_{(1)}\, {}^3e^{(WSW) s}_{(3)}\,
{}^3\nabla_r\, {}^3e^{(WSW)}_{(2)s} +{}^3e^{(WSW) r}_{(3)}\,
 {}^3e^{(WSW) s}_{(2)}\, {}^3\nabla_r\, {}^3e^{(WSW)}_{(1)s}+\nonumber \\
 &+&{}^3e^{(WSW) r}_{(2)}\, {}^3e^{(WSW) s}_{(1)}\, {}^3\nabla_r\,
{}^3e^{(WSW)}_{(3)s}=0.
 \label{a18}
 \end{eqnarray}

The WSW tetrads of Eq.(201) of Ref.\cite{3} are

\begin{eqnarray}
{}^4_{(\Sigma )}{\check {\tilde E}}^{(WSW)}{}^A_{(o)}&=& {1\over
{-\epsilon +n}} (1; -n^r),\nonumber \\
  {}^4_{(\Sigma )}{\check {\tilde E}}^{(WSW)}{}^A_{(a)} &=& (0; {}^3e^{(WSW)}{}^r_{(a)}),
  \nonumber \\
  &&{}\nonumber \\
  {}^4_{(\Sigma )}{\check E}^{(WSW)}{}^{\mu}_{(o)} &=& l^{\mu},\nonumber \\
  {}^4_{(\Sigma )}{\check E}^{\mu}_{(a)} &=& b^{\mu}_s\,\,\, {}^3e^{(WSW)}{}^s_{(a)}.
 \label{a19}
 \end{eqnarray}

The embedding of the WSW hyper-surfaces of oue gauge into the
space-time, given in Eq.(208) of Ref.\cite{3}, are

\begin{eqnarray}
z^{\mu}_{(WSW)}(\tau ,\vec \sigma)&=& \delta^{\mu}_{(\mu )}
x^{(\mu )}_{(\infty )}(0)+ l^{\mu}(\tau ,\vec \sigma ) \tau
+\epsilon^{\mu}_r(\tau ,\vec \sigma ) \sigma^r =\nonumber \\
 &=&x^{\mu}_{(\infty )}(0)+ l^{\mu}(\tau ,\vec \sigma ) \tau + b^{\mu}_s(\tau ,\vec \sigma )
 \, {}^3e^{(WSW)}{}^s_{(a)}(\tau ,\vec \sigma ) \delta_{(a)r} \sigma^r=\nonumber \\
 &=& x^{\mu}_{(\infty )}(0) +b^{\mu}_A(\tau ,\vec \sigma ) F^A(\tau ,\vec \sigma ),\nonumber \\
  &&{}\nonumber \\
  &&F^{\tau}(\tau ,\vec \sigma )= {{\tau}\over {-\epsilon +n(\tau ,\vec \sigma )}},
  \nonumber \\
  && F^s(\tau ,\vec \sigma )={}^3e^{(WSW)}{}^s_{(a)}(\tau ,\vec \sigma ) \delta
  _{(a)r} \sigma^r -{{n^s(\tau ,\vec \sigma )}\over {-\epsilon +n(\tau ,\vec \sigma )}} \tau ,
 \label{a20}
 \end{eqnarray}

\noindent with the transition coefficients $b^{\mu}_A$ determined
by  Eqs.(209) of Ref.\cite{3}

\begin{eqnarray}
b^{\mu}_A &=& {{\partial z^{\mu}_{(WSW)}}\over {\partial
\sigma^A}}= b^{\mu}_B {{\partial F^B}\over {\partial \sigma^A}} +
{{\partial b^{\mu}_B}\over {\partial \sigma^A}} F^B,\nonumber \\
 &&{}\nonumber \\
 &&A_A{}^B = \delta_A^B - {{\partial F^B}\over {\partial \sigma^A}},\nonumber \\
 &&{}\nonumber \\
 F^B {{\partial b^{\mu}_B}\over {\partial \sigma^A}} &=& A_A{}^B b^{\mu}_B,\nonumber \\
 &&or\quad b^{\mu}_b = (A^{-1})_B{}^a F^C {{\partial b^{\mu}_C}\over {\partial \sigma^A}}.
 \label{a21}
 \end{eqnarray}

\vfill\eject

\section{The Kernels $F^{(o)r}_{(a)(b)}$.}

As shown in Ref.\cite{3} the kernels $F^{(o)r}_{(a)(b)}$'s, zero
curvature limit of the kernels $F^r_{(a)(b)}$'s of Eq.(\ref{I4}),
are given by the equations

\bea
 F^{(o)s}_{(a)(b)}(\vec \sigma ,{\vec \sigma}_1)&=&{1\over
 2}\Big[\epsilon_{(a)(s)(b)}\delta^3(\vec \sigma ,{\vec
 \sigma}_1)-\sum_{u,r}\epsilon_{(u)(r)(b)}{{\partial
 G^{(o)su}_{(a)}(\vec \sigma ,{\vec \sigma}_1)}\over {\partial
 \sigma_1^r}}\Big],\nonumber \\
 &&{}\nonumber \\
F^{(o) r}_{(a)(1)}({\vec \sigma}_1 ,{\vec \sigma};\tau ) &=&
 {{\partial G^{(o) r3}_{(a)}({\vec \sigma}_1 ,{\vec \sigma};\tau )}\over
 {\partial \sigma_1^2}} - \delta_{(a)3} \delta^r_2 \delta^3(\vec \sigma ,
 {\vec \sigma}_1) =\nonumber \\
 &=& \delta_{(a)2} \delta^r_3 \delta^3(\vec \sigma , {\vec \sigma}_1)
  - {{\partial G^{(o) r2}_{(a)}({\vec \sigma}_1 ,{\vec \sigma};\tau )}
 \over {\partial \sigma_1^3}},\nonumber \\
 F^{(o) r}_{(a)(2)}({\vec \sigma}_1 ,{\vec \sigma};\tau ) &=&
 {{\partial G^{(o) r1}_{(a)}({\vec \sigma}_1 ,{\vec \sigma};\tau )}\over
 {\partial \sigma_1^3}} - \delta_{(a)1} \delta^r_3 \delta^3(\vec \sigma ,
 {\vec \sigma}_1) =\nonumber \\
 &=& \delta_{(a)3} \delta^r_1 \delta^3(\vec \sigma , {\vec \sigma}_1)
  -{{\partial G^{(o) r3}_{(a)}({\vec \sigma}_1 ,{\vec \sigma};\tau )}
 \over {\partial \sigma_1^1}},\nonumber \\
 F^{(o) r}_{(a)(3)}({\vec \sigma}_1 ,{\vec \sigma};\tau ) &=&
 {{\partial G^{(o) r2}_{(a)}({\vec \sigma}_1 ,{\vec \sigma};\tau )}\over
 {\partial \sigma_1^1}} - \delta_{(a)2} \delta^r_1 \delta^3(\vec \sigma ,
 {\vec \sigma}_1) =\nonumber \\
 &=& \delta_{(a)1} \delta^r_2 \delta^3(\vec \sigma , {\vec \sigma}_1)
  -{{\partial G^{(o) r1}_{(a)}({\vec \sigma}_1 ,{\vec \sigma};\tau )}
 \over {\partial \sigma_1^2}},
\label{b1}
 \eea

\noindent so that from Eqs.(\ref{II12}) we get

\bea
 F^{(o) a}_{(a)(b)}({\vec \sigma} ,{\vec \sigma}_1) &=& 0,\qquad a,b = 1,2,3,\nonumber \\
 &&{}\nonumber \\
 F^{(o) 3}_{(2)(1)}({\vec \sigma} ,{\vec \sigma}_1) &=&-
 F^{(o) 2}_{(3)(1)}({\vec \sigma} ,{\vec \sigma}_1) =
 F^{(o) 1}_{(3)(2)}({\vec \sigma} ,{\vec \sigma}_1) =-
 F^{(o) 3}_{(1)(2)}({\vec \sigma} ,{\vec \sigma}_1) =\nonumber \\
 &=& F^{(o) 2}_{(1)(3)}({\vec \sigma} ,{\vec \sigma}_1) =-
 F^{(o) 1}_{(2)(3)}({\vec \sigma} ,{\vec \sigma}_1) = {1\over 2}
 \delta^3(\vec \sigma ,{\vec \sigma}_1),\nonumber \\
 &&{}\nonumber \\
 F^{(o) 2}_{(1)(1)}({\vec \sigma} ,{\vec \sigma}_1) &=&
 F^{(o) 1}_{(2)(1)}({\vec \sigma} ,{\vec \sigma}_1) =\nonumber \\
 &=& -{1\over 2} {{\partial}\over {\partial \sigma_1^3}}
  \int_{-\infty}^{\sigma_1^1} dw_1^1 \delta^3(\vec \sigma , w_1^1\sigma_1^2\sigma_1^3
 )=\nonumber \\
 &=& -{1\over 2} \theta (\sigma_1^1,\sigma^1) \delta (\sigma^2,\sigma^2_1)
 {{\partial \delta (\sigma^3,\sigma^3_1)}\over {\partial \sigma_1^3}},\nonumber \\
  F^{(o) 3}_{(2)(3)}({\vec \sigma} ,{\vec \sigma}_1) &=&
 F^{(o) 2}_{(3)(3)}({\vec \sigma} ,{\vec \sigma}_1) =\nonumber \\
 &=& {1\over 2} {{\partial}\over {\partial \sigma_1^1}}
  \int_{-\infty}^{\sigma_1^3} dw_1^3 \delta^3(\vec \sigma , \sigma_1^1\sigma_1^2w_1^3
 )=\nonumber \\
 &=& {1\over 2} {{\partial \delta (\sigma^1,\sigma^1_1)}\over {\partial \sigma_1^1}}
 \delta (\sigma^2,\sigma^2_1) \theta (\sigma_1^3,\sigma^3),\nonumber \\
 F^{(o) 3}_{(1)(1)}({\vec \sigma} ,{\vec \sigma}_1) &=&
 F^{(o) 1}_{(3)(1)}({\vec \sigma} ,{\vec \sigma}_1) =\nonumber \\
 &=& {1\over 2} {{\partial}\over {\partial \sigma_1^2}}
  \int_{-\infty}^{\sigma_1^1} dw_1^1 \delta^3(\vec \sigma , w_1^1\sigma_1^2\sigma_1^3
 )=\nonumber \\
 &=& {1\over 2} \theta (\sigma_1^1,\sigma^1) {{\partial \delta (\sigma^2,\sigma^2_1)}\over
 {\partial \sigma_1^2}} \delta (\sigma^3,\sigma^3_1),\nonumber \\
 F^{(o) 3}_{(2)(2)}({\vec \sigma} ,{\vec \sigma}_1) &=&
 F^{(o) 2}_{(3)(2)}({\vec \sigma} ,{\vec \sigma}_1) =\nonumber \\
 &=& -{1\over 2} {{\partial}\over {\partial \sigma_1^1}}
  \int_{-\infty}^{\sigma_1^2} dw_1^2 \delta^3(\vec \sigma , \sigma_1^1w_1^2\sigma_1^3
 )=\nonumber \\
 &=& -{1\over 2} {{\partial \delta (\sigma^1,\sigma^1_1)}\over {\partial \sigma_1^1}}
 \theta (\sigma_1^2,\sigma^2) \delta (\sigma^3,\sigma^3_1),\nonumber \\
 F^{(o) 2}_{(1)(2)}({\vec \sigma} ,{\vec \sigma}_1) &=&
 F^{(o) 1}_{(2)(2)}({\vec \sigma} ,{\vec \sigma}_1) =\nonumber \\
 &=& {1\over 2} {{\partial}\over {\partial \sigma_1^3}}
  \int_{-\infty}^{\sigma_1^2} dw_1^2 \delta^3(\vec \sigma , \sigma_1^1w_1^2\sigma_1^3
 )=\nonumber \\
 &=& {1\over 2} \delta (\sigma^1,\sigma^1_1) \theta (\sigma_1^2,\sigma^2)
 {{\partial \delta (\sigma^3,\sigma^3_1)}\over {\partial \sigma_1^3}},\nonumber \\
 F^{(o) 3}_{(1)(3)}({\vec \sigma} ,{\vec \sigma}_1) &=&
 F^{(o) 1}_{(3)(3)}({\vec \sigma} ,{\vec \sigma}_1) =\nonumber \\
 &=& -{1\over 2} {{\partial}\over {\partial \sigma_1^2}}
  \int_{-\infty}^{\sigma_1^3} dw_1^3 \delta^3(\vec \sigma , \sigma_1^1\sigma_1^2w_1^3
 )=\nonumber \\
 &=& -{1\over 2} \delta (\sigma^1,\sigma^1_1) {{\partial \delta (\sigma^2,\sigma^2_1)}\over
 {\partial \sigma_1^2}} \theta (\sigma_1^3,\sigma^3).
 \label{b2}
 \eea

\vfill\eject

\section{Fourier Transforms.}

By assuming the validity of the Fourier transform on the
linearized WSW CMC-hyper-surfaces, which are assimilated to flat
$R^3$ surfaces at the lowest level, for the real functions
$r_{\bar a}(\tau ,\vec \sigma )$, $\pi_{\bar a}(\tau ,\vec \sigma
)$ we have

\bea
 r_{\bar a}(\tau ,\vec \sigma ) &=& \int {{d^3k}\over {(2\pi
)^3}} {\tilde r}_{\bar a}(\tau ,\vec k) e^{i\vec k\cdot \vec
\sigma},\qquad {\tilde r}^*_{\bar a}(\tau ,\vec k) = {\tilde
r}_{\bar a}(\tau , -\vec k), \nonumber \\
 \pi_{\bar a}(\tau ,\vec \sigma ) &=& \int {{d^3k}\over {(2\pi )^3}}
{\tilde \pi}_{\bar a}(\tau ,\vec k) e^{i\vec k\cdot \vec
\sigma},\qquad {\tilde \pi}^*_{\bar a}(\tau ,\vec k) = {\tilde
\pi}_{\bar a}(\tau , -\vec k), \nonumber \\
 &&{}\nonumber \\
 &&{\tilde r}_{\bar a}(\tau ,\vec k ) = \int
d^3{\sigma} r_{\bar a}(\tau ,\vec \sigma) e^{-i\vec k\cdot \vec
\sigma}, \qquad
 {\tilde \pi}_{\bar a}(\tau ,\vec k ) = \int
d^3{\sigma} {\pi}_{\bar a}(\tau ,\vec \sigma) e^{-i\vec k\cdot
\vec \sigma},\nonumber \\
 &&{}\nonumber \\
 \{{\tilde r}_{\bar a}(\tau ,\vec k),{\tilde \pi}_{\bar b}(\tau
,\vec k_{1})\}&=&\{\int d^3{\sigma} r_{\bar a}(\tau ,\vec \sigma)
e^{-i\vec k\cdot \vec \sigma},\int d^3{\sigma_{1}} \pi_{\bar
b}(\tau ,\vec \sigma_{1}) e^{-i\vec k_{1}\cdot \vec
\sigma_{1}}\}\nonumber\\
 &&=\delta_{\bar a \bar b}\int
d^3{\sigma_{1}}\int d^3{\sigma} e^{-i\vec k_{1}\cdot \vec
\sigma_{1}} e^{-i\vec k\cdot \vec \sigma}\delta(\vec \sigma - \vec
\sigma_{1})\nonumber\\
 &&=\delta_{\bar a \bar b}\int d^3{\sigma}
e^{-i(\vec k+\vec k_{1})\cdot \vec \sigma}=(2\pi )^3\delta_{\bar a
\bar b}\delta(\vec k+\vec k_{1}).
 \label{c1}
 \eea

The conditions (\ref{II15}) and (\ref{II16}) become

\bea
 && {\tilde \pi}_{\bar a}(\tau , 0 k^2 k^3) = {\tilde \pi}_{\bar a}(\tau ,
 k^1 0  k^3) = {\tilde \pi}_{\bar a}(\tau , k^1 k^2 0) =
 0,\nonumber \\
 &&{}\nonumber \\
 &&{\tilde r}_{\bar a}(\tau , 0 k^2 k^3) = {\tilde r}_{\bar a}(\tau ,
 k^1 0  k^3) = {\tilde r}_{\bar a}(\tau , k^1 k^2 0) =
 0.
 \label{c2}
 \eea

\bigskip

Some useful relations are

\bea
 && \theta (x) = lim_{\epsilon \rightarrow 0} \int {{dk}\over {2\pi
i}} {{e^{ikx}}\over {k-i\epsilon}},\quad \int {{d^3k}\over {(2\pi
)^3}}\, {{e^{i\vec k\cdot \vec \sigma }}\over {|\vec k|^2}} \, =\,
{1\over {4\pi |\vec \sigma |}},
 \nonumber \\
  &&\psi (x) = \int dy \theta (x-y) f(y)\quad \Rightarrow \quad
 \tilde f(k) =ik \tilde \psi (k),\nonumber \\
 &&{}\nonumber \\
 &&\Box r_{\bar a}(\tau ,\vec \sigma ) = \int {{d^3k}\over {(2\pi
 )^3}} [\partial^2_{\tau}+|\vec k|^2] {\tilde r}_{\bar a}(\tau
 ,\vec k)  e^{i\vec k\cdot \vec \sigma},\nonumber \\
 &&{}\nonumber \\
 &&\int d^3\sigma\, e^{-i\vec k\cdot \vec
\sigma }\,  \int d^3\sigma_1 \, {{f({\vec \sigma}_1)}\over
 {4\pi |\vec \sigma -{\vec \sigma}_1|}}\, =\, {{\tilde f(\vec
 k)}\over {|\vec k|^2}},\nonumber \\
 &&{}\nonumber \\
 &&\int d^3\sigma_1
\, {{r_{\bar a}(\tau ,{\vec \sigma}_1)}\over
 {4\pi |\vec \sigma -{\vec \sigma}_1|}}\, =\, \int {{d^3k}\over
 {(2\pi )^3}}\,\, {{{\tilde r}_{\bar a}(\tau ,\vec k)}\over {|\vec
 k|^2}}\, e^{i\vec k\cdot \vec \sigma},\nonumber \\
 &&{}\nonumber \\
 &&\int d\sigma_1^u \theta (\sigma^u,\sigma_1^u) \int d\sigma_2^u
 \theta (\sigma_2^u,\sigma_1^u)
 e^{ik^u\sigma_2^u}={{e^{ik^u\sigma^u}}\over {k_u^2}},\quad
 \int d\sigma_1^r
 \theta (\sigma_1^r,\sigma^u)
 e^{ik^r\sigma_1^r}=-{{e^{ik^r\sigma^r}}\over {i k_r}}, \nonumber \\
 &&{}\nonumber \\
 &&since\nonumber \\
 &&{}\nonumber \\
 &&{{\partial^2}\over {(\partial \sigma^u)^2}}\Big(
 {{e^{-ik^u\sigma^u}}\over {k_u^2}}\Big) =
 e^{ik^u\sigma^u}={{\partial^2}\over {(\partial \sigma^u)^2}}
\int^{\sigma^u}_{-\infty} d\sigma_1^u \int_{\sigma_1^u}^{\infty}
d\sigma_2^u\, e^{ik^u\sigma_2^u},\nonumber \\
 &&{}\nonumber \\
 &&{}\nonumber \\
&&\int^{\sigma^u}_{-\infty} d\sigma_1^u
\int^{\sigma_1^u}_{-\infty} d\sigma_2^u\, \int^{\infty}_{\sigma^v}
d\sigma_1^v \int^{\infty}_{\sigma_1^v} d\sigma_2^v
 \int d^3\sigma_3\, {{\partial^2_{3r}\partial^2_{3s}\partial^2_{3w}\partial^2_{3t}
 r_{\bar c}(\tau ,{\vec \sigma}_3)}\over {4\pi |{\vec
 \sigma}_2-{\vec \sigma}_3|}}{|}_{{\vec \sigma}_2=(\sigma_2^u
 \sigma_2^v \sigma^{k\not= u,v})}=\nonumber \\
 &&= \int {{d^3k}\over {(2\pi )^3}}\, {{k^2_rk^2_sk^2_wk^2_t
 }\over {k_u^2k_v^2|\vec k|^2}} {\tilde r}_{\bar c}(\tau ,\vec
 k) e^{i\vec k\cdot \vec \sigma}.
 \label{c3}
 \eea

\bigskip

Let us consider Eqs.(\ref{V2}). Their Fourier transform is

 \beq
 \partial_{\tau}{\tilde r}_{\bar a}(\tau ,\vec k) = \sum_{\bar b} A_{\bar a\bar b}(\vec k)
{\tilde \pi}_{\bar b}(\tau ,\vec k),
 \label{c4}
 \eeq

\noindent where the matrix $A_{\bar a\bar b}(\vec k)$ is

\bea
 A_{\bar a\bar b}(\vec k) &=& A_{\bar b\bar a}(\vec k) =
 A_{\bar a\bar b}(- \vec k) = {{24\pi G}\over {c^3}}
 \Big[ \delta_{\bar a\bar b} + {1\over 4} \sum_{r,s,u,v}
 \gamma_{\bar ar}\gamma_{\bar bs}\nonumber \\
 &&(1-\delta_{uv})[1-2(\delta_{ur}+\delta_{vr})] [1-2(\delta_{us}+\delta_{vs})]
 \Big( {{k^r k^s}\over {k^u k^v}} \Big)^2 \Big],\nonumber \\
 &&{}\nonumber \\
 && det\, \Big( A_{\bar a\bar b}(\vec k) \Big) = {{192\pi^2 G^2}\over {c^6}}
 {{|\vec k|^6}\over {(k^1k^2k^3)^2}}.
 \label{c5}
 \eea

\bigskip

Eq.(\ref{c2}) implies that  Eq.(\ref{c4}) is well defined, even if
the matrix $A_{\bar a\bar b}(\vec k)$ diverges for $k^r
\rightarrow 0$, if ${\tilde \pi}_{\bar a}(\tau ,\vec k)$ {\it
vanishes for $\vec k \rightarrow 0$ at least as} $(k^1 k^2 k^3)^{2
+ \epsilon}$, $\epsilon > 0$.

\bigskip

For $\vec k \not= 0$ its inverse is

\bea
 A^{-1}_{\bar a\bar b}(\vec k) &=&  A^{-1}_{\bar b\bar a}(\vec k)
 =  A^{-1}_{\bar a\bar b}(- \vec k) =\nonumber \\
 &=& {1\over {det\, A(\vec k)}} \sum_{\bar c\bar d}
 \epsilon_{\bar a\bar c} \epsilon_{\bar b\bar d} A_{\bar c\bar d}(\vec k)=
   {{c^3}\over {8\pi G}} {1\over {|\vec k|^6}} \Big[ (k^1k^2k^3)^2 \delta
 _{\bar a\bar b}+\nonumber \\
 &+&{1\over 2} \sum_{r,s,t,\bar c,\bar d} \epsilon_{\bar a\bar c}\gamma_{\bar cr}
 \epsilon_{\bar b\bar d}\gamma_{\bar ds} (2\delta_{tr}-1)(2\delta_{ts}-1)
 (k^r k^s k^t)^2 \Big].
 \label{c6}
 \eea

\bigskip

The Fourier transform of Eq.(\ref{V5}) is

\beq
 \partial_{\tau}{\tilde \pi}_{\bar a}(\tau ,\vec k) =
  \sum_{\bar b} B_{\bar a\bar b}(\vec k)
{\tilde r}_{\bar b}(\tau ,\vec k),
 \label{c7}
 \eeq

\noindent where $B_{\bar a\bar b}(\vec k)$ is the matrix

\bea
 B_{\bar a\bar b}(\vec k) &=& B_{\bar b\bar a}(\vec k) =
 B_{\bar a\bar b}(- \vec k) = \nonumber \\
 &=& {{c^3}\over{24 \pi G}}\Big[-\delta_{\bar a\bar b}\sum_r k^2_r +
\sum_{rs}\gamma_{\bar ar}\gamma_{\bar bs}\, \Big( 2 \delta_{rs}
k^2_r-{{k^2_r k^2_s}\over{2|\vec k|^2}}\Big) \Big], \nonumber \\
 && det\, \Big( B_{\bar a\bar b}(\vec k) \Big) =
{{c^6\, (k^1k^2k^3)^2}\over{192\pi^2 G^2|\vec k|^2}},
 \label{c8}
 \eea

\noindent satisfying

\bea
 \sum_{\bar b}A _{\bar a\bar b}(\vec k)B_{\bar b\bar c}(\vec
k) &=& -|\vec k|^2 \delta_{\bar a \bar c}=\sum_{\bar b}B _{\bar
a\bar b}(\vec k)A_{\bar b\bar c}(\vec k),\nonumber \\
 &&{}\nonumber \\
 \Rightarrow && A^{-1}(\vec k) = - |\vec k|^2\, B(\vec k).
 \label{c9}
 \eea

\bigskip

But this implies that the Fourier transform of Eq.(\ref{V7})
reduces to the wave equation

\bea
 {\ddot {\tilde r}}_{\bar a}(\tau ,\vec k) &{\buildrel \circ
 \over =}& \sum_{\bar b} \Big( -|\vec k|^2 \delta_{\bar a\bar b}
 +2 \sum_r\gamma_{\bar ar}\gamma_{\bar br} \, k_r^2-
 {1\over 2} \sum_{rs}\gamma_{\bar ar} {{k_r^2k_s^2}\over {|\vec
 k|^2}}\Big){\tilde r}_{\bar b}(\tau
 ,\vec k)+\nonumber \\
&+&{1\over 8}\sum_{\bar c}\Big( \sum_{rs}\sum_{tw}\sum_{uv}\Big
[-2\gamma_{\bar ar}\gamma_{\bar cs}
 +4\gamma_{\bar ar}\gamma_{\bar ct}(\delta_{ts}-{1\over 3})
 -\gamma_{\bar ar}\gamma_{\bar cw}(\delta_{ts}-{1\over 3})
 \Big](1-\delta_{uv})\nonumber \\
&&[1-2(\delta_{ur}+\delta_{vr})][1-2(\delta_{us}+\delta_{vs})]{{k_t^2k_w^2
 k_r^2k_s^2}\over {k_u^2k_v^2|\vec k|^2}} \Big){\tilde r}_{\bar c}(\tau
 ,\vec k)  =\nonumber \\
 &=& -|\vec k|^2{\tilde r}_{\bar a}(\tau ,\vec k),
 \label{c10}
 \eea

\noindent whose solutions

\bea
 r_{\bar a}(\tau ,\vec{\sigma})&=& \int {{d^3k}\over
{(2\pi )^3}}\, \Big(C_{\bar a}(\vec k) e^{-i (|\vec
 k|\, \tau - \vec k \cdot \vec \sigma )}+C^{\ast}_{\bar a}(\vec k)e^{i(|\vec
k|\tau-\vec k\cdot \vec \sigma)}\Big),\nonumber \\
 &&{}\nonumber \\
 && {\tilde r}_{\bar a}(\tau ,\vec k) = C_{\bar a}(\vec k)\,
 e^{-i\, |\vec k|\, \tau} + C^*_{\bar a}(-\vec k)\, e^{i\, |\vec
 k|\, \tau},
 \label{c11}
 \eea

\noindent  have the arbitrary functions $C_{\bar a}(\vec k)$
satisfying Eq.(\ref{c2}). Then Eqs.(\ref{V4}) imply the following
form of the momenta restricted to the solutions

 \bea
  \pi_{\bar a}(\tau ,\vec \sigma ) &=& - i \int {{d^3k}\over {(2\pi
 )^3}}\, \sum_{\bar b}\, A^{-1}_{\bar a\bar b}(\vec k)\, |\vec
 k|\, \Big[ C_{\bar b}(\vec k)\, e^{-i (|\vec
 k|\, \tau - \vec k \cdot \vec \sigma )} - C^{\ast}_{\bar a}(\vec k)e^{i(|\vec
k|\tau-\vec k\cdot \vec \sigma)}\Big],\nonumber \\
 &&{}\nonumber \\
 {\tilde \pi}_{\bar
a}(\tau ,\vec k)&=&\sum_{\bar b}A^{-1}_{\bar a \bar b}(\vec k)\,
\partial_{\tau}{\tilde r}_{\bar b}(\tau ,\vec k) =\nonumber \\
 &=&\sum_{\bar b}A^{-1}_{\bar a \bar b}(\vec k)\, \Big[-i|\vec
k|C_{\bar b}(\vec k)e^{-i|\vec k|\tau}+i|\vec k|C_{\bar
b}^{\ast}(-\vec k)e^{i|\vec k|\tau}\Big].
 \label{c12}
  \eea

\noindent To have ${\tilde \pi}_{\bar a}(\tau ,\vec k)$ vanishing
for $\vec k \rightarrow 0$ at least as $(k^1 k^2 k^3)^{2 +
\epsilon}$, $\epsilon > 0$, we must require that the functions
$C_{\bar a}(\vec k)$  {\it vanish at least as} $(k^1\, k^2\, k^3
)^{\epsilon}$ for $k^r \rightarrow 0$.

\bigskip

The Fourier transform of the ADM energy (\ref{III1}) is

\bea
 {\hat E}_{ADM}&=&{{12\pi G}\over{c^3}}\int {{d^3k}\over
{(2\pi )^3}} \sum_{\bar a}{\tilde \pi}_{\bar a}(\tau ,\vec
k){\tilde \pi}_{\bar a}(\tau ,-\vec k)+ \nonumber \\
  &+& {{c^3}\over{48 \pi G}}\int
{{d^3k}\over {(2\pi )^3}} \sum_{\bar a r}k^2_r {\tilde r}_{\bar
a}(\tau ,\vec k){\tilde r}_{\bar a}(\tau ,-\vec k) +\nonumber \\
 &&{}\nonumber \\
 &+&{{3\pi G}\over {c^3}}\int {{d^3k}\over {(2\pi )^3}}\sum_{\bar a \bar b}
 \Big[ \sum_{r,s,u,v} \gamma_{\bar ar}\gamma_{\bar
 bs}(1-\delta_{uv})[1-2(\delta_{ur}+\delta_{vr})]\nonumber \\
 &&[1-2(\delta_{us}+\delta_{vs})]
 \Big( {{k^r k^s}\over {k^u k^v}} \Big)^2 {\tilde \pi}_{\bar a}(\tau ,\vec k){\tilde \pi}_{\bar
b}(\tau ,-\vec k)\Big] -\nonumber\\
 &-&{{c^3}\over{24 \pi G}}\int {{d^3k}\over {(2\pi )^3}} \sum_{\bar
a \bar b r}\gamma_{\bar ar}\gamma_{\bar
 b r}k^2_r {\tilde r}_{\bar a}(\tau ,\vec
k){\tilde r}_{\bar b}(\tau ,-\vec k)+\nonumber \\
 &+&{{c^3}\over{96
\pi G}}\int {{d^3k}\over {(2\pi )^3}} \sum_{\bar a \bar b r
u}\gamma_{\bar ar}\gamma_{\bar
 b u}{{k^2_r k^2_u}\over{|\vec k|^2}} {\tilde r}_{\bar a}(\tau ,\vec
k){\tilde r}_{\bar b}(\tau ,-\vec k).
 \label{c13}
  \eea

As a check we can recover the Fourier transforms  (\ref{c4}) and
(\ref{c7}) of the Hamilton equations by using this form of the
weak ADM energy and the Poisson brackets (\ref{c1}).

\bigskip

The Fourier transform of the shift functions (\ref{III5}) is
\bigskip

\beq
 {\tilde n}_r(\tau ,\vec k ) =-i\sqrt{3} {{2\pi G}\over {c^3}}
 \sum_{\bar bvuc} \gamma_{\bar bv}(1-{\delta}_{uc})
 [1-2(\delta_{uv}+\delta_{cv})][1-2(\delta_{ur}+\delta_{cr})]
 {{k_r k^2_v}\over{k^2_u k^2_c}} {\tilde \pi}_{\bar b}(\tau ,\vec k
 ),
 \label{c14}
 \eeq

\noindent while the Fourier transform of the extrinsic curvature
is
\bigskip

\beq
 {}^3{\tilde {\hat K}}_{rs}(\tau ,\vec k ) =\epsilon \sqrt{3}
{{4\pi G}\over {c^3}} \sum_{\bar a} \Big[ 2 \delta_{rs}
 \gamma_{\bar a s} -(1-\delta_{rs})\sum_w\gamma_{\bar aw}
  [1-2(\delta_{rw}+\delta_{sw})]{{k^2_w}\over{k_r k_s}}\Big]{\tilde \pi}_{\bar a}(\tau ,\vec k
  ),
  \label{c15}
 \eeq

\bigskip

The following Fourier transforms are used in the study of the
geodesic deviation equation

\bea
 {}^4{\tilde {\hat {\Gamma}}}_{\tau s}^r(\tau ,\vec k )&=& i
k_s {\tilde n}_r(\tau ,\vec k )+ \epsilon {}^3{\tilde {\hat
K}}_{rs}(\tau ,\vec k ),  \nonumber \\
 &&{}\nonumber \\
 {}^4{\tilde {\hat R}}_{\tau
s \tau}^r(\tau ,\vec k )&=& - \epsilon
\partial_{\tau}{}^3{\tilde {\hat K}}_{rs}(\tau ,\vec k ).
 \label{c16}
 \eea
\bigskip

The Fourier transforms of the exponent $q(\tau ,\vec \sigma )$
($\sqrt{{}^4{\hat g}}\approx -\epsilon (1+3q)$) of the conformal
factor of the 3-metric, given in  Eq.(\ref{II2}), and of the
inverse 3-metric (\ref{III7}) are

\beq
 {\tilde q}(\tau ,\vec k)={1\over{2\sqrt3}}\sum_{\bar a
u}\gamma_{\bar a u}{{k^2_u}\over{|\vec k|^2}}{\tilde r}_{\bar
a}(\tau ,\vec k),
 \label{c18}
 \eeq

\noindent and

\beq
 {}^3{\tilde{\hat{\gamma}}}^{rs}(\tau ,\vec k)=\delta^{rs}
\Big(1-{2\over{\sqrt{3}}}\sum_{\bar a}\Big[\gamma_{\bar ar}{\tilde
r}_{\bar d}(\tau ,\vec k)+{1\over 2}\sum_{u}\gamma_{\bar
au}{{k^2_u}\over{|\vec k|^2}}{\tilde r}_{\bar a}(\tau ,\vec
k)\Big]\Big).
 \label{c19}
 \eeq

\vfill\eject

\section{Linearized 3- and 4-Tensors.}

\subsection{Linearized 3-Tensors on the WSW Hyper-Surfaces
$\Sigma_{\tau}$}

>From Eqs.(\ref{III8}) and  (\ref{a7}) we get the following results
for the linearized Riemannian structure of the WSW hyper-surfaces
$\Sigma_{\tau}^{(WSW)}$.

The  Christoffel symbols are

\bea
 {}^3{\hat \Gamma}^r_{uv}(\tau ,\vec \sigma )&=&{1\over
 {\sqrt{3}}}\sum_{\bar a}\Big( -\delta_{uv}\Big[ \gamma_{\bar au}
 \partial_r r_{\bar a}(\tau ,\vec \sigma )-\nonumber \\
 &-&{1\over 2} \sum_w\gamma_{\bar aw} \partial_r
  \int d^3\sigma_1\, {{\partial^2_{1w}r_{\bar a}(\tau ,{\vec
 \sigma}_1)}\over {4\pi |\vec \sigma -{\vec
 \sigma}_1|}}\Big]+\nonumber \\
 &+&\delta_{ru}\Big[ \gamma_{\bar au}
 \partial_v r_{\bar a}(\tau ,\vec \sigma )-\nonumber \\
 &-&{1\over 2} \sum_w\gamma_{\bar aw} \partial_v
  \int d^3\sigma_1\, {{\partial^2_{1w}r_{\bar a}(\tau ,{\vec
 \sigma}_1)}\over {4\pi |\vec \sigma -{\vec
 \sigma}_1|}}\Big]+\nonumber \\
 &+&\delta_{rv} \Big[ \gamma_{\bar av}
 \partial_u r_{\bar a}(\tau ,\vec \sigma )-\nonumber \\
 &-&{1\over 2} \sum_w\gamma_{\bar aw} \partial_u
  \int d^3\sigma_1\, {{\partial^2_{1w}r_{\bar a}(\tau ,{\vec
 \sigma}_1)}\over {4\pi |\vec \sigma -{\vec
 \sigma}_1|}}\Big] \Big) + O(r^2_{\bar a}),\nonumber \\
 &&{}\nonumber \\
 &&{}\nonumber \\
 \sum_v\, {}^3{\hat \Gamma}^v_{uv}(\tau ,\vec \sigma
 )&=&-{{\sqrt{3}}\over 2} \sum_{\bar aw}\gamma_{\bar aw}
 \partial_u  \int d^3\sigma_1\, {{\partial^2_{1w}r_{\bar a}(\tau ,{\vec
 \sigma}_1)}\over {4\pi |\vec \sigma -{\vec
 \sigma}_1|}}  + O(r^2_{\bar a}),\nonumber \\
 \sum_u \partial_{(s}\, {}^3{\hat \Gamma}^u_{r)u}(\tau ,\vec
 \sigma )&=&- {{\sqrt{3}}\over 2} \sum_{\bar aw}\gamma_{\bar aw}
 \partial_r\partial_s  \int d^3\sigma_1\, {{\partial^2_{1w}r_{\bar a}(\tau ,{\vec
 \sigma}_1)}\over {4\pi |\vec \sigma -{\vec
 \sigma}_1|}}  + O(r^2_{\bar a}),\nonumber \\
 \sum_r\partial_r\, {}^3{\hat \Gamma}^r_{uv}(\tau ,\vec \sigma
 )&=&{1\over {\sqrt{3}}} \sum_{\bar a}\Big(\Big[ -\delta_{uv}\Big(
 \gamma_{\bar au} \triangle +{1\over 2}\sum_w\gamma_{\bar
 aw}\partial_w^2\Big) +(\gamma_{\bar au}+\gamma_{\bar
 av})\partial_u\partial_v\Big] r_{\bar a}(\tau ,\vec \sigma
 )-\nonumber \\
 &-& \sum_w\gamma_{\bar aw} \partial_u\partial_v  \int d^3\sigma_1\,
 {{\partial^2_{1w}r_{\bar a}(\tau ,{\vec
 \sigma}_1)}\over {4\pi |\vec \sigma -{\vec
 \sigma}_1|}} \Big)  + O(r^2_{\bar a}),\nonumber \\
 \sum_{ur} \partial_r\, {}^3{\hat \Gamma}^u_{ru}(\tau ,\vec \sigma
 )&=& \sum_{ru}\partial_r\, {}^3{\hat \Gamma}^r_{uu}(\tau ,\vec
 \sigma )= {{\sqrt{3}}\over 2} \sum_w\gamma_{\bar aw} \partial_w^2
 r_{\bar a}(\tau ,\vec \sigma )  + O(r^2_{\bar a}),
 \label{d1}
 \eea

\noindent while the spin connection and the field strength are

\bea
 {}^3{\hat \omega}_{r(a)}(\tau ,\vec \sigma ) &=&
  {1\over {\sqrt{3}}}\,  \sum_u \epsilon_{(a)(r)(u)}
   \sum_{\bar a}\Big[ \gamma_{\bar ar}
  \partial_ur_{\bar a}(\tau ,\vec \sigma )-\nonumber \\
 &-& {1\over 2}\, \sum_{v} \gamma_{\bar av}{{\partial}\over {\partial
\sigma^u}} \int d^3\sigma_1 {{\partial^2_{1u}r_{\bar a}(\tau
,{\vec \sigma}_1)} \over {4\pi |\vec \sigma -{\vec
\sigma}_1|}}\Big] + O(r^2_{\bar a}),
 \label{d2}
 \eea

\bea
 {}^3{\hat \Omega}_{rs(a)}(\tau ,\vec \sigma )&=& \partial_r\,
 {}^3{\hat \omega}_{s(a)}(\tau ,\vec \sigma )-\partial_s\,
 {}^3{\hat \omega}_{r(a)}(\tau ,\vec \sigma ) +O(r^2_{\bar
 a})=\nonumber \\
 &=&{1\over {\sqrt{3}}}\sum_{\bar au} \Big( \epsilon_{(a)(s)(u)}\Big[
 \gamma_{\bar as} \partial_u\partial_r r_{\bar a}(\tau ,\vec
 \sigma )-\nonumber \\
 &-&{1\over 2} \sum_v\gamma_{\bar av} \partial_u\partial_r \int
 d^3\sigma_1\, {{\partial^2_{1v}r_{\bar a}(\tau ,{\vec
 \sigma}_1)}\over {4\pi |\vec \sigma -{\vec
 \sigma}_1|}}\Big]-\nonumber \\
 &-&\epsilon_{(a)(r)(u)}\Big[
 \gamma_{\bar ar} \partial_u\partial_s r_{\bar a}(\tau ,\vec
 \sigma )-\nonumber \\
 &-&{1\over 2} \sum_v\gamma_{\bar av} \partial_u\partial_s \int
 d^3\sigma_1\, {{\partial^2_{1v}r_{\bar a}(\tau ,{\vec
 \sigma}_1)}\over {4\pi |\vec \sigma -{\vec
 \sigma}_1|}}\Big] \Big)  + O(r^2_{\bar a}).
 \label{d3}
 \eea

Finally the Riemann and Ricci tensors and the curvature scalars
are

\bea
 {}^3{\hat R}_{rsuv}(\tau ,\vec \sigma )&=&\epsilon_{(r)(s)(a)}\,
 {}^3{\hat \Omega}_{uv(a)}(\tau ,\vec \sigma )+O(r^2_{\bar
 a})=\nonumber \\
 &=&{1\over {\sqrt{3}}}\sum_{\bar a} \Big( \Big[\gamma_{\bar
 ar}(\delta_{rv}\partial_s\partial_u-\delta_{ru}\partial_s\partial_v)-\gamma_{\bar
 as}(\delta_{sv}\partial_r\partial_u-\delta_{su}\partial_r\partial_v)\Big]
 r_{\bar a}(\tau ,\vec \sigma )-\nonumber \\
 &-&{1\over 2} \sum_t\gamma_{\bar at}\Big[
 \delta_{rv}\partial_s\partial_u-\delta_{ru}\partial_s\partial_v-
 (\delta_{sv}\partial_r\partial_u-\delta_{su}\partial_r\partial_v)\Big]\nonumber \\
 &&\int d^3\sigma_1\, {{\partial^2_{1t}r_{\bar a}(\tau ,{\vec
 \sigma}_1)}\over {4\pi |\vec \sigma -{\vec
 \sigma}_1|}} \Big)  + O(r^2_{\bar a}),\nonumber \\
 &&{}\nonumber \\
 {}^3{\hat R}_{rs}(\tau ,\vec \sigma )&=& \sum_u\, {}^3{\hat
 R}_{urus}(\tau ,\vec \sigma )+O(r^2_{\bar a})=\nonumber \\
 &=&{1\over {\sqrt{3}}} \sum_{\bar a}\Big( \Big[ (\gamma_{\bar
 ar}+\gamma_{\bar as})\partial_r\partial_s
 -\delta_{rs}(\gamma_{\bar ar}\triangle +{1\over
 2}\sum_t\gamma_{\bar at} \partial^2_t)\Big] r_{\bar a}(\tau ,\vec
 \sigma )-\nonumber \\
 &-&{1\over 2}\, \int d^3\sigma_1\, \sum_t\gamma_{\bar at} \partial_r\partial_s
  d^3\sigma_1\, {{\partial^2_{1t}r_{\bar a}(\tau ,{\vec
 \sigma}_1)}\over {4\pi |\vec \sigma -{\vec
 \sigma}_1|}} \Big)  + O(r^2_{\bar a}),\nonumber \\
 &&{}\nonumber \\
 {}^3{\hat R}(\tau ,\vec \sigma )&=& \sum_r\, {}^3{\hat
 R}_{rr}(\tau ,\vec \sigma )+O(r^2_{\bar a})={1\over {\sqrt{3}}}
 \sum_{\bar av} \partial^2_v r_{\bar a}(\tau ,\vec \sigma
 )  + O(r^2_{\bar a}).
 \label{d4}
 \eea

\subsection{The 4-Christoffel Symbols, the 4-Riemann Tensor and Einstein
Equations.}

In this Subsection we will give the linearized form of the main
4-tensors of our space-time.

By using the parametrization of the 4-metric ${}^4g_{AB}$ given in
footnote 6 and Eqs.(\ref{IV1}), with the ${}^3{\hat
\Gamma}^u_{rs}$'s of Eq.(\ref{d1}), we get the following
4-Christoffel symbols [$N = -\epsilon + n$, $N^r = n_r +
O(r^2_{\bar a})$]

\bea
 {}^4{\hat \Gamma}^{\tau}_{\tau\tau}(\tau ,\vec \sigma )&=&
\Big( {1\over N}[\partial_{\tau}N+N^r\partial_rN-N^rN^s\,
{}^3{\hat K}_{rs}]\Big) (\tau ,\vec \sigma )=\nonumber \\
 &=& 0 +O(r^2_{\bar a}),\nonumber \\
 &&{}\nonumber \\
 {}^4{\hat \Gamma}^{\tau}_{r\tau}(\tau ,\vec \sigma )&=&\Big(
 {1\over N} [\partial_rN- {}^3{\hat K}_{rs}N^s]\Big) (\tau ,\vec
 \sigma )=0 +O(r^2_{\bar a}),\nonumber \\
 &&{}\nonumber \\
 {}^4{\hat \Gamma}^{\tau}_{rs}(\tau ,\vec \sigma )&=&-\Big(
 {1\over N} \, {}^3{\hat K}_{rs}\Big) (\tau ,\vec \sigma
 )=\nonumber \\
 &=& {{4\pi G}\over
{c^3}}{}^3G_{o(a)(b)(c)(d)}\delta_{(a)r}\delta_{(b)s}
\sum_u\delta_{(c)u} {}^3{\check {\tilde \pi}}^u_{(d)}(\tau ,\vec
\sigma ) + O(r^2_{\bar a})=\nonumber \\ &=& \sqrt{3} {{4\pi
G}\over {c^3}} \sum_{\bar a} \Big[ 2 \delta_{rs}
 \gamma_{\bar a r} \pi_{\bar a}(\tau ,\vec \sigma ) +\nonumber \\
 &+&  [1-\delta_{rs}]\sum_w\gamma_{\bar aw} [1-2(\delta_{rw}+\delta_{sw})] \nonumber \\
 &&{{\partial^2}\over {\partial (\sigma^w)^2}}  \int^{\infty}_{\sigma^r} d\sigma_1^r
 \int^{\infty}_{\sigma^s} d\sigma_1^s\,\, \pi_{\bar a}(\tau
 ,\sigma_1^r\sigma_1^s\sigma^{k\not= r,s}) \Big] + O(r^2_{\bar
 a}),\nonumber \\
 &&{}\nonumber \\
 {}^4{\hat \Gamma}^u_{\tau\tau}(\tau ,\vec \sigma )&=&\Big(
 \partial_{\tau}N^u -{{N^u}\over N} \partial_{\tau}N+({}^3{\hat
 g}^{uv}-{{N^uN^v}\over {N^2}}) N\partial_vN+\nonumber \\
 &+&N^u{}_{|v} N^v- 2N ({}^3{\hat g}^{uv}-{{N^uN^v}\over {N^2}})
 {}^3{\hat K}_{vr}N^r\Big) (\tau ,\vec \sigma )=\nonumber \\
 &=& \partial_{\tau} n_u(\tau ,\vec \sigma )+O(r^2_{\bar a})=\nonumber \\
 &=&  \sqrt{3} {{2\pi G}\over {c^3}}
 \sum_{\bar at} \gamma_{\bar at}
  \sum_{mn}(1-\delta_{mn})[1-2(\delta^u_m+\delta^u_n)]
 [1-2(\delta_{tm}+\delta_{tn})] \nonumber \\
 &&{{\partial}\over {\partial \sigma^u}} \int^{\sigma^m}_{-\infty}
 d\sigma_1^m \int^{\sigma^n}_{-\infty} d\sigma_1^n
 \int^{\infty}_{\sigma_1^m} d\sigma_2^m
 \int^{\infty}_{\sigma_1^n}d\sigma_2^n
 {{\partial^2 \partial_{\tau} \pi_{\bar a}(\tau ,\sigma_2^m
 \sigma_2^n \sigma_2^{k\not= m,n})}\over {(\partial \sigma_2^t)^2}}
 {|}_{\sigma_2^k=\sigma^k}+O(r^2_{\bar a}),\nonumber \\
 &&{}\nonumber \\
 {}^4{\hat \Gamma}^u_{r\tau}(\tau ,\vec \sigma )&=& \Big(
 N^u{}_{|r} -{{N^u}\over N} \partial_rN -N ({}^3{\hat
 g}^{uv}-{{N^uN^v}\over {N^2}}) {}^3{\hat K}_{vr}\Big) (\tau ,\vec
 \sigma )=\nonumber \\
 &=& \partial_rn_u(\tau ,\vec \sigma ) +\epsilon \, {}^3{\hat
 K}_{ur}(\tau ,\vec \sigma )+O(r^2_{\bar a})=\nonumber \\
 &=& \partial_rn_u(\tau ,\vec \sigma )+{{4\pi
G}\over {c^3}}{}^3G_{o(a)(b)(c)(d)}\delta^u_(a)\delta_{(b)r}
\sum_w\delta_{(c)w} {}^3{\check {\tilde \pi}}^w_{(d)}(\tau ,\vec
\sigma ) + O(r^2_{\bar a})=\nonumber \\ &=&\sqrt{3} {{4\pi G}\over
{c^3}} \sum_{\bar a} \Big[ 2 \delta^u_r
 \gamma_{\bar a r} \pi_{\bar a}(\tau ,\vec \sigma )
 +  [1-\delta^u_r]\sum_t\gamma_{\bar at} [1-2(\delta^u_t+\delta_{rt})] \nonumber \\
 &&{{\partial^2}\over {\partial (\sigma^t)^2}}  \int^{\infty}_{\sigma^u}
 d\sigma_1^u \int^{\infty}_{\sigma^r} d\sigma_1^r\,\, \pi_{\bar a}(\tau
 ,\sigma_1^u\sigma_1^r\sigma^{k\not= u,r})+\nonumber \\
 &+&{1\over 2}\sum_{\bar at} \gamma_{\bar at}
  \sum_{mn}(1-\delta_{mn})[1-2(\delta^u_m+\delta^u_n)]
 [1-2(\delta_{tm}+\delta_{tn})] \nonumber \\
 &&{{\partial^2}\over {\partial \sigma^r \partial \sigma^u}} \int^{\sigma^m}_{-\infty}
 d\sigma_1^m \int^{\sigma^n}_{-\infty} d\sigma_1^n
 \int^{\infty}_{\sigma_1^m} d\sigma_2^m
 \int^{\infty}_{\sigma_1^n}d\sigma_2^n
 {{\partial^2 \pi_{\bar a}(\tau ,\sigma_2^m
 \sigma_2^n \sigma_2^{k\not= m,n})}\over {(\partial \sigma_2^t)^2}}
 {|}_{\sigma_2^k=\sigma^k}\Big]+\nonumber \\
 &+& O(r^2_{\bar a}),\nonumber \\
 &&{}\nonumber \\
 {}^4{\hat \Gamma}^u_{rs}(\tau ,\vec \sigma )&=& \Big( {}^3{\hat
 \Gamma}^u_{rs} +{{N^u}\over N}\, {}^3{\hat K}_{rs}\Big) (\tau
 ,\vec \sigma ) =\nonumber \\
 &=& {}^3{\hat \Gamma}^u_{rs}(\tau ,\vec \sigma ) +O(r^2_{\bar
 a})=\nonumber \\
 &=&{1\over
 {\sqrt{3}}}\sum_{\bar a}\Big( -\delta_{uv}\Big[ \gamma_{\bar au}
 \partial_r r_{\bar a}(\tau ,\vec \sigma )-{1\over 2} \sum_w\gamma_{\bar aw} \partial_r
  \int d^3\sigma_1\, {{\partial^2_{1w}r_{\bar a}(\tau ,{\vec
 \sigma}_1)}\over {4\pi |\vec \sigma -{\vec
 \sigma}_1|}}\Big]+\nonumber \\
 &+&\delta_{ru}\Big[ \gamma_{\bar au}
 \partial_v r_{\bar a}(\tau ,\vec \sigma )-{1\over 2} \sum_w\gamma_{\bar aw} \partial_v
  \int d^3\sigma_1\, {{\partial^2_{1w}r_{\bar a}(\tau ,{\vec
 \sigma}_1)}\over {4\pi |\vec \sigma -{\vec
 \sigma}_1|}}\Big]+\nonumber \\
 &+&\delta_{rv} \Big[ \gamma_{\bar av}
 \partial_u r_{\bar a}(\tau ,\vec \sigma )-{1\over 2} \sum_w\gamma_{\bar aw} \partial_u
  \int d^3\sigma_1\, {{\partial^2_{1w}r_{\bar a}(\tau ,{\vec
 \sigma}_1)}\over {4\pi |\vec \sigma -{\vec
 \sigma}_1|}}\Big] \Big)+O(r^2_{\bar a}).
 \label{d5}
 \eea

Since $\varphi_{(a)}(\tau ,\vec \sigma ) =0$ in our gauge, the
space-time spin connection ${}^4\omega_A{}^{(\alpha )}{}_{(\beta
)}  = {}^4E^{(\alpha )}_B[\partial_A\, {}^4E^B_{(\beta
)}+{}^4\Gamma^B_{AC}\, {}^4E^C_{(\beta )}]$ can be evaluated with
the formula

\begin{equation}
{}^4{\buildrel \circ \over {\omega}}_A{}^{(\alpha )}{}_{(\beta
)}={}^4 _{(\Sigma )}{\check {\tilde E}}^{(\alpha )}_B[\partial_A\,
{}^4_{(\Sigma )} {\check {\tilde E}}^B_{(\beta
)}+{}^4\Gamma^B_{AC}\, {}^4_{(\Sigma )}{\check {\tilde
E}}^C_{(\beta )}],
 \label{d6}
 \end{equation}

\noindent by using Eqs.(\ref{III9}) and (\ref{d5}).

For the 4-Riemann tensor  ${}^4R^A{}_{BCD}={1\over 2}\,
{}^4g^{AE}\, (\partial_B\partial_D\,
{}^4g_{EC}+\partial_E\partial_C \,
{}^4g_{BD}-\partial_E\partial_D\,
{}^4g_{BC}-\partial_B\partial_C\, {}^4g _{ED})+
 {}^4g^{AL}\, {}^4g_{EF}\, ({}^4\Gamma^E_{LC}\,
{}^4\Gamma^F_{BD}-{}^4\Gamma^E_{LD}\, {}^4\Gamma^F_{BC})$ we get
\bigskip

 \bea
 {}^4{\hat R}^A{}_{BCD} &=& \partial_C\, {}^4{\hat \Gamma}^A_{BD}
 - \partial_D\, {}^4{\hat \Gamma}^A_{CB} +O(r^2_{\bar
 a}),\nonumber \\
 &&{}\nonumber \\
 {}^4{\hat R}^{\tau}{}_{\tau CD}(\tau
,\vec \sigma ) &=&
 0+O(r^2_{\bar a}),\nonumber \\
 {}^4{\hat
 R}^{\tau}{}_{r\tau s}(\tau
,\vec \sigma )&=&\partial_{\tau}\, {}^4{\hat
 \Gamma}^{\tau}_{rs}+O(r^2_{\bar a})=\epsilon \partial_{\tau}\, {}^3{\hat
 K}_{rs}+O(r^2_{\bar a})=\nonumber \\
 &=&\sqrt{3} {{4\pi G}\over {c^3}} \sum_{\bar a} \Big[ 2
\delta_{rs}
 \gamma_{\bar a r} \pi_{\bar a}(\tau ,\vec \sigma ) +\nonumber \\
 &+& [1-\delta_{rs}]\sum_w\gamma_{\bar aw} [1-2(\delta_{rw}+\delta_{sw})] \nonumber \\
 &&{{\partial^2}\over {\partial (\sigma^w)^2}}  \int^{\infty}_{\sigma^r} d\sigma_1^r
 \int^{\infty}_{\sigma^s} d\sigma_1^s\,\,\partial_{\tau} \pi_{\bar a}(\tau
 ,\sigma_1^r\sigma_1^s\sigma^{k\not= r,s}) \Big] + O(r^2_{\bar
 a}),\nonumber \\
{}^4{\hat
 R}^{\tau}{}_{ruv}(\tau
,\vec \sigma )&=&\partial_u \, {}^4{\hat
 \Gamma}^{\tau}_{rv}-\partial_v\, {}^4{\hat
 \Gamma}^{\tau}_{ur}=+\epsilon(\partial_u \, {}^3{\hat
 K}_{rv}-\partial_v\, {}^3{\hat K}_{ru})=\nonumber \\
&=&\sqrt{3} {{4\pi G}\over {c^3}} \sum_{\bar a} \Big[
2\gamma_{\bar
ar}(\delta_{rv}\partial_u-\delta_{ru}\partial_v)\pi_{\bar a}(\tau
,\vec \sigma )+\sum_t \gamma_{\bar at} \nonumber \\
 &&\Big((1-\delta_{rv})[1-2(\delta_{rt}+\delta_{vt})]
 {{\partial^3}\over {\partial \sigma^u \partial (\sigma^t)^2}}  \int^{\infty}_{\sigma^r} dw^r
 \int^{\infty}_{\sigma^v} dw^v\,\pi_{\bar a}(\tau
 ,w^rw^v\sigma^{k\not= r,v})-\nonumber \\
 &-&(1-\delta_{ru})[1-2(\delta_{rt}+\delta_{ut})]
 {{\partial^3}\over {\partial \sigma^v \partial (\sigma^t)^2}}  \int^{\infty}_{\sigma^r} dw^r
 \int^{\infty}_{\sigma^u} dw^u\,\pi_{\bar a}(\tau
 ,w^r w^u\sigma^{k\not= r,u})\Big)\Big]+\nonumber \\
 &+&O(r^2_{\bar
 a}),\nonumber \\
{}^4{\hat R}^r{}_{\tau s\tau}(\tau ,\vec \sigma )&=&-\epsilon
\partial_{\tau}\, {}^3{\hat
 K}_{rs}(\tau
,\vec \sigma )+O(r^2_{\bar a})=\nonumber \\
 &=&-\sqrt{3} {{4\pi
G}\over {c^3}} \sum_{\bar a} \Big[ 2 \delta_{rs}
 \gamma_{\bar a r}\partial_{\tau} \pi_{\bar a}(\tau ,\vec \sigma ) +\nonumber \\
 &+&  [1-\delta_{rs}]\sum_w\gamma_{\bar aw} [1-2(\delta_{rw}+\delta_{sw})] \nonumber \\
 &&{{\partial^2}\over {\partial (\sigma^w)^2}}  \int^{\infty}_{\sigma^r} d\sigma_1^r
 \int^{\infty}_{\sigma^s} d\sigma_1^s\,\,\partial_{\tau} \pi_{\bar a}(\tau
 ,\sigma_1^r\sigma_1^s\sigma^{k\not= r,s}) \Big] + O(r^2_{\bar
 a}),\nonumber \\
  {}^4{\hat
 R}^r{}_{\tau uv}(\tau ,\vec \sigma )&=&{}^4{\hat
 R}^{\tau}{}_{ruv}(\tau ,\vec \sigma ),\nonumber \\
 {}^4{\hat
 R}^r{}_{suv}(\tau ,\vec \sigma )&=&\partial_u\, {}^4{\hat \Gamma}^r_{sv}-\partial_v\,
 {}^4{\hat \Gamma}^r_{us}+O(r^2_{\bar a})=\nonumber \\
 &=&{1\over{\sqrt 3}} \sum_{\bar a} \Big( \delta_{su}\Big [
 \gamma_{\bar ar}\partial_v \partial_r r_{\bar a}(\tau ,\vec \sigma
 )-{1\over2}\sum_w\gamma_{\bar aw}{{\partial^2}\over{\partial
 \sigma^v \partial \sigma^r}}\int
 d^3\sigma_1{{(\partial^2_{1w}r_{\bar a})(\tau ,\vec \sigma
 )}\over{4\pi|\vec \sigma -{\vec \sigma}_1|}}\Big]-\nonumber \\
 &-&\delta_{sv}\Big [
 \gamma_{\bar as}\partial_u \partial_r r_{\bar a}(\tau ,\vec \sigma
 )-{1\over2}\sum_w\gamma_{\bar aw}{{\partial^2}\over{\partial
 \sigma^u \partial \sigma^r}}\int
 d^3\sigma_1{{(\partial^2_{1w}r_{\bar a})(\tau ,\vec \sigma
 )}\over{4\pi|\vec \sigma -{\vec \sigma}_1|}}\Big]+\nonumber \\
 &+&\delta_{rv}\Big [
 \gamma_{\bar av}\partial_u \partial_s r_{\bar a}(\tau ,\vec \sigma
 )-{1\over2}\sum_w\gamma_{\bar aw}{{\partial^2}\over{\partial
 \sigma^u \partial \sigma^s}}\int
 d^3\sigma_1{{(\partial^2_{1w}r_{\bar a})(\tau ,\vec \sigma
 )}\over{4\pi|\vec \sigma -{\vec \sigma}_1|}}\Big]-\nonumber \\
 &-&\delta_{ru}\Big [
 \gamma_{\bar au}\partial_v \partial_s r_{\bar a}(\tau ,\vec \sigma
 )-{1\over2}\sum_w\gamma_{\bar aw}{{\partial^2}\over{\partial
 \sigma^v \partial \sigma^s}}\int
 d^3\sigma_1{{(\partial^2_{1w}r_{\bar a})(\tau ,\vec \sigma
 )}\over{4\pi|\vec \sigma -{\vec \sigma}_1|}}\Big]\Big)+O(r^2_{\bar a}).
 \label{d7}
 \end{eqnarray}

Then we get the following linearization of the Ricci tensor and of
the curvature scalar

 \bea
 {}^4{\hat R}_{AB}&=& {}^4{\hat R}^E{}_{AEB}={}^4{\hat
 R}^{\tau}{}_{A\tau B}+{}^4{\hat R}^r{}_{ArB},\nonumber \\
 &&{}\nonumber \\
 {}^4{\hat R}_{\tau\tau}&=& \sum_r\, {}^4{\hat R}^r{}_{\tau
 r\tau}=-\epsilon \sum_r \partial_{\tau}\, {}^3{\hat K}_{rr}+O(r^2_{\bar a})
 \cir 0 +O(r^2_{\bar a}),\nonumber \\
 {}^4{\hat R}_{\tau u}&=&\sum_r\, {}^4{\hat R}^r{}_{\tau
 ru}=\epsilon \sum_r(\partial_r\, {}^3{\hat K}_{ru}-\partial_u\,
 {}^3{\hat K}_{rr})+O(r^2_{\bar a})=\nonumber \\
 &=& \epsilon \sum_r \partial_r\, {}^3{\hat K}_{ru}+O(r^2_{\bar
 a}) \cir 0+O(r^2_{\bar a}),\nonumber \\
 {}^4{\hat R}_{rs}&=& {}^4{\hat R}^{\tau}{}_{r\tau
 s}+\sum_u\, {}^4{\hat R}^u{}_{rus}=\nonumber \\
 &=&\epsilon \, \partial_{\tau}\, {}^3{\hat K}_{rs}+\sum_u
 (\partial_u\, {}^3{\hat \Gamma}^u_{rs}-\partial_{(s}\, {}^3{\hat
 \Gamma}^u_{r)u})+O(r^2_{\bar a}),\nonumber \\
 &&{}\nonumber \\
 {}^4\hat R&=& {}^4\eta^{AB}\, {}^4{\hat R}_{AB}+O(r^2_{\bar
 a})=\epsilon ({}^4{\hat R}_{\tau\tau}-\sum_r\, {}^4{\hat
 R}_{rr})+O(r^2_{\bar a})=\nonumber \\
 &=& -\epsilon \sum_{ur}(\partial_u\, {}^3{\hat
 \Gamma}^u_{rr}-\partial_r\, {}^3{\hat \Gamma}^u_{ru})+O(r^2_{\bar
 a})= 0 +O(r^2_{\bar a}).
 \label{d8}
 \eea

We see that the Einstein's equations ${}^4{\hat R}_{\tau A} \cir
0$ (corresponding to the super-hamiltonian and super-momentum
constraints) are satisfied as a consequence of the results of
Section II.
\bigskip

The spatial Einstein equations ${}^4{\hat R}_{rs}\, {\buildrel
\circ \over =}\, 0$ are independent from the shift functions $n_r$
and, after having used the Hamilton equations (\ref{V6}) to
eliminate $ \partial_{\tau} \pi_{\bar a}(\tau
 ,\vec \sigma )$, it can be checked with a  long but straightforward
calculation  that they are satisfied

 \bea
 {}^4{\hat R}_{rs}(\tau ,\vec \sigma )&=&\epsilon \,
 \partial_{\tau}\, {}^3{\hat K}_{rs}+\sum_u
 (\partial_u\, {}^3{\hat \Gamma}^u_{rs}-\partial_{(s}\, {}^3{\hat
 \Gamma}^u_{r)u})(\tau ,\vec \sigma )+O(r^2_{\bar a})=\nonumber \\
 &=&\sqrt{3} {{4\pi G}\over {c^3}} \sum_{\bar a} \Big[
 2\delta_{rs} \gamma_{\bar as} \partial_{\tau} \pi_{\bar a}(\tau
 ,\vec \sigma )+\nonumber \\
 &+&(1-\delta_{rs})\sum_w\gamma_{\bar
 aw}[1-2(\delta_{rw}+\delta_{sw})] \nonumber \\
 &&\partial^2_w \int^{\infty}_{\sigma^r} d\sigma_1^r
 \int^{\infty}_{\sigma^s} d\sigma_1^s\,\, \partial_{\tau}\,
 \pi_{\bar a}(\tau ,\sigma_1^r \sigma_1^s \sigma^{k\not=
 r,s})\Big] +\nonumber \\
 &+&{1\over {\sqrt{3}}} \sum_{\bar a}\Big( \Big[ -\delta_{rs}
 (\gamma_{\bar ar} \triangle +{1\over 2} \sum_w\gamma_{\bar aw}
 \partial_w^2)+(\gamma_{\bar ar}+\gamma_{\bar
 as})\partial_r\partial_s\Big] r_{\bar a}(\tau ,\vec \sigma
 )-\nonumber \\
 &-&\sum_w\gamma_{\bar aw} \partial_r\partial_s \int d^3\sigma_1
 {{\partial_{1w}^2 r_{\bar a}(\tau ,{\vec \sigma}_1)}\over {4\pi
 |\vec \sigma -{\vec \sigma}_1|}} \Big) +\nonumber \\
 &+& {{\sqrt{3}}\over 2} \sum_{\bar aw}\gamma_{\bar aw}
 \partial_r\partial_s  \int d^3\sigma_1
 {{\partial_{1w}^2 r_{\bar a}(\tau ,{\vec \sigma}_1)}\over {4\pi
 |\vec \sigma -{\vec \sigma}_1|}} \cir 0+O(r^2_{\bar a}).
 \label{d9}
 \eea

\bigskip

Then we can obtain the field strength in our gauge by using the
equation  ${}^4{\buildrel \circ \over {\Omega}}_{AB(\alpha )(\beta
)}= {}^4_{(\Sigma )}{\check {\tilde E}}^C_{(\alpha )}\,
{}^4_{(\Sigma )}{\check {\tilde E}}^D_{(\beta )}\,
{}^4R_{CDAB}=\partial_A\, {}^4{\buildrel \circ \over
{\omega}}_{B(\alpha )(\beta )}-\partial_B\, {}^4{\buildrel \circ
\over {\omega}}_{A(\alpha )(\beta )}+{}^4{\buildrel \circ \over
{\omega}}_{A(\alpha )(\gamma )}\,$ ${}^4{\buildrel \circ \over
{\omega}}_{B(\beta )}^{(\gamma )}- {}^4{\buildrel \circ \over
{\omega}}_{B(\alpha )(\gamma )}\, {}^4{\buildrel \circ \over
{\omega}}_{A(\beta )}^{(\gamma )}$.

\subsection{The 4-Weyl and 4-Bel-Robinson Tensors and the
Kretschmann Invariant.}

In this Section we will study the radiation gauge form of others
4-tensors.

Let us evaluate the 4-Weyl tensor
${}^4C_{ABCD}={}^4R_{ABCD}+{1\over 2}({}^4R_{AC}\,
{}^4g_{BD}-{}^4R_{BC}\, {}^4g_{AD}+{}^4R_{BD}\,
{}^4g_{AC}-{}^4R_{AD}\, {}^4g_{BC})+ {1\over 6}({}^4g_{AC}\,
{}^4g_{BD}-{}^4g_{AD}\, {}^4g_{BC})\, {}^4R\, {\buildrel \circ
\over =}\, {}^4R_{ABCD}$. We get

\begin{eqnarray*}
 {}^4C_{rstu}&=&{}^4R_{rstu}+{1\over 2}(-\epsilon
{}^4R_{rt}\delta_{su}+\epsilon {}^4R_{st}\delta_{ru}-\epsilon
{}^4R_{su}\delta_{rt}+\epsilon {}^4R_{ru}\delta_{st})=\nonumber \\
&=&-\epsilon{1\over{\sqrt 3}} \sum_{\bar a} \Big( \delta_{st}\Big[
 \gamma_{\bar ar}\partial_u \partial_r r_{\bar a}(\tau ,\vec \sigma
 )-{1\over2}\sum_w\gamma_{\bar aw}{{\partial^2}\over{\partial
 \sigma^u \partial \sigma^r}}\int
 d^3\sigma_1{{(\partial^2_{1w}r_{\bar a})(\tau ,\vec \sigma
 )}\over{4\pi|\vec \sigma -{\vec \sigma}_1|}}\Big]-\nonumber \\
 &-&\delta_{su}\Big [
 \gamma_{\bar as}\partial_t \partial_r r_{\bar a}(\tau ,\vec \sigma
 )-{1\over2}\sum_w\gamma_{\bar aw}{{\partial^2}\over{\partial
 \sigma^t \partial \sigma^r}}\int
 d^3\sigma_1{{(\partial^2_{1w}r_{\bar a})(\tau ,\vec \sigma
 )}\over{4\pi|\vec \sigma -{\vec \sigma}_1|}}\Big]+\nonumber \\
 &+&\delta_{ru}\Big [
 \gamma_{\bar au}\partial_t \partial_s r_{\bar a}(\tau ,\vec \sigma
 )-{1\over2}\sum_w\gamma_{\bar aw}{{\partial^2}\over{\partial
 \sigma^t \partial \sigma^s}}\int
 d^3\sigma_1{{(\partial^2_{1w}r_{\bar a})(\tau ,\vec \sigma
 )}\over{4\pi|\vec \sigma -{\vec \sigma}_1|}}\Big]-\nonumber \\
 &-&\delta_{rt}\Big [
 \gamma_{\bar at}\partial_u \partial_s r_{\bar a}(\tau ,\vec \sigma
 )-{1\over2}\sum_w\gamma_{\bar aw}{{\partial^2}\over{\partial
 \sigma^u \partial \sigma^s}}\int
 d^3\sigma_1{{(\partial^2_{1w}r_{\bar a})(\tau ,\vec \sigma
 )}\over{4\pi|\vec \sigma -{\vec
 \sigma}_1|}}\Big]\Big)+\nonumber \\
 &+&{1\over2}\Big(-\epsilon \delta_{su}\sqrt{3} {{4\pi G}\over {c^3}} \sum_{\bar a} \Big[
 2\delta_{rt} \gamma_{\bar at} \partial_{\tau} \pi_{\bar a}(\tau
 ,\vec \sigma )+\nonumber \\
 &+&(1-\delta_{rt})\sum_w\gamma_{\bar
 aw}[1-2(\delta_{rw}+\delta_{tw})] \nonumber \\
 &&\partial^2_w \int^{\infty}_{\sigma^r} d\sigma_1^r
 \int^{\infty}_{\sigma^t} d\sigma_1^t\,\, \partial_{\tau}\,
 \pi_{\bar a}(\tau ,\sigma_1^r \sigma_1^s \sigma^{k\not=
 r,t})\Big] +\nonumber \\
 &+&{1\over {\sqrt{3}}} \sum_{\bar a}\Big( \Big[ -\delta_{rt}
 (\gamma_{\bar ar} \triangle +{1\over 2} \sum_w\gamma_{\bar aw}
 \partial_w^2)+(\gamma_{\bar ar}+\gamma_{\bar
 at})\partial_r\partial_t\Big] r_{\bar a}(\tau ,\vec \sigma
 )-\nonumber \\
 &-&\sum_w\gamma_{\bar aw} \partial_r\partial_t \int d^3\sigma_1
 {{\partial_{1w}^2 r_{\bar a}(\tau ,{\vec \sigma}_1)}\over {4\pi
 |\vec \sigma -{\vec \sigma}_1|}} \Big) +\nonumber \\
 &+& {{\sqrt{3}}\over 2} \sum_{\bar aw}\gamma_{\bar aw}
 \partial_r\partial_t  \int d^3\sigma_1
 {{\partial_{1w}^2 r_{\bar a}(\tau ,{\vec \sigma}_1)}\over {4\pi
 |\vec \sigma -{\vec \sigma}_1|}}+\nonumber \\
 &+&\epsilon \delta_{ru}\sqrt{3} {{4\pi G}\over {c^3}} \sum_{\bar a} \Big[
 2\delta_{st} \gamma_{\bar at} \partial_{\tau} \pi_{\bar a}(\tau
 ,\vec \sigma )+\nonumber \\
 &+&(1-\delta_{st})\sum_w\gamma_{\bar
 aw}[1-2(\delta_{sw}+\delta_{tw})]\times \nonumber \\
 &&\partial^2_w \int^{\infty}_{\sigma^s} d\sigma_1^s
 \int^{\infty}_{\sigma^t} d\sigma_1^t\,\, \partial_{\tau}\,
 \pi_{\bar a}(\tau ,\sigma_1^s \sigma_1^t \sigma^{k\not=
 s,t})\Big] +\nonumber \\
 &+&{1\over {\sqrt{3}}} \sum_{\bar a}\Big( \Big[ -\delta_{st}
 (\gamma_{\bar as} \triangle +{1\over 2} \sum_w\gamma_{\bar aw}
 \partial_w^2)+(\gamma_{\bar as}+\gamma_{\bar
 as})\partial_s\partial_t\Big] r_{\bar a}(\tau ,\vec \sigma
 )-\nonumber \\
 &-&\sum_w\gamma_{\bar aw} \partial_s\partial_t \int d^3\sigma_1
 {{\partial_{1w}^2 r_{\bar a}(\tau ,{\vec \sigma}_1)}\over {4\pi
 |\vec \sigma -{\vec \sigma}_1|}} \Big) +\nonumber \\
 &+& {{\sqrt{3}}\over 2} \sum_{\bar aw}\gamma_{\bar aw}
 \partial_s\partial_t  \int d^3\sigma_1
 {{\partial_{1w}^2 r_{\bar a}(\tau ,{\vec \sigma}_1)}\over {4\pi
 |\vec \sigma -{\vec \sigma}_1|}}-
\end{eqnarray*}

\begin{eqnarray*}
 &-&\epsilon \delta_{rt}\sqrt{3} {{4\pi G}\over {c^3}} \sum_{\bar a} \Big[
 2\delta_{su} \gamma_{\bar au} \partial_{\tau} \pi_{\bar a}(\tau
 ,\vec \sigma )+\nonumber \\
 &+&(1-\delta_{su})\sum_w\gamma_{\bar
 aw}[1-2(\delta_{sw}+\delta_{uw})] \nonumber \\
 &&\partial^2_w \int^{\infty}_{\sigma^s} d\sigma_1^s
 \int^{\infty}_{\sigma^u} d\sigma_1^u\,\, \partial_{\tau}\,
 \pi_{\bar a}(\tau ,\sigma_1^s \sigma_1^u \sigma^{k\not=
 s,u})\Big] +\nonumber \\
 &+&{1\over {\sqrt{3}}} \sum_{\bar a}\Big( \Big[ -\delta_{su}
 (\gamma_{\bar as} \triangle +{1\over 2} \sum_w\gamma_{\bar aw}
 \partial_w^2)+(\gamma_{\bar as}+\gamma_{\bar
 au})\partial_s\partial_u\Big] r_{\bar a}(\tau ,\vec \sigma
 )-\nonumber \\
 &-&\sum_w\gamma_{\bar aw} \partial_s\partial_u \int d^3\sigma_1
 {{\partial_{1w}^2 r_{\bar a}(\tau ,{\vec \sigma}_1)}\over {4\pi
 |\vec \sigma -{\vec \sigma}_1|}} \Big) +\nonumber \\
 &+& {{\sqrt{3}}\over 2} \sum_{\bar aw}\gamma_{\bar aw}
 \partial_s\partial_u  \int d^3\sigma_1
 {{\partial_{1w}^2 r_{\bar a}(\tau ,{\vec \sigma}_1)}\over {4\pi
 |\vec \sigma -{\vec \sigma}_1|}}+\nonumber \\
 &+&\epsilon \delta_{st}\sqrt{3} {{4\pi G}\over {c^3}} \sum_{\bar a} \Big[
 2\delta_{rs} \gamma_{\bar au} \partial_{\tau} \pi_{\bar a}(\tau
 ,\vec \sigma )+\nonumber \\
 &+&(1-\delta_{ru})\sum_w\gamma_{\bar
 aw}[1-2(\delta_{rw}+\delta_{uw})] \nonumber \\
 &&\partial^2_w \int^{\infty}_{\sigma^r} d\sigma_1^r
 \int^{\infty}_{\sigma^u} d\sigma_1^u\,\, \partial_{\tau}\,
 \pi_{\bar a}(\tau ,\sigma_1^r \sigma_1^u \sigma^{k\not=
 r,u})\Big] +\nonumber \\
 &+&{1\over {\sqrt{3}}} \sum_{\bar a}\Big( \Big[ -\delta_{ru}
 (\gamma_{\bar ar} \triangle +{1\over 2} \sum_w\gamma_{\bar aw}
 \partial_w^2)+(\gamma_{\bar ar}+\gamma_{\bar
 as})\partial_r\partial_u\Big] r_{\bar a}(\tau ,\vec \sigma
 )-\nonumber \\
 &-&\sum_w\gamma_{\bar aw} \partial_r\partial_u \int d^3\sigma_1
 {{\partial_{1w}^2 r_{\bar a}(\tau ,{\vec \sigma}_1)}\over {4\pi
 |\vec \sigma -{\vec \sigma}_1|}} \Big) +\nonumber \\
 &+& {{\sqrt{3}}\over 2} \sum_{\bar aw}\gamma_{\bar aw}
 \partial_r\partial_u  \int d^3\sigma_1
 {{\partial_{1w}^2 r_{\bar a}(\tau ,{\vec \sigma}_1)}\over {4\pi
 |\vec \sigma -{\vec \sigma}_1|}}\Big)+O(r^2_{\bar a}),\nonumber\\
\end{eqnarray*}

\bea
 {}^4C_{\tau ruv}&=&{}^4R_{\tau ruv}+O(r^2_{\bar a})= \nonumber \\
 &=& \epsilon\sqrt{3} {{4\pi G}\over {c^3}} \sum_{\bar a} \Big[
2\gamma_{\bar ar} (\delta_{rv}
\partial_u - \delta_{ru}\partial_v)\pi_{\bar a}(\tau ,\vec \sigma
)+\sum_t \gamma_{\bar at} \nonumber \\
&&\Big((1-\delta_{rv})[1-2(\delta_{rt}+\delta_{vt})]
 {{\partial^3}\over {\partial \sigma^u \partial (\sigma^t)^2}}  \int^{\infty}_{\sigma^r} dw^r
 \int^{\infty}_{\sigma^v} dw^v\,\pi_{\bar a}(\tau
 ,w^r w^v\sigma^{k\not= r,v})-\nonumber \\
 &-&(1-\delta_{ru})[1-2(\delta_{rt}+\delta_{ut})]
 {{\partial^3}\over {\partial \sigma^v \partial (\sigma^t)^2}}  \int^{\infty}_{\sigma^r} dw^r
 \int^{\infty}_{\sigma^u} dw^u\,\pi_{\bar a}(\tau
 ,w^r w^u\sigma^{k\not= r,u})\Big)\Big]+\nonumber \\
 &+&O(r^2_{\bar a}),\nonumber \\
 &&{}\nonumber \\
{}^4C_{\tau r \tau s}&=&{}^4R_{\tau r\tau s}+{1\over 2}\epsilon
{}^4R_{rs}= \nonumber \\
 &=&\epsilon \sqrt{3} {{4\pi G}\over {c^3}}
\sum_{\bar a} \Big[ 2 \delta_{rs}
 \gamma_{\bar a r} \pi_{\bar a}(\tau ,\vec \sigma ) +\nonumber \\
 &+&  [1-\delta_{rs}]\sum_w\gamma_{\bar aw} [1-2(\delta_{rw}+\delta_{sw})] \nonumber \\
 &&{{\partial^2}\over {\partial (\sigma^w)^2}}  \int^{\infty}_{\sigma^r} d\sigma_1^r
 \int^{\infty}_{\sigma^s} d\sigma_1^s\,\,\partial_{\tau} \pi_{\bar a}(\tau
 ,\sigma_1^r\sigma_1^s\sigma^{k\not= r,s}) \Big]+\nonumber \\
 &+&{\epsilon \over2}\sqrt{3} {{4\pi G}\over {c^3}} \sum_{\bar a} \Big[
 2\delta_{rs} \gamma_{\bar as} \partial_{\tau} \pi_{\bar a}(\tau
 ,\vec \sigma )+\nonumber \\
 &+&(1-\delta_{rs})\sum_w\gamma_{\bar
 aw}[1-2(\delta_{rw}+\delta_{sw})] \nonumber \\
 &&\partial^2_w \int^{\infty}_{\sigma^r} d\sigma_1^r
 \int^{\infty}_{\sigma^s} d\sigma_1^s\,\, \partial_{\tau}\,
 \pi_{\bar a}(\tau ,\sigma_1^r \sigma_1^s \sigma^{k\not=
 r,s})\Big] +\nonumber \\
 &+&{1\over {\sqrt{3}}} \sum_{\bar a}\Big( \Big[ -\delta_{rs}
 (\gamma_{\bar ar} \triangle +{1\over 2} \sum_w\gamma_{\bar aw}
 \partial_w^2)+(\gamma_{\bar ar}+\gamma_{\bar
 as})\partial_r\partial_s\Big] r_{\bar a}(\tau ,\vec \sigma
 )-\nonumber \\
 &-&\sum_w\gamma_{\bar aw} \partial_r\partial_s \int d^3\sigma_1
 {{\partial_{1w}^2 r_{\bar a}(\tau ,{\vec \sigma}_1)}\over {4\pi
 |\vec \sigma -{\vec \sigma}_1|}} \Big) +\nonumber \\
 &+&{{\sqrt{3}}\over 2} \sum_{\bar aw}\gamma_{\bar aw}
 \partial_r\partial_s  \int d^3\sigma_1
 {{\partial_{1w}^2 r_{\bar a}(\tau ,{\vec \sigma}_1)}\over {4\pi
 |\vec \sigma -{\vec \sigma}_1|}}+O(r^2_{\bar
 a}),\nonumber \\
 &&{}\nonumber \\
{}^4C_{\tau s\tau \tau }&=&{}^4R_{\tau s \tau \tau}=0,\nonumber \\
 {}^4C_{\tau \tau \tau \tau}&=&{}^4R_{\tau \tau \tau \tau}=0.
 \label{d10}
 \eea

On the solutions of the Hamilton equations the Weyl tensor
${}^4{\hat C}^A{}_{BCD}(\tau ,\vec \sigma )$ coincides with the
Riemann tensor.\bigskip

Let us consider the congruence of time-like observers whose
4-velocities are given by the unit normals to
$\Sigma^{(WSW)}_{\tau}$, viz. ${}^4_{(\Sigma )}{\check {\tilde
E}}^A_{(o)}(\tau ,\vec \sigma )= \epsilon l^A(\tau ,\vec \sigma )
= -(1; \delta^{rs}n_s(\tau ,\vec \sigma ))$, ${}^4_{(\Sigma
)}{\check {\tilde E}}^{(o)}_A(\tau ,\vec \sigma )= l_A(\tau ,\vec
\sigma ) = -\epsilon (1; 0)$ [see Eqs.(\ref{II2})]. The {\it
electric and magnetic parts of the Weyl tensor} with respect to
these time-like observers are \footnote{$*{}^4{\hat
C}_{ABCD}={1\over 2}\epsilon_{AB}{}^{EF}\, {}^4{\hat
C}_{EFCD}={1\over 2} \epsilon_{CD}{}^{EF}\, {}^4{\hat C}_{ABEF}$
is, due to the Lanczos identity, the unique Hodge dual of the Weyl
tensor \cite{47} (the Riemann tensor has different left and right
duals).}

 \bea
  {}^4{\hat E}_{AB}(\tau ,\vec \sigma ) &=& {}^4{\hat
E}_{BA}(\tau ,\vec \sigma ) =[{}^4{\hat C}_{ACBD}\, l^C\,
l^D](\tau ,\vec \sigma ),\nonumber \\
 &&{}\nonumber \\
 &&[{}^4{\hat E}_{AB}\, l^B](\tau ,\vec \sigma )=
 -\Big[ {}^4{\hat E}_{A\tau}+\, {}^4{\hat E}_{Ar} n_r\Big]
 (\tau ,\vec \sigma )=0,\quad {}^4{\hat E}^A{}_A(\tau ,\vec \sigma )=0,\nonumber \\
 {}^4{\hat E}_{rs}(\tau ,\vec \sigma ) &=&
 [{}^4{\hat C}_{r\tau s\tau}+ ({}^4{\hat C}_{r\tau su}+{}^4{\hat C}_{s\tau ru}) n_u
 +{}^4{\hat C}_{rusv} n_un_v](\tau ,\vec \sigma )=\nonumber \\
 &=& {}^4{\hat C}_{r\tau s\tau} + O(r^2_{\bar a}),\nonumber \\
 {}^4{\hat E}_{r\tau}(\tau ,\vec \sigma ) &=&
 [{}^4{\hat C}_{r\tau\tau u} n_u +{}^4{\hat C}_{ru\tau v} n_un_v](\tau
 ,\vec \sigma )=-[{}^4{\hat E}_{rs} n_s](\tau ,\vec \sigma )=\nonumber \\
 &=& 0 + O(r^2_{\bar a}),\nonumber \\
 {}^4{\hat E}_{\tau\tau}(\tau ,\vec \sigma ) &=&
 [{}^4{\hat C}_{\tau r\tau s}n_rn_s](\tau ,\vec \sigma ) =
 -[{}^4{\hat E}_{\tau s}n_s](\tau ,\vec \sigma )=\nonumber \\
 &=& 0 +O(r^2_{\bar a}), \nonumber \\
 &&{}\nonumber \\
 {}^4{\hat H}_{AB}(\tau ,\vec \sigma ) &=& {}^4{\hat H}_{BA}(\tau ,\vec \sigma )=
 {1\over 2}\epsilon_{BE}{}^{CD} \,[{}^4{\hat C}_{AFCD}\, l^E\, l^F](\tau ,\vec \sigma )=
 [*{}^4{\hat C}_{ABCD} l^C l^D](\tau ,\vec \sigma ),\nonumber \\
 &&{}\nonumber \\
 &&[{}^4{\hat H}_{AB}\, l^B](\tau ,\vec \sigma ) =
 -\Big[ {}^4{\hat H}_{A\tau}+\, {}^4H_{Ar} n_r\Big]
 (\tau ,\vec \sigma )=0,\quad {}^4{\hat H}^A{}_A(\tau ,\vec \sigma ) =0,\nonumber \\
 {}^4{\hat H}_{rs}(\tau ,\vec \sigma )&=&
 [*{}^4{\hat C}_{r\tau s\tau}+ (*{}^4{\hat C}_{r\tau su}+*{}^4C_{s\tau ru}) n_u
 +*{}^4{\hat C}_{rusv} n_un_v](\tau ,\vec \sigma )=\nonumber \\
 &=& *{}^4{\hat C}_{r\tau s\tau} + O(r^2_{\bar a}),  \nonumber \\
 {}^4{\hat H}_{r\tau}(\tau ,\vec \sigma ) &=&
 [*{}^4{\hat C}_{r\tau\tau u} n_u +*{}^4{\hat C}_{ru\tau v} n_un_v](\tau
 ,\vec \sigma )=-[{}^4{\hat H}_{rs} n_s](\tau ,\vec \sigma )=\nonumber \\
 &=& 0 + O(r^2_{\bar a}),\nonumber \\
 {}^4{\hat H}_{\tau\tau}(\tau ,\vec \sigma ) &=&
 [*{}^4{\hat C}_{\tau r\tau s}n_rn_s](\tau ,\vec \sigma ) =
 -[{}^4{\hat H}_{\tau s}n_s](\tau ,\vec \sigma )=\nonumber \\
 &=& 0 + O(r^2_{\bar a}).
 \label{d11}
 \eea

Both ${}^4{\hat E}_{AB}$ and ${}^4{\hat H}_{AB}$ have five
independent components. See Ref.\cite{47} for an analogous
decomposition of the Riemann tensor.
\bigskip

The four eigenvalues $\Lambda_{\alpha}$, $\alpha =1,..,4$, of the
Weyl tensor are

\bea &&{}^4{\hat C}_{ABCD}\, {}^4{\hat C}^{ABCD} = O(r^2_{\bar a})
,\nonumber \\
 &&{}^4{\hat C}_{AB}{}^{CD}\, \epsilon_{CDEF}\, {}^4{\hat C}^{ABEF}
  = O(r^2_{\bar a}),\nonumber \\
 && {}^4{\hat C}^{AB}{}_{CD}\, {}^4{\hat C}^{CD}{}_{EF}\, {}^4{\hat C}^{EF}{}_{AB}
  = O(r^3_{\bar a}), \nonumber \\
 &&{}^4{\hat C}^{ABCD}\, \epsilon_{CDEF}\, {}^4{\hat C}^{EF}{}_{UV}\,
{}^4{\hat C}^{UV}{}_{AB}  = O(r^3_{\bar a}).
 \label{d12}
  \eea

\noindent As shown in Ref.\cite{15}, the 4-coordinate system
$\sigma^A = \{ \tau ,\vec \sigma \}$, corresponding to our
completely fixed 3-orthogonal gauge, is identifiable by means of 4
gauge-fixing constraints $\sigma^A - F^{\bar
A}[\Lambda_{\alpha}(\tau ,\vec \sigma )] \approx 0$, where the
$F^{\bar A}$ are 4 suitable scalar functions of the 4 eigenvalues
of the Weyl tensor such that in the linearized theory we have
$F^{\bar A} = F_{(L)}^{\bar A} + O(r^2_{\bar a})$ with $F^{\bar
A}_{(L)} = O(r_{\bar a})$.

The {\it Kretschmann invariant and pseudo-invariant} are
respectively [$*{}^4{\hat R}^{ABCD}={1\over 2}\epsilon^{ABEF}\,
{}^4{\hat R}_{EF}{}^{CD}$]

  \bea
&&{\bf \hat R}\cdot {\bf \hat R}={}^4{\hat R}_{ABCD}\, {}^4{\hat
R}^{ABCD}  = O(r^2_{\bar a}),\nonumber \\
 &&{\bf *\hat R}\cdot {\bf \hat R}=
*{}^4{\hat R}^{ABCD}\, {}^4{\hat R}_{ABCD}  = O(r^2_{\bar a}).
 \label{d13}
 \eea

\noindent In Ref.\cite{16} it is suggested that in presence of
matter a coordinate-independent characterization of
gravito-magnetism is ${\bf *\hat R}\cdot {\bf \hat R}\not= 0$.

\bigskip

The {\it Bel-Robinson and Bel tensors} are respectively

\bea {}^4{\hat {\cal T}}_{ABCD} &=& {}^4{\hat C}_{AECF}\,
{}^4{\hat C}_B{}^E{}_D{}^F + {}^4{\hat C}_{AEDF}\, {}^4{\hat
C}_B{}^E{}_C{}^F- \nonumber \\
 &-&{1\over 2} {}^4g_{AB}\, {}^4{\hat C}_{EFCG}\, {}^4{\hat C}^{EF}{}_D{}^G-
 {1\over 2} {}^4g_{CD}\, {}^4{\hat C}_{AEFG}\, {}^4{\hat C}_B{}^{EFG}
 +\nonumber \\
 &+& {1\over 8} {}^4g_{AB}\, {}^4g_{CD}\, {}^4{\hat C}_{EFGH}\, {}^4{\hat C}^{EFGH}
  = O(r^2_{\bar a}), \nonumber \\
 {}^4{\hat {\cal B}} &=& {}^4{\hat R}_{AECF}\, {}^4{\hat
R}_B{}^E{}_D{}^F + {}^4{\hat R}_{AEDF}\, {}^4{\hat
R}_B{}^E{}_C{}^F- \nonumber \\
 &-&{1\over 2} {}^4g_{AB}\, {}^4{\hat R}_{EFCG}\, {}^4{\hat R}^{EF}{}_D{}^G-
 {1\over 2} {}^4g_{CD}\, {}^4{\hat R}_{AEFG}\, {}^4{\hat R}_B{}^{EFG}
 +\nonumber \\
 &+& {1\over 8} {}^4g_{AB}\, {}^4g_{CD}\, {}^4{\hat R}_{EFGH}\, {}^4{\hat R}^{EFGH}
  = O(r^2_{\bar a}), \nonumber \\
 &&{}\nonumber \\
 &&{}^4{\hat {\cal B}}_{ABCD} = {}^4{\hat {\cal T}}_{ABCD} + {}^4{\hat {\cal M}}_{ABCD}
 + {}^4{\hat {\cal Q}}_{ABCD},
 \label{d14}
 \eea

\noindent where ${}^4{\hat {\cal Q}}_{ABCD}={1\over 6} {}^4\hat R
\, [{}^4{\hat C}_{ACBD}+ {}^4{\hat C}_{ADBC}]$ is the
matter-gravity coupling tensor and ${}^4{\hat {\cal M}}_{ABCD}$ is
the pure matter gravitational super-energy tensor (see
Ref.\cite{47} for its expression).

\vfill\eject

\end{document}